%% file: TESE.tex
\documentclass[
12pt,		
openright,	
twoside,  
a4paper,			
chapter=TITLE,		
english,			
french,				
spanish,			
brazil				
]{USPSC-classe/USPSC}
\usepackage[T1]{fontenc}		
\usepackage[utf8]{inputenc}		
\usepackage{lmodern}			
\usepackage{lastpage}			
\usepackage{indentfirst}		
\usepackage{color}				
\usepackage{graphicx}			
\usepackage{float} 				
\usepackage{chemfig}            
\usepackage{chemmacros}         
\usepackage{tikz}				
\usetikzlibrary{positioning}
\usepackage{pdfpages}
\usepackage{makeidx}            
\usepackage{hyphenat}          
\usepackage[absolute]{textpos} 
\usepackage{eso-pic}           
\usepackage{makebox}           
\usepackage{pdfpages}

\newcommand{\upA}{\uparrow}
\newcommand{\downA}{\downarrow}

\newcommand{\e}{\varepsilon}

\newcommand{\PR}[1]{\ensuremath{\left[#1\right]}}
\newcommand{\PC}[1]{\ensuremath{\left(#1\right)}}
\newcommand{\Mrm}{\mathrm}

\newcommand{\Sou}{{ \newline  Source: By the author.}}   
\newcommand{\Rb}{{$\mathfrak{a}_0$}}
\input{newcommands}

\usepackage{cite}              
\usepackage[num, abnt-emphasize=bf, abnt-thesis-year=both, abnt-repeated-author-omit=no, abnt-last-names=abnt, abnt-etal-cite=3, abnt-etal-list=3, abnt-etal-text=it, abnt-and-type=e, abnt-doi=doi, abnt-url-package=none, abnt-verbatim-entry=no]{abntex2cite} 
\renewcommand{\footnotesize}{\small} 

\usepackage{lipsum}				

\usepackage{multicol}	
\usepackage{multirow}	
\usepackage{longtable}	
\usepackage{threeparttablex}    
\usepackage{array}

\usepackage{USPSC-classe/ABNT6023-10520}

\include{CAPA_INFOR}

\programa
\definecolor{blue}{RGB}{41,5,195}

\makeatletter
\hypersetup{
	pdftitle={\@title}, 
	pdfauthor={\@author},
	pdfsubject={\imprimirpreambulo},
	pdfcreator={LaTeX with abnTeX2},
	pdfkeywords={abnt}{latex}{abntex}{USPSC}{trabalho acadêmico},
	colorlinks=true,       		
	linkcolor=blue,	          	
	citecolor= red,        		
	filecolor=black,      		
	urlcolor=black,
	bookmarksdepth=4	
}
\makeatother

\makeindex

\begin{document}

\selectlanguage{english}

\frenchspacing 

\renewcommand{\ABNTEXchapterfontsize}{\fontsize{12}{12}\bfseries}
\renewcommand{\ABNTEXsectionfontsize}{\fontsize{12}{12}\bfseries}
\renewcommand{\ABNTEXsubsectionfontsize}{\fontsize{12}{12}\normalfont}
\renewcommand{\ABNTEXsubsubsectionfontsize}{\fontsize{12}{12}\normalfont}
\renewcommand{\ABNTEXsubsubsubsectionfontsize}{\fontsize{12}{12}\normalfont}

\imprimircapa
\imprimirfolhaderosto*

\setcounter{page}{1}
\includepdf{CAP/PaginaEmBranco.pdf}
\includepdf[pages=1]{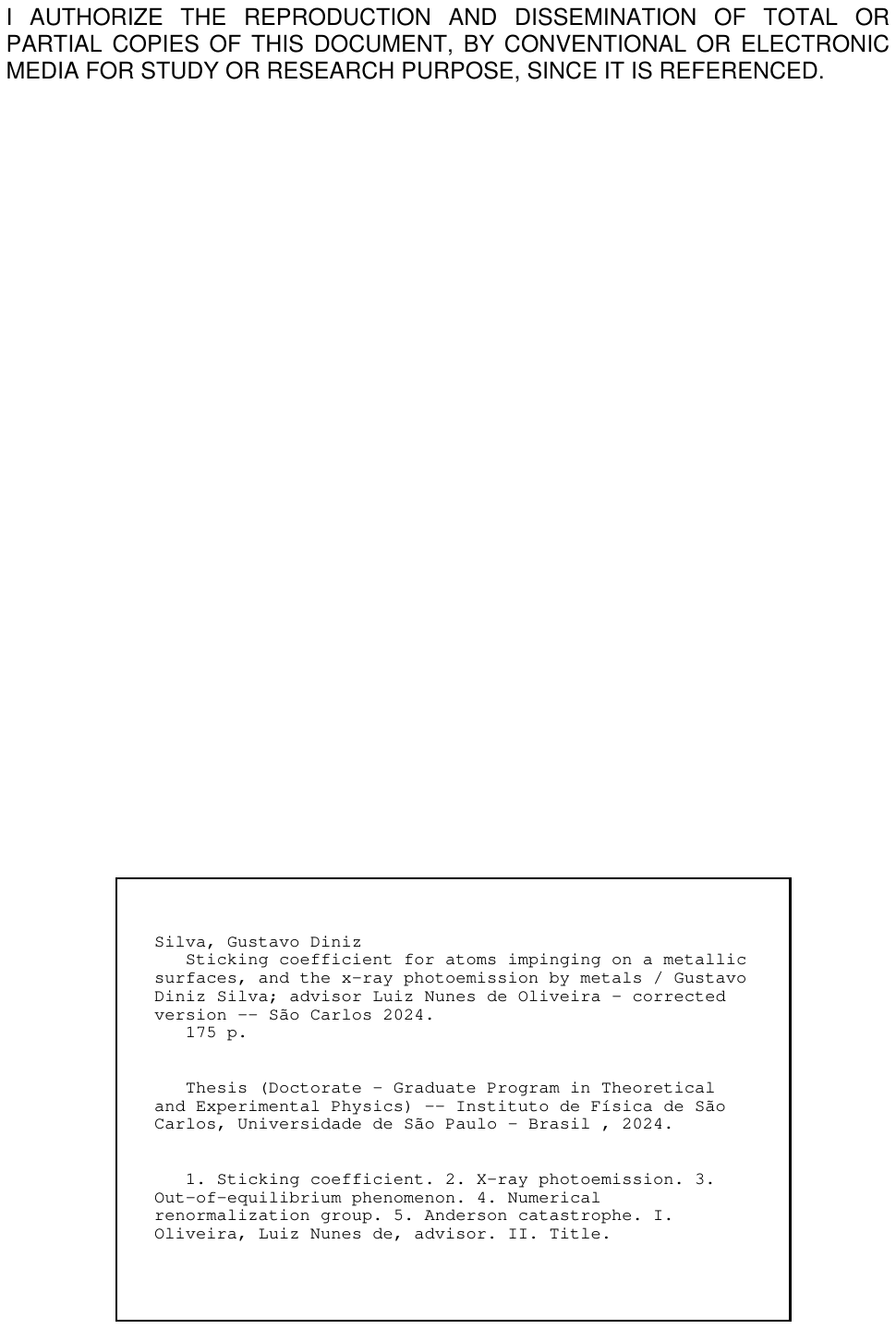}











\include{CAP/Agradecimentos}


\include{CAP/Abstract}

%
\include{CAP/Resumo}

\pdfbookmark[0]{\listfigurename}{lof}
\listoffigures*
\cleardoublepage

\pdfbookmark[0]{\listtablename}{lot}
\listoftables*
\cleardoublepage



\pdfbookmark[0]{\contentsname}{toc}
\tableofcontents*
\cleardoublepage
\textual

\include{CAP/Cap1}

\include{CAP/Cap2}

\include{CAP/Cap3}

\include{CAP/Cap4}

\include{CAP/Cap5}

\include{CAP/Cap6}

\include{CAP/Cap7}

\include{CAP/Cap8}


\postextual

\bibliography{TESE}

%
%

\include{CAP/Apendices}




\end{document}

%% file: newcommands.tex
\usepackage[utf8]{inputenc}
\usepackage[T1]{fontenc}
\usepackage{fixltx2e}
\usepackage{graphicx}
\usepackage{longtable}
\usepackage{float}
\usepackage{wrapfig}
\usepackage{rotating}
\usepackage[normalem]{ulem}
\usepackage{amsmath}
\usepackage{textcomp}
\usepackage{marvosym}
\usepackage{wasysym}
\usepackage{amssymb}
\usepackage{hyperref}
\usepackage{xcolor}
\usepackage{graphicx}
\usepackage{tikz}
\usepackage{fp}
\usepackage{relsize}
\usepackage{fancybox}
\usetikzlibrary{decorations.pathmorphing, patterns,angles,quotes}
\usepackage{siunitx}
\usepackage{amsmath}

\newcommand{\cd}[1]{c_{#1}^{\dagger}}
\newcommand{\cn}[1]{c_{#1}^{\phantom{\dagger}}}
\newcommand{\fd}[1]{f_{#1}^{\dagger}}
\newcommand{\fn}[1]{f_{#1}^{\phantom{\dagger}}}

\newcommand{\ket}[1]{\vert #1\rangle}

\newcommand{\bra}[1]{\langle#1\vert}

\newcommand{\braket}[2]{\langle#1\vert#2\rangle}

\newcommand{\hc}{\mbox{h.~c.}}

%% file: CAPA_INFOR.tex
\instituicao{Instituto de F\'isica de S\~ao Carlos, Universidade de S\~ao Paulo}
\unidade{INSTITUTO DE F\'ISICA DE S\~AO CARLOS}
\unidademin{Instituto de F\'isica de S\~ao Carlos}
\universidademin{Universidade de S\~ao Paulo}

\notafolharosto{\hspace{3cm} Corrected version\newline (Original version available on the Program Unit)}
 
\universidade{UNIVERSIDADE DE S\~AO PAULO}
\titulo{Sticking coefficient for atoms impinging on a metallic surfaces, and the x-ray photoemission by metals}

\titleabstract{Sticking coefficient for atoms impinging on a metallic surfaces, and the x-ray photoemission by metals}
\tituloresumo{Coeficiente de adesão de um átomo colidindo com uma superfície metálica, e o problema da fotoemissão por metais}
\autor{Gustavo Diniz Silva}
\autorficha{Silva, Gustavo D.}
\autorabr{SILVA, G. D.}

\cutter{S856m}

\local{S\~ao Carlos}
\data{2024}
\renewcommand{\orientadorname}{Orientador:}
\orientador{Prof. Dr. Luiz Nunes de Oliveira}
\orientadorcorpoficha{orientador Luiz Nunes de Oliveira}
\orientadorficha{Oliveira, L. N.}

\notaautorizacao{AUTORIZO A REPRODU\c{C}\~AO E DIVULGA\c{C}\~AO TOTAL OU PARCIAL DESTE TRABALHO, POR QUALQUER MEIO CONVENCIONAL OU ELETR\^ONICO PARA FINS DE ESTUDO E PESQUISA, DESDE QUE CITADA A FONTE.}
\notabib{Ficha catalogr\'afica revisada pelo Servi\c{c}o de Biblioteca e Informa\c{c}\~ao Prof. Bernhard Gross, com os dados fornecidos pelo(a) autor(a)}

\newcommand{\programa}{
	\renewcommand{\areaname}{Concentration area:}
	\renewcommand{\opcaoname}{Option:}
	\renewcommand{\orientadorname}{Advisor:}
	\tipotrabalho{Tese (Doutorado em Ci\^encias)}
	\tipotrabalhoabs{Thesis (Doctor in Science)}
	\area{Theoretical Physics}
	\preambulo{Thesis presented to the Graduate Program in Physics at the Instituto de F\'isica de S\~ao Carlos da Universidade de S\~ao Paulo, to obtain the degree of Doctor in Science.}
	\notaficha{Thesis (Doctorate - Graduate Program in Theoretical and Experimental Physics)}
}

%% file: CAP/Agradecimentos.tex
\begin{agradecimentos}
First and foremost, to my mother, Ms. Girlene Firmina Diniz, my sister Yngrid Martins Diniz, and my girlfriend Ms. Amanda Aparecida Vieira Dias, the most important people in my life who provided essential emotional support during this process. To my uncle Leomar José da Silva. To my foster grandmother Augusta Maria Godinho and my late foster grandfather Sebastião Martins Godinho, for all the affection throughout all these years.

To my advisor, Dr. Luiz Nunes de Oliveira, for the opportunity to work in this interesting and challenging project, as well as for all the guidance, knowledge, and assistance provided throughout the PhD process.

To my friends and research colleagues, including Mr. Felipe D. Picoli, Israel C. Ribeiro, and Marino P. Lenzarini, who, in addition to shared moments, good stories, bar outings, and great post-lunch coffees, contributed greatly to this project both emotionally and through discussions of results, simulations, and sharing relevant data for the research conducted. And also to my friends Mr. Gabriel dos Reis Trindade, Mr. Guilherme Nogueira, the group from 61, and Tiago D. R. dos Santos, who, beyond emotional support, coffee, and good times shared, also helped with discussions and opinions about the project.

To Dr. Irene D'Amico, whose discussions and expertise greatly enriched our work.

To Dr. Edson Vernek, Dr. Luiz H. B. Guessi, and Dr. Krissia Zawadzki for productive discussions about the project and possible approaches and perspectives that helped in the final outcome.

To Dr. Celso Ricardo Rêgo for the parameters obtained via DFT.

To Prof. Dr. Renato V. Gonçalves and the Nanomaterials and Advanced Ceramics research group (NaCA) for the experimental XPS results on silver.

To all my other friends and family members (I won't name them all, as there are many) and colleagues from IFSC. And also to all my friends inside and outside the University of São Paulo.

To the University of São Paulo and the Institute of Physics of São Carlos for the opportunity to pursue this course.

This study was financed in part by the Coordenação de Aperfeiçoamento de Pessoal de Nível Superior – Brasil (CAPES) – Finance Code 88887.495890/2020-00.

Research developed using computational resources from the Center for Mathematical Sciences Applied to Industry (CeMEAI), funded by FAPESP (proc. 2013/07375-0).

\end{agradecimentos}

%% file: CAP/Abstract.tex
\begin{resumo}[Abstract]
 \begin{otherlanguage*}{english}
	\begin{flushleft} 
		\setlength{\absparsep}{0pt} 
 		\SingleSpacing  		\imprimirautorabr~~\textbf{\imprimirtitleabstract}.	\imprimirdata.  \pageref{LastPage} p. 
		\imprimirtipotrabalhoabs~-~\imprimirinstituicao, \imprimirlocal, 	\imprimirdata. 
 	\end{flushleft}
	\OnehalfSpacing

Out-of-equilibrium electron-gas systems contain rich physics. We discuss the time evolution of three such systems. Our first subject is photoemission from metals, a problem traditionally studied in the frequency domain. We find unexpected features in the time dependence of the photoemission rate.  The rate oscillates at relatively high frequency as it decays, and the amplitude of the oscillations decay faster than the average current. We combine analysis with numerical data to trace  the oscillatory behavior to the interference between two excitation processes, one of which decays according to the Doniach-Sunjic power law while the other decays faster, following the Nozières-De Dominicis power law. We expect XPS experiments focused on this feature to identify the corresponding peak in the frequency domain. As our second problem, with a view to quantifying adiabaticity, we consider an electron gas subject to a localized potential that ramps up from zero to a maximum at constant rate. Again on the basis of analytical and numerical results, we identify the region of the parametric space of the model in which the system behaves adiabatically. In contrast with the Quantum Adiabatic Criterion, which associates adiabaticity with small ramp-up rates, our results show that the number of energy scales participating in the screening of the localized potential determines whether non-adiabaticity emerges. The object of our final study is the collision between an initially neutral hydrogen atom and a copper surface, represented by a half-filled conduction band. As the atom approaches the surface, the overlap between the atomic and surface orbitals allows electron transfer to and negative ionization of the H atom. The ionization switches on a image-charge potential, which pulls the ion towards the surface. We define a spinless model that captures the physics of the collision and, on the basis of numerical treatment, follow the evolution of the atomic wave packet and compute the sticking coefficient, that is, the probability that the atom remain close to the surface after a long time. Plotted as function of the incident energy, the sticking coefficient has a maximum around 0.3 eV. Assisted by the experience gained with first two problems, we interpret the peak as a compromise between the contribution of non-adiabatic processes, which grows with the initial energy, and the time the atom takes to traverse the region where such processes occur. The numerical results are in semi-quantitative agreement with the available experimental data.

   \vspace{\onelineskip}
 
 
\end{otherlanguage*}
\end{resumo}

%% file: CAP/Resumo.tex
\setlength{\absparsep}{18pt} 
\begin{resumo}
	\begin{flushleft} 
			\setlength{\absparsep}{0pt} 
			\SingleSpacing 
			\imprimirautorabr~~\textbf{\imprimirtituloresumo}.	\imprimirdata. \pageref{LastPage} p. 
			\imprimirtipotrabalho~-~\imprimirinstituicao, \imprimirlocal, \imprimirdata. 
 	\end{flushleft}
\OnehalfSpacing

Sistemas de gás de elétrons fora de equilíbrio contêm uma rica física. Discutimos a evolução temporal de três desses sistemas. Nosso primeiro tema é a fotoemissão de metais, um problema tradicionalmente estudado no domínio da frequência. Encontramos características inesperadas na dependência temporal da taxa de fotoemissão. A taxa oscila em frequência relativamente alta enquanto decai, e a amplitude das oscilações decai mais rapidamente do que a corrente média. Combinamos análise com dados numéricos para rastrear o comportamento oscilatório até a interferência entre dois processos de excitação, um dos quais decai de acordo com a lei de potência de Doniach-Sunjic, enquanto o outro decai mais rapidamente, seguindo a lei de potência de Nozières-De Dominicis. Esperamos que experimentos de XPS focados nessa característica identifiquem o pico correspondente no domínio da frequência. Como nosso segundo problema, com o intuito de quantificar a adiabaticidade, consideramos um gás de elétrons sujeito a um potencial localizado que aumenta de zero até um máximo a uma taxa constante. Novamente, com base em resultados analíticos e numéricos, identificamos a região do espaço paramétrico do modelo na qual o sistema se comporta adiabaticamente. Em contraste com o Critério Adiabático Quântico, que associa a adiabaticidade a baixas taxas de aumento do potencial, nossos resultados mostram que o número de escalas de energia participantes na blindagem do potencial localizado determina se a não-adiabaticidade emerge. O objeto de nosso estudo final é a colisão entre um átomo de hidrogênio inicialmente neutro e uma superfície de cobre, representada por uma banda de condução meio preenchida. À medida que o átomo se aproxima da superfície, a sobreposição entre os orbitais atômicos e os da superfície permite a transferência de elétrons e a ionização negativa do átomo de H. A ionização ativa um potencial de carga-imagem, que puxa o íon em direção à superfície. Definimos um modelo sem spin que captura a física da colisão e, com base em um tratamento numérico, seguimos a evolução do pacote de ondas atômicas e calculamos o coeficiente de adesão, ou seja, a probabilidade de o átomo permanecer próximo à superfície após um longo tempo. Quando plotado em função da energia incidente, o coeficiente de adesão tem um máximo em torno de 0.3 eV. Assistidos pela experiência adquirida com os dois primeiros problemas, interpretamos o pico como um compromisso entre a contribuição de processos não adiabáticos, que aumenta com a energia inicial, e o tempo que o átomo leva para atravessar a região onde tais processos ocorrem. Os resultados numéricos estão em acordo semi-quantitativo com os dados experimentais disponíveis.

 \textbf{Palavras-chave}: Coeficiente de adesão. Fotoemissão de raio-X. Fenômeno fora do equilíbrio. Grupo de renormalização numérica. Catástrofe de Anderson.
\end{resumo}

%% file: CAP/Cap1.tex

\chapter[Introdcution]{Introduction}
\label{Introdução}

Unraveling the physics of an interacting many-body system is always a challenge. Despite the simplicity of the mutual interactions between its constituents, collectively, the system often behaves in unexpected ways. This intriguing aspect of nature has been eloquently discussed in a seminal paper, "More is Different", by P. W. Anderson~\cite{Anderson393}. These characteristics have been observed in some of the most interesting and difficult problems in condensed matter physics, such as the Kondo problem \cite{RevModPhys.47.773,PhysRevB.109.155127,PhysRevB.91.085101}, superconductivity \cite{PhysRev.108.1175,PhysRevResearch.6.033218}, heavy fermions \cite{RevModPhys.56.755}, topological materials \cite{PhysRevLett.101.010504,PhysRevB.107.045121}, and many others \cite{PICOLI2022169062,PhysRevB.110.064404}.

Recently, the time-dependent properties of strongly correlated systems have attracted significant attention due to their crucial relevance in emerging technologies \cite{960386,Do_2014,4408808}. However, extracting the precise physics from dynamical properties requires solving the time-dependent Schrödinger equation for a large system, which is impractical. One way to bypass this is to explore alternative approaches that simplify the problem and drastically reduce the number of variables. Several methods have been developed in this direction, such as the Time-Dependent Numerical Renormalization Group (TDNRG) \cite{PhysRevLett.95.196801}, Time-Dependent Density Functional Theory (TDDFT) \cite{PhysRev.140.A1133,PhysRevLett.52.997}, and the Density Matrix Renormalization Group (DMRG) \cite{PhysRevLett.69.2863, PhysRevLett.93.076401}. However, even with these tools, comprehending the real-time evolution of many-body fermionic systems poses a formidable challenge.

In this work, our objective is to delve into the physics of out-of-equilibrium many-body interacting time-dependent systems. Specifically, we will focus on studying the time evolution of atoms impinging on metallic surfaces and the x-ray photoemission problem. For the first problem, our aim is to comprehend the processes involved in atom-surface interactions and compute the sticking coefficient. In the second problem, our goal is to calculate the time-dependent transition probability associated with x-ray photoemission. In the following sections, we will explain more about both problems and their connection to each other and discuss our contributions to understanding these phenomena.

\newpage
\section{Sticking coefficient for atoms impinging on metallic surfaces}

The adsorption of atomic particles on metallic surfaces holds significant importance in both theoretical research and practical applications \cite{SupCu,10.1063/1.470836,PhysRevLett.39.1417,10.1093/mnras/213.2.295}. Catalysis and corrosion are two prominent themes in this domain \cite{EOM20132804,GRUNDMEIER1998165,BECKER1955135}. From a scientific perspective, a key unresolved issue is the physical mechanism underlying these phenomena, particularly in the dynamics of the collision between an atom or molecule and a metallic surface. For this reason, theoretical and experimental researchers have studied this type of collision using hydrogen atoms for the past 80 years \cite{Smith1938-vs,PhysRev.118.158,BORISOV199299,10.1126-science.aad4972,doi:10.1021-acs.jpca.1c00361}. But, even for the simplest atom in the periodic table, the problem remains a challenge.

\begin{figure}[hbt!]
		\begin{center}
		\includegraphics[scale=0.56]{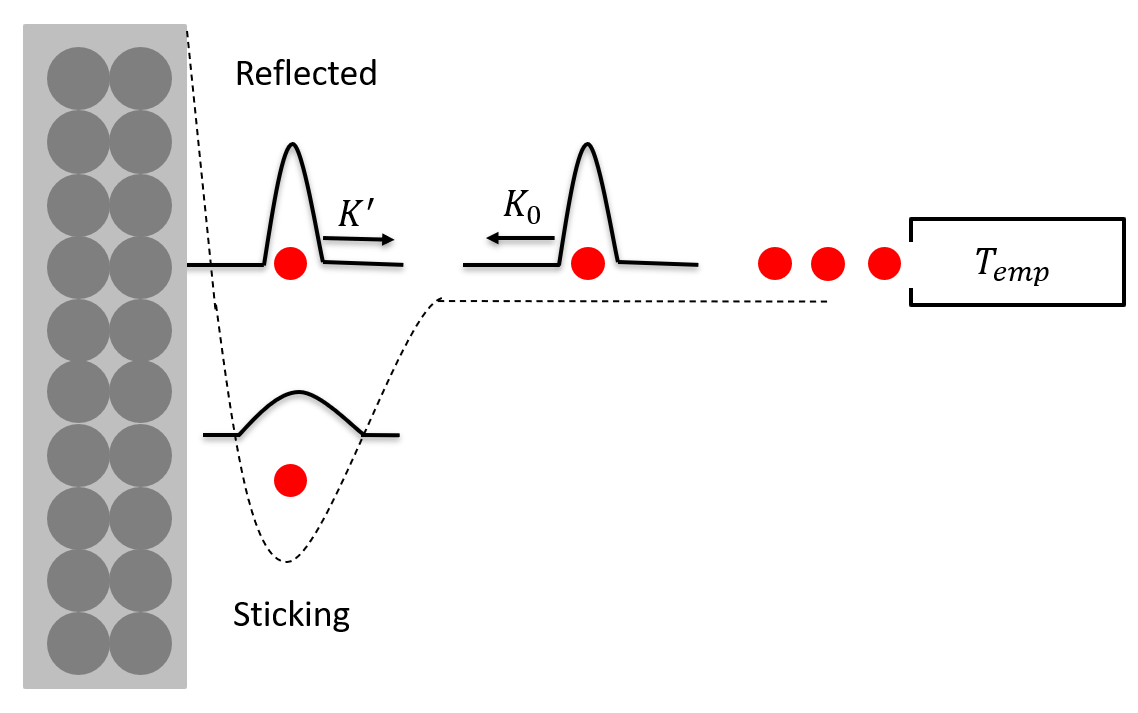}
		\end{center}
		\vspace{-0.3cm}
		\caption{\footnotesize Schematic representation of the experiment. A beam of neutral particles is emitted with high kinetic energy $K_0$ from a hot furnace towards the surface at temperature $T_{emp}$. The dashed line qualitatively represents the atomic potential. Close to the surface, the atom interacts with atoms on the surface and can subsequently be ionized, lose energy, or even become stuck. $K'$ represents the kinetic energy after the collision. \newline Source: By the author.
}\label{EXPE}
\end{figure}

A schematic representation of a real-life experiment is shown in Fig. \ref{EXPE}. A beam of neutral particles is emitted from a furnace at high temperatures, giving an initial kinetic energy $K_0 \sim k_B T_{emp}$ to the particles in the direction of the surface, where $T_{emp}$ is the furnace temperature and $k_B$ is the Boltzmann constant. The subsequent interactions between the particles and the surface give rise to a rich array of phenomena, including ionization \cite{PhysRev.118.158,BORISOV199299,Smith1938-vs}, nuclear quantum effects \cite{Jiang2021-rs}, magnetic interactions \cite{PhysRevLett.93.223201}, electronic friction \cite{Lecroart2021-fi}, and sticking \cite{Makoto,TICHMANN2012291}. These phenomena can be observed indirectly by measuring the final energy and momentum spectra of the beam and the total number of reflected particles.

Solving this problem not only offers insights into physical and chemical phenomena, but also represents a substantial advancement in understanding dynamic processes involving the nuclear movement of atoms or molecules coupled to electronic states. In such cases, non-adiabatic processes can become important, and the traditional Born-Oppenheimer (BO) approximation may fail, especially when nuclear states couple to electronic states separated by small energy differences. These nonadiabaticities were experimentally observed by computing the energy loss for hydrogen-germanium surface collisions \cite{Kruger2023-dg}, and the authors showed that the BO approximation fails to explain this observation.

\begin{figure}[hbt!]
		\begin{center}
		\includegraphics[scale=0.57]{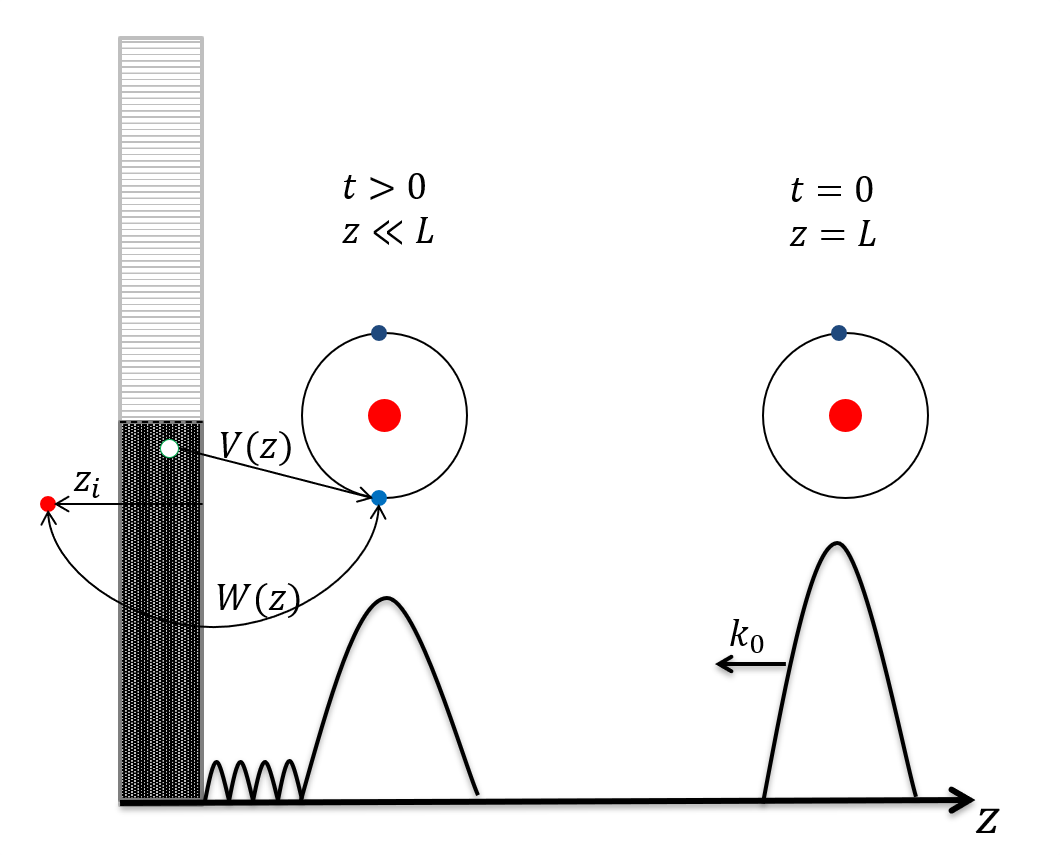}
		\end{center}
        \vspace{-0.3cm}
		\caption{\footnotesize Schematic representation of the collision dynamics: The hydrogen atom is represented by a single level with energy $\e_d$. The distance between the atom and the surface is $z$. The metal is represented by its half-filled conduction band. The transfer of charge between the atom and the surface, which is allowed by the coupling term $V(z)$, ionizes the atom, producing the attractive image-charge potential $W(z)$. The nuclear part of the wave function is initially represented by a Gaussian centered at the initial position, moving towards the surface with momentum $k_0$, as shown qualitatively.  \newline  Source: By the author. }\label{H_MS}
\end{figure}

Qualitatively, the problem can be described in a few lines. An initially neutral atom approaches the metallic surface. The overlap between the atomic orbitals and the orbitals of the ions on the surface increases as the particle approaches the metal, allowing for the transfer of electrons between them. When an electron is transferred, the particle becomes ionized; consequently, an attractive image-charge potential appears, which accelerates the particle towards the surface, as shown in Fig. \ref{H_MS}, which considers a hydrogen atom. In the subsequent collision, the generation of phonons and electron-hole pairs in the metal steals energy from the incident particle, which may become trapped in the attractive potential. Therefore, there is non-zero probability that the particle ends up attached to the surface. The theoretical challenge is to calculate the probability that this will happen, which is called the sticking coefficient ($\mathcal{S}$).

The contribution of phonons to this process is reasonably well understood \cite{D2CP01144A,10.1063/1.463233,doi:10.1080/00107518508210991}. By contrast, calculating the contribution of electron-hole pairs to $\mathcal{S}$ as a function of the initial kinetic energy of incidence is still an open question. Recently, in collaboration with researchers at the Max Planck Institute in Halle, significant progress has been made in studying this problem \cite{2019APS..MARH20006R}. In particular, they have shown that the Born-Oppenheimer approximation, traditionally employed to treat this type of problem, is unreliable and that a complete calculation is necessary using precise numerical treatment of the time-dependent wave function. However, this preliminary treatment relied on a mean-field approximation to describe the interaction between the atom and the electrons in the metallic conduction band, as well as the Coulomb repulsion.

To improve the calculation, we replaced the mean-field approximation with the iterative diagonalization strategy of the Numerical Renormalization Group (NRG) method \cite{RevModPhys.47.773}. For the computational solution procedure, the distance between the particle and the surface is discretized as $z=z_m$, and for each coordinate the electronic energy spectrum is calculated using the iterative diagonalization method of the NRG \cite{Pinto_2014}. In the initial state, the particle is several Bohr radius away from the surface, and its wave function $\Psi(z,{\ell},t=0)$ (where ${\ell}$ represents the array of electronic states) is the product of the electronic ground state of the surface with one electron in hydrogen and a Gaussian centered at an initial position, describing the initial nuclear part of the wave function. The Crank-Nicolson \cite{crank_nicolson_1947} (CN) method then allows the calculation of the wave function $\Psi(z, {\ell},t=\Delta t)$, where $\Delta t$ is the interval used to discretize the time axis. Repeated applications of the same procedure determine the temporal evolution of the wave function until, after the collision, the function splits into a part localized near the surface and another part moving away from it. The spatial integral of the squared modulus of the first part determines the sticking coefficient $\mathcal{S}$.

Consider, now, the practical aspects of carrying out this numerical calculations. To compute the dynamical simulation of the collision, we need to discretize the $z$ axis into approximately $10^{3}$ intervals. Considering only $30$ electronic states, the Crank-Nicolson method requires the inversion of a $\left(3 \times 10^{4}\right) \times \left(3 \times 10^{4}\right) $ complex matrix, typically occupying $32$ GB of RAM memory. This number grows very rapidly as we increase the number of electronic states in the simulation. Additionally, the high-energy electronic states contribute significantly to the dynamics. Cutting off the spectrum at a predefined energy is hence a poor approximation. Therefore, even when using the NRG method, the total number of electronic states needed for a complete basis calculation is huge, usually on the order of $10^{4}$, requiring for the collision the inversion of a $\left(  10^{7}\right) \times \left( 10^{7}\right) $ complex matrix,  which is impractical.

\begin{figure}[H]
		\begin{center}
		\includegraphics[scale=0.48]{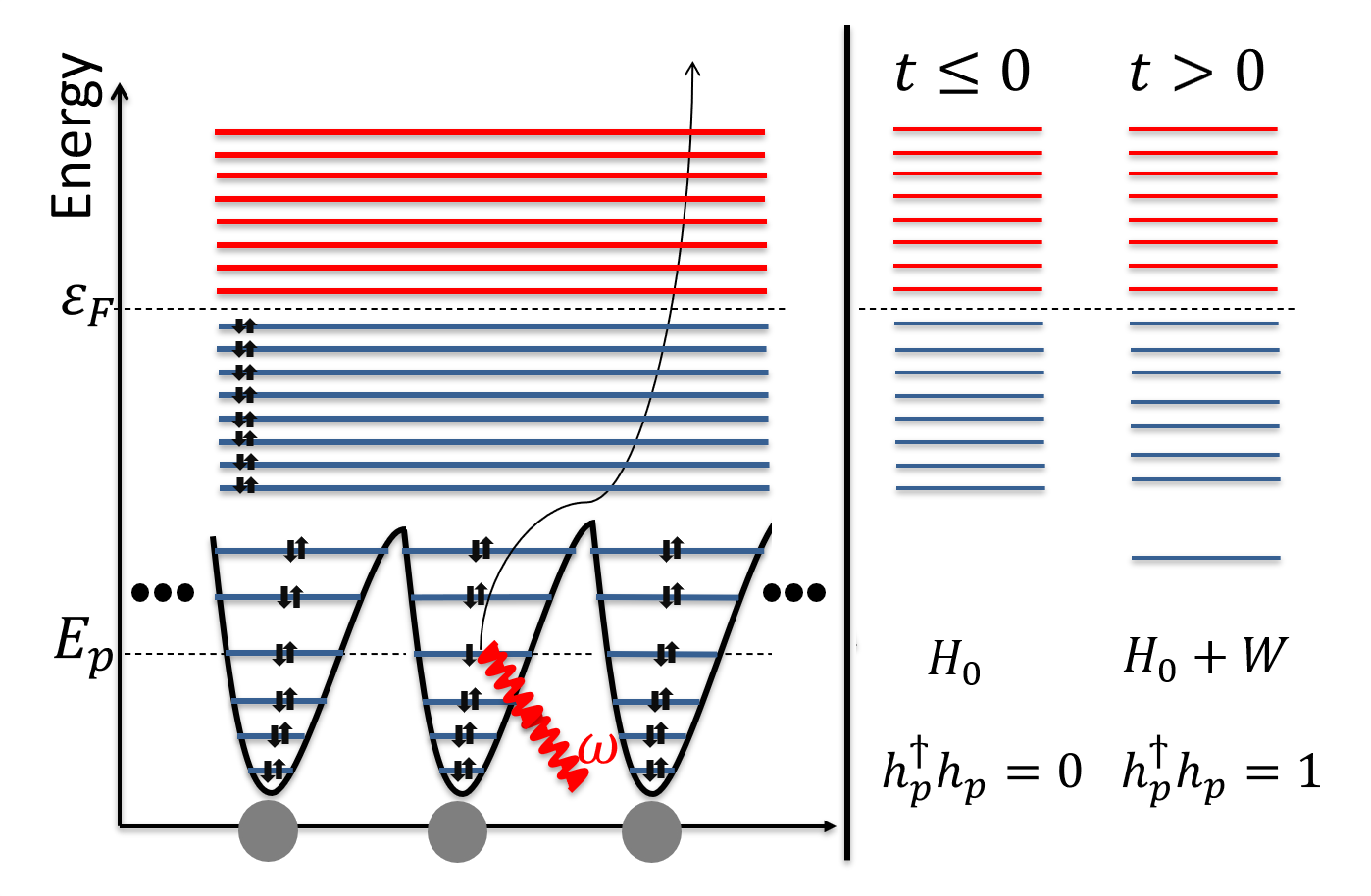}
        		\vspace{-0.4cm}
		\end{center}
		\caption{\footnotesize Schematic representation of x-ray photoemission by a metal. The interaction with the x-ray beam of energy $\omega$ excites a deep core state with energy level $E_p$. This interaction ejects an electron from the core, exciting it to a state $\e_k$ above the Fermi level $\e_F$. The operator $h_p^\dagger$ represents the hole in this core state. The absence of this electron in the electronic core (deep hole) generates an attractive potential $W$, which behaves as a scattering potential for the conduction electrons. The presence of this potential changes the energy levels and creates a bound state (right).  \newline  Source: By the author.  }
		\label{x_Ray}
\end{figure}

Fortunately, there is another interesting problem that exhibits electronic physics similar to collisions and allows for similar treatment with lower computational cost, namely the x-ray photoemission by a metal. The Figure \ref{x_Ray} shows a schematic representation of this problem. By studying the photoemission, we have learned how to select the important electronic states and important energy scales of the problem, which has allowed us to to drastically reduce the number of electronic states needed to yield trustworthy results.

\section{X-ray photoemission in metals}

An x-ray photon striking a metal can interact with the inner-shell electrons, causing eventual emission as photoelectrons, - the x-ray photoemission phenomenon schematically shown in Fig. \ref{x_Ray}. As the x-ray photon ejects one deep core electron, the electrons in the metallic band perceive this hole as a localized attractive scattering potential $W$. For $t>0$, this scattering potential shakes the conduction electrons and shift the energy levels, thus creating many pair of particle-hole excitations. Under these conditions, the Anderson orthogonality catastrophe (AOC) emerges \cite{Anderson_1967}, and the initial and final ground states are orthogonal. This problem has been extensively studied for decades \cite{Mahan, Nozieres,Doniach_1970,Nozieres71,oliveira1985,PhysRevB.24.4863,ohtaka1990,libero1990}; it is a strongly out-of-equilibrium problem exhibiting numerous interesting properties.

The photoelectric effect \cite{Einstein} occurs when electrons are excited to external states. Singularities in x-ray emission and absorption were first studied by Mahan in 1967 and later, in 1969, by Nozières and De Dominicis (ND) \cite{Nozieres} in the asymptotic limit. Those authors found the power-law behavior observed in experiments, which arises from the readjustment of the ground state of the entire Fermi gas due to the presence of the effective potential of the hole. Both works examined the response of conduction band electrons to a sudden change in the scattering potential caused by the hole potential created by interaction with the x-ray beam. Doniach and Sunjic \cite{Doniach_1970}, in 1970, studied the photoemission of electrons due to this sudden change in the scattering potential for long times and finite lifetimes of the deep state excitation. Notwithstanding the voluminous literature on this subject, this phenomenon is only completely understood for weak scattering potentials.

For strong scattering potentials, the presence of an excitonic bound state is expected, involving the electron in the band and the deep hole. In 1971, Combescot and Nozières \cite{Nozieres71} demonstrated, in the context of x-ray emission and absorption spectra, that excitations from this bound state can influence the absorption spectrum, leading to the emergence of an additional peak with a specific decay in the spectrum when such a bound state is present. Latter, Ohtaka and Tanabe \cite{ohtaka1990} showed that these extra peak also is present in the x-ray photoemission spectra, and showed how to estimate numerically the relative contribution of this peak. However, even after these instrumental works, some aspects of these extra peaks for strong scattering potentials remain unexplored, such as deriving an analytical expression to estimate these peaks, as well as understanding why and how they emerge as the potential increases. We aim to address some of these questions in Chapter 3.

The power-law behavior is crucial for understanding one of the most important methods for characterizing materials: x-ray photoemission spectroscopy (XPS) \cite{GRECZYNSKI2020100591,Baer2019-tt}. XPS involves irradiating a sample with x-rays and measuring the kinetic energy and number of emitted photoelectrons. Each peak found in the experiment can be associated with a characteristic binding energy of the material, typically associated with atomic deep core electronic transitions. For some materials, however, the presence of satellite peaks is common \cite{PAULY2016317,D1CP04886D,GROSVENOR20061771,https://doi.org/10.1002/sia.740230705,Slaughter_Weber_Güntherodt_Falco_1992,SILVERSMIT2004167,Steiner1980-im}. Determining the source of these peaks is a difficult problem because the associated binding energy involves not only deep core transitions but also interactions with the band \cite{https://doi.org/10.1002/sia.740230705,Slaughter_Weber_Güntherodt_Falco_1992,PhysRevA.15.1486}. Therefore, a better understanding of x-ray photoemission represents an advance not only in theoretical research, but also in materials science.

Also interesting from our viewpoint, x-ray photoemission by a metal can simulate some aspects of collisions. As already discussed, when an atom is near the surface, an electron can transfer from the surface to the atomic orbitals. This leads to the creation of an image-charge potential, causing a sudden change in the scattering potential for conduction electrons. This situation is similar to the ejection of an electron from a core state. In both cases, we expect the system to behave according to the Anderson orthogonality catastrophe \cite{Anderson_1967}, which states that if a fermionic system undergoes a sudden change in the Hamiltonian, then the final ground state is orthogonal to the initial ground state. In other words, at a very small time after this sudden change, the system can only be found in its excited states, which result in many-particle-hole excitations.

When the atom is far from the surface, the coupling between the atom's orbital and the surface orbitals is small and varies slowly; consequently, the probability of an electron being exchanged between the atom and the metal is low; i.e., it is unlikely that the image-charge term become significant. Therefore, during the temporal evolution, we expect the system to remain in the ground state while the coupling term are very small, and the image-charge term is not significant, - an adiabatic process. If the Hamiltonian changes very slowly, we can use the adiabatic theorem \cite{Born}, which states that a system initially in the ground state evolves, adiabatically, always remaining in the ground state. We can study the behavior of the electronic states of the metal using the x-ray photoemission problem as a toy model, but with a time-dependent scattering potential that slowly increases. Naturally, as the atom approaches the surface, the variation in the coupling increases, along with the image-charge term, and thus, the Anderson orthogonality catastrophe competes with the adiabatic behavior. One of our goals is to understand the regimes where each of these effects become more important.

During the temporal evolution of a fermionic system in response to a rapidly increasing scattering potential, a high-energy state may become populated, and latter decay onto other high-energy states. Under these circumstance, it is difficult to predict the time evolution of the system. If this occurs, these excited state becomes crucial for the temporal evolution and cannot be ignored. Of course, considering all electronic excited states is impractical. Therefore, understanding the dynamics of the temporal evolution for these system is essential for correctly selecting the relevant states. This is significant in solving both the sticking and the x-ray photoemission problems.

Since the photoemission follows similar physics and can simulate some aspects of the collisions, we will use this problem to formulate strategies on how to select the most relevant electronic states for these types of problems. We can calculate the transition probability from the initial ground state to the states of the final Hamiltonian (after the sudden creation of the hole scattering potential). In this situation, the Anderson catastrophe occurs, resulting in low transition probabilities to the final ground state. Thus, excited states can have significant importance to the system dynamics, but the energy scale of these important excitations is difficult to predict.

Therefore, we first want to find the transition probability for the x-ray photoemission problem with a time-dependent potential, and use the knowledge obtained from this problem to help us to find the sticking coefficient. To do this, we will adopt a numerical solution strategy similar to the collision. At the instant $t=0$, the system is in the ground state, i.e., all energy levels below the Fermi level ($\e_F$) are occupied, and it is represented by the wave function $\Psi({\ell}, 0)$ (where ${\ell}$ represents the array of electronic states). We will use the NRG to diagonalize the Hamiltonian $H(t)$ for any $t$, and the Crank-Nicolson procedure to compute the function at time $\Psi({\ell},t=\Delta t)$. The wave function at any time $t$ is obtained by successively applying this method.

%% file: CAP/Cap2.tex

\chapter{Modeling the phenomena}

In this Chapter, we will find the Hamiltonian to describe each phenomenon. Section I considers an impurity-metal system. Section II discusses the interaction between the x-ray and the atoms, which is an important aspect of the x-ray photoemission problem. Finally, Section III describes the model for the hydrogen-surface collision.

A metal is characterized by its band structure, formed by the overlap of the wave functions of electronic states farthest from the nuclei of the metallic ions. For the properties studied in this work, these electronic states primarily contribute to the physical properties of the metal. The simplest model is one dimensional, with a single electronic level for each atom, coupled only to the nearest neighbors. The coupling is defined by a parameter $\tau$. This model is known as the tight-binding model. In second quantization, this model can be written as:
\begin{equation}\label{TBM}
\begin{aligned}
{H}_{B} \equiv -\tau \sum_\sigma\sum_{n=0}^{N-1} \left(a_{n\sigma}^{\dagger}a_{n+1 \sigma}+\mathrm{h.c.}\right).
\end{aligned}
\end{equation}
Here, the operators $a_{n\sigma}^{\dagger}$ creates one electron in the metallic site $n$ with a spin projection $\sigma = \{ \upA, \downA \}$, and $N$ is the number of sites. In momentum space, the tight-binding Hamiltonian reads
\begin{equation}\label{TBM_M}
\begin{aligned}
{H}_{B} \equiv \sum_{k\sigma}\varepsilon_{k}\tilde{a}^\dagger_{k\sigma}\tilde{a}_{k\sigma}.
\end{aligned}
\end{equation}
Where $\varepsilon_k = 2\tau\sin\left(\frac{\pi k}{N}\right)$ and $k$ an integer such that $-N/2 \leq k \leq  N/2$ and $\tilde{a}_{k\sigma}^\dagger = \frac{1}{\sqrt{N}}\sum_n a_{n\sigma}^\dagger \exp\left(-i\frac{\pi k}{N} n\right)$, given that we used the periodic boundary condition.  To avoid confusion, note that we shifted the phase of \(\varepsilon_{k'} = 2\tau\cos\left(\frac{\pi \left( k' - N/2\right)}{N} + \frac{\pi}{2}\right)\), resulting in band levels odd symmetric under \(k = k' - N/2 \). For small values of $|k|$, $\varepsilon_k \approx \frac{2\tau\pi}{N} k$ and the smaller energy gap $\Delta\varepsilon_k \approx \Delta\varepsilon = \frac{2\tau\pi}{N}$.

\section{Anderson model}

To understand the behavior of a quantum impurity coupled to the conduction band, we start out with the single impurity Anderson model (SIAM) \cite{PhysRev.124.41}, described by the Hamiltonian
\begin{eqnarray}\label{AM}
 H_{A} \equiv \left[ \varepsilon_d \left(n_{d\uparrow} + n_{d\downarrow} \right) + U n_{d\uparrow}n_{d\downarrow} \right] 
+ H_B + H_{\mathrm{hyb.}}.
\end{eqnarray}
The first term within brackets on the right-hand side is the contribution of impurity, where $n_{d\sigma} = d_\sigma^\dagger d_\sigma$, and $d_{\sigma}$ represents an impurity level with energy $\varepsilon_d$ and spin $\sigma$. The Coulomb repulsion term penalizes double occupancy. Consequently, the impurity can have one of three energies, depending on the impurity occupancy (single, empty, and double). $H_B$ represents the contribution of the conduction band, and the hybridization term $H_{\mathrm{hyb.}}$ is the coupling between the impurity and the metal, allowing electron transfer between the impurity and the metal.

\begin{figure}[htb!]
		\begin{center}
		\includegraphics[scale=0.47]{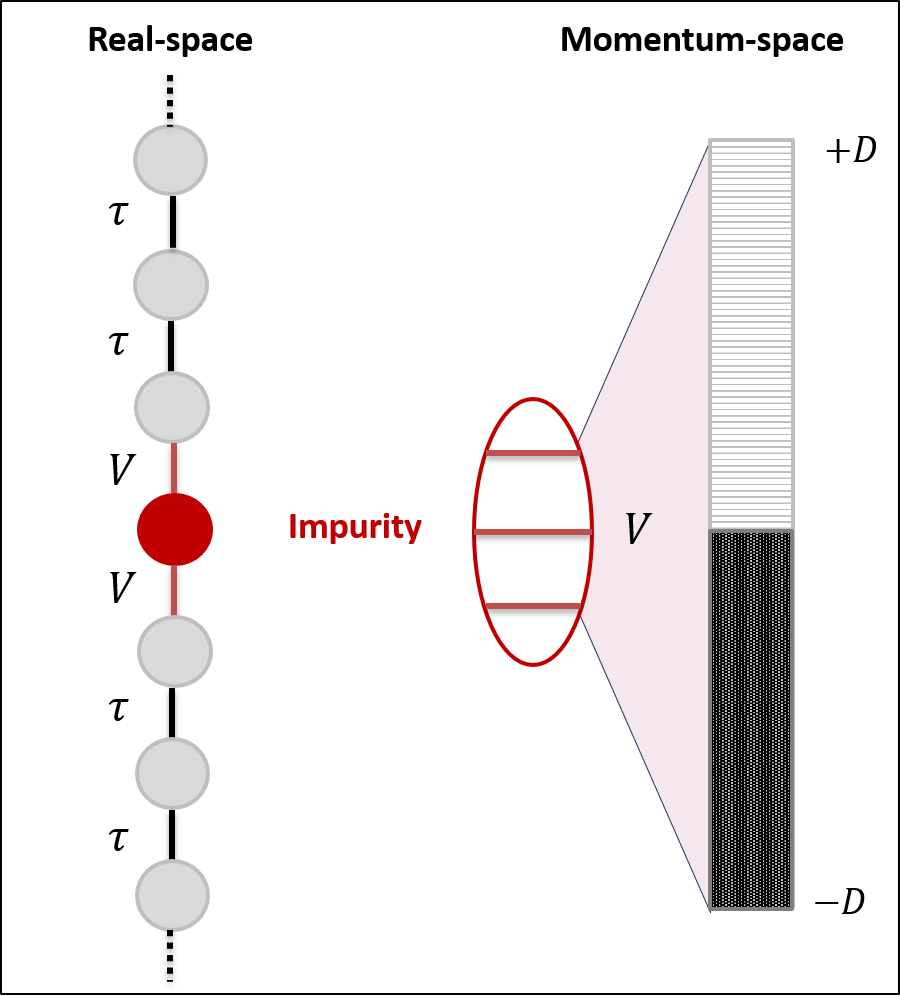}
		\end{center}
		\vspace{-0.2cm}
		\caption{\footnotesize Schematic representation of the SIAM. On the left, the real-space representation of the impurity-metal system, where there is a coupling only in the first neighbors. On the right, the momentum-space representation of the impurity-metal system. \Sou }
	\label{Anderson_Model}
\end{figure}

Figure \ref{Anderson_Model} shows the schematic representation of the SIAM. In real space, the impurity couples only to the neighboring metallic atoms. By exploring the symmetry of the chain, we can map the chain into a semi-chain, and the hybridization term can be written as:
\begin{eqnarray}\label{Hyb.RS}
 H_{\mathrm{hyb.}} \equiv \sqrt{2} V\left(a_0^\dagger d + d^\dagger a_0 \right).
\end{eqnarray}
 Here, $a_0$ represents the orbital of the first metallic site and $V$ is the coupling amplitude. For simplicity, after this point, the sum over the spin projection will be omitted. In cases where the impurity is connected only with one side of the chain shown in Fig. \ref{Anderson_Model}, the factor $\sqrt{2}$ needs to be removed from Eq. \eqref{Hyb.RS}. In momentum space; the coupling is constant, and the hybridization term becomes:
\begin{eqnarray}
 H_{\mathrm{hyb.}} \equiv \frac{V}{\sqrt{N}} \sum_k\left(\tilde{a}_k^\dagger d + d^\dagger \tilde{a}_k \right).
\end{eqnarray}

The SIAM has rich physics. In spite of its simplicity, it offers insight into many phenomena, such as the Kondo problem \cite{RevModPhys.47.773,PhysRevB.101.125115}, heavy fermions \cite{RevModPhys.56.755}, and others. For more general problems beyond, the tight-binding model, a more general band structure can be used in Eq. \ref{TBM_M}, characterized by the density of states $\rho(\epsilon)$. The physics of SIAM is parameterized by the impurity energies $\varepsilon_d$ and $2\varepsilon_d + U$, the hybridization energy $\Gamma = \pi \rho V^2$, and the band structure $\rho$. The hybridization energy $\Gamma$ represents the energy width of the impurity level, due to its coupling to the conduction band. Here, we will use this model to describe the metal-hydrogen interaction in atomic-surface collisions or to consider an impurity in the x-ray photoemission problem.

\section{X-ray - metal interactions}

The x-ray photoemission problem was qualitatively discussed in the introduction, and the schematic representation of this problem is shown in Fig. \ref{x_Ray}. Even tough we are interested in the conduction states, the electronic core of the ions cannot be ignored, because they can interact with x-ray photons. Thus, as a consequence of this interaction, an electron can be ejected from a core level to a free electron state outside the metal. If this happens, the atom will be positively charged, generating an attractive Coulomb potential $W$ (known as a core-hole potential) interacting with the electrons in the conduction band. This interaction adds an attractive term $ W a_0^\dagger a_0 $ to the Hamiltonian, with $ W < 0 $. Since the scattering potential is due to the core hole, it would be more rigorous to add $ W a_0^\dagger a_0 h_p^\dagger h_p$ to the Hamiltonian instead, where the operator $ h_p^\dagger $ creates a core hole.

Now, we can propose an appropriate model to study the metal when an x-ray photon strikes a core level. As discussed in the last section, the metal system is well-described by the tight-binding model. Therefore, we can write the Hamiltonian in real-space as follows:
\begin{align}\label{H0}
H = H_B + E_p h_p h_p^\dagger + W a_0^\dagger a_0h_p^\dagger h_p.
\end{align}
Here, the states in the band are represented by $H_B$, the core level energy is $E_p$, and the last term is the hole potential, non-zero and attractive when $h_p^\dagger h_p > 0$.

The x-ray photon and deep core level interaction is modeled via perturbation theory, enabling electrons to transfer from the core level to a free electron state outside the metal. Then, at $ t = 0 $, one electron from the core level is expelled from the metal, resulting in $ h_p^\dagger h_p = 0 $ for $ t < 0 $ and $ h_p^\dagger h_p = 1 $ for $ t \ge 0 $. The number of electrons in the conduction band is conserved in the photoemission phenomena. Additionally, we consider that the lifetime of the core hole is infinite, which simplifies the problem.


In summary, the core level is initially occupied, and the scattering potential term $W a_0^\dagger a_0h_p^\dagger h_p$ is null, since $h_p^\dagger h_p = 0$. But, at $t=0$, the absorption of an x-ray photon removes an electron from this deep core state ($h_p^\dagger h_p  \rightarrow 1$) and switches on the attractive core-hole potential ($W<0$). The many-body state $\ket{\Psi(t)}$ then evolves under the final Hamiltonian
\begin{align}\label{H_photo_MI}
H_{p}(t) = H_B + W(t) a_0^\dagger a_0.
\end{align}

The time scale of the interaction between x-rays and inner-shell electrons, and consequently the ejection of deep core electrons, is extremely short. Consequently, the scattering potential appears as a sudden quench $W(t) = W \cdot \Theta(t)$, where $\Theta(x)$ is the Heaviside step function. However, to study the electronic behavior of the band in a more general situation, we will let $W(t)$ a time-dependent function. This make it possible to simulate the collision conditions and to study adiabaticity.

\subsection{The photoemission critical behavior}

Once the Hamiltonian in Eq. \eqref{H_photo_MI} represents a Fermi gas with a localized scattering potential \cite{Anderson_1967,Nozieres,PhysRevB.24.4863}, the potential shifts the phase of the conduction electrons by
\begin{eqnarray}\label{phase_shift}
&\delta \equiv \mathrm{atan} \left(-\pi\rho W\right).
\end{eqnarray}

X-ray photoelectron spectroscopy (XPS) measures the Fourier transform of the square of the projection $\braket{\phi_0}{\Psi(t)}$, where $\ket{\Psi(t)}$ is the system wave function and $\ket{\phi_0} = \left(\prod_{\epsilon_k < 0}\tilde{a}_k^\dagger \right) \ket{0}$ is the initial ground state, before the x-ray photon remove an electron from the deep core. Once we know the system start in the initial unperturbed ground state $\ket{\Psi(0)} =  \ket{\phi_0}$, the projection $\braket{\phi_0}{\Psi(t)}$ quantifies how much of the system remains in the initial ground state and what is the probability of finding the system in one of the excited states. If there were no scattering potential, once the system would remains in the initial ground state with energy $\tilde{E}_0$, the projection $\braket{\phi_0}{\Psi}_{t} = 1$ or $\braket{\phi_0}{\Psi}_{\hbar\omega} = \delta(\hbar \omega - \tilde{E}_0)$ in the frequency domain.

If the scattering potential is non-zero, the switching-on of the potential creates particle-hole excitations in the band. But, due to energy conservation, only transitions for excited states with very small energies should have significance. By consequence, the expected behavior for $\braket{\phi_0}{\Psi}_{\hbar\omega}$ in the frequency domain is a sharp peak, resembling a Lorentzian $\braket{\phi_0}{\Psi}_{\hbar\omega} \sim \frac{\eta / \pi}{\eta^2 + (\hbar\omega -  \tilde{E}_0)^2}$, where $\eta$ controls how sharp the peak is.

However, this oversimplified perspective misses two important physical aspects of this out-of-equilibrium Fermi gas system. First, in the presence of a localized scattering potential, the Anderson orthogonality catastrophe arises \cite{Anderson_1967}, stating that, for a large number of electrons in the band, the ground state before and after the introduction of the scattering potential are orthogonal. Second, for finite times, the Heisenberg uncertainty principle ($\Delta \epsilon \cdot t \sim \hbar$) permits short-time violations of energy conservation. As time progresses, the particle-hole excitations must have smaller energies.

Therefore, as discussed before, Doniach and Sunjic \cite{Doniach_1970} pointed out that, instead of a sharp peak in the frequency domain, the projection has the form $\langle \phi_0 | \Psi \rangle_{\hbar\omega} \sim (\hbar \omega - \tilde{E}_0)^{-1 + \left(\frac{\delta}{\pi}\right)^2}$, on the basis of the analogous result for x-ray absorption, first derived by Nozières and De Dominicis \cite{Nozieres}. This power law behavior, a consequence of the many small energy particle-hole excitations that arise in the transient regime of the system due to the re-adjustment of the ground state \cite{Doniach_1970}, is observed in XPS experiments. In the time domain, this is equivalent to $\langle \phi_0 | \Psi \rangle_{t} \sim t^{-\left(\frac{\delta}{\pi}\right)^2}$, at long times. The projection approaches zero for all $\delta \ne 0$.

This behavior combines the Heisenberg uncertainty principle with Anderson's orthogonality. Let us start from $t=0$, when the scattering potential suddenly appears and changes the energy levels, leading to Anderson's orthogonality catastrophe and creating many low-energy particle-hole excitations. The system wave function represented by $\Psi(t)$ is initially in the initial ground state, and it takes time to change. For $t>0$, the excitations with energy $\epsilon \sim \frac{\hbar}{t}$ become important and contribute to the projection, a consequence of the Heisenberg uncertainty principle, which results in many small energy particle-hole excitations arising as time passes and the quantity $\langle \phi_0 | \Psi \rangle_{t}$ dropping by the factor $\sim t^{-\left(\frac{\delta}{\pi}\right)^2}$, known as the Doniach-Sunjic power law.

Additionally, one important point to note about photoemission is the presence of the bound state. To understand this, let us first consider a situation in which the localized scattering potential is very strong, $ |W| \gg 2\tau $, with the band described by the tight-binding model in Eq \eqref{TBM}. In this case, an electron at the site 0 (represented by the operator $ a_0^\dagger $) cannot flow from it to the neighboring sites, causing this level to decouple from the rest of the metallic chain. This decoupled level, becomes the lowest energy level in the Hamiltonian \eqref{H_photo_MI}, with an energy $W$, since $W<0$. This level is a bound state, that is, it correspond to an electron bounded to the site $0$. This bound state also occurs for smaller potentials. As we will discuss in detail in Chapter 3, this level is extremely important to describe the photoemission accurately.

\subsection{Continually increasing scattering potential}

Another interesting case occurs when the scattering potential amplitude $|W(t)|$ in Eq. \eqref{H_photo_MI} is continuously increasing in time. This situation allow us to study adiabaticity in the Fermi gas system and also help us understand the atomic-surface behavior when the atom is far from the surface. We will consider $W(t)$ growing linearly up to a maximum amplitude $\bar{K}$, on a time scale $T \leq T_m = \frac{\hbar}{\Delta\varepsilon}$. Clearly, when $T=0$ the sudden quench is recovered. This time-dependence allows us to study all dynamical regimes, from non-adiabatic to adiabatic \cite{GDiniz}. Furthermore, we have two time scales in this system. The first is defined by the smallest energy scale in the system: $ T_m = \frac{\hbar}{\Delta \varepsilon} $ is the time necessary for all electronic levels in the system with energy $ \epsilon > \Delta \varepsilon $ respond to the attractive potential at the first site. The second is the time scale $ T $ controlling the ramp up from $W=0$ to $W=\bar K$.

If $T$ is very small, the system responds to a rapid change, analogous to the sudden quench in the previous subsection. The time-dependent behavior is governed by the Doniach-Sunjic power law, and the probability to find the system in the instantaneous ground state ($\ket{\varphi_0(t)}$) is governed by the Anderson catastrophe
\begin{equation}
|\braket{\varphi_0 }{\Psi}|^2 \approx \left(T_m D/ \hbar \right) ^{-2\left(\frac{\delta}{\pi}\right)^2}.
\end{equation}
Here, $D = \Delta\varepsilon \cdot N_e$ is the bandwidth. For a large number $N_e$ of electrons, numerous particle-hole excitations emerge from the ground state, indicating that the system is far from adiabaticity.

In contrast, if the time scale $T$ is large, the picture changes, and only small energy excitations become impossible. Using a quasi-adiabatic approximation, in which only one particle-hole excitations are allowed, we will show in Chapter 4 that the probability of the system remaining in the instantaneous ground state during its evolution is given by
\begin{equation}
\begin{aligned}
|\braket{\varphi_0(T) }{\Psi(T) }|^2 \approx \left({T_m}/{T} \right)^{-\left(\frac{{\delta}}{\pi} \right)^2 - \left(\frac{{\delta} }{\pi} \right)^4}.
\end{aligned}
\end{equation}
The time behavior still follows a power law, parameterized by the phase shift, but the exponent is different.

If the maximum scattering potential $\bar{K}$ is small and $T$ is large enough, the system remains in the instantaneous ground state, and the evolution is adiabatic. We are specially interested in the competition between the adiabatic phase and the non-adiabatic regime, which depends on both time scales $T_m$, $T \leq T_m$, and on $\bar{K}$.

Our study led to two important conclusions, which will be discussed in Chapter 4. The first is that the traditional adiabatic criterion \cite{Avron1999-qt} is an unreliable criterion, under the studied circumstances. Additionally, we discovered that if this Fermi gas system is truly gapless, there exists a critical $\bar{K}_\mathrm{max}$ such that for any $\bar{K} > \bar{K}_\mathrm{max}$, the system behaves non-adiabatically for any $T \leq T_m$.

\section{Atom - surface collision}

Now, let us focus on the atomic sticking problem. Qualitatively, the dynamics of the collision between a particle and a metallic surface has already been briefly discussed in the introduction, and it is schematically represented in Fig. \ref{H_MS}.  Here, we want to further explore and discuss this system in detail and propose a suitable model.  For simplicity, we will consider the incident particle as a hydrogen atom represented by $\mathrm{H^0}$ for the neutral particle and $\mathrm{H^{- , +}}$ when it is ionized.

If the incident atom were adsorbed on the metallic surface, it would behave like an impurity in the metal. Therefore, in this case, the system would be effectively described by the SIAM. However, the atom is in motion towards the surface, initially neutral and at a distance $z_0$. As the atom approaches the surface, the overlap between the wave functions of the electronic states on the surface and the hydrogen orbital increases, and so does the coupling term $V(z)$, leading to a higher probability of electron transfer between the atom and the surface. The kinetic energy of the incident particle must be represented by an operator $\dfrac{P_z^2}{2M}$, where $M$ is the nuclear mass, and the position $z$ defines the degree of freedom of the atomic wave function.

The only missing ingredient now is the image-charge potential. This potential takes the form
\begin{equation}
    W(z) = -\frac{1}{4\pi\varepsilon_0} \frac{e^2}{4z}
\end{equation}
for a hydrogen atom \cite{Zangwill_2012}. Similarly as in the case of x-ray photoemission, this localized image-charge creates a scattering potential attracting the electrons in the conduction band. When the atom is ionized, this potential is represented by the term $W(z) f_0^\dagger f_0$, where $f_0^\dagger$ is the Wannier orbital of the metal given by 
\begin{eqnarray}
    f_{0}^\dagger \equiv \dfrac1{\sqrt {N}} \sum_k \tilde{a}_k^\dagger.
\end{eqnarray}
This potential is only created when the atom is ionized, regardless of the sign of the atomic charge. Therefore, in the context of the collision problem, this scattering potential can be modeled as 
\begin{eqnarray}
    W(z) f_0^\dagger f_0 (n_d -1)^2,
\end{eqnarray}
where $n_d$ represents the occupation of the hydrogen orbital.

Finally, we can write the Hamiltonian describing the collision as
\begin{align}\label{H_Colision}
H_c(z) = \dfrac{P_z^2}{2M} + \left\{ H_d +\sum_{k}\varepsilon_{k}\tilde{a}^\dagger_{k}\tilde{a}_{k} + V(z)\left(\cd{d}\fn{0}+\hc \right)+ W(z)(n_{d}-1)^2\fd{0}\fn{0} \right\},
\end{align}
where $H_d = \e_d d^\dagger d + U n_{d\upA}n_{d\downA}$ represents the hydrogen levels, $d$ denotes the atomic orbital with energy $\e_d$, $U$ is the Coulomb interaction between the electrons in the atom and $z$ is the distance between the atom and the surface.

The Hamiltonian format is ready, but additional details are needed to describe the physics of the collision. Let us start with the coordinate $z$, which measures the distance between the surface and the atom. The surface at $z=0$ does not correspond to the position of the metallic atoms. Instead, we define $z=0$ as the point where the surface potential is significantly large, making it nearly impossible for the hydrogen atom to penetrate the surface with the relatively small energies, as represented in Fig.\ref{EXPE}. The image-charge potential can be then written as:
\begin{align}
\label{eq:3}
W(z) \equiv -\dfrac{W_0}{4(z+z_{\text{im}})},
\end{align}
where $W_0 = \frac{e^2}{4\pi\varepsilon_0} = 27.21$ eV.Bohr, and $z_{\text{im}}$ indicates the asymmetry of the image position.

Another term that requires closer examination is the hydrogen-surface coupling $V(z)$. This term represents the overlap between the metallic orbitals and the hydrogen orbitals. Typically, the spatial distribution of the atomic orbitals decreases exponentially as $\exp(-z/\mathfrak{a}_0)$, where $\mathfrak{a}_0$ is the Bohr radius. Therefore, we will consider the hydrogen-surface coupling as
\begin{align}
\label{eq:2}
V(z) \equiv V_0\exp(-z/r),
\end{align}
The constant $V_0$ represents the coupling at $z=0$, and $r$ is the range of the potential.

We consider that the incident atom is initially at the position $z_0$, with kinetic energy $K_0 = \frac{\hbar^2k_0^2}{2M}$ towards the surface. At the time $t=0$, the wave function of the incident particle is the product of the electronic state $\lvert \Phi_0 \rangle$ (the metallic ground state and a neutral hydrogen atom) with a Gaussian spatial distribution centered at $z=z_0$, moving to the left of the $z$ axis with momentum $k_0$, i.e.,
\begin{align}
\label{eq:5}
\ket{{\Psi}(z,0)}= Be^{-{(z-z_0)^2}/{2 \eta }} e^{-ik_0 z} \lvert \Phi_0 \rangle,
\end{align}
where $\eta$ controls how sharp the distribution is initially and $B \approx (\pi\eta)^{-1/4}$ is the normalization constant.

For the hydrogen atom $1s^1$ orbital, the energies are $\varepsilon_d = -13.6$ eV and $U = 12.8$ eV, which result in the negatively ionized $\mathrm{H}^-$ being more stable than the neutral hydrogen $\mathrm{H}^0$. Both $\mathrm{H}^-$ and $\mathrm{H}^0$ are more stable than $\mathrm{H}^+$. Even when the atom is far from the surface, the minimum energy occurs when the hydrogen is negatively ionized. Due to the enormous energy difference, the positively ionized state contributes insignificantly to the sticking.

Then, in this model, for a region near the surface with length $L_z$, the sticking coefficient
\begin{eqnarray}
   \mathcal{S} \equiv \lim_{t \rightarrow \infty} \int_0^{L_{z}} dz \braket{{\Psi}(z,t)}{{\Psi}(z,t)} 
\end{eqnarray}
is a function of $\mathcal{S}\left(D,\epsilon_F,\varepsilon_d,U,V_0,r,z_{\text{im}},K_0,M\right)$, where $D$ and $\epsilon_F$ are surface parameters, $\varepsilon_d$, $U$ and $M$ are atomic parameters, $V_0$, $r$ and $z_{\text{im}}$ depends on the interaction atom-surface and $K_0 \sim k_B T_{emp}$ depends on the initial atomic thermal energy, that is, the furnace temperature.

Although qualitatively simple, this problem is complex. As mentioned earlier, atomic states are coupled to electronic states since, depending on each possible electronic distribution, the atom behaves differently. For example, hypothetically, for an electronic configuration where the atom remains neutral during the trajectory, the image charge potential does not appear, and consequently, the particle simply collides with the metallic surface and is reflected. Conversely, for an electronic configuration where the atom is ionized, depending on its initial kinetic energy and energy loss during the collision process, it may end up stuck on the surface.

\subsection{Physics of the collision}

Here we discuss in more detail the physics of the atomic-surface collision. This problem has several time scales. The first one is the time that the atom interacts with the surface, defined by $ T_{\mathrm{kin}} \sim { z_c}{\sqrt{\frac{2M}{K_0}}} $, where $ z_c $ defines the region where the atomic-surface interaction is significant. The second is $ T_{\mathrm{ele}} \sim \frac{\hbar}{\Delta\varepsilon} $, which is the time scale of the small energy excitations in the band. The last one is associated with the hybridization $T_{\mathrm{hyb}} \sim \frac{\hbar}{V}$ and it represents the necessary time to hybridize the neutral and ionized levels.

By the Heisenberg uncertainty principle, only electronic energy states where $ T_{\mathrm{ele}} \leq T_{\mathrm{kin}} $ will participate in the electronic transitions. This is because electronic levels separated by smaller energy differences than $ \Delta \varepsilon' < \frac{\hbar}{T_{\mathrm{kin}}} $ do not have enough time to participate in the atomic-surface collision dynamics. Additionally, as the hybridization between the electronic levels needs to occur during the collision, so $ T_{\mathrm{hyb}} \le T_{\mathrm{kin}} $. Clearly, the sticking mechanism depends on these three time scales being comparable.

\begin{figure}[hbt!]
		\begin{center}
        \begin{tabular}{ll}
        \hspace{-0.1cm}
        \includegraphics[scale=0.32]{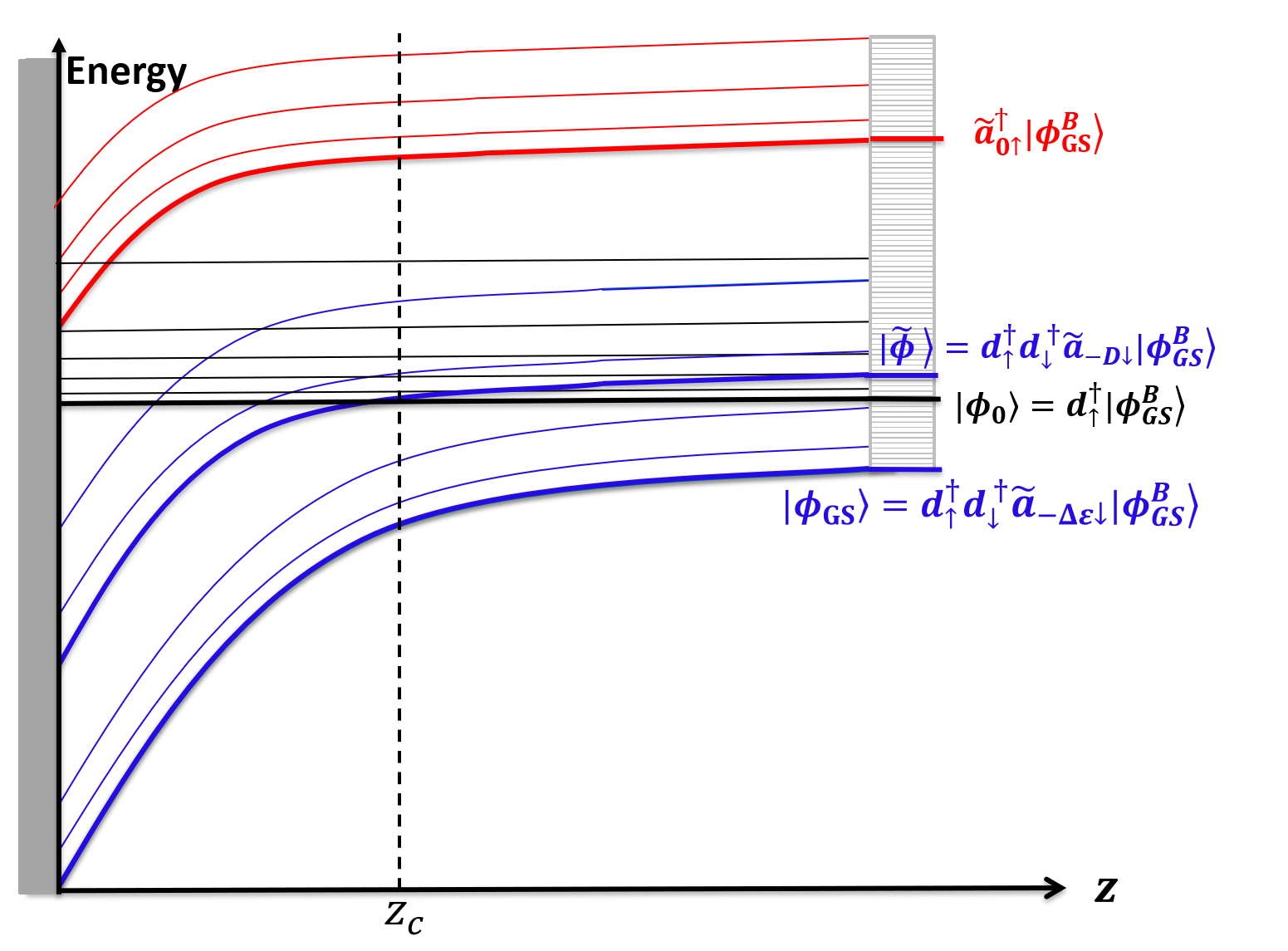}
        & \hspace{-0.2cm}
        \includegraphics[scale=0.3]{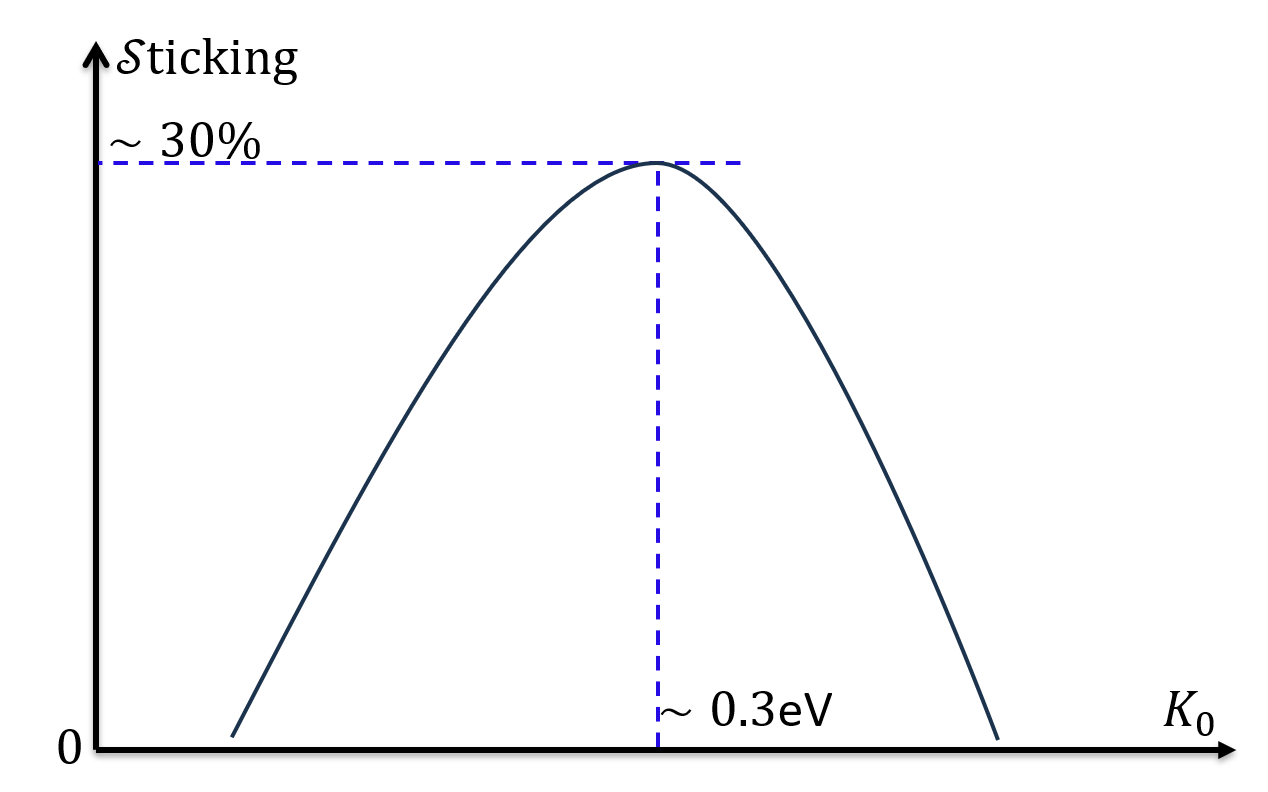}
        \end{tabular}
    \vspace{-0.3cm}
	\end{center}
	\caption{\footnotesize (Left panel) Potential energy surfaces (schematic). The colors are associated with the charge of the hydrogen atom: black, blue and red represent $\mathrm{H}^0$, $\mathrm{H}^-$ and $\mathrm{H}^+$, respectively. The ket $\ket{\phi^B_{GS}}$ stands for half-filled conduction band. The thick line on the graph represents the indicated states constructed by the creation operators of the band, \(\tilde{a}_k^\dagger\), and the hydrogen levels, \(d^\dagger\), acting on the metallic ground state $\ket{\phi^B_{GS}}$. The thin line corresponds to particle-hole excitations of these discussed states. Since $2\varepsilon_d + U < \varepsilon_d$, in the total ground state the hydrogen is negatively ionized. (Right panel) Sticking coefficient as a function of the initial kinetic energy (schematic). \Sou  }
    \label{Draw_Collision}
\end{figure}

Another important point is the non-adiabaticity created by the atomic movement, which steals energy from the incident atom. As Fig. \ref{Draw_Collision} (Left panel) shows, as the atom approaches the surface, the ionized levels are displaced by the scattering potential,  which lowers their energy, and the hybridization grows, allowing charge transfer between the atom and the surface. Since the atomic motion is coupled to the electronic states, and since the image-charge potential $W(z)$ and the hybridization $V(z)$ depend on the atomic position, fast movement close to the surface causes sudden changes in the Hamiltonian. The displacement of the electronic levels creates particle-hole excitations, which can steal a considerable part of the kinetic energy, as a result, the ion may be bound to the surface.

On the other hand, if the atom moves slowly close to the surface, the Hamiltonian changes gradually, and only small energy particle-hole excitations will be significant.  Since there is no dissipative term in the Hamiltonian, the atom will merely accelerate and collide with the surface, rebound, and lose the kinetic energy gained from the image-charge potential. No interesting physics emerges from this reversible collision, and the sticking coefficient approaches zero, as the right panel of Fig. \ref{Draw_Collision} shows. Conversely, if the initial kinetic energy is too high, with $ T_{\mathrm{hyb}} > T_{\mathrm{kin}} $, the system does not have enough time to hybridize; the resulting sticking coefficients are equally small. Only at intermediate energy scales, around $ K_0 \sim 0.3 $ eV, does the  probability of binding to the surface become significant.

To discuss the collision from another viewpoint, consider the avoided crossing levels in the left panel of Fig. \ref{Draw_Collision} (at $z = z_c$). Initially, when the atom is far from the surface, the hybridization is nearly zero, making transitions from the initial neutral configuration to any ionized state improbable. As the atom approaches the surface, the hybridization gradually increases and small-energy transitions become possible. The probability becomes appreciable if the atom remains in this region for a considerable time. Since the magnitude of the scattering potential grows as the atom approaches the surface, the energies of the ionized levels decrease and may become smaller than those of the neutral configuration. The resulting avoided crossings signal hybridization and, consequently, a significant transition probability between the neutral and ionized configurations.

When the atom is far from the surface, the hybridization is very small ($V(z) \sim 0$), and we can write the electronic part of the Hamiltonian in Eq. \eqref{H_Colision} as
\begin{eqnarray}
    H_e(z) = \left[\varepsilon_d n_d + U n_{d\upA} n_{d\downA}  \right] + H_B + W(z)(n_d - 1)^2 f_0^\dagger f_0. 
\end{eqnarray}
Within this approximation, the charge $n_d$ is conserved. In the ionized sector ($n_{d} = 2$) the Hamiltonian coincides with the final Hamiltonian in the photoemission problem, Eq. \eqref{H_photo_MI}. In the neutral sector ($n_d = 1$), which contains the initial state, the conduction band coincides with the initial ground state in the photoemission problem. Therefore, the photoemission problem can not only  mimic the collision process qualitatively, as discussed in the introduction, but it shares its initial and final Hamiltonians with the $V=0$ limit of the electronic Hamiltonian for the atom-surface problem.

Additionally, in analogy with photoemission, the localized scattering potential makes the near-surface conduction-band ground state nearly 
orthogonal to the initial ground state, a consequence of the Anderson orthogonality catastrophe. Excited states are therefore important and cannot be disregarded in accurate computations of the sticking coefficient. Dropping excited states above a predefined cutoff may therefore be a poor approximation. Hence, we aim to keep all relevant states.

For that we will need an efficient way to select these states. The simplest approach is to identify the leading contributions to photoemission, on the basis of analytical or numerical treatment. For this reason, the next two chapters will deal with photoemission. After that, we will return to the collision.

%% file: CAP/Cap3.tex
\chapter{Photoemission from simple metals}


To understand the  physics underlying the photoemission, let us start with the Hamiltonian in Eq. \eqref{H_photo_MI} without spin. To simplify it, we can consider a flat band (flat density of states, or linear energy dispersion) and write the localized scattering potential in momentum space. This transforms $W a_0^\dagger a_0 \rightarrow \frac{W}{N} \sum_{k,q} \tilde{a}_k^\dagger\tilde{a}_q $. The Hamiltonian is now
\begin{align}\label{H_photo}
H = \sum_{k}\varepsilon_{k}\tilde{a}^\dagger_{k}\tilde{a}_{k} + \frac{W}{N} \sum_{k,q} \tilde{a}_k^\dagger\tilde{a}_q.
\end{align}
Where $\varepsilon_{k} = k.\Delta\varepsilon$. Here, $\Delta\varepsilon$ is the energy gap and $k$ an integer such that $-N/2 \leq k \leq N/2$.

The Hamiltonian in the Eq. \eqref{H_photo} is quadratic and can be diagonalized analytically. With this in mind, we employ the Sommerfeld-Watson transformation \cite{sommerfeld1949partial}, as detailed in Appendix \ref{Appendix_Analytical_Diagonalization}. The energies of the resulting Hamiltonian are simply given by 
\begin{equation}\label{Band_energies}
\epsilon_l = \varepsilon_l - \frac{\Delta\varepsilon}{\pi}\delta_l,
\end{equation}
corresponding to a small shift from the initial energy levels $\varepsilon_l$, where $\delta_l \approx \delta $ is the phase shift in Eq. \eqref{phase_shift}. As shown in Appendix \ref{Appendix_Analytical_Diagonalization}, the eigenstate of the final Hamiltonian with energy $\epsilon_l$ can be written as a linear combination of the unperturbed band operators as
\begin{equation}
g_l^\dagger = \sum_k \alpha_{l,k} \tilde a_k^\dagger, 
\end{equation}
where the coefficients $\alpha_{l,k}$ are
\begin{eqnarray}\label{Band_proj}
\alpha_{l,k}  = -\frac{\sin \delta_l }{\pi} \frac{\Delta\varepsilon}{\varepsilon_l - \varepsilon_k - \frac{\delta_l}{\pi}\Delta\varepsilon}.
\end{eqnarray}
Clearly, the overlap between a final eigenstate with energy $\epsilon_l = \varepsilon_l - \frac{\delta_l}{\pi}\Delta\varepsilon$ and an initial eigenstate with energy $\varepsilon_k$ will be very small if $|\epsilon_l - \varepsilon_k| \gg \Delta\varepsilon$.

To describe the photoemission, even at the qualitative level, we have to examine the ground-state of the initial Hamiltonian, which can be constructed from the initial single-particle eigenstates bellow Fermi level, and the many-body eigenstates that  can be constructed from the final single-particle Hamiltonian  \eqref{H_photo}. Algebraically, the final many-body eigenstates $\{ \ket{\varphi_n}\}$, with energy $E_n$, can be constructed as particle-hole excitations from the many-body ground state $\ket{\varphi_0} = \prod_{\epsilon_q < 0}g_q^\dagger \ket{0}$, where $\ket{0}$ is the vacuum. We will use the $\{ \ket{\varphi_n}\}$ as the basis states to expand $\ket{\Psi(t)}$.

As already mentioned, the quantity $\braket{\phi_0}{\Psi(t)}$ gives us information about the photoemission. Since $\ket{\Psi(0)} = \ket{\phi_0} = \left(\prod_{\epsilon_k < 0}\tilde{a}_k^\dagger \right) \ket{0}$, the projection $\braket{\phi_0}{\Psi(t)}$ is the amplitude of finding the system in the initial ground state. For this reason, in this chapter, we will focus on computing the fidelity $\mathcal{F}(t)$ of the quantum state of the system $\ket{\Psi(t)}$ with respect to the initial ground state $\ket{\phi_0}$
\begin{equation}
    \mathcal{F}(t) \equiv |\braket{\phi_0}{\Psi(t)}|^2 = \left| \sum_n |\braket{\phi_0}{\varphi_n}|^2 \exp\left(-i\frac{E_n t}{\hbar}\right)  \right|^2.
\end{equation}

To compute the fidelity $ \mathcal{F}(t)$, we fist need $\braket{\phi_0}{\varphi_0}$, given by the Slater determinant
\begin{eqnarray}\label{Slater}
\braket{\phi_0}{\varphi_0} = \mathrm{det}
\begin{pmatrix}
\{\tilde a_{-1},g_{-1}^\dagger\} & ... & \{\tilde a_{-1},g_{-N_e}^\dagger\} \\
\vdots & ... & \vdots \\
\{\tilde a_{-N_e},g_{-1}^\dagger\} & ... & \{\tilde a_{-N_e},g_{-N_e}^\dagger\} \\
\end{pmatrix}.
\end{eqnarray}
Here, $N_e$ is the number of electrons and $\{\tilde a_{k},g_{l}^\dagger\}$ can be obtained from Eq. \eqref{Band_proj}. 
Letting $\delta$ be constant, we are led to the Anderson catastrophe \cite{Anderson_1967}
\begin{eqnarray}\label{AOC EQ}
\braket{\phi_0}{\varphi_0} \sim N_e^{-\left(\frac{\delta}{\pi}\right)^2}.
\end{eqnarray}
As typically $N_e \sim 10^{23}$, the right-hand side approaches zero, even for small values of $\delta$. As a consequence, a large number of particle-hole excitations must contribute to $\ket{\Psi(t)}$.

Asymptotically, Eqs. \eqref{Band_energies}, \eqref{Band_proj}, and \eqref{AOC EQ} yield the Doniach-Sunjic power law behavior \cite{Doniach_1970}, a result that was established five decades ago. Little attention has been given, however, to the contributions of the bound state created by the attractive potential.

\section{The bound state}
            
As discussed by elementary textbooks on Quantum Mechanics, attractive potentials applied to one-dimensional systems create bound states. Our final Hamiltonian is no exception. Appendix \ref{Appendix_Analytical_Diagonalization} shows that the bound state energy is
\begin{eqnarray}\label{Bond_State_energy}
\epsilon_{B} = -D.\mathrm{coth}\left(  \frac{1}{-2\rho W} \right),
\end{eqnarray}
with the coefficients $\alpha_{B,k}$ expressing the corresponding  eigenoperator $g_B$ on the the basis of the conduction operators $\tilde a_k$ are
\begin{eqnarray}\label{Bond_State_proj}
\alpha_{B,k} = \sqrt{\frac{\Delta\varepsilon}{2D}} \frac{\sqrt{(-\epsilon_B+\Delta\varepsilon)^2 - D^2}}{(-\epsilon_B+\Delta\varepsilon) + \varepsilon_k}.
\end{eqnarray}

Although the well-known expressions  Eqs. \eqref{phase_shift} and \eqref{Band_proj} \cite{Mahan,Nozieres,Doniach_1970,PhysRevB.24.4863},  we did not find Eqs. \eqref{Bond_State_energy} and \eqref{Bond_State_proj} in the literature, apparently because has traditionally been focused on states near the Fermi level. As we shall see, however, the bound state affects the time dependence of the fidelity dramatically and adds structure to the photoemission spectra. In a experimental system, a core-hole potential tends to be screened; treating it as localized is therefore a good approximation, and the conclusions drawn from our results may be checked.


The energy difference between the initial and final ground state energies is given by the equality
\begin{equation}
    E_0 - \tilde E_0 = \sum_{\epsilon_k < 0} \epsilon_k - \sum_{\varepsilon_k < 0} \varepsilon_k, 
\end{equation}
where the initial ground state energy is the sum of all energies $\varepsilon_k$ up to the Fermi level ($\varepsilon_F = 0$), and the final ground state energy is the sum of all energies $\epsilon_k < 0$. By using the Eqs. \eqref{Band_energies} and \eqref{Bond_State_energy}, we find the equality
\begin{equation}
E_0 - \tilde E_0 = \epsilon_B + D -  \frac{\delta}{\pi}\sum_{k<0} \Delta \epsilon,
\end{equation}
which results in
\begin{equation}\label{Binding_energy_shift}
    E_0 - \tilde E_0  \approx  - D\left(\coth\left( \frac{-1}{2\rho W}\right)- 1 + \frac{\delta}{\pi} \right),
\end{equation}
a negative energy, as expected. 



\begin{figure}[hbt!]
		\centering
        \begin{tabular}{ll}
        \hspace{-0.55cm}
        \includegraphics[scale=0.515]{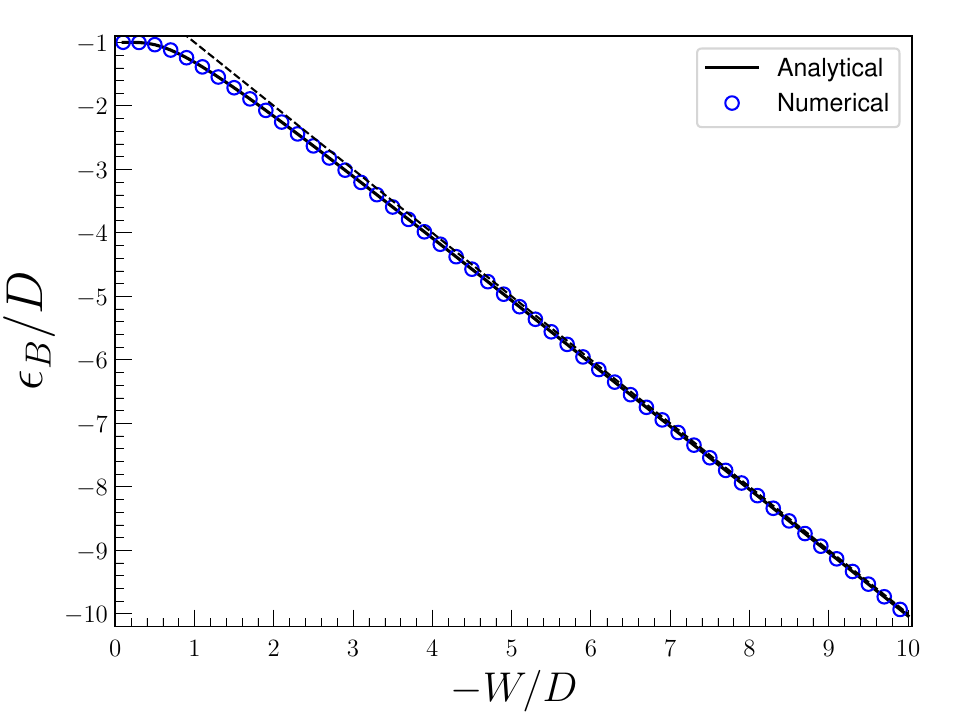}
        & \hspace{-0.8cm}
        \includegraphics[scale=0.515]{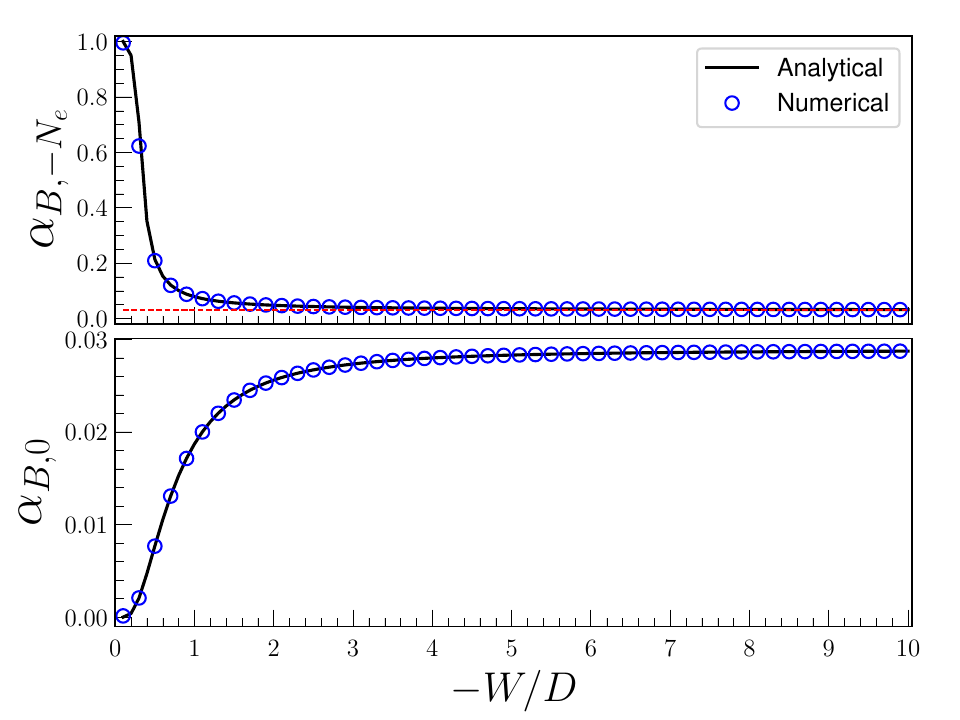}
        \end{tabular}
    \vspace{-0.5cm}
	\caption{\footnotesize  (Left panel) Bound state energy $\epsilon_{B}$ as function of the scattering potential. The open circles display the results of numerical diagonalization for $N=1200$ sites, while the solid line represents the Eq. \eqref{Bond_State_energy}. (Right panel) Commutators $\{\tilde a_0, g_B^\dagger \}$ and $\{\tilde a_{-N_e}, g_B^\dagger \}$  as function of the scattering potential, with same symbols as in (Left panel).  \Sou}
        \label{AR-1}
\end{figure}

Fig. \ref{AR-1} compares Eqs. \eqref{Bond_State_energy} (left panel) and \eqref{Bond_State_proj} (right panel) with numerical values obtained by direct diagonalization of the Hamiltonian \eqref{H_photo}, using $N=1200$ and $0 \le -W/D \le 10$. The coefficients $\alpha_{B,-N/2}$ and $\alpha_{B,0}$ correspond to, the bottom of the band and the first level above the Fermi level, respectively. It is evident from Fig. \ref{AR-1} that the numerical and analytical results are in excellent agreement. As $|W|$ increases, the bound state level $g_B^\dagger$ shifts from being a delocalized level $\tilde a^\dagger_{-N/2}$ to becoming strongly localized at the first real site $a_0^\dagger $.

\newpage 

An accurate approximation for the projection between the initial and final ground state is obtained in the Appendix \ref{AOC proof}, on the basis of the Cauchy determinant formula \cite{cauchy1841exercices}:
\begin{eqnarray}\label{AOC EQ II}
|\braket{\phi_0}{\varphi_0}| \approx \left(\frac{\sin\delta}{\delta} \right)^{-1+\frac{\delta}{\pi} + 1.25\left( \frac{\delta}{\pi} \right)^4 } \left(\frac{\exp\left(2C_{\mathrm{EM}} + \frac{\pi^2}{6}\right) }{600}\right)^{\frac{1}{2}\left(\frac{\delta}{\pi}\right)^2} N_e^{-\frac{1}{2}\left(\frac{\delta}{\pi}\right)^2}.
\end{eqnarray}
where $C_{\mathrm{EM}}$ is the Euler-Mascheroni constant. 

Observe that in Eq. \eqref{AOC EQ II}, the exponent $-\frac{1}{2}\left(\frac{\delta}{\pi}\right)^2$ differs from the result obtained by Anderson in Eq. \eqref{AOC EQ}, but it aligns with more modern and rigorous works on the orthogonality catastrophe \cite{Gebert2014-fu}. If the phase shift were constant for all single-level energies, the exponent would match Anderson's prediction exactly. However, since the phase shift depends on the energy, as shown in Appendix \ref{Appendix_Analytical_Diagonalization}, it acquires this $\frac{1}{2}$ factor.

The bound state is occupied in the final ground state. A class of high energy excited final states of special interest are the single particle-hole excitations from the bound state to levels above the Fermi energy. Simplest among them is the state $\ket{\bar\varphi} = g_{0}^\dagger g_{B} \ket{\varphi_0}$. The overlap $\braket{\phi_0}{\bar\varphi}$ can be determined using a Slater determinant similar to Eq. \eqref{Slater}, with the last column replaced by the coefficients $\alpha_{B,h} \rightarrow \alpha_{p,h}$. Appendix \ref{AOC proof} then finds the following approximate expression for the projections between this and the initial ground state:
\begin{eqnarray}\label{Bound_State_MB_proj}
 {|\bra{\phi_0} g_{0}^\dagger g_{B} \ket{\varphi_0}|} \approx \frac{2}{\pi} \frac{{|\epsilon_B + D|}}{|\epsilon_B|} \left(\frac{\sin\delta}{\delta}\right)^{2\left(1-{\delta}-\left(\frac{\delta}{\pi}\right)^4 \right)} \gamma^{2\frac{\delta}{\pi}}\left(1+\frac{\delta}{\pi}\right) \left( \frac{N_e}{600}\right)^{-\frac{1}{2} + \frac{\delta}{\pi} } |\braket{\phi_0}{\varphi_0}|, ~~
\end{eqnarray} 
where $\gamma(x)$ is the well known Gamma function.

\begin{figure}[hbt!]
		\centering
        \begin{tabular}{ll}
        \hspace{-0.6cm}
        \includegraphics[scale=0.52]{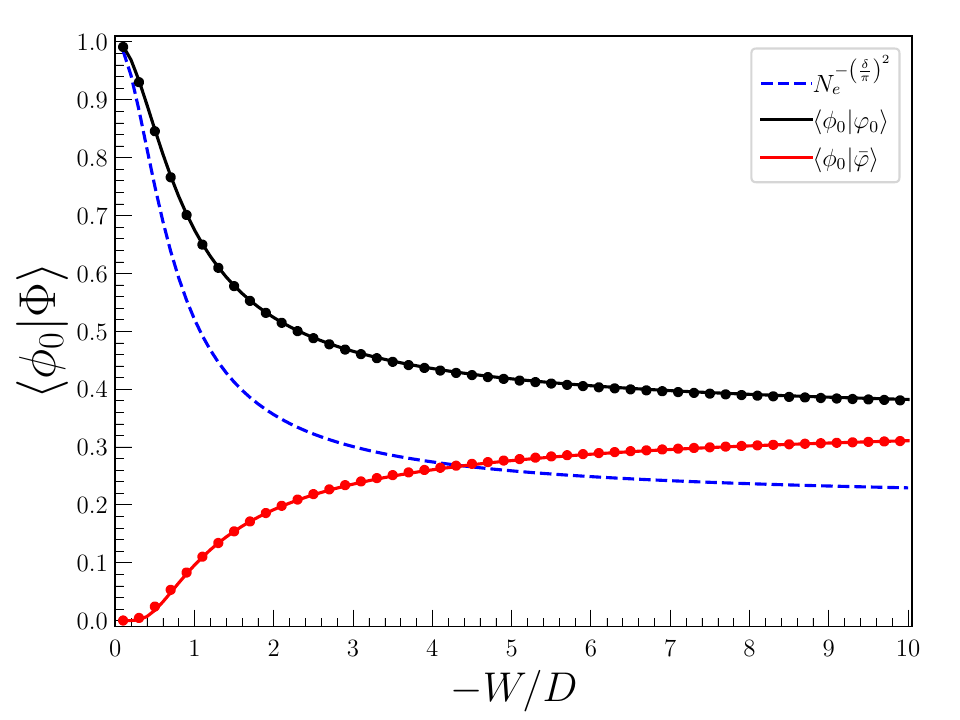}
        & \hspace{-0.8cm}
        \includegraphics[scale=0.52]{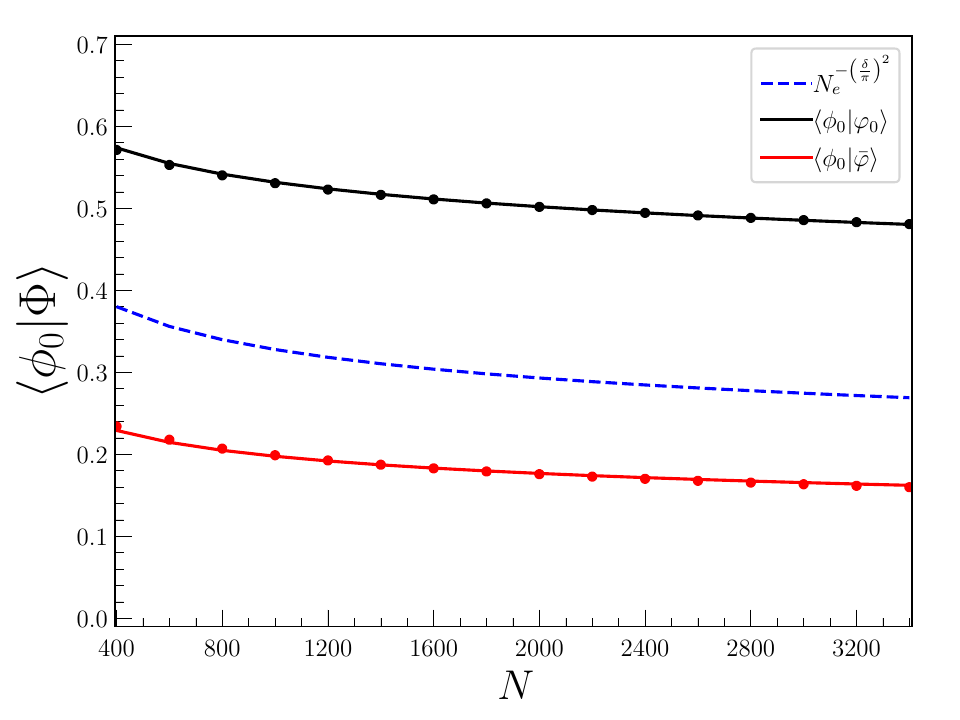}
        \end{tabular}
        \vspace{-0.4cm}
		\caption{\footnotesize (Left panel) Projections as function of $W$ for a fixed $N=1200$. (Right panel) Projections as function of $N$ for a fixed $W/D=-2$. For both panels, the black circles and the black line represent the numerical and analytical results for $\braket{\phi_0}{\varphi_0}$, respectively, while the red circles and the red line represent the numerical and analytical results for $\braket{\phi_0}{\bar\varphi}$, respectively. \Sou}\label{MB Proj.}
\end{figure}

Figure \ref{MB Proj.}, compares Eqs. \eqref{AOC EQ II} and \eqref{Bound_State_MB_proj} with the numerical results for different $W$ (left panel) and $N$ (right panel). The black circles and the solid black line represent the numerical and analytical results for $\braket{\phi_0}{\varphi_0}$, respectively, while the red circles and the solid red line represent the numerical and analytical results for $\braket{\phi_0}{\bar\varphi}$, respectively. Clearly, Eqs. \eqref{AOC EQ II} and \eqref{Bound_State_MB_proj} are good approximations for these projections.
 
In Fig. \ref{MB Proj.}, we observe that exists a highly energetic excited state (with energy $-\epsilon_B$) that has a significant projection onto the initial ground state. This result suggests that these excitations from the bound state level are crucial for accurately describe the photoemission.

\section{Bond state and the Doniach-Sunjic law}

As we already discussed, the behavior of a system described by the Hamiltonian \eqref{H_photo} is well understood for small values of $\delta$, or small $W$, as first explained by Doniach-Sunjic \cite{Doniach_1970}. For long times, the projection $|\langle \phi_0 | \Psi \rangle| $ obeys the Doniach-Sunjic power law \cite{Doniach_1970} $|\langle \phi_0 | \Psi \rangle| \sim t^{-\left(\frac{\delta}{\pi} \right)^2}$. This behavior is a consequence of the generation of a large number of low-energy electron-hole pairs, which occurs after the sudden creation of the scattering potential shakes up the electrons occupying the levels above the Fermi level in the conduction band.


Here, we aim to investigate the influence of these high-energy excitations on the physics of this system. To achieve this, we will initially consider the system in the unperturbed ground state, denoted $\ket{\Psi(0)} = \ket{\phi_0}$. Then, we will suddenly introduce the scattering potential at $t > 0$. The wave function at time $t>0$ is then given by 
\begin{equation}\label{wave_function_photo}
\ket{\Psi(t)} = \sum_{n=0} \braket{\varphi_n}{\phi_0}~e^{-iE_n\frac{t}{\hbar}} \ket{\varphi_n}, 
\end{equation}
or equivalently for all times
\begin{equation}
\braket{\Psi(t)}{\phi_0}= \sum_{n=0} |\braket{\varphi_n}{\phi_0}|^2e^{+iE_n\frac{t}{\hbar}}\Theta(t) + \Theta(-t)
\end{equation}
where $\Theta(t)$ is the Heaviside step function. Now,  defining the correlation function $G(t) = i \braket{\Psi(t)}{\Psi(0)} \Theta(t) $  and applying the Fourier transform, we find that:
\begin{eqnarray}\label{Gw}
G(\omega) = \sum_{n=0} |\braket{\varphi_n}{\phi_0}|^2 \left( i\pi \delta(\omega - \omega_n) + \mathcal{P}\frac{1}{\omega-\omega_n} \right),
\end{eqnarray}
where 
\begin{equation}
    G(\omega) = i \int_{-\infty}^\infty dt ~ e^{-i\omega t} \braket{\Psi(t)}{\phi_0} \Theta(t) ,
\end{equation}
and we used that
\begin{equation}
\int_{-\infty}^\infty dt ~ e^{-i\omega t} \left( e^{+iE_n\frac{t}{\hbar}} \Theta(t) \right) = \pi \delta(\omega - \omega_n) -i \mathcal{P} \frac{1}{\omega-\omega_n}.
\end{equation}
Here, $\omega_n = E_n/\hbar$, $\delta(x)$ is the Dirac delta function and $\mathcal{P}$ represents the Cauchy principal part.

Before computing $G(\omega)$, we will partition the many-body basis to isolate the contributions from the low-energy states to $G(\omega)$ from those arising from high-energy excitations of the bound state level. To accomplish this, we will define the set $\{\ket{\varphi_n'}\}$ as the final many-body states where the bound state level remains occupied (plugged states), and the set $\{\ket{\bar\varphi_n}\}$ as the final many-body states where the bound state is vacant (unplugged states). Clearly, the complete final many-body basis is the union of these two sets
\begin{equation}
  \{\ket{\varphi_n}\} = \{\ket{\varphi_n '}\} \cup \{\ket{\bar\varphi_n}\}.  
\end{equation}

We can now express the plugged states $\ket{\varphi_n'}$ as a product of pairs of particle-hole excitations from the final ground state
\begin{equation}
    \ket{\varphi_n'} = \prod_{j_n} g_{p_{j_n}}^\dagger g_{h_{j_n}} \ket{\varphi_0},
\end{equation}
Here, $\varepsilon_{p_{j_n}}$ and $\varepsilon_{h_{j_n}}$ represent the energies of the particle and hole levels, respectively. The set $\{ p_{j_n} \}$ indicates the levels with energy above the Fermi level that are occupied in the $\ket{\varphi_n'}$ configuration, while the set $\{ h_{j_n} \}$ represents the empty levels below the Fermi level. The index $j_n \in \{1, \dots, \text{number of pairs}\}$ thus arbitrarily organizes the particles $\{ p_{j_n} \}$ and holes $\{ h_{j_n} \}$ into pairs and counts the total number of particle-hole pairs.

For example, in the many-body state $ \ket{\varphi_n'} = g_{2}^\dagger g_{1}^\dagger g_{-1} g_{-2} \ket{\varphi_0} $, one possible choice to represent this state is the pairs $(g_{2}^\dagger, g_{-2})$ ($j_n = 1$) and $(g_{1}^\dagger, g_{-1})$ ($j_n = 2$). In this case, $p_1 = 2$, $h_1 = -2$, $p_2 = 1$, and $h_2 = -1$. A different choice of pairing for this state will only change the sign of the state.

Similarly, we can express the unplugged states $\ket{\bar\varphi_n}$ as a product of pairs of particle-hole excitations from the special state $\ket{\bar\varphi} = g_0^\dagger g_B \ket{\varphi_0}$ as
\begin{equation}
\ket{\bar\varphi_n} = \prod_{j_n} g_{p_{j_n}}^\dagger g_{h_{j_n}} \ket{\bar\varphi}.
\end{equation}

The energy of the plugged state $\ket{\varphi_n'}$ is 
\begin{equation}
    E_n' = \sum_{j_n} (\varepsilon_{p_{j_n}} - \varepsilon_{h_{j_n}}),
\end{equation}
and the energy of unplugged state $\ket{\bar\varphi_n}$ is 
\begin{equation}
    \bar E_n = -\epsilon_B + \sum_{j_n} (\varepsilon_{p_{j_n}} - \varepsilon_{h_{j_n}}),
\end{equation} 
as there is already one particle-hole excitation from the bound level.

\newpage

\begin{figure}[htb!]
		\centering
		\includegraphics[scale=0.55]{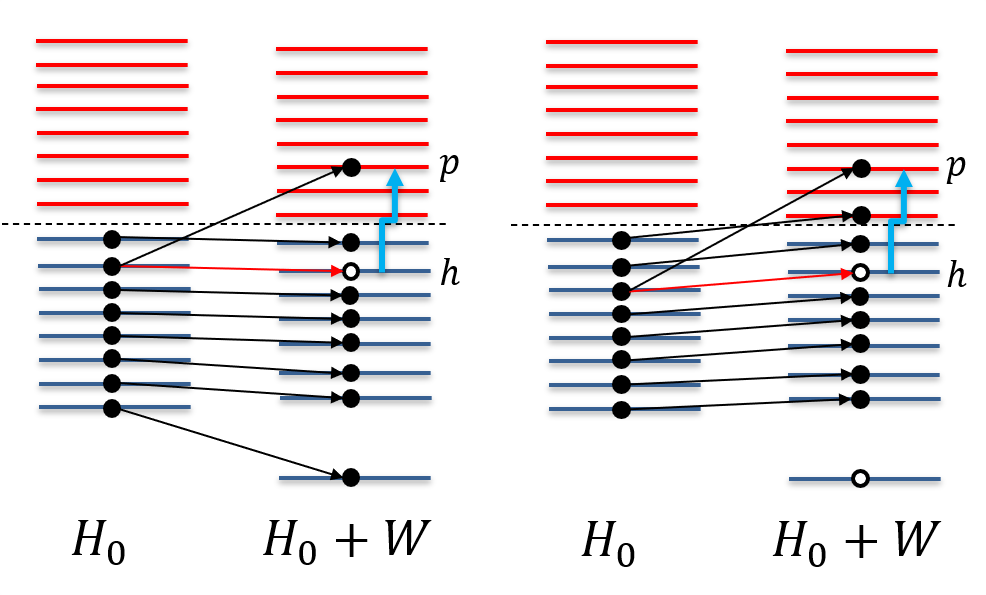}
        \vspace{-0.4cm}
    \caption{\footnotesize (Left panel) Schematic representation of a one-particle-hole excitation from $\ket{\varphi_0}$ where the bound state level remains occupied. (Right panel) Schematic representation of a one-particle-hole excitation from $\ket{\bar{\varphi}}$. The black arrows indicate the diagonal overlap elements of the Slater determinant, following Eq. \eqref{Slater}. The blue line represents the one-particle-hole excitation, and the red arrow indicates the diagonal element before the excitation. \Sou}
	   \label{Change_levels_energy}
\end{figure}

From Eq. \eqref{wave_function_photo}, the probability of finding the system in the final state $\ket{\varphi_n}$ is given by $ |\braket{\varphi_n}{\Psi_t}|^2 =  |\braket{\varphi_n}{\phi_0}|^2 $, which can be determined using the Slater determinant. This computation is necessary to find both the fidelity and the Green function $ G(\omega)$. Computing this probability for all many-body states, however, would require a very large number of Slater determinants, which is impractical.

One way to avoid this is by approximating the value of the Slater determinant, as schematically shown in Fig. \ref{Change_levels_energy} (Left panel). Starting with the projection $\braket{\phi_0}{\varphi_0}$ in Eq. \eqref{Slater}, the diagonal elements of the determinant are $\{\tilde a_k, g_k^\dagger \} = \frac{\sin \delta}{\delta}$, except for the bound state column. Now, to compute the projection $\bra{\phi_0}g_p^\dagger g_h \ket{\varphi_0}$, where the excited state is composed of one particle-hole excitation from the final ground state, we need to replace in Eq. \eqref{Slater} the elements of $\{\tilde a_k, g_h^\dagger \}$ with $\{\tilde a_k, g_p^\dagger\}$. Considering that the diagonal terms $ \alpha_{h,h} \rightarrow \alpha_{p,h} $ are more relevant to the determinant, this leads to the left case as:
\begin{eqnarray}
{\braket{\phi_0}{\varphi_{p,h}'}} \approx \left(\frac{\delta}{\pi} \right) \frac{\Delta\varepsilon}{\varepsilon_p - \varepsilon_h - \frac{\delta}{\pi} \Delta\varepsilon}  {\braket{\phi_0}{\varphi_0}}.
\end{eqnarray}

The projection $\braket{\phi_0}{\tilde \varphi}$ can be obtained similarly to $\braket{\phi_0}{\varphi_0}$. However, before changing the elements $\{\tilde a_k, g_B^\dagger \}$ to $\{\tilde a_k, g_0^\dagger \}$, we can translate the bound state column to the first column of the matrix, resulting in
\begin{eqnarray}\label{Slater_II}
\braket{\phi_0}{\tilde \varphi} = \mathrm{det}
\begin{pmatrix}
\{\tilde a_{-1},g_{0}^\dagger\} & \{\tilde a_{-1},g_{-1}^\dagger\} & ... & \{\tilde a_{-1},g_{-(N_e-1)}^\dagger\} \\
\{\tilde a_{-2},g_{0}^\dagger\} & \{\tilde a_{-2},g_{-1}^\dagger\} & ... & \{\tilde a_{-2},g_{-(N_e-1)}^\dagger\} \\
\{\tilde a_{-3},g_{0}^\dagger\} & \{\tilde a_{-3},g_{-1}^\dagger\} & ... & \{\tilde a_{-3},g_{-(N_e-1)}^\dagger\} \\
\vdots & ... & \vdots \\
\{\tilde a_{-N_e},g_{0}^\dagger\} & \{\tilde a_{-N_e},g_{-1}^\dagger\} & ... & \{\tilde a_{-N_e},g_{-(N_e-1)}^\dagger\} \\
\end{pmatrix}, 
\end{eqnarray}
as schematically shown in Fig. \ref{Change_levels_energy} (Right panel). Thus, in the case in which the bound state is empty, the diagonal terms are $\alpha_{k-1,k}$. After replacing $\alpha_{h-1,h} \rightarrow \alpha_{p,h}$, the determinant $\bra{\phi_0} g_p^\dagger g_h \ket{\tilde \varphi}$ can be approximated as
\begin{eqnarray}
{\braket{\phi_0}{\bar\varphi_{p,h}}} \approx \left[ \left(\frac{\pi-\delta}{\pi} \right) \frac{\Delta\varepsilon}{\varepsilon_p - \varepsilon_h - \frac{\pi-\delta}{\pi} \Delta\varepsilon} \right] {\braket{\phi_0}{\bar\varphi}}.
\end{eqnarray}

Extending these ideas to any number of particle-hole excitation, by using this approximation that elements of the Slater determinant $\bra{\phi_0}\prod_{j_n} g_{p_{j_n}}^\dagger g_{h_{j_n}} \ket{\varphi_0}$ on the diagonal are dominant, using the Eq. \eqref{Band_proj} we can write
\begin{eqnarray}\label{Proj_II}
{\braket{\phi_0}{\varphi_n'}} \approx \left[\prod_{j_n} \left(\frac{\delta}{\pi} \right) \frac{\Delta\varepsilon}{\varepsilon_{p_{j_n}} - \varepsilon_{h_{j_n}} - \frac{\delta}{\pi} \Delta\varepsilon} \right] {\braket{\phi_0}{\varphi_0}},
\end{eqnarray}
\begin{eqnarray}\label{Proj_III}
{\braket{\phi_0}{\bar\varphi_n}} \approx \left[\prod_{j_n} \left(\frac{\pi - \delta}{\pi} \right) \frac{\Delta\varepsilon}{\varepsilon_{p_{j_n}} - \varepsilon_{h_{j_n}} - \frac{\pi - \delta}{\pi} \Delta\varepsilon} \right] {\braket{\phi_0}{\bar\varphi}}.
\end{eqnarray}
Here, $\varepsilon_{p_{j_n}}$ and $\varepsilon_{h_{j_n}}$ represent the energies of the particle and hole level, respectively, where this parameter ${j_n}$ depends on the many-body state $\ket{\varphi_n}$ and represents the pairs of particle-hole excitations which compose this state.

Now, let us focus on the imaginary part of $G(\omega)$, as this part is proportional to the XPS experimental spectra. After partitioning the energy spectrum into plugged states ${\ket{\varphi_n'}}$ and unplugged states ${\ket{\bar\varphi_n}}$, we can rewrite Eq. \eqref{Gw} as:
\begin{eqnarray}\label{GwII}
\mathrm{Im}G(\omega)=\pi\sum_{n=0} |\braket{\varphi_n'}{\phi_0}|^2 \delta(\omega - \omega_n') +  \pi \sum_{n=0} |\braket{\bar\varphi_n}{\phi_0}|^2 \delta(\omega - \omega_B - \omega_n'),
\end{eqnarray}
where $\hbar\omega_n' = \sum_{{j_n}} (\varepsilon_{p_{j_n}} - \varepsilon_{h_{j_n}})$ represents the energy of the plugged many-body excitations from the ground state, and $\hbar \omega_B = |\epsilon_B|$ represents the energy of the bound level. Then, we can use the Eqs. \eqref{Proj_II} and \eqref{Proj_III} to rewrite Eq. \eqref{GwII} as
\begin{eqnarray}\label{ImGw}
\mathrm{Im}G(\hbar\omega) &=& \pi |\braket{\varphi_0}{\phi_0}|^2 \sum_{n=0} \Omega_n^2(\delta) \delta\left(\hbar\omega - \sum_{{j_n}} (\varepsilon_{p_{j_n}} - \varepsilon_{h_{j_n}}) \right) \nonumber \\
&+&  \pi |\braket{\bar\varphi}{\phi_0}|^2 \sum_{n=0} \Omega_n^2(\pi-\delta) \delta\left(\hbar\omega - |\epsilon_B| - \sum_{{j_n}} (\varepsilon_{p_{j_n}} - \varepsilon_{h_{j_n}}) \right).
\end{eqnarray}
Here, the product
\begin{equation}
    \Omega_n^2 (\delta) = \prod_{j_n} \left(\frac{\delta}{\pi} \right)^2 \frac{{\Delta\varepsilon}^2}{\left(\varepsilon_{p_{j_n}} - \varepsilon_{h_{j_n}} - \frac{\delta}{\pi} \Delta\varepsilon\right)^2}
\end{equation}
was obtained by Eqs. \eqref{Proj_II} and \eqref{Proj_III}, and the projections $|\braket{\varphi_0}{\phi_0}|^2$ and $|\braket{\bar\varphi}{\phi_0}|^2$, from Eqs. \eqref{AOC EQ II} and \eqref{Bound_State_MB_proj}.

The imaginary part of the Green's function in Eq. \eqref{ImGw} can be divided into two parts: $\mathrm{Im}G(\hbar\omega) = \mathrm{Im}G^{\mathrm{ND}}(\hbar\omega) + \mathrm{Im}G^{\mathrm{BS}}(\hbar\omega)$, where the first one refers to plugged states, and the second one to unplugged states. The first part, 
\begin{equation}\label{AUX_NZ_I}
    \mathrm{Im}G^{\mathrm{ND}}(\hbar\omega)= \pi |\braket{\varphi_0}{\phi_0}|^2 \sum_{n=0} \Omega_n^2(\delta) \delta\left(\hbar\omega - \sum_{{j_n}} (\varepsilon_{p_{j_n}} - \varepsilon_{h_{j_n}}) \right) \propto \left( {\hbar\omega}\right)^{-1+ \left(\frac{\delta}{\pi}\right)^2} ,
\end{equation}
gives rise to the Doniach-Sunjic power law \cite{Doniach_1970} with the characterized decay $\omega^{-1+\left(\frac{\delta}{\pi}\right)^2}$ in the frequency domain, and  $t^{-\left(\frac{\delta}{\pi}\right)^2}$ in the time domain. A detailed proof of $\mathrm{Im}G^{\mathrm{ND}}(\hbar\omega)$ is given in Appendix \ref{DN law} with the pertinent prefactor.

Clearly, if the phase shift ${\delta}$ or $|W|$ is small, then $|\braket{\bar\varphi}{\phi_0}|^2 \ll |\braket{\varphi_0}{\phi_0}|^2$ as shown in the Fig. \ref{MB Proj.}, and the first part of the $\mathrm{Im}G(\hbar\omega)$ is dominant. If $|W|$ is strong, however, excitations from the bound state become important, and both terms are comparable. Luckily, it is not difficult to show that 
\begin{eqnarray}\label{AUX_NZ_II}
\mathrm{Im}G^{\mathrm{BS}}(\hbar\omega >|\epsilon_B| ) = \frac{|\braket{\bar\varphi}{\phi_0}|^2}{|\braket{\varphi_0}{\phi_0}|^2} \left[ \mathrm{Im}G^{\mathrm{ND}}(\hbar\omega - |\epsilon_B|) \right]_{\delta \rightarrow (\pi - \delta)} \propto (\hbar\omega- |\epsilon_B|)^{-2\frac{\delta}{\pi}+ \left(\frac{\delta}{\pi}\right)^2},~
\end{eqnarray}
with the Nozières and Dominicis \cite{Nozieres} power law for the x-ray absorption problem.

For strong scattering potentials, when a bound state is present, Combescot and Nozières \cite{Nozieres71} demonstrated, in the context of x-ray emission and absorption spectra, the exactly same decays described by Eqs. \eqref{AUX_NZ_I} and \eqref{AUX_NZ_II}. Additionally, Ohtaka and Tanabe \cite{ohtaka1990} showed that this extra peak also appears in x-ray photoemission spectra with the same decays. In the following we will explore this problem even further and more generally using the approximations in Eq. \eqref{Proj_II} and \eqref{Proj_III}. With these calculations, we not only reproduce these decays but also provide formulas for each contribution for any scattering potential amplitude, determine the position of the extra peak, and discuss the behavior of the phenomenon in the time domain, without being restricted to the long-time behavior or particular small/large amplitudes of the scattering potential.

Using our calculations shown in Appendix \ref{DN law}, we can write $\mathrm{Im}G(\hbar\omega) $ as
\begin{eqnarray}\label{ImGw_}
  \mathrm{Im}G(\hbar\omega) &\approx& \pi |\braket{\varphi_0}{\phi_0}|^2 \delta(\hbar\omega) + \mathcal{K}(\delta) \left( {\hbar\omega}\right)^{-1+\left(\frac{\delta}{\pi}\right)^2}\Theta(\hbar\omega) + \pi |\braket{\bar\varphi}{\phi_0}|^2 \delta(\hbar\omega - |\epsilon_B|) \nonumber \\
 &+& r\mathcal{K}(\pi-\delta) (\hbar\omega- |\epsilon_B|)^{-2\frac{\delta}{\pi}+\left(\frac{\delta}{\pi}\right)^2}\Theta(\hbar\omega- |\epsilon_B|),
\end{eqnarray}
where 
\begin{equation}
 \mathcal{K}(\delta)= \pi \left(\frac{\delta}{\pi}\right)^2 |\braket{\varphi_0}{\phi_0}|^2 (\Delta\varepsilon)^{-\left(\frac{\delta}{\pi}\right)^2}
\end{equation}
and the ratio 
\begin{equation}\label{Relative_Contribution_BS}
 r(\delta) \equiv \frac{|\braket{\bar\varphi}{\phi_0}|^2}{|\braket{\varphi_0}{\phi_0}|^2}.
\end{equation}
Note that this ratio $r(\delta)$ can be obtained analytically using our expressions for each projection in the Eqs. \eqref{AOC EQ II} and \eqref{Bound_State_MB_proj}, and it depends only on the phase shift and the number of electrons.

\newpage
The excitations from the bound state resemble the Nozières-Dominicis x-ray absorption spectra \cite{Nozieres}. Physically, when an electron is promoted from the bound state to the conduction band, the process is very similar to the x-ray absorption problem, the levels near the Fermi energy have one extra electron, resulting in the same exponent. In the limit $\delta \rightarrow \frac{\pi}{2}$, the Doniach-Sunjic and Nozières-Dominicis exponents approach.

Our calculation measures the energies from that of the final ground state. The difference between the initial and final ground states energies is given by Eq. \eqref{Binding_energy_shift}, which yields $E_0 - \tilde E_0  = - D\left(\coth\left( \frac{-1}{2\rho W}\right)- 1 + \frac{\delta}{\pi} \right)$. Since $E_0 - \tilde E_0 < 0$, we can see that the core-hole potential shifts all energies in the photoemission to lower values.

\begin{figure}[htb!]
		\centering
		\includegraphics[scale=0.52]{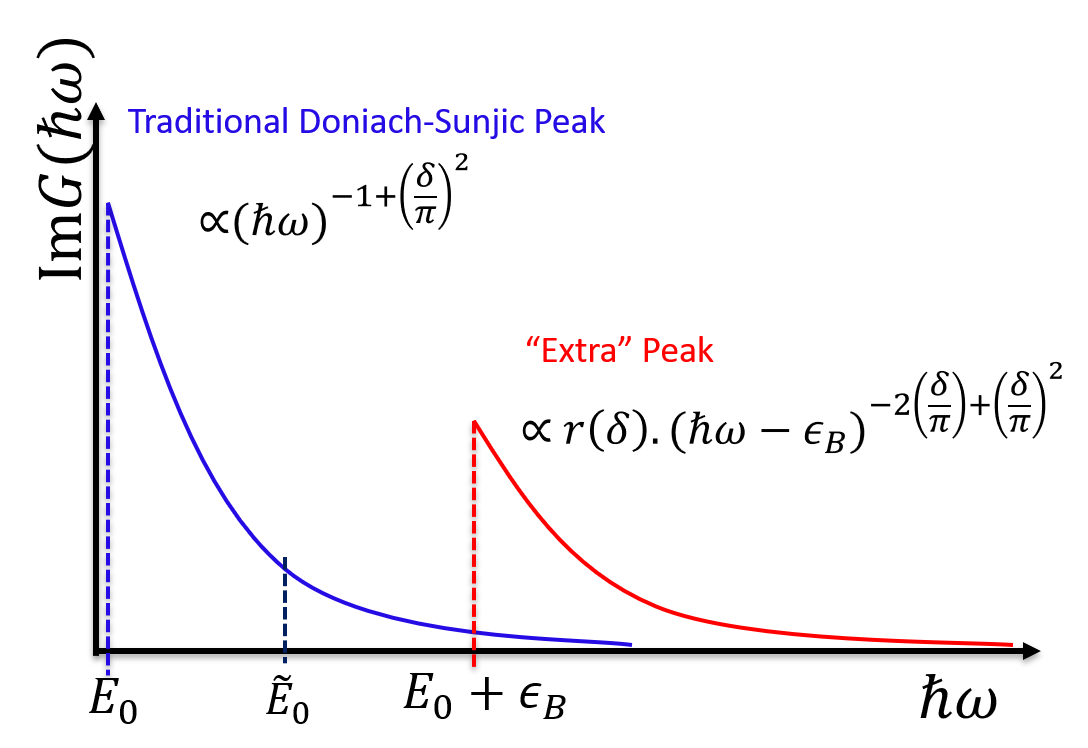}
        \vspace{-0.3cm}
	    \caption{\footnotesize Schematic representation of $\mathrm{Im}G(\hbar\omega)$. The 
            divergence at $E_0$ represents the well-known Doniach-Sunjic peak. If the scattering potential is sufficiently strong, so that the ratio $r(\delta)$ becomes significant, then a second peak emerges associated with the unplugged states. The energy $\tilde E_0$ represents the initial ground state energy. \Sou }
	   \label{Doniach_Sunjic_Peaks}
\end{figure}


Figure \ref{Doniach_Sunjic_Peaks} represents  Eq. \eqref{ImGw_} schematically. The peak at $E_0$ represents the Doniach-Sunjic peak. If $|W|/D  \sim 1$ or bigger, the ratio $r(\delta) = \frac{|\braket{\bar\varphi}{\phi_0}|^2}{|\braket{\varphi_0}{\phi_0}|^2}$ becomes significant (see Fig. \ref{MB Proj.}), a second peak emerges, due to particle-hole excitations from the bound state level. Satellite peaks observed in XPS experiments \cite{PAULY2016317,D1CP04886D,GROSVENOR20061771,https://doi.org/10.1002/sia.740230705}, ones that cannot be explained as Auger excitations \cite{https://doi.org/10.1002/sia.740230705,Slaughter_Weber_Güntherodt_Falco_1992} and lie above the threshold photoemission energy, are likely to be due to particle-hole excitations from the bound states. 


A time-dependent signature of such excitations from the bound state are oscillations with frequency $|\epsilon_B|/\hbar$. To show this we have computed the Fidelity, $\mathcal{F}(t) = |\langle \phi_0 | \Psi(t) \rangle|^2$:
\begin{eqnarray}\label{Fidelity_time}
\mathcal{F}(t) \approx \frac{\left(1 + {\mathcal{R}(t)}^2 + 2 \mathcal{R}(t) \cos\left(\frac{\epsilon_B t}{\hbar}\right) \right)}{(1+r)^2} e^{ -2\left(\frac{\delta}{\pi}\right)^2 \left(\ln \left(\frac{D}{\Delta\varepsilon}\right) + \mathrm{CI}\left(\frac{\Delta\varepsilon  t}{\hbar} \right) - \mathrm{CI}\left( \frac{D t}{\hbar} \right) \right)}, 
\end{eqnarray}
with 
\begin{eqnarray}\label{TDR}
\mathcal{R}(t) = r \times e^{ -\left(1-2\left(\frac{\delta}{\pi}\right) \right).\left(\ln \left(\frac{D}{\Delta\varepsilon}\right) + \mathrm{CI}\left(\frac{\Delta\varepsilon  t}{\hbar} \right) - \mathrm{CI}\left( \frac{D t}{\hbar} \right) \right)}, 
\end{eqnarray}
Here, $\mathrm{CI}(x)$ is the cosine integral function. The derivation of this expression is presented in Appendix \ref{DN law}, following the same strategy and approximations used in deriving $\mathrm{Im}G(\hbar\omega)$.

In Eq. \eqref{Fidelity_time}, there are three contributions. The first term on the right-hand side represents the contribution only from the plugged states, given by
\begin{equation}
    \mathcal{F}_p(t) = \frac{1}{(1+r)^2} e^{ -2\left(\frac{\delta}{\pi}\right)^2 \left(\ln \left(\frac{D}{\Delta\varepsilon}\right) + \mathrm{CI}\left(\frac{\Delta\varepsilon  t}{\hbar} \right) - \mathrm{CI}\left( \frac{D t}{\hbar} \right) \right)}.
\end{equation}
At long times
\begin{equation}
    \mathcal{F}_p(t) \approx  \frac{1}{(1+r)^2} t^{-2 \left(\frac{\delta}{\pi}\right)^2},
\end{equation}
the Doniach-Sunjic power law in the time domain, as expected. Here, we have used that
\begin{equation}
   \left[ e^{-\left(\ln \left(\frac{D}{\Delta\varepsilon}\right) + \mathrm{CI}\left(\frac{\Delta\varepsilon  t}{\hbar} \right) - \mathrm{CI}\left( \frac{D t}{\hbar} \right) \right)}\right]_{t \gg 1} \approx t^{-1}.
\end{equation}

The second term on the right-hand side of Eq. \eqref{Fidelity_time} represents the contribution only from the unplugged states, given by
\begin{equation}
    \mathcal{F}_u(t) \approx \frac{r^2}{(1+r)^2}e^{ -2\left(1 -\frac{\delta}{\pi}\right)^2 \left(\ln \left(\frac{D}{\Delta\varepsilon}\right) + \mathrm{CI}\left(\frac{\Delta\varepsilon  t}{\hbar} \right) - \mathrm{CI}\left( \frac{D t}{\hbar} \right) \right)}. 
\end{equation}
Here, we used Eq. \eqref{TDR} to explicitly shows the decay of this contribution over time. At long times, the expression approaches
\begin{equation}
    \mathcal{F}_u(t) \approx  \frac{r^2}{(1+r)^2} t^{-2 \left(1 - \frac{\delta}{\pi}\right)^2},
\end{equation}
which is the Nozières-Dominicis power law. 

At last, the third term on the right-hand side of Eq. \eqref{Fidelity_time} is
\begin{equation}
    \mathcal{F}_{pu}(t) \approx \frac{2r \cos\left(\frac{\epsilon_B t}{\hbar}\right) }{(1+r)^2}   e^{- \left[ \left(1 -\frac{\delta}{\pi}\right)^2 + \left(\frac{\delta}{\pi}\right)^2 \right]  \left(\ln \left(\frac{D}{\Delta\varepsilon}\right) + \mathrm{CI}\left(\frac{\Delta\varepsilon  t}{\hbar} \right) - \mathrm{CI}\left( \frac{D t}{\hbar} \right) \right)},
\end{equation}
and combines the contribution of plugged and unplugged states for the fidelity. As expected, Eq. \eqref{Fidelity_time} shows oscillations with frequency $|\epsilon_B|/\hbar$ and amplitude proportional to $r$, but the amplitude slowly decays with time. These oscillations arise from the phase difference between the plugged and unplugged states.

\begin{figure}[htb!]
        \centering
        \includegraphics[scale=0.88]{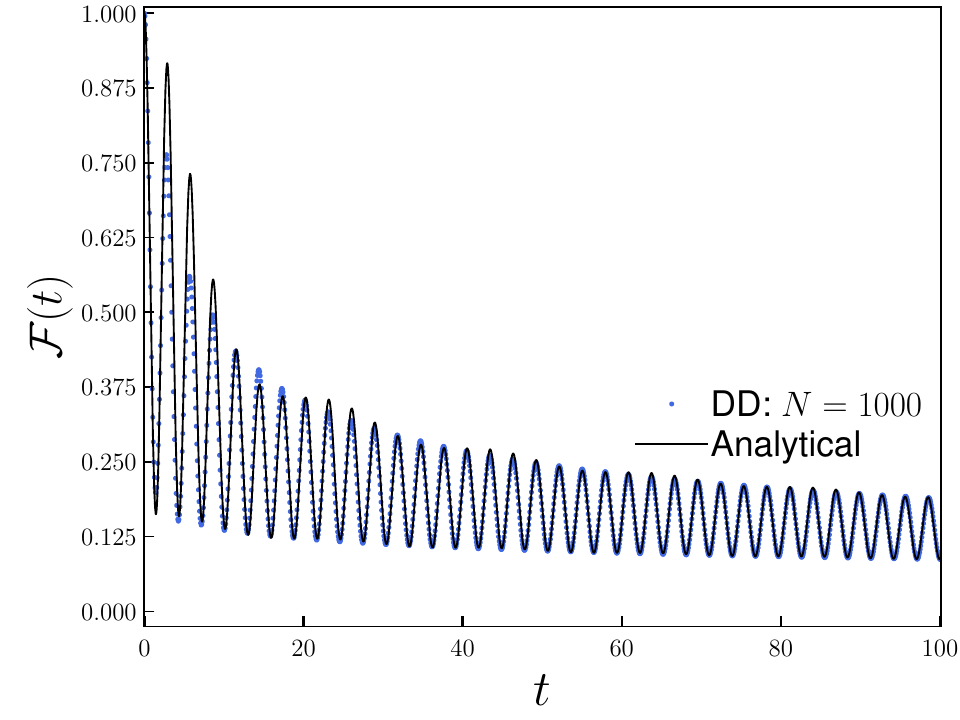}
        \includegraphics[scale=0.88]{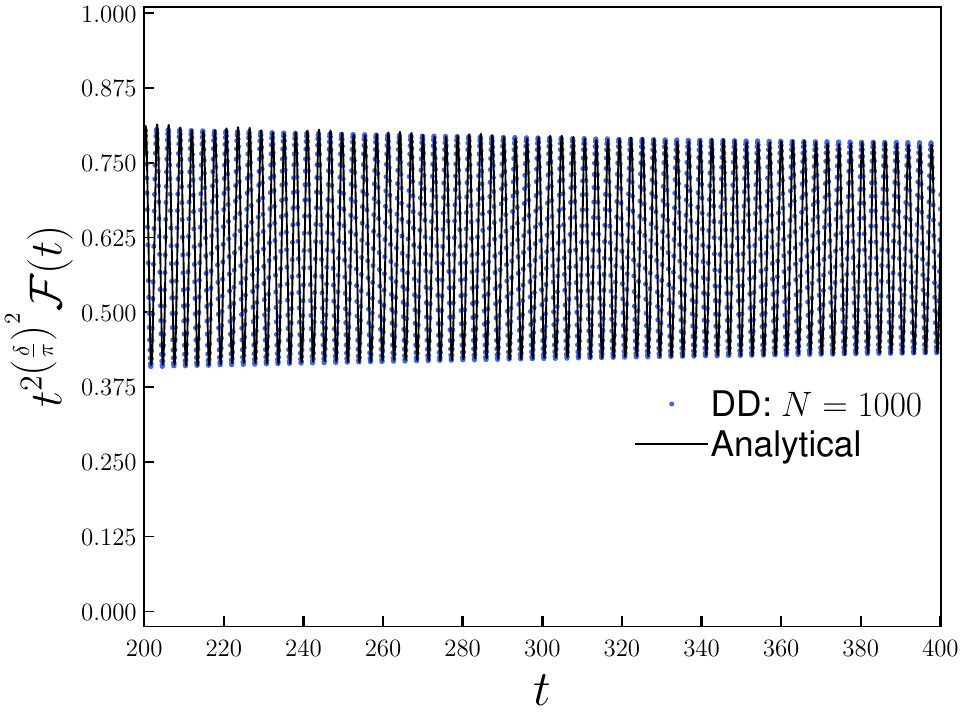}
        \vspace{-0.55cm}
	    \caption{\footnotesize  (Top plot)  Fidelity as a function of time. (Bottom plot) Long time behavior of the fidelity multiplied by the Doniach-Sunjic factor.  Here, we compare the analytical results (solid black line) from Eq. \eqref{Fidelity_time} with results from direct diagonalization (blue circular dots) using $W=-2D$ and $N=1000$. \Sou}
        \vspace{-0.5cm}
	   \label{Doniach_Sunjic_time}
\end{figure}

To derive the Eqs. \eqref{Fidelity_time} and \eqref{ImGw_}, we have used approximations, such as the energy independence of the phase shift and the diagonal dominance in the Slater determinant. These approximations are reliable near the Fermi levels, but not so very well for the high energy levels. Notwithstanding these approximations, the average value of the right-hand side of Eq. \eqref{Fidelity_time} approaches the Doniach-Sunjic power law, for long times. Moreover, it reproduces the results of brute-force numerical diagonalization of the Hamiltonian semi quantitatively at short times and very well at long times - Fig. \ref{Doniach_Sunjic_time} offers an illustration.

~
~

~
\newpage

\begin{figure}[htb!]
        \centering
        \includegraphics[scale=0.6]{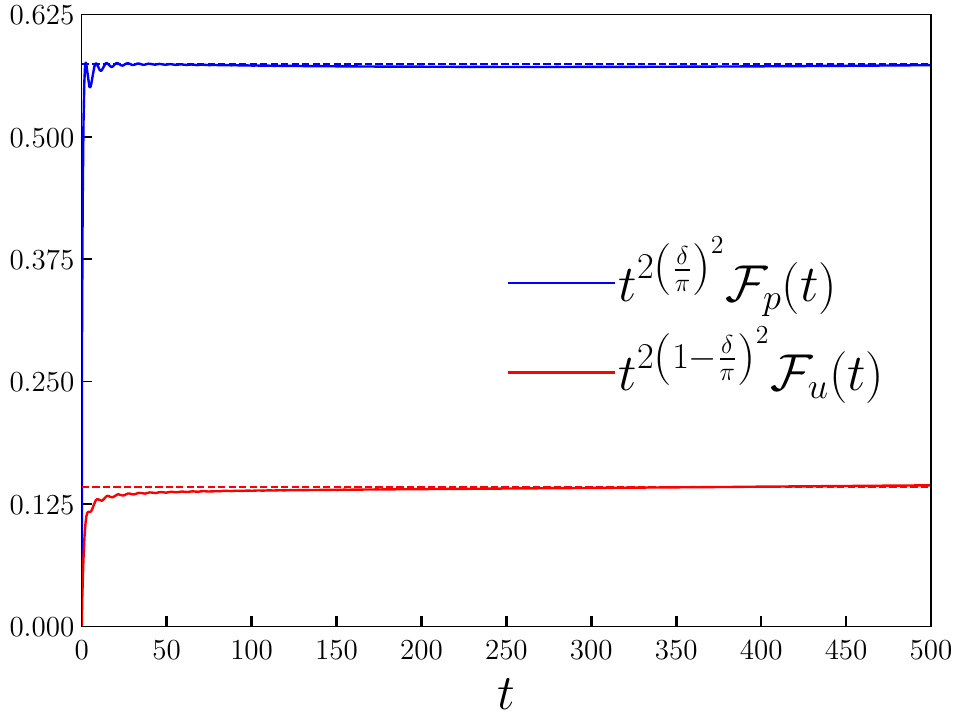}
        \vspace{-0.5cm}
	    \caption{\footnotesize  Plugged and unplugged contributions for the fidelity as function of time. The results was obtained from direct diagonalization using $\frac{W}{D}=-2$ and $N=1000$. \Sou}
	   \label{Doniach_Sunjic_time_2}
\end{figure}

In summary, the Doniach-Sunjic power law emerges from the contributions of the plugged states, as the sudden introduction of the potential creates many particle-hole excitations around the Fermi level. However, particle-hole excitations from the bound state (unplugged states) are also significant in describing the phenomenon. Their contributions introduce a secondary behavior that follows the Nozières-de Dominicis power law. These two contributions to the fidelity are illustrated in Fig. \ref{Doniach_Sunjic_time_2}. Since the contributions from the unplugged states carry a phase associated with the bound state energy, the phase difference with the plugged states induces oscillations with a frequency of $ |\epsilon_B|/\hbar $.

\section{Improvement of the photoemission calculations}

In the last section, we attempted to extract analytical solutions from the Hamiltonian \eqref{H_photo_MI}. To do so, we had to relay on several approximations, such as a flat band for the tight-binding, using an energy-independent phase shift, and considering the dominant terms in the Slater determinant only. Now we wish to improve the results by treating the Hamiltonian
\begin{equation}\label{H_photo_TTB_}
\begin{aligned}
\mathcal{H} = -\tau \sum_{n=0}^{N-1} \left(a_{n}^{\dagger}a_{n+1}+\mathrm{h.c.}\right) + W a_0^\dagger a_0,
\end{aligned}
\end{equation}
more accurately. Where $N$ is the number of lattice sites.

We start out with a discussion of the time and energy scales. Measured from the Fermi level $\epsilon_F$, the single-particle eigenvalues of the unperturbed tight-binding Hamiltonian have the form  $\varepsilon_k = 2\tau \sin\left(\frac{\pi}{N} k\right)$. Near the Fermi energy, the energy differences between successive levels are given by
\begin{equation}
    |\Delta\varepsilon_k| = \frac{2 \tau\pi}{N} \cos\left(\frac{\pi k}{N}\right) + \mathcal{O} \left( N^{-3} \right) \approx \frac{2\pi \tau}{N},
\end{equation}
and hence vanish as $N \rightarrow \infty$. For a half-filled band, the number of levels below the Fermi level is $N_e = \frac{N}{2} = \frac{\pi\tau}{\Delta\varepsilon}$.

The improvement we will discuss requires a numerical diagonalization of $\mathcal{H}$. We will have to deal with finite lattices, and, consequently, with finite gaps. If one observes the system for a measurement time $T_m$, according to Heisenberg's uncertainty principle only energy differences above $\frac{\hbar}{T_m}$ will be discernible \cite{Cui2014, PhysRevX.9.011044, PhysRevB.59.10935, PhysRevLett.109.087401}. For this reason, the behavior of the finite system with a gap
\begin{equation}
    \Delta\varepsilon = \frac{\hbar}{T_m}
\end{equation}
is similar to that of a gapless system up to the time scale of $T_m$. This condition tells us that the energy gap $\Delta\varepsilon$ of $\mathcal{H}$ determines the time scale $\frac{\hbar}{\Delta\varepsilon}$ up to when the behavior of the system can faithfully simulate the behavior of a metallic band. For  $t > \frac{\hbar}{\Delta\varepsilon}$, the finite size effects become too important and this approximation fails.

As a consequence, as time grows a bigger lattice, or a smaller gap $\Delta \varepsilon$, becomes necessary to accurately represent the metallic band:
\begin{equation}
    N \ge  2\pi \frac{\tau . T_m}{\hbar}.
\end{equation}
Clearly, this task becomes impractical for a large chain because the many-body basis grows exponentially with the chain size $N$. To address this problem, we have implemented the Real-Space Numerical Renormalization Group method (eNRG), as explained in the following section.

\section{Renormalization-group approach}\label{eNRG method}

\begin{figure}[htb!]
\centering
\includegraphics[scale=0.42]{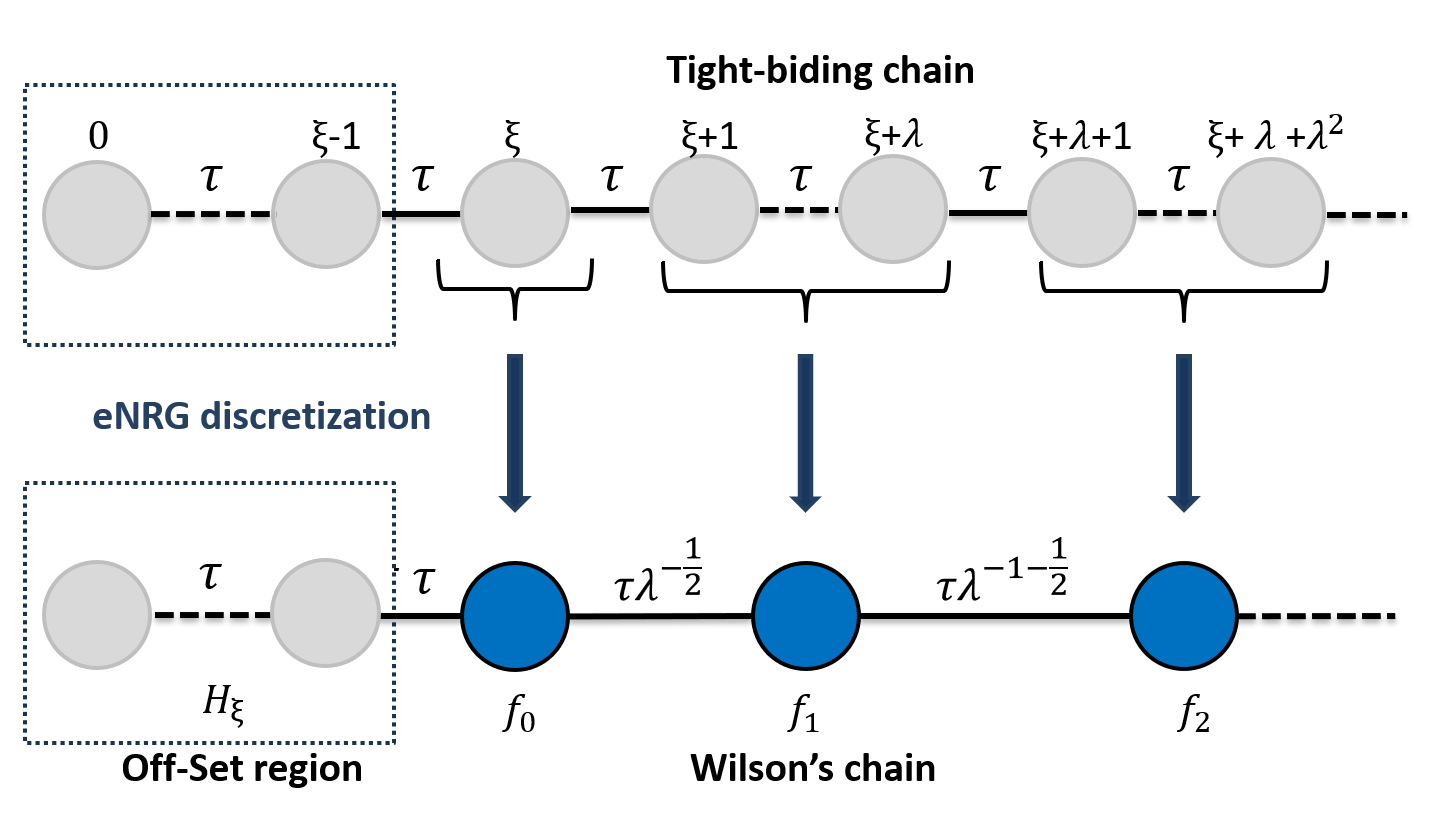}
\vspace{-0.1cm}
\caption{\footnotesize Schematic representation of the eNRG method. The procedure initiates with a 1D tight-binding chain (gray circles). Then, it involves grouping $\lambda^n$ sites $(n=0,1,2...)$, where $\lambda>1$, into a single operator $f_n$ (blue circles), and the new hopping parameter becomes $\tau_n=\tau\lambda^{-n-1/2}$. A second discretization parameter $\xi$ (offset) is introduced, where the initial $\xi$ sites are treated individually. \Sou}
\label{eNRG}
\end{figure}

The eNRG procedure \cite{PhysRevB.106.075129} is a new technique based on the traditional NRG \cite{RevModPhys.47.773} procedure, but it is more flexible. The procedure start with a tight-binding description of the conduction band, as in Eq. \eqref{TBM}, then it groups $\lambda^n$ tight-binding sites operator into a single operator $f_n$, where $\lambda > 1$ is a dimensionless discretization parameter. The $\xi \ge 0$ (offset) first sites are preserved (see Fig.\ref{eNRG}). After this procedure, we can write the metallic band as:
\begin{equation}\label{H_ENRG}
\begin{aligned}
{H}_{\tilde N}^{\mathrm{eNRG}} = \sum_{n=0}^{\xi-1} \tau \left( a_n^{\dagger}a_{n+1}+\mathrm{h.c.}\right) + \sum_{n=0}^{\tilde N-1}\tau_n\left( f_n^{\dagger}f_{n+1}+\mathrm{h.c.}\right).
\end{aligned}
\end{equation}
The operators $f_n^{\dagger}$ represent the eNRG sites, with the condition $f_0^\dagger = a_\xi^\dagger$  on the eNRG basis, the hopping $\tau_n=\tau\lambda^{-n-1/2}$ decrease exponentially with $n$, and $\tilde N$ is the Wilson chain size, here defined by the criterion that the smallest energy gap be minute compared to the measurement time $T_m$, that is, $\Delta\varepsilon T_m \ll \hbar$, where 
\begin{eqnarray}
    \Delta\varepsilon \approx 4\tau\lambda^{-(\tilde N - 0.5)}.
\end{eqnarray}

\newpage

\begin{figure}[hbt!]
		\centering
        \includegraphics[scale=0.42]{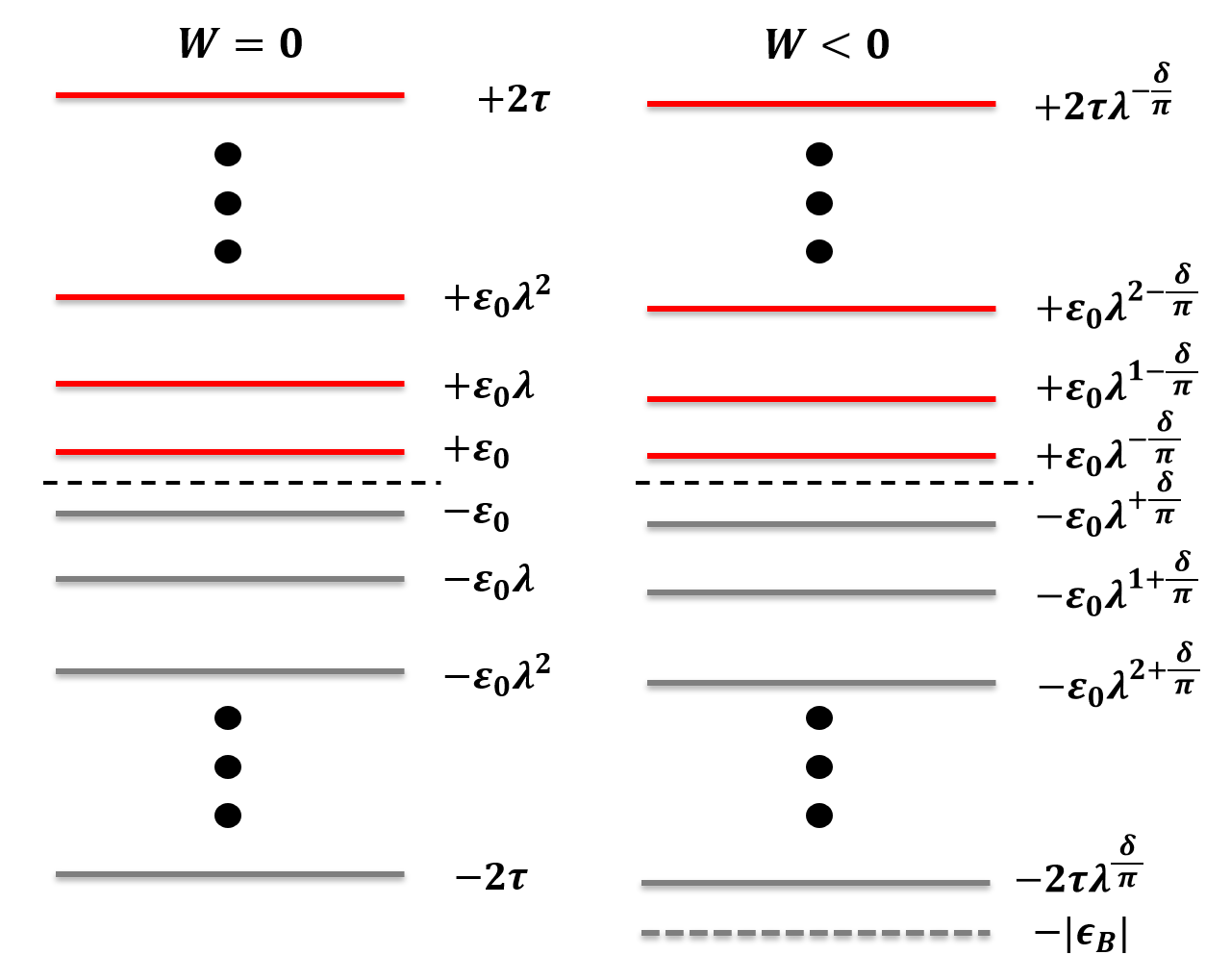}
        \vspace{-0.25cm}
	    \caption{\footnotesize Energy levels of the eNRG Hamiltonian in \eqref{H_photo_eNRG} for $W=0$ (Left) and $W<0$ (Right). The presence of the scattering potential shifts the energy levels by the factor $\lambda^{-\frac{\delta}{\pi}}$, where $\delta$ is the phase shift. The dashed line represents the bound state, which splits off the conduction band for large $|W|$. \Sou}
        \label{Fig: eNRG_Energy}
\end{figure}

The eNRG procedure shares similarities with the NRG, including the numerical diagonalization procedure. The exponential coupling between sites results in a logarithmic discretization of the energy spectra, similar to the NRG spectra, as show in Fig. \ref{Fig: eNRG_Energy}. Recently, a smoothing procedure has been developed for the eNRG method designed to eliminate non-physical oscillations in the results due to the discretization \cite{Picoli}. In essence, this procedure amounts to choosing two successive values $\xi$, typically $\xi=3$ and $4$, followed by changing the hopping parameter by a small transformation $\tau_n\rightarrow \tau_n \lambda^{\theta}$, and averaging over a uniform distribution $\theta\in[-1,1]$, analogous to the z-trick \cite{oliveira1994}. This construction is described in Ref. \cite{Picoli}.

In the x-ray photoemission problem numerically, the final Hamiltonian becomes:
\begin{equation}\label{H_photo_eNRG}
\begin{aligned}
\mathcal{H}_{\tilde N} = \sum_{n=0}^{\xi-1} \tau \left( a_n^{\dagger}a_{n+1}+\mathrm{h.c.}\right) + \sum_{n=0}^{\tilde N-1}\tau_n\left( f_n^{\dagger}f_{n+1}+\mathrm{h.c.}\right)+ W a_0^{\dagger}a_0.
\end{aligned}
\end{equation}

Since the Hamiltonian in Eq. \eqref{H_photo_eNRG} is quadratic, it can be diagonalized numerically. The resulting single-particle spectrum of the final Hamiltonian is shown in Fig. \ref{Fig: eNRG_Energy}. Here the minimum energy above the Fermi level is $\varepsilon_0 = 2 \tau \lambda^{0.5 - \tilde {N}}$ and $\Delta\varepsilon = 2 \varepsilon_0$. In the presence of the attractive localized scattering potential, the initial energy levels are shifted by the factor of $\lambda^{-\frac{\delta}{\pi}}$, where $\delta$ is the phase shift. For strong scattering potentials $W > -2\tau$, the bond state splits off the conduction band  with energy $\epsilon_B \sim - W$.

\newpage

Finally, since $\Delta\varepsilon \approx 4\tau \lambda^{-\tilde {N}} $, by the the Heisenberg uncertainty principle, to accurately represent the metallic band up to the time $T_m$, an eNRG chain of size
\begin{equation}
    \tilde N \ge   \frac{\ln \left(4 \frac{\tau . T_m}{\hbar}\right)}{\ln \lambda},
\end{equation}
 is necessary. The chain size grows with $\ln T_m$ instead of $T_m$, resulting in a much smaller chain and many-body matrices. This makes the problem numerically treatable even for very large $T_m$. Additionally, the procedure can compute any observable and can be applied to interacting Hamiltonians, allowing study of the Kondo problem among other applications.

\section{Numerical results}



We now go back to the photoemission problem. Our previous discussion considered a flat band. To adjust the discussion for the tight-binding Hamiltonian, we only have to let $D \rightarrow 2\tau$, $\rho \rightarrow \frac{1}{\pi\tau}$ and $N_e = \frac{D}{\Delta\varepsilon} \rightarrow \frac{\pi \tau}{\Delta \varepsilon}$, and then substitute the expressions
\begin{eqnarray}\label{phase_Shift_ttb}
\tan\delta = - {W}/{\tau} ,
\end{eqnarray}
\begin{eqnarray}\label{Bond_State_energy_ttb}
\epsilon_{B} \approx -2\tau.\mathrm{coth}\left(\frac{2\tau}{|W|} \right),
\end{eqnarray}
\begin{eqnarray}\label{Fidelity_time_ttb}
\mathcal{F}(t) \approx \frac{\left(1 + \mathcal{R}(t)^2 + 2\mathcal{R}(t) \cos\left(\frac{\epsilon_B t}{\hbar}\right) \right)}{(1+r)^2} e^{ -2\left(\frac{\delta}{\pi}\right)^2 \left(\ln \left(\frac{\pi \tau }{\Delta\varepsilon}\right) + \mathrm{CI}\left(\frac{\Delta\varepsilon  t}{\hbar} \right) - \mathrm{CI}\left( \frac{\pi \tau t}{\hbar} \right) \right)},
\end{eqnarray}
for the phase shift, the bound state energy and the fidelity, respectively. Here, the ratio 
\begin{equation}
    r = \frac{|\braket{\phi_0}{\tilde \varphi}|^2}{|\braket{\phi_0}{\varphi_0}|^2},
\end{equation}
can be computed numerically by the eNRG procedure.

\begin{figure}[hbt!]
		\centering
        \begin{tabular}{ll}
        \hspace{-0.65cm}
        \includegraphics[scale=0.545]{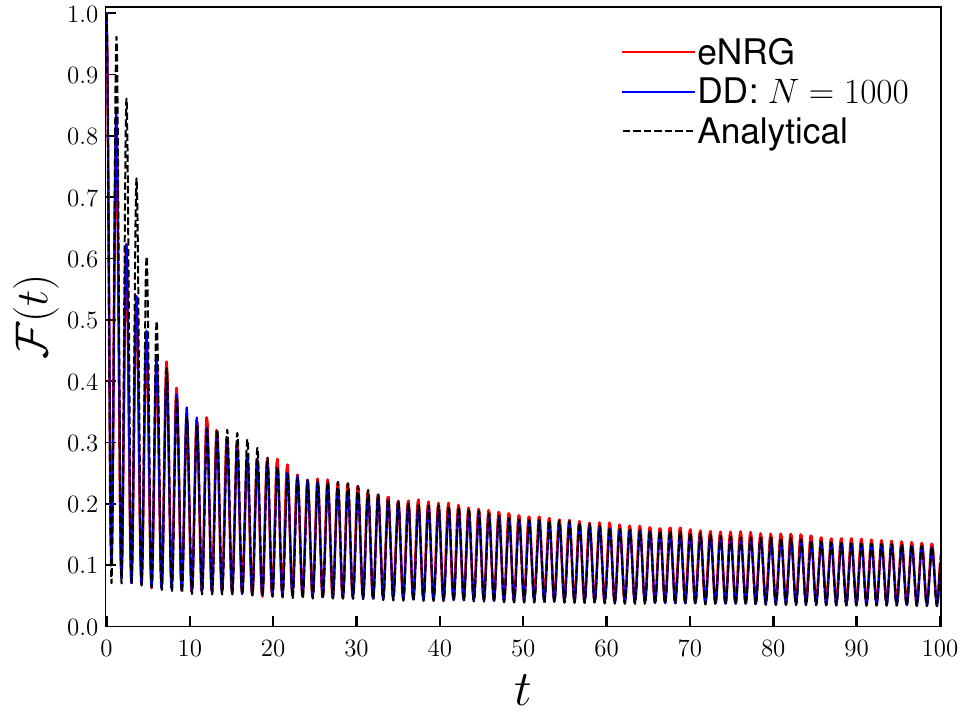}
        & \hspace{-0.55cm}
        \includegraphics[scale=0.45]{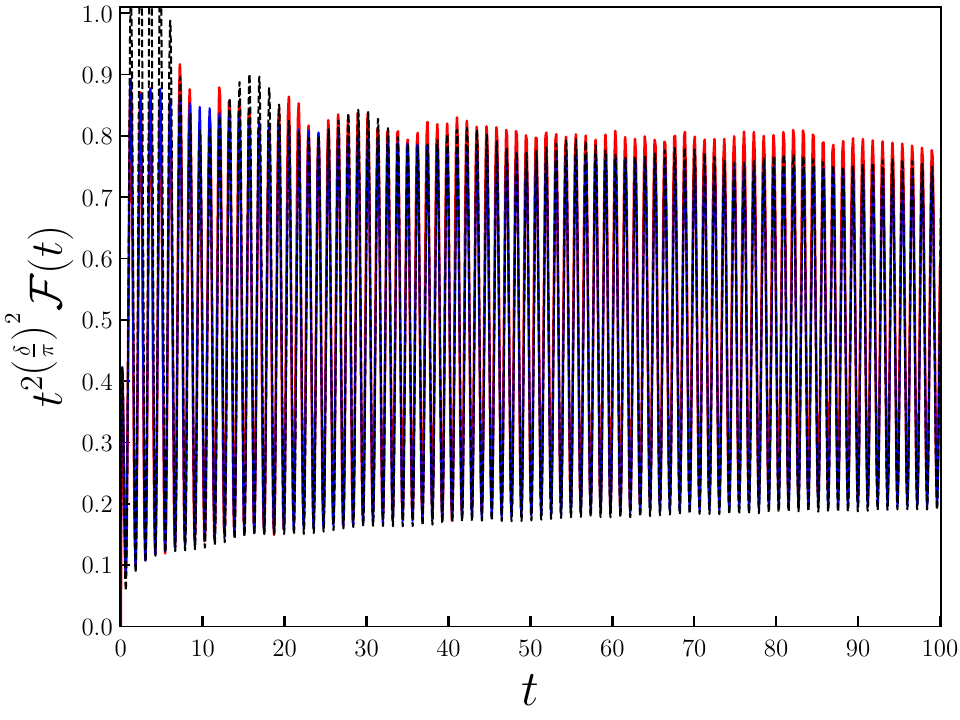}
        \end{tabular}
        \vspace{-0.3cm}
	    \caption{\footnotesize (Left panel)  Numerical results for the fidelity computed by the eNRG (solid blue line), analytical results (dashed black lines) and direct diagonalization (DD) using $750$ tight-binding sites (red solid line) for $\frac{W}{\tau} = -5.0$. For the eNRG results we used $\lambda=1.5$, $\tilde N = 12$, $\xi = 3 \mathrm{~and~} 4$ and $10$  $\theta \mathrm{s} \in [-1,1] $ uniformly distributed. (Right panel) The same numerical results shown in the left panel multiplied by $t^{2\left(\frac{\delta}{\pi}\right)^2}$. \Sou }
        \label{Fig: Fidelity}
\end{figure}
\begin{figure}[hbt!]
		\centering
        \begin{tabular}{ll}
        \hspace{-0.65cm}
        \includegraphics[scale=0.545]{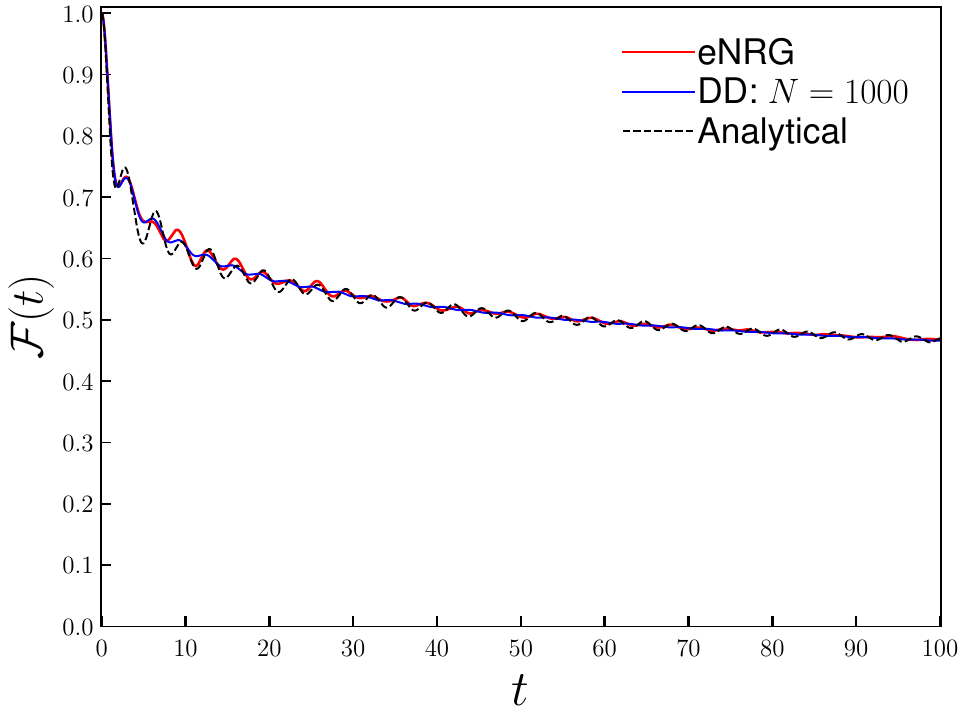}
        & \hspace{-0.55cm}
        \includegraphics[scale=0.45]{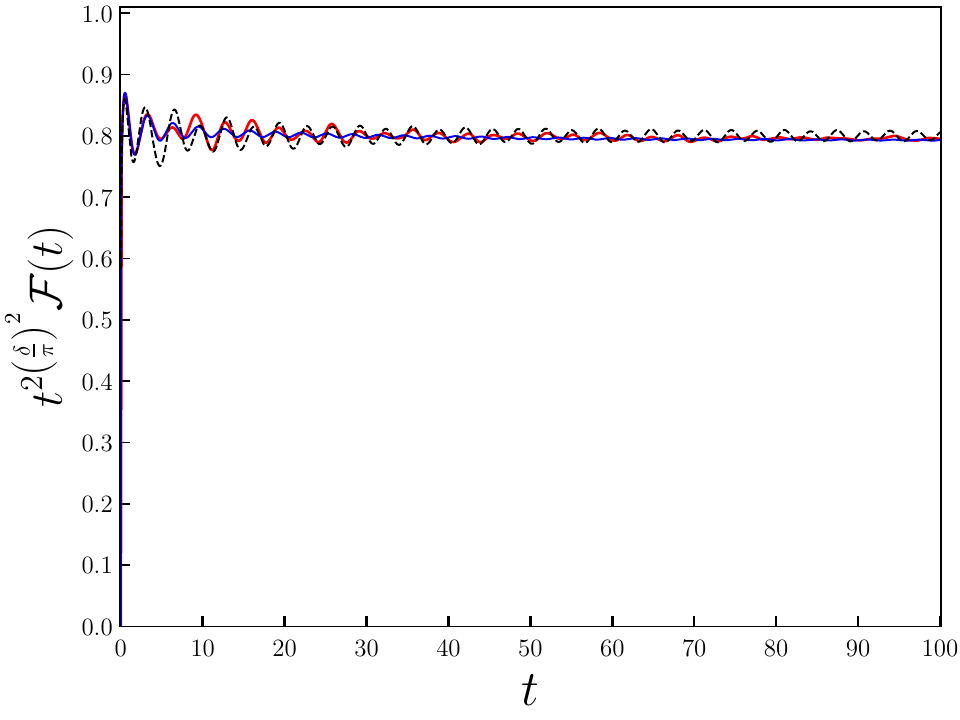}
        \end{tabular}
        \vspace{-0.3cm}
	    \caption{\footnotesize (Left panel)  Numerical results for the fidelity computed by the eNRG (solid blue line), analytical results (dashed black lines) and direct diagonalization (DD) using $750$ tight-binding sites (red solid line) for $\frac{W}{\tau} = -1.0$. For the eNRG results we used $\lambda=1.5$, $\tilde N = 12$, $\xi = 3 \mathrm{~and~} 4$ and $10$  $\theta \mathrm{s} \in [-1,1] $ uniformly distributed. (Right panel) The same numerical results shown in the left panel multiplied by $t^{2\left(\frac{\delta}{\pi}\right)^2}$. \Sou}
        \label{Fig: Fidelity - 2}
\end{figure}

Figures Fig. \ref{Fig: Fidelity} and \ref{Fig: Fidelity - 2} show the computed fidelities for $W=-5\tau$ and $W=-1\tau$, respectively. The solid blue line shows the fidelity computed with the eNRG method, and we compare this result with Eq. \eqref{Fidelity_time_ttb} (dashed black line) and with the fidelity from the direct diagonalization of the tight-binding 
Hamiltonian \eqref{H_photo_eNRG} with $N=750$ sites. For the eNRG fidelity were computed with $\lambda = 1.5$, $\tilde{N} = 12$, $\xi = 3 \mathrm{~and~} 4$, and $10$ values of $\theta \in [-1,1]$ uniformly distributed.

\begin{figure}[hbt!]
		\centering
        \begin{tabular}{ll}
        \hspace{-0.6cm}
        \includegraphics[scale=0.505]{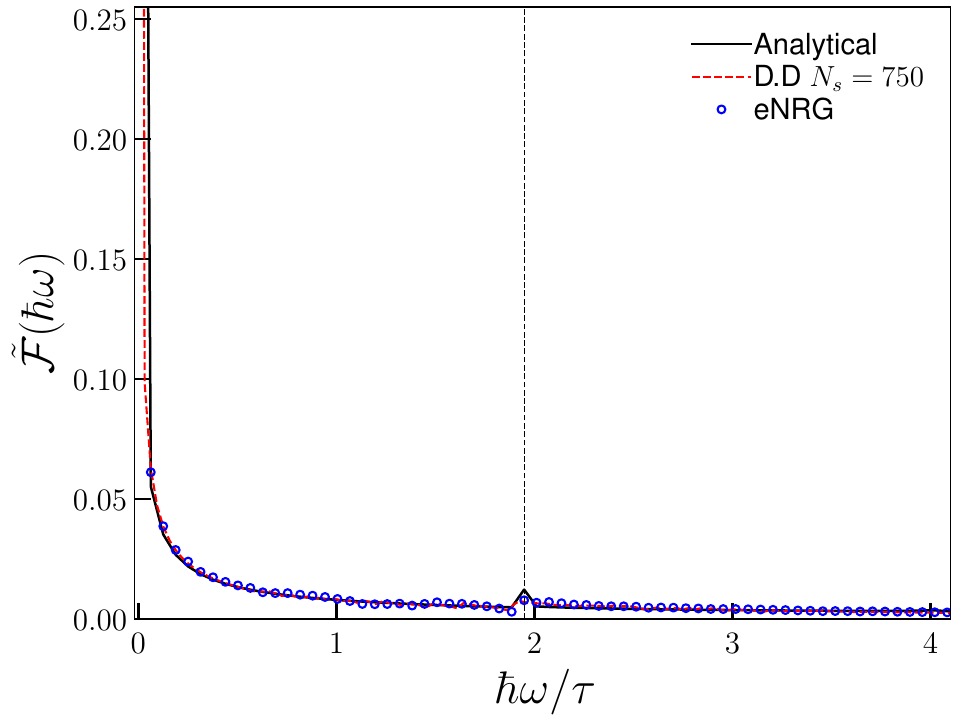}
        &\hspace{-0.6cm}
        \includegraphics[scale=0.505]{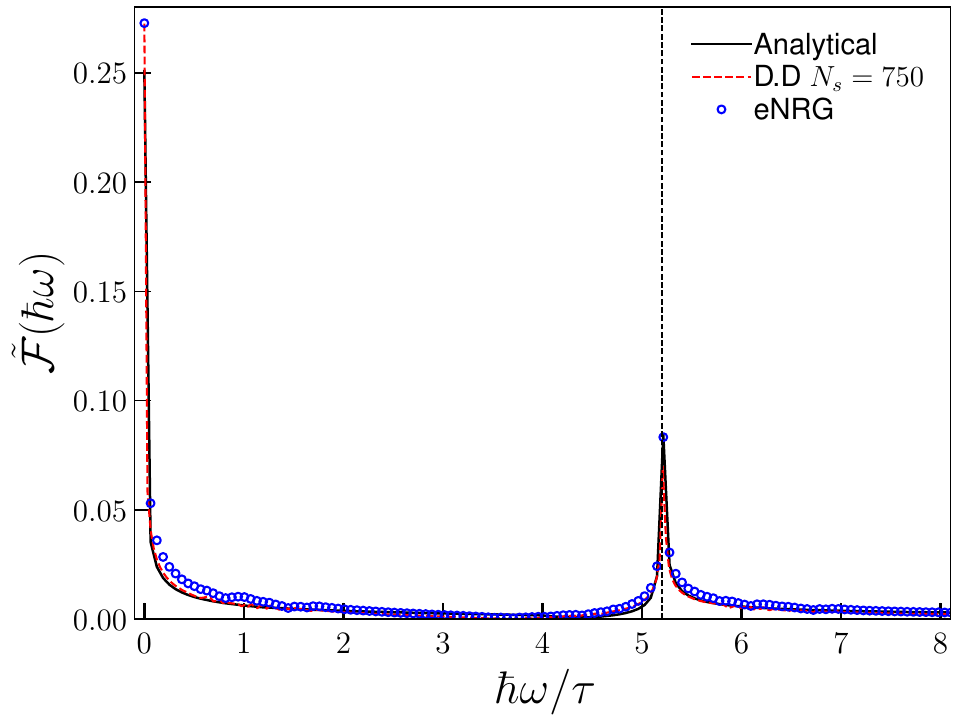}
        \end{tabular}
        \vspace{-0.1cm}
	    \caption{\footnotesize Numerical results for the fidelity in the frequency domain computed by the eNRG (blue circular dots), DD (dashed red line), and analytical (black solid line) for ${W} = -1.0\tau$ (Left panel) and ${W} = -5.0\tau$ (Right panel). The dashed line represent the minimum energy of an excitation from the bound state. For the eNRG results we used $\lambda=1.5$, $\tilde N = 12$, $\xi = 3,4$ and $10$ values of $\theta$ uniformly distributed. \Sou}
        \label{Fig: Fidelity_FFT}
\end{figure}

The results validate our conclusions. At long times, the averages of the computed fidelities follow the Doniach-Sunjic law, as shown in the right panels of Figs. \ref{Fig: Fidelity} and \ref{Fig: Fidelity - 2}. The results also exhibit an oscillation with a frequency of $ \epsilon_B/\hbar $, which is more clearly observed in the frequency-domain plots in Fig. \ref{Fig: Fidelity_FFT}. This oscillation arises from pairs of particle-hole excitations that include the bound state (unplugged states). For all computed fidelities, the eNRG results show excellent agreement with direct diagonalization (DD) and also semi-quantitative agreement with the analytical solution.

We conclude that with only $10$ different values of $\theta$ and  $\tilde N = 17$ is sufficient to accurately capture the photoemission behavior up to the time scale of $T_m \sim 200 \hbar/\tau$. The small differences can be reduced further by using more values of $\theta$. The eNRG procedure can faithfully simulates the metallic behavior, the smoothing procedure proposed in \cite{Picoli} indeed removes the non-physical artifacts that arise from the discretization.


The numerical data in Figs. \ref{Fig: Fidelity}, \ref{Fig: Fidelity - 2}, and \ref{Fig: Fidelity_FFT} show significant deviations from Eq. \eqref{Fidelity_time_ttb} at short times but maintain good agreement at long times. These discrepancies arise from differences in the band structures: while the numerical results are based on a tight-binding model, the analytical treatment assumes a flat band. In renormalization group language, the different dispersion relations introduce irrelevant operators that affect the short-time behavior—or the large-energy dependence—of physical properties. The low-energy levels, however, are well represented by the flat band approximation with a constant phase shift, leading to accurate results from Eq. \eqref{Fidelity_time_ttb} at long times.

Another interesting quantity to compute is the expected occupation of each tight-binding site $\langle n_l \rangle = \langle \Psi |  a_l^\dagger a_l | \Psi \rangle$. After some manipulations, the occupation number of the $l$-th site can be easily computed by
\begin{eqnarray}\label{OccupatioNite_eNRG}
        \langle n_l(t) \rangle =\sum_{\substack{p,h}}\sum_{\substack{k: \varepsilon_k<\epsilon_F}}\{  g_p,  a_l^\dagger  \} \{ a_l ,   g_h^\dagger \} \{g_p, \tilde a_k^\dagger  \} \{\tilde a_k,  g_h^\dagger  \}  e^{-i(\epsilon_p-\epsilon_h)\frac{t}{\hbar}}\;\;.
\end{eqnarray}
Here, $g_q^\dagger$ is the creation operator of a single level of the final Hamiltonian ($W\ne 0$), with energy $\epsilon_q$. $\tilde a_k^\dagger$ is the creation operator of a level of the initial Hamiltonian ($W=0$) and $a_l^\dagger$ represents the site $l$.

As it is evident so far, finding properties of the system using the many-body wave function is challenging, even when the Hamiltonian is quadratic, because the number of states increases very quickly. In contrast, calculating the local density is typically easier, especially for calculations performed using electronic density approaches such as Density Functional Theory (DFT) \cite{PhysRev.140.A1133}. The Runge–Gross theorem \cite{PhysRevLett.52.997} plays an important role here, as it states that the information provided by the time-dependent wave function and the local density is the same. Thus, by using the local density $\langle n_l \rangle$ for each site at time $t$, it is possible, in principle, to compute the fidelity $\mathcal{F}(t)$. Although we will not explore this idea further here, it motivated us to compute the local density.

\begin{figure}[hbt!]
		\centering
        \begin{tabular}{ll}
        \hspace{-0.65cm}
        \includegraphics[scale=0.58]{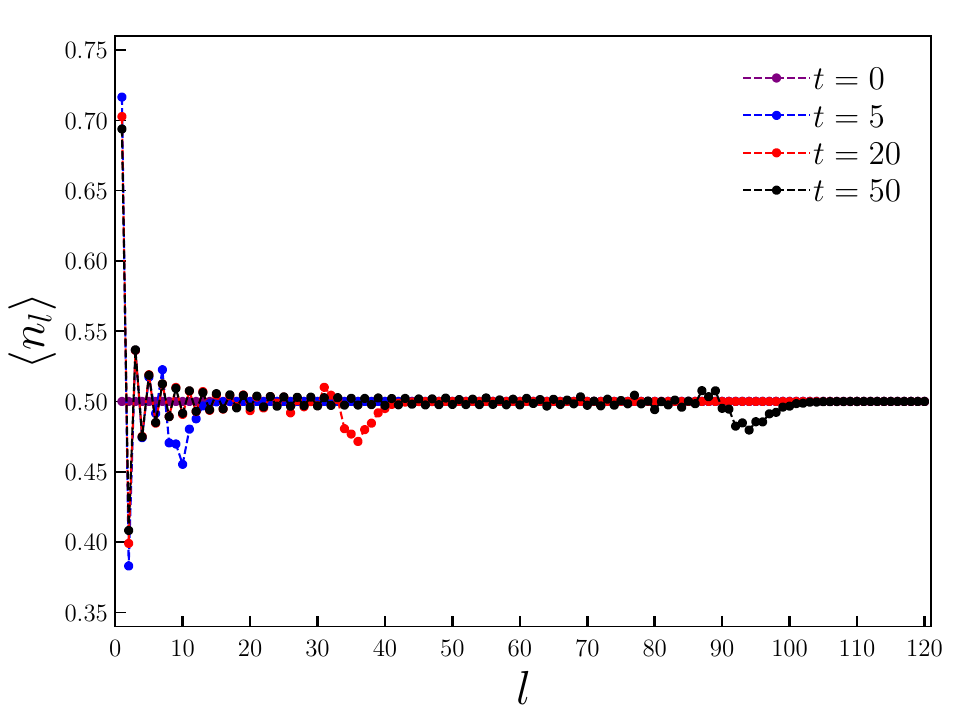}
        & \hspace{-0.7cm}
        \includegraphics[scale=0.43]{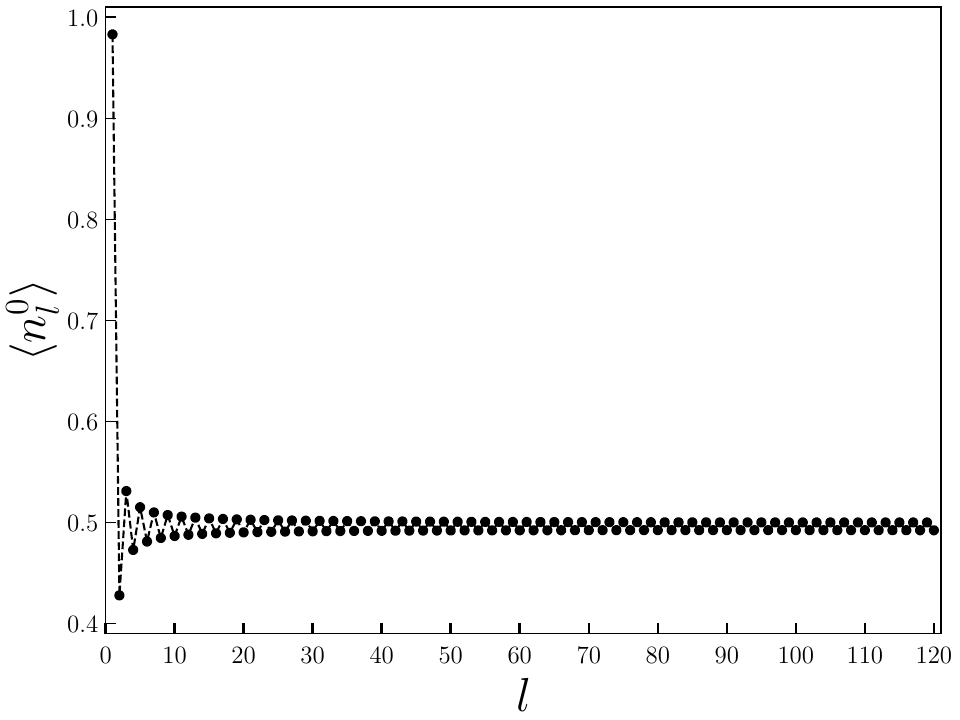}
        \end{tabular}
        \vspace{-0.3cm}
	    \caption{\footnotesize (Left panel) Numerical results of the site occupation $n_l$ along the chain $l$ for different times (in unities of $\frac{\hbar}{\tau}$). (Right panel) The local density configuration of the final ground state $\langle n_l^0 \rangle$. For this plot we used $W=-5$ and $ N = 120$. \Sou }
        \label{Fig: Occupation}
\end{figure}

\newpage

Fig. \ref{Fig: Occupation} (left panel) shows the occupation of each site along the chain ($N = 120$) at different times, alongside the final ground state occupation of each site (right panel). Note that from Eq. \eqref{OccupatioNite_eNRG}, computing the local density $\langle n_l \rangle$ requires only one-particle projections, making it simpler and numerically cheaper than computing the fidelity. This allows us to work directly with the tight-binding Hamiltonian.

At $t=0$, the system is purely in the ground state. For a half-filled tight-binding band, this implies that the occupation of each site is $1/2$. For $t>0$, the attractive localized scattering potential shifts the energy level of the first site and disturbs the surrounding sites, creating distortions in the final ground state local density (as shown in Fig. \ref{Fig: Occupation} (Right)). Once $W < 0$, the electrons tends to preferentially occupy the first site, breaking the translation symmetry. {However, as $n_i(t)$ is an observable and takes time to change, the presence of this attractive potential in the first site} creates a wave propagating along the chain, reorganizing the occupation of each site closer to the final ground state configuration, as shown in Fig. \ref{Fig: Occupation} (Left). This behavior of the occupation on the chain is associated with Friedel oscillations, with a more detailed discussed in \cite{Picoli,friedel1952}.

Also, from Fig. \ref{Fig: Occupation}, we observe that the wave propagating in the chain creates a valley from the initial value of 0.5, and this valley propagates with the wave. This occurs because the electrons in the chain are being pushed in the direction of the first site, creating an electrical current in the chain flowing from the left to the right (by convention). The propagation velocity of the wave seems to be approximately constant, as indicated by the ratios $\frac{\mathrm{site}}{\mathrm{time}}$ of the position of the valley over time: $ \frac{9}{5} = 1.8 $, $ \frac{36}{20} = 1.80 $, and $ \frac{92}{50} = 1.84 $.

\section{Qualitative comparison with experimental results}

As mentioned in section 3.2, experimental signatures of theses excitations from the bound state are three. The first is the shift in the binding energy $\Delta E_{\mathrm{bin}} = |E_0 - \tilde{E_i}| = D\left(\coth\left(\frac{-1}{2\rho W}\right) - 1 + \frac{\delta}{\pi}\right)$. The second is the peak associated with the excitations from the bound state, with energy $E_0 + D\coth\left(\frac{-1}{2\rho W}\right)$ and obeying the Nozières-De Dominicis's power law, as shown in Fig. \ref{Fig: Fidelity_FFT}.  Also, this extra peak in the frequency domain fidelity can explain satellite peaks observed in experimental XPS data, which are beyond explanation by other mechanisms. The last part is the electrical current propagating in the chain, a consequence of the reorganization of the electrons in the chain due to the presence of the localized potential at the first level.

\begin{figure}[hbt!]
		\centering
        \includegraphics[scale=0.6]{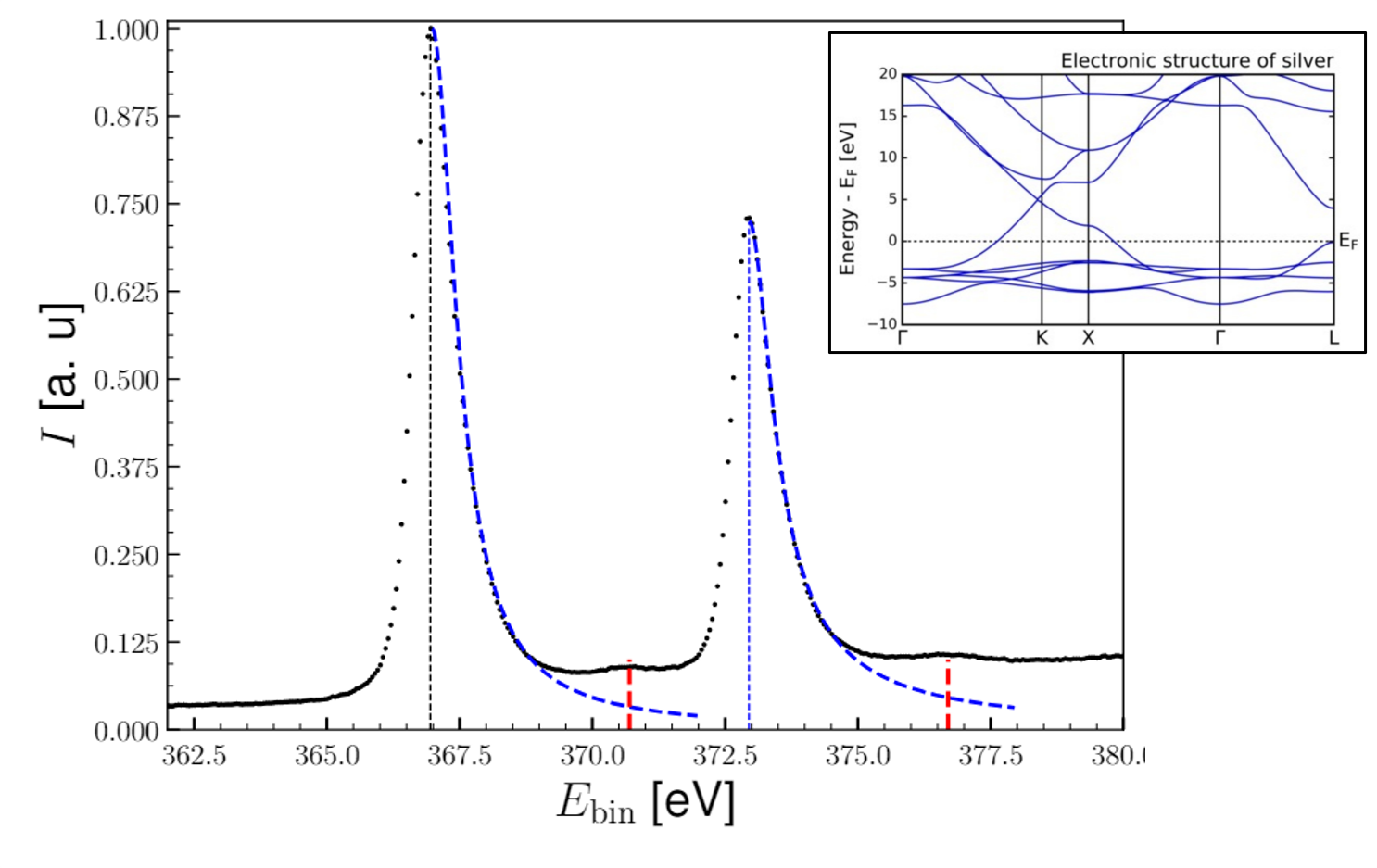}
        \vspace{-0.5cm}
	    \caption{\footnotesize Experimental XPS data of the 3d transition from silver (bulk). The two characteristic peaks of these transitions are the blue dashed lines and follow the Doniach-Sunjic power law. The red dashed vertical lines show small satellite peaks, compatible with the energy of a bound state $\epsilon_B \approx 3.5$ eV, below the bottom of the conduction band at the $\Gamma$ point (in the top inset) of the band structure ($D \approx 3.3$ eV). \Sou}
        \label{Fig:Experimental}
\end{figure}

Figure \ref{Fig:Experimental} shows experimental photoemission spectra around the 3d transitions in bulk silver \cite{AgXPSEXP}. The two characteristic peaks of the multiplet 3d transitions are represented by the blue dashed lines and follow the Doniach-Sunjic power law \cite{Doniach_1970}. The red dashed lines show satellite peaks. Each such peak is nearly 3.5 eV away from the corresponding threshold, approximately equal to the energy of the bottom of the conduction band at the $\Gamma$ point measured from the Fermi level, where a weakly bound state would be expected. Our results predict that the decay of these bound state peak candidates will follow the Nozières-De Dominicis law, with the same $\delta$ that fits the main peaks.

Unfortunately, the large width of the satellite peak in Fig. \ref{Fig:Experimental} makes it impossible to verify this prediction. We have been unable to find results with sufficient resolution to allow for a quantitative comparison, and we hope that future experiments will be conducted with this goal in mind.

Besides the interesting physics we have just discussed, important conclusion from our work comes from the success of the eNRG procedure in reproducing the tight-binding results, even better than the analytical solutions. To compare the computational performances, using the eNRG with $\xi = 3$ and $\tilde{N} = 12$, we needed to consider 12870 many-body states. In contrast, the brute force diagonalization of the tight-binding model, truncating the many-body states up to three particle-hole excitations from the ground state, required dealing with 562500 many-body states (a number that satisfies the condition $\sum_n |\braket{\phi_0}{\varphi_n}|^2 > 0.998 $), 44 times larger than the eNRG requirement. Clearly, averaging over 10 $\theta$s is a small price to pay for the accurate results we obtained. Motivated by this success, we will use the same procedure to discuss a similar system in the following chapter: a Fermi gas system subject to a time-dependent localized potential.

%% file: CAP/Cap4.tex
\chapter{Quasi-adiabatic Fermi gas system}

Even more interesting physics arises when a continuous function of time $W(t)$ is substituted for the scattering potential in Eq. \eqref{H_photo_MI}, instead of the sudden creation discussed in Chapter 3. Consider a conduction band with uniform density of states and write the Hamiltonian in momentum space:
\begin{align}\label{H_photo_CCT}
H(t) = \sum_{k}\varepsilon_{k}\tilde{a}^\dagger_{k}\tilde{a}_{k} + \frac{W(t)}{N} \sum_{k,q} \tilde{a}_k^\dagger\tilde{a}_q.
\end{align}
This modification allows us to study adiabaticity and helps to understand the atom-surface behavior when the hydrogen is far from the surface. Therefore, it is convenient work with the instantaneous basis. The instantaneous basis $\{\ket{\varphi_n (t)}\}$ is constituted by the eigenstates of the instantaneous Hamiltonian:
\begin{equation}
    H(t) \ket{\varphi_n (t)}=E_n(t) \ket{\varphi_n (t)}
\end{equation}
at each $t$, with the instantaneous eigenenergies  $E_n(t)$. Here, we will consider an evolution starting from the ground state of the initial Hamiltonian $H_0$ and evolving under $H(t)$. The expansion of the time-dependent wave function on the instantaneous basis is 
\begin{equation}
    \ket{\Psi(t)} = \sum_{k} c_k(t) \ket{\varphi_k (t)}.
\end{equation}
Our main goal is to find an expression for the coefficients $\{ c_k(t) \}$ of each many-body state $\ket{\varphi_k (t)}$.

To track their time evolution, we need to solve the Schrödinger equation, which, after the transformation 
\begin{equation}
    \tilde{c}_n(t) = {c}_{n}(t) e^{\frac{i}{\hbar}\int_0^t dt' E_n(t')},
\end{equation}
yields
\begin{eqnarray}\label{SE_}
\frac{d\tilde{c}_n(t)}{dt}=-\sum_m \tilde{c}_m(t)\bra{\varphi_n(t)} {\partial _t} \ket{\varphi_m(t)} e^{-\frac{i}{\hbar}\int_0^t (E_m(t') - E_n(t')) dt' }.
\end{eqnarray}
To solve this system of $L$ coupled differential equations, we therefore have to compute the matrix elements of $\bra{\varphi_n(t)} {\partial _t} \ket{\varphi_m(t)}$.

Using the analytical diagonalization of the Hamiltonian \eqref{H_photo} in the Appendix \ref{Appendix_Analytical_Diagonalization}, combined with the derivative computed by finite differences, the Appendix \ref{Model_Continuous_W} shows that:
\begin{eqnarray}\label{Coupling_SE}
&\bra{\varphi_n(t)}\partial_t ~ g_p^\dagger(t) g_h(t) \ket{\varphi_n(t)} \approx -\dfrac{1}{\pi}\dfrac{\Delta \delta}{\Delta t} \dfrac{\Delta\varepsilon}{\varepsilon_p - \varepsilon_h} \nonumber, \\
&\bra{\varphi_n(t)}\partial_t ~ g_p^\dagger(t) g_h(t)  g_{p'}^\dagger(t) g_{h'}(t) \ket{\varphi_n(t)} \approx -\dfrac{1}{\pi^2}\dfrac{\Delta \delta^2}{\Delta t} \dfrac{\Delta\varepsilon}{\varepsilon_p - \varepsilon_h} \dfrac{\Delta\varepsilon}{\varepsilon_{p'} - \varepsilon_{h'}}.
\end{eqnarray}

\begin{figure}[hbt!]
    \centering
    \includegraphics[scale=0.51]{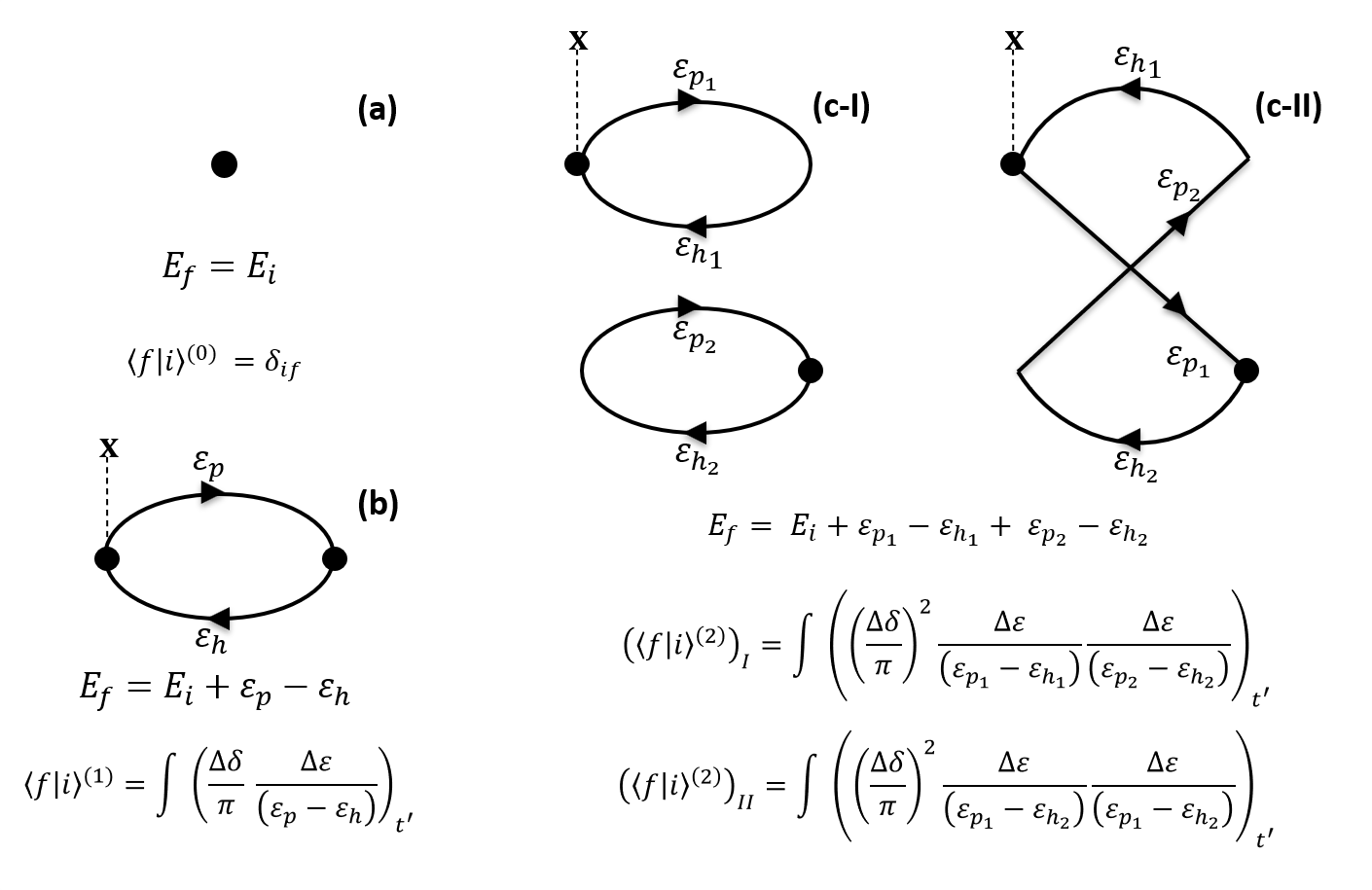}
    \vspace{-0.45cm}
\caption{\footnotesize Diagrams of Eq. \eqref{SE_} after a small time interval $\Delta t$. The left side represents the unperturbed system at $t = 0$ in its initial configuration. The X indicates the scattering potential. (a) Zeroth-order diagram illustrates the transition to the final state with the same configuration. (b) First-order diagram illustrates the transition to a final many-body state with a particle-hole excitation from the initial configuration. (c-I) Second-order diagram illustrates the direct transition to a final many-body state with two particle-hole excitations from the initial configuration. (c-II) Second-order diagram illustrates the indirect transition to a final many-body state with two particle-hole excitations from the initial configuration. \Sou}
    \label{fig:Feynman}
\end{figure}

Even through we will not solve the Eq. \eqref{SE_} using perturbation theory, the diagrammatic representation of the Dyson series \cite{Mahan2010-xj} for Eq. \eqref{SE_} proves instructive, to identify the most important transitions.  Figure \ref{fig:Feynman} shows the diagrams in the time domain (see also Appendix \ref{diagrams_dayson}). The vacuum of the diagram represents the ground state of the instantaneous Hamiltonian. The other electronic states are represented by particle-hole excitations from the ground state, depicted by lines with arrows indicating the creation of electrons and holes. Thus, after the initial electronic configuration interacts with the potential, it may create one (or more) pairs of particle-hole excitations from this configuration. The intermediate states along the timeline can be observed by making a vertical cut at any point before the final point in the diagram.

Let us consider that the system at $t=0$ is in the initial configuration $\ket{\varphi_i}$. At $t = \Delta t$, the potential changes by $\Delta W$, resulting in a phase shift of $\Delta \delta$. We want to know what is the possibles final configuration $\ket{\varphi_f}$. In Figures \ref{fig:Feynman}, we illustrate the principal transitions for this system, after a small $\Delta t$, arranged by the number of particle-hole excitations and the contribution of each diagram to the final states using Eq. \eqref{Coupling_SE}. In the zero-order expansion, the system remains in the same initial configuration $\braket{\varphi_f}{\varphi_i}^{(0)} \sim 1$. Meanwhile, the one- and two-pairs of particle-hole excitations contributions are $\braket{\varphi_f}{\varphi_i}^{(1)} \sim \frac{\Delta\delta}{\pi}$ and $\braket{\varphi_f}{\varphi_i}^{(2)} \sim \left(\frac{\Delta\delta}{\pi}\right)^2$, respectively. 

\newpage
Since the scattering potential changes continuously in time, we can express the change in the phase shift as $\Delta \delta = \frac{d\delta}{dt} \Delta t$. In this case, the contribution of the two-pair of particle-hole excited states becomes $\mathcal{O}(\Delta t^2)$, implying that only contributions from the zero and one-pair diagrams are important, up to first order in $\Delta t$. Since contributions for transition $\sim \mathcal{O}(\Delta t^2)$ or higher vanish in the integral, it is sufficient to consider up to one-pair contributions. Then, the only way to reach a final state with two particle-hole excitations is by transitioning through an intermediate state with one particle-hole excitation. This simplification would be inappropriate if $W(t)$ had discontinuities.

Appendix \ref{Model_Continuous_W} shows that substituting Eq. \eqref{Coupling_SE} into Eq. \eqref{SE_} leads to the evolution of the coefficients $ c_k(t) $, described by
\begin{equation}\label{SEFM_}
\begin{aligned}
\frac{d\tilde{c}_n(t)}{dt} = -\frac{1}{\pi} \frac{d\delta(t)}{dt}\sum_p\sum_{h\neq p} \tilde{c}_{n,p,h}(t) \frac{\Delta\varepsilon}{\varepsilon_p - \varepsilon_h} e^{-i(\varepsilon_p - \varepsilon_h)\frac{t}{\hbar}}.
\end{aligned}
\end{equation}
Here, $\tilde{c}_{n,p,h}(t)$ denotes the coefficient of the state $g_p^\dagger g_h \ket{\varphi_n}$, which corresponds to one particle-hole excitation on the many-body state $\ket{\varphi_n}$. For continuous $W(t)$, only many-body states differing by a single particle-hole excitation are directly coupled to each other.

\begin{figure}[hbt!]
    \centering
    \includegraphics[scale=0.56]{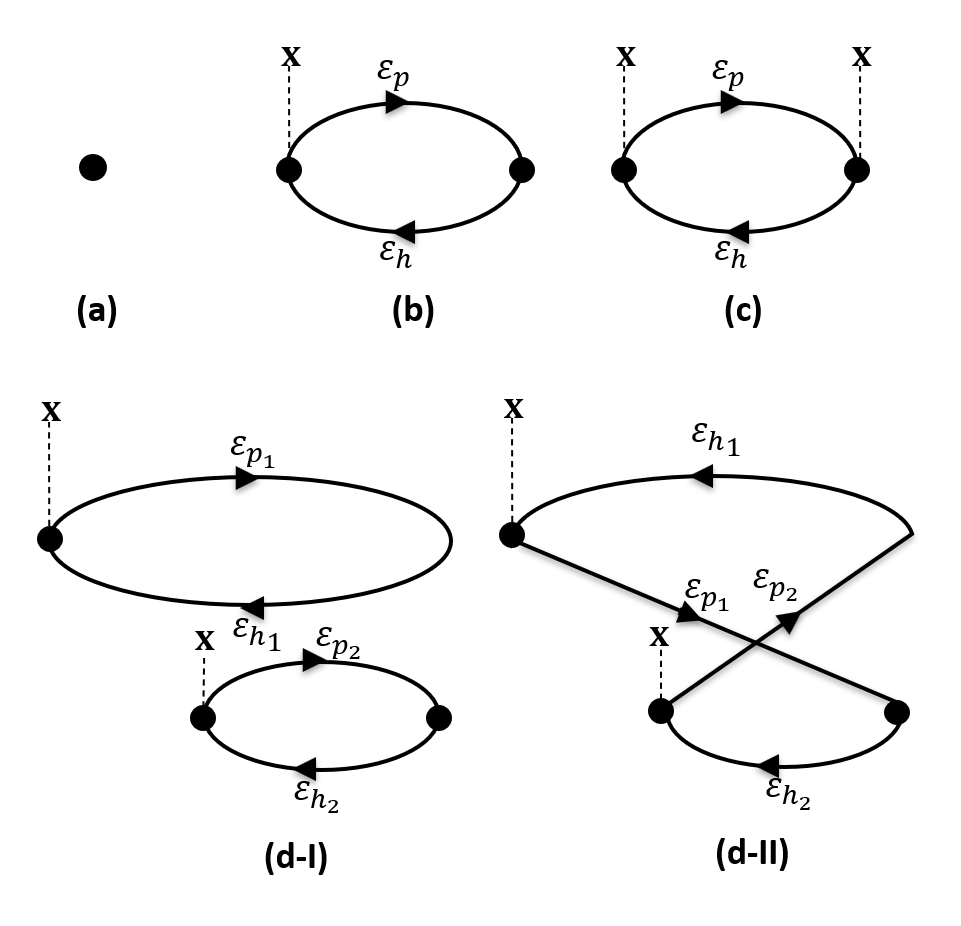}
    \vspace{-0.4cm}
    \caption{\footnotesize Diagrams of Eq. \eqref{SEFM_} up to second order. (a) Zeroth-order diagram: the system remains in the ground state. (b) First-order diagram: the transition to a final many-body state with a one-particle-hole excitation. (c) Second-order diagram: the transition back to the ground state via an intermediate state with one particle-hole excitation. (d-I) Second-order diagram: the direct transition to a final state with two-particle-hole excitations via an intermediate state with one particle-hole excitation. (d-II) Second-order diagram: the indirect transition to a final state with two-particle-hole excitations via an intermediate state with one particle-hole excitation. The X indicates the presence of the time-dependent potential. \Sou}
    \label{fig:Feynman_}
\end{figure}

\newpage

Even though a state cannot transitioned instantaneously to another state that differs by more than a single particle-hole excitation, this process can occur indirectly, mediated by single particle-hole excitation states. In practice, this means that all many-body eigenstates of the instantaneous Hamiltonian with $N_e$ particles are indirectly coupled to each other. However, the coupling between a state $\ket{\varphi_n}$ and another state with $m$ additional (or fewer) particle-hole excitations is of the order $\mathcal{O} \left( \frac{1}{\pi} \frac{d\delta}{dt} \right)^m$. Since the system is at the ground state at $t=0$, some time must pass before a substantial number of particle-hole excitations builds up. Depending on how fast the potential grows, this delay may range from very small to effectively infinity.


To be more explicit, Fig. \ref{fig:Feynman_} shows the diagrammatic representation of Eq. \eqref{SEFM_}, up to second order. Now, the zero- and first-order contributions were discussed with Fig. \ref{fig:Feynman}. It is not difficult to show that $\braket{\varphi_f}{\varphi_i}^{(0)} \sim 1 $ and $\braket{\varphi_f}{\varphi_i}^{(1)} \sim \frac{1}{\pi} \frac{d \delta}{d t}$, respectively. Now, the second order contribution represents a final state with two-particle-hole excitations from the initial state and the contribution is $\braket{\varphi_f}{\varphi_i}^{(2)} \sim  \left( \frac{1}{\pi}\frac{d \delta}{d t}\right)^2$.

From Eq.~\eqref{phase_shift} it follows that
\begin{equation}
    \frac{d\delta(t)}{dt} = \frac{\pi\rho}{1+(\pi\rho W(t))^2}\frac{dW(t)}{dt}.
\end{equation}
Therefore, we can consider three possible scenarios, depending on how quickly the scattering potential changes.

When $W(t)$ changes rapidly, the one-particle-hole excited states steal probability from the instantaneous ground state, the latter feed the 
two-particle-hole excitations (and feed back the ground state) and so on. This diffusion-like process rapidly reduces the probability to find the system on the ground state, which amounts to non-adiabatic behavior. Accurate description of states with a large number of particle-hole excitations becomes necessary.


If $ W(t) $ changes slowly, then $ \left( \frac{1}{\pi} \frac{d \delta}{d t} \right) $ is small, and $ \left( \frac{1}{\pi} \frac{d \delta}{d t} \right)^2 \ll 1 $. In this case, the probability of finding the system in a state with two or more particle-hole excitations is very low, meaning that only one-particle-hole excitations from the ground state will be significant. We can refer to this situation as quasi-adiabatic evolution. If the change is slow enough, such that $ \left( \frac{1}{\pi} \frac{d \delta}{d t} \right) \ll 1 $, the system remains in the instantaneous ground state, and the dynamics are adiabatic. Additionally, a smaller $\Delta\varepsilon$ makes it easier for excited states to appear, resulting in less adiabatic behavior.

To accurately simulate the time evolution of a half-filled band numerically, accounting for all possible many-body states, one must construct a matrix with dimensions of $ \frac{N!}{[(N/2)!]^2} $, where $ N $ is the total number of single-particle levels, for each time step in the simulation. If the calculation is restricted to one-particle-hole excitations, this dimension is drastically reduced to $ \left(\frac{N}{2}\right)^2 $. Incorporating two-particle-hole excitations requires an increase in the matrix size by $ \left[\frac{1}{2} \left(\frac{N}{2}\right) \left(\frac{N}{2} - 1\right)\right]^2 $. This number grows exponentially with the number of particle-hole excitations.

Slowly-changing scattering potentials, which correspond to quasi-adiabatic behavior, lead to physics that is analogous to that of the electronic states in the early phase of atom-surface collisions, when the incident particle is still far from the surface. In the quasi-adiabatic regime, only the ground states and single-particle-hole excitations require attention. The set of differential Eqs. \eqref{SEFM_} then reduces to
\begin{equation}\label{EDM}
\begin{aligned}
&\frac{d\tilde{c}_0(t)}{dt} = - \frac{1}{\pi}\frac{d\delta(t)}{dt} \sum_{p,h} \frac{\Delta\varepsilon}{\varepsilon_p - \varepsilon_h }  e^{-i (\varepsilon_p - \varepsilon_h)\frac{t}{\hbar} } \tilde{c}_{p,h}(t), \\
&\frac{d\tilde{c}_{p,h}(t)}{dt} \approx  \frac{1}{\pi} \frac{d\delta(t)}{dt} \frac{\Delta\varepsilon}{\varepsilon_p - \varepsilon_h } e^{+i(\varepsilon_p - \varepsilon_h)\frac{t}{\hbar}} \tilde{c}_{0}(t).
\end{aligned}
\end{equation}
Here, $\tilde{c}_{p,h}(t)$ is the coefficient for the state $\ket{\varphi_{p,h} (t)} = g_p(t)^\dagger g_h(t) \ket{\varphi_0(t)}$.

\begin{figure}[hbt!]
    \centering
    \includegraphics[scale=0.69]{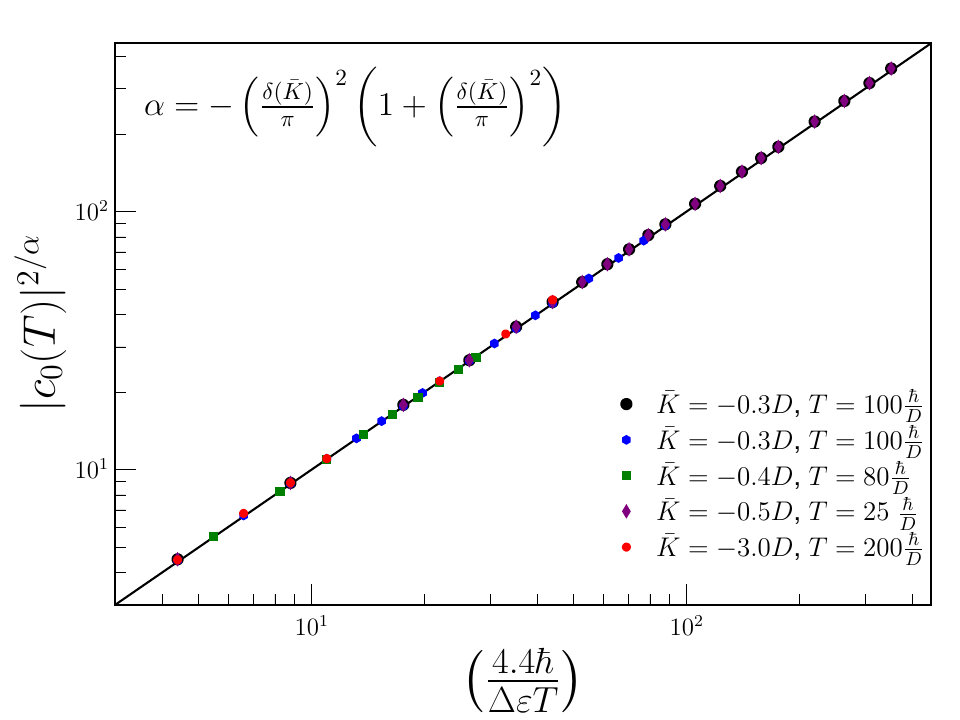}
    \vspace{-0.3cm}
    \caption{\footnotesize Universal behavior of $|c_0(T)|^2$. The coefficient was computed from Eq. \eqref{EDM} for the displayed maximum potentials $\bar K$ and ramp-up times $T$. The black solid line shows that Eq. \eqref{AS} fits the numerical data virtually perfectly. \Sou}
    \label{fig:RPTxN}
\end{figure}

To be specific, we will consider the simplest time dependence, a scattering potential that grows at a constant rate until a maximum value of  $\bar K$: $W(t) = -\frac{\bar K}{T} t$, for $0 \le t \le T$. The time scale $T$ controls the rate of growth. We used Eq. \eqref{EDM} to compute $|c_0(t=T)|^2$  for different combinations of $\bar{K}$, $T \le \frac{\hbar}{\Delta\varepsilon}$, and $\Delta\varepsilon$, within the ranges $-3 \leq \bar{K}/D <0 $, $10 \le T \frac{D}{\hbar} \le 200$ and $ 10^{-2} \le \frac{\Delta\varepsilon}{D} \le 10^{-4}$. Figure \ref{fig:RPTxN} shows the results. Motivated by the calculations in Appendix \ref{|C_0|}, where we found that $ |c_0(T)|^2 \sim \left( \frac{\hbar}{\Delta\varepsilon T} \right)^{- \left(\frac{\delta}{\pi}\right)^2} $, we scaled the axes as $ y = |c_0(T)|^{2/\alpha} $ and $ x = \left( \frac{4.4\hbar}{\Delta\varepsilon T} \right) $, where $ \alpha = -\left(\frac{{\delta}(\bar{K})}{\pi}\right)^2 \left( 1 + \left(\frac{{\delta}(\bar{K})}{\pi}\right)^2 \right) $. This scaling reveals a universal behavior. Under these conditions, $ |c_0(T)|^2 $ can be expressed as a function of $ \bar{K} $, $ T $, and $ \Delta\varepsilon $ as follows:
\begin{equation}\label{AS}
\begin{aligned}
|c_0(T)|^2 = \left( \frac{4.4\hbar}{\Delta\varepsilon T} \right)^{- \left(\frac{1}{\pi} {\delta}(\bar{K})\right)^2\left( 1 + \left(\frac{1}{\pi} {\delta}(\bar{K}) \right)^2 \right)}.
\end{aligned}
\end{equation}

For a very fast ramp ups, we expect the final ground-state coefficients to be close to those discussed in Chapter 3, namely, $|\braket{\Psi(0)}{\varphi_0}|^2 \sim N_e^{-2\left(\frac{\delta}{\pi}\right)^2}$, where $N_e \equiv \frac{D}{\Delta \epsilon}$ is the number of electrons, that is, the number of single-particle energies bellow the Fermi level. As $T$ is reduced, the system is driven into the sudden-quench regime when $\hbar/T $ becomes comparable to the energy $N_e D$, needed to excite all particles to single-particle states above the Fermi level. Substitution of $T = \frac{\hbar}{N_e D}$ and $\Delta \varepsilon = \frac{D}{N_e} $ the right-hand side of Eq. \eqref{AS} yields
\begin{equation}
    \left|c_0\left(T = \frac{\hbar}{D N_e} \right)\right|^2 \approx \left(\sqrt{4.4} N_e\right)^{-2\left(\frac{\delta}{\pi}\right)^2 \left( 1 + \left(\frac{\delta}{\pi}\right)^2 \right) } \sim N_e^{-2\left(\frac{\delta}{\pi}\right)^2},
\end{equation}
recovering qualitatively the sudden quench case, except by the difference in the exponent.

One interesting property of Eq. \eqref{AS} is that it depends only on the maximum scattering potential $ \bar{K} $ and the product $ \Delta\varepsilon \cdot \frac{T}{\hbar} $. This means that for a fixed $ \bar{K} $ and constant $ \Delta\varepsilon \cdot \frac{T}{\hbar} $, the value of $ |c_0(T)| $ remains unchanged. For a fixed $ \bar{K} $, a larger product $ \Delta\varepsilon \cdot \frac{T}{\hbar} $ results in more adiabatic behavior, and vice versa. If $ T = T_m = \frac{\hbar}{\Delta\varepsilon} $, only very low-energy excitations are significant. On the other hand, if $ T = \frac{\hbar}{N_e D} $, in the sudden quench case, excitations across all energy levels will appear.

One possible interpretation of Eq. \eqref{AS} is that the timescale $ T $ defines an energy scale $ \frac{\hbar}{T} $, which acts as a cutoff for possible excitations. The fraction of participating electrons is, therefore, approximately $ \frac{\hbar}{T \Delta\varepsilon} $. However, Eq. \eqref{AS} provides a more precise estimate, indicating that the fraction of electrons is actually $ \frac{4.4\hbar}{T \Delta\varepsilon} $. Consequently, we can identify two regimes depending on the value of $ T $. If $ T \leq \frac{\hbar}{D N_e} $, the system behaves as in the sudden quench case, where all possible electronic transitions are influenced by the scattering potential, resulting in a strongly non-adiabatic response. On the other hand, if $ \frac{\hbar}{D N_e} < T \leq T_m $, only a fraction of the possible particle-hole excitations is affected by the increasing potential. As $ T $ increases, this fraction decreases, leading to fewer particle-hole excitations and a more adiabatic evolution of the system.

To close this section, we emphasize that, although limited to single-particle-hole excitations, our treatment is non-perturbative and yields accurate results in the quasi-adiabatic regime, even for strong scattering potentials. In the following section, we will rely on the eNRG method described in the previous chapter to compare the trustworthy eNRG results from with the analytical calculations. This will moreover allow us to explore the adiabatic regime.


\newpage 

\section{Time scales and adiabatic threshold}
Two important time scales are present in our problem: the first one is $T_m=\frac{\hbar}{\Delta\varepsilon}$ the maximum time  at which the system can be observed, and the other one is the ramp-up time $T$. These time scales reflect the number of electronic levels in the tight-binding chain and the speed at which the scattering potential grows, respectively.

In \cite{Skelt}, the concept of an adiabatic threshold was introduced: for practical purposes, a system is said to evolve adiabatically if the Bures distance between its quantum state and its instantaneous ground state remains under a specified fraction of the maximum Bures separation.
Reference \cite{Skelt} chose this threshold to be 10\%. Here, we will adopt the trace distance $D_{\rho}^{T}(t)$  \cite{wilde_2013} to monitor deviations from adiabatic behavior and let $\eta$ denote the threshold. The trace distance between two pure states is associated with the overlap between them  \cite{wilde_2013}. In our notation, on the instantaneous basis, this relation reads 
\begin{eqnarray}\label{TR}
D_{\rho}^{T}(t) = \sqrt{1 - |c_0(t)|^2}.
\end{eqnarray}
Here $c_0(t) = \braket{\Psi(t)}{\varphi_0(t)}$ and $\ket{\varphi_0(t)}$ is the instantaneous ground state.

From Eq. \eqref{AS}, it is straightforward to compute the trace distance. However, Eq. \eqref{AS} is only reliable under quasi-adiabatic conditions, a regime where the trace distance is necessarily small. To study systems with highly non-adiabatic dynamics, it is necessary to compute $ |c_0(t)| $ numerically.

\section{Numerical results and adiabaticity}

Before turning to rapidly varying potentials, it seems appropriate to check the accuracy of both procedures. With this purpose in mind, we have carried out eNRG computations of $|c_0(t)|^2$. These computations follows the procedure described in Chapter 3, with $W(t)$ substituted for the constant $W$. At each time t, instead of projected upon the ground state of the initial Hamiltonian ($W=0$), the resulting many-body eigenstates are projected upon those of the Hamiltonian at the previous time step. This yields the set of coefficients $c_n(t)$, out of which we select the ground state coefficient, $c_0(t)$.

\begin{figure}[hbt!]
		\centering
        \includegraphics[scale=0.74]{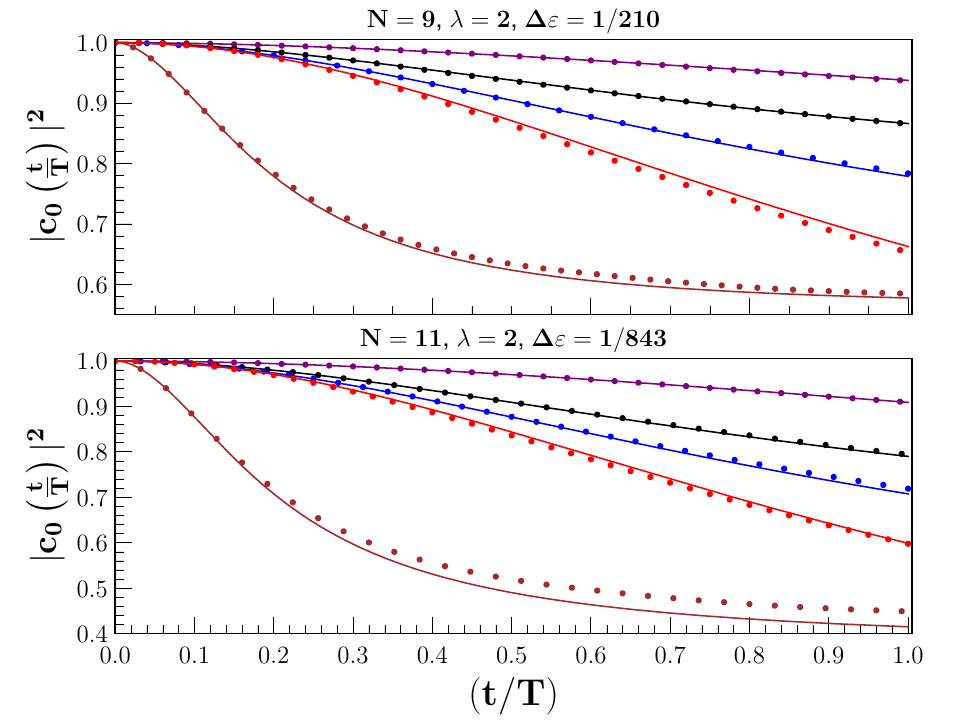}
        \vspace{-0.25cm}
	    \caption{\footnotesize $|c_0(t)|^2$ computed with the Eq. \eqref{EDM} (solid lines) and the eNRG (circular dots) method for different combinations of $\bar{K}$ and $T$: purple ($\bar{K}=-0.5$; $T=50$),  black ($\bar{K}=-1.0$; $T=100$), blue ($\bar{K}=-1.0$; $T=20$), brown ($\bar{K}=-5.0$; $T=100$) and red ($\bar{K}=-1.0$; $T=1$). The title of each panel lists the eNRG parameters and the resulting gap. \Sou }
        \label{Fig: PT}
\end{figure}

Figure \ref{Fig: PT} compares the numerical results for various model parameters with the analytical expression, Eq. \eqref{AS}. The gap $\Delta\epsilon$ one the right-hand side of that equality is equivalent to the eNRG gap $\Delta\varepsilon = 2\tau\lambda^{-N+0.5}$. The data in the top and bottom panels of Fig. \ref{Fig: PT} were obtained with $\tilde N=11$ and $\tilde N=9$, respectively. The purple curves in the two panels lie close to $|c_0|^2 = 1$  and ere hence in the quasi-adiabatic regime; under such conditions Eq. \eqref{AS} yields accurate coefficients. The excellent agreement between the dots and the solid lines therefore attests to the accuracy of the eNRG computations. Since the eNRG method has uniform accuracy over all parametric space, all dots in the figure represent essentially exact results. They can also be regarded as bench marks against the accuracy of the black, blue, brown and red solid lines can be checked.

As the trace distance depends only on $|c_0(t)|^2$, this quantity will be the focus of discussions. In Fig. \ref{Fig: PT}, we show $|c_0(t)|^2$ over time using Eq. \eqref{EDM} (solid lines) and eNRG data (circular dots) for different combinations of $\bar{K}$ and $T$, for $\Delta = 1/210$ (top panel) and $1/843$ (bottom panel). The results exhibit very different behaviors for combinations of $(\bar{K}, T)$ with the same ratio $\bar{K}/T$ . For example, each pair of lines (black, purple) and (blue, brown) corresponds to the same $\bar{K}/T$, but as the magnitude of $|\bar{K}|$ increases, the system becomes less adiabatic, meaning $|c_0(t)|^2$ decreases more rapidly, and quite dramatically so for the second pair. In contrast to the traditional expectation, the amplitude of $\bar{K}$ is more critical for tracking adiabaticity than $\bar{K}/T$. 
For a fixed $\bar{K}$, however, a longer time scale $T$ indeed leads to a more adiabatic dynamic.

Comparison between any two points belonging to the upper and lower panels, with the same ramp-up period T, the same maximum potential $\bar K$, and the same scaled time $t/T$ shows that $|c_0(t)|^2$ become smaller as the gap $\Delta\varepsilon$ is reduced. This can be readily understood: smaller gapes favor creation of particle-hole pairs, which rob probability from the ground state.


\newpage
Figure \ref{Fig: PT} shows that the results from Eq. \eqref{EDM} and eNRG methods align not just near adiabaticity, when $|c_0(t/T)|^2 \geq 0.90$, but even when the evolution is far from adiabaticity, up to $|c_0(t/T)|^2 \approx 0.6$. Since Eq. \eqref{AS} accounts accurately for single-particle-hole excitations, the deviations measure the significance of the two or more particle-hole pairs excitations from the ground state. Figure \ref{Fig: PT} demonstrates that fairly strong potentials, with $ |\bar{K}| $ substantially greater than unity, are required to consider two-particle-hole excited states in the model to accurately represent the evolution. The beown plots in the bottom panel offers a clear example.



Near adiabaticity, Eq. \eqref{AS} is accurate. When combined with Eqs. \eqref{phase_shift}, \eqref{TR}, and \eqref{AS}, it yields the following inequality relating the maximum scattering potential, the ramp-up timescale, and the band gap to the adiabatic threshold parameter:
\begin{eqnarray}\label{DA:RUC}
 \rho{|\bar{K}|}\leq {\eta}\left(\log\left( \frac{4.4\hbar}{\Delta\varepsilon T} \right)\right)^{-1/2}.
\end{eqnarray}

\begin{figure}[hbt!]
		\centering
        \includegraphics[scale=0.77]{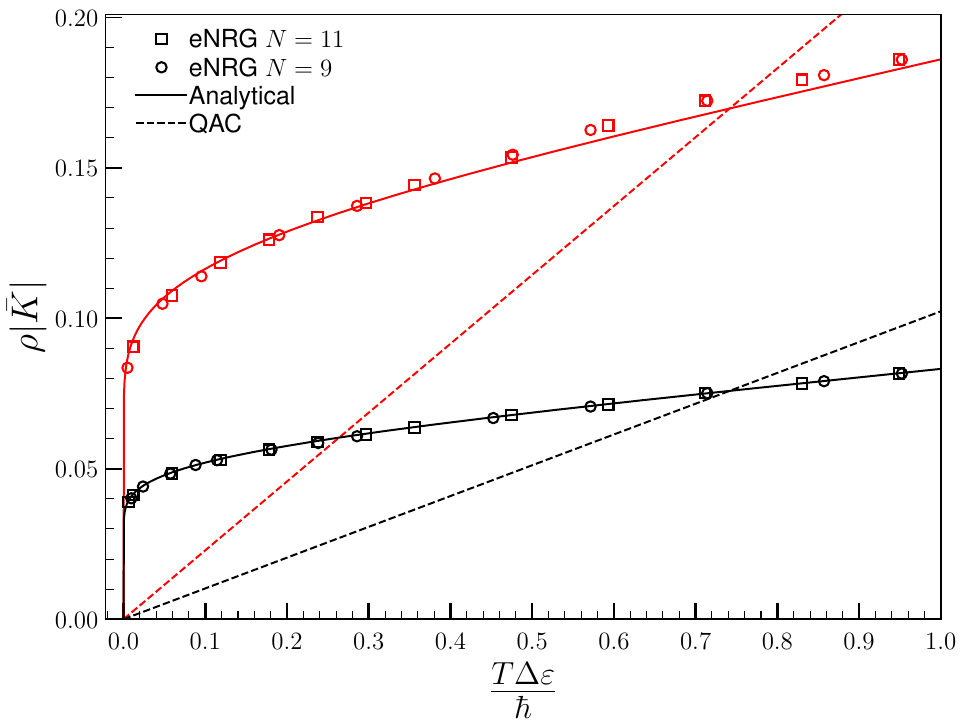}
        \vspace{-0.2cm}
	    \caption{\footnotesize Adiabatic diagram for $N=11$, $\lambda=2$ and $\Delta\varepsilon=1/843$ (black and red square dots), and $N=9$ , $\lambda=2$ and $\Delta\varepsilon = 1/210$ (black and red circular dots). The dots $(\bar{K},T)$ represent the threshold conditions for adiabaticity defined by the trace distance and the threshold $\eta$, for two different values of the threshold, $\eta = 0.1$ (black curves) and $\eta = 0.2236$ (red curves). The QAC criterion is represented by the dashed black and red lines. \Sou}
        \label{Fig: Phase_Diagram}
\end{figure} 

The black lines and dots in  Fig. \ref{Fig: Phase_Diagram} show the region in $\{(\bar{K}, T)\}$ space satisfying the adiabatic condition $D_{\rho}^{T}(t) = \eta = 0.1$ (with threshold $\eta = 0.1$). The region above each line is non-adiabatic, and the region below, adiabatic. 
The solid line represents Eq. \eqref{DA:RUC}, while the circular dots were obtained from the eNRG computations of $|c_0(t)|^2$ for the parameters in the top panel in Fig \ref{Fig: PT}. The curves and circles are in excellent agreement. To compare with the quantum adiabatic criterion (QAC) (black dashed line), we used Eq. \eqref{QAC} and considered only the first excited state $\ket{\varphi_1(t)}$ with $|E_1 - E_0| = \Delta\varepsilon$. Since $\Delta \dot{H} = \frac{\bar{K}}{T} a_0^\dagger a_0$, the condition reduces to
\begin{equation}
    |\bar{K}| \leq \frac{\eta\Delta\varepsilon}{|\bra{\varphi_0} a_0^\dagger a_0 \ket{\varphi_1}|} \frac{T \Delta\varepsilon}{\hbar}.
\end{equation}

From Fig. \ref{Fig: Phase_Diagram}, we can see that compared with Eq. \eqref{DA:RUC} the QAC amounts to a strikingly different condition. For one thing, the QAC depends on the ratio $\frac{|\bar{K}|}{T}$ and not on $(\bar{K}, T)$ separately. Figs. \ref{fig:RPTxN}, \ref{Fig: PT}, and \ref{Fig: Phase_Diagram} are much more sensitive to the strength of the potential $|\bar{K}|$. For this reason, the QAC can be tight, even for weak potentials, provided that $T$ to be small; or it can loose, even for strong potentials, as long as the ramp-up is slow. Equation \eqref{DA:RUC} is consistent with the eNRG results, explicitly showing correctly the dependence of $\bar{K}$ on $T$, for each threshold $\eta$. 
As the black plots shows, for $\frac{\Delta\epsilon T}{\hbar}<\frac{160}{210} = 0.762$, the QAC underestimates the region of adiabaticity, and for $\frac{\Delta\epsilon T}{\hbar} > 0.762$, it attributes adiabaticity to points with nonadiabatic dynamics. Clearly, the QAC fails to track adiabaticity for this problem.

Keeping the discussions about the results in Fig. \ref{Fig: Phase_Diagram}, we compare now Eq. \eqref{DA:RUC} and the eNRG results for different adiabatic thresholds. The circular dots correspond to eNRG results with $N=9$, $\lambda = 2$, and $\Delta\varepsilon = \frac{1}{210}$; the square dots, to $N=11$, $\lambda=2$, and $\Delta\varepsilon=\frac{1}{843}$. For $\eta = 0.1$ (black solid line), the agreement between the eNRG and analytical results is excellent. It follows from Eqs.\eqref{AS} and \eqref{DA:RUC}, that the shape of the curve corresponding to the adiabatic threshold in Fig. \ref{Fig: Phase_Diagram} is the same for all $\frac{\hbar}{\Delta\varepsilon}$. For the higher threshold $\eta = 0.2236$ (red solid line, corresponding to $|c_0|^2 \ge 0.95$), the agreement between both sets of eNRG results shows that the dependence on $\frac{T\Delta\varepsilon}{\hbar}$ persists; the comparison with the analytical results although inferior to that on the black curves, remains very good.

Near adiabaticity, Eq. \eqref{DA:RUC} performed well when compared with the eNRG results. This suggests that Eq. \eqref{DA:RUC} defines the adiabatic regime for any combinations of $\bar{K}$, $T$ and $\frac{\hbar}{\Delta\varepsilon}$. This is confirmed by the universality of the two curves in Fig. \ref{Fig: Phase_Diagram}. We expect therefore this expression to hold as $\Delta\varepsilon \rightarrow 0$, the continuum limit.

An interesting consequence of Eq. \eqref{DA:RUC} is that when the two time scales of the system coincide $T=T_m = \frac{\hbar}{\Delta\varepsilon}$ and $\rho|\bar{K}| \leq \frac{\eta}{\sqrt{4.4}}$, the evolution is always adiabatic. On the other hand, if $\rho|\bar{K}| > \frac{\eta}{\sqrt{4.4}}$, as long as $T \le \frac{\hbar}{\Delta\varepsilon}$, the dynamics is nonadiabatic even for a small ratio $\frac{|\bar{K}|}{T}$, which seems to violate the adiabatic theorem. This characterizes a critical scattering potential
\begin{equation}
    \rho|\bar{K}_c| \equiv \frac{\eta}{\sqrt{4.4}}.
\end{equation}
 This counter intuitive result arises because in a truly gapless fermionic system, the apparent gap $\Delta\varepsilon = \frac{\hbar}{T_m}$ depends on $T_m$, and $T \le \frac{\hbar}{\Delta\varepsilon}$.

However, if the spectrum is discrete with a small real gap $\Delta$, then we need to consider two regimes: when $T_m \ge \frac{\hbar}{\Delta}$ and when $T_m < \frac{\hbar}{\Delta}$. In the first case, when $T_m \geq \frac{\hbar}{\Delta}$, the system has had enough time for even the small excitations with energy $\Delta$ to become perceptible. In this case, $\Delta\varepsilon = \Delta$ and we can modify $\frac{\hbar}{\Delta\varepsilon T}$ to $\frac{\hbar}{T \Delta}$ in Eq. \eqref{DA:RUC}, resulting in $\rho{|\bar{K}|} = {\eta}\left(\log\left( \frac{4.4\hbar}{\Delta ~ T} \right)\right)^{-1/2}$. Since $\frac{\hbar}{T \Delta}$ is no bound by unity, this implies that there exists a $T$ large enough to ensure adiabaticity for any finite $|\bar{K}|$, thus recovering the Adiabatic Theorem. In the case when $T_m < \frac{\hbar}{\Delta}$, the artificial gap is larger than the real gap, $\Delta\varepsilon > \Delta$, and $\frac{\hbar}{\Delta\varepsilon T} \leq 1$, an inequality we have already discussed.

Now that we have a good understanding of the electronic behavior of the system in the presence of the scattering potential, both after a sudden change in the potential and when the scattering potential grows from $0$ to its maximum amplitude, we are ready to return to the hydrogen-surface collision problem. However, unlike the previously discussed situations, the electronic part of the Hamiltonian is not quadratic and cannot be diagonalized into single-particle levels. To circumvent this problem, various approximations are typically applied. However, as we will discuss in the next chapter, these approximations often fail to accurately capture the phenomena we are interested in for this work.

%% file: CAP/Cap5.tex
\chapter{Standard Approximations}

Now, we will return our focus to the atomic-surface collision problem, which, unlike the photoemission problem, cannot be solved analytically or numerically on the basis of single-particle concepts, because the Hamiltonian is not quadratic. We will have to deal with many-body states and start out with a discussion of approximations typically adapted in treatments of many-body Hamiltonians. Unfortunately, in atom-surface collisions, these approximations only provide insights, numerical treatment being required to uncover the physics of the problem. In the following subsections, we will present those approximations and explain why they fail to solve the atom-surface collision problem.

\section{Perturbation Theory}

Let us begin with the perturbation theory, a golden fruit of the 19th century. This theory applies when the Hamiltonian can be written as 
\begin{eqnarray}
 H = H_0 + V,
\end{eqnarray}
where the eigenvalues ${ \ket{\phi_n} }$ and eigenstates ${ E^{(0)}_n}$ of $H_0$ are known, and the perturbation $V$ is weak. ${ \ket{\phi_n} }$ form a complete basis, upon which any state can be projected. Rewrite, the above equation as $H=H_0 + \alpha V$, where $\alpha$ is an accounting parameter, set to unity at the end. The eigenstates and eigenvalues of $H$ can be approximated by
\begin{equation}
\ket{\varphi_n } = \ket{\phi_n} + \alpha\Delta \ket{\varphi_n^{(1)}} + \alpha^2 \Delta \ket{\varphi_n^{(2)}} + ... 
\end{equation}
 and 
\begin{equation}
E_n = E^{(0)}_n + \alpha \Delta E^{(0)}_n + \alpha^2 \Delta E^{(2)}_n + ....
\end{equation}
If $V$ is a weak perturbation, the main contributions to the pair $(\ket{\varphi_n}; E_n)$ come from $(\ket{\phi_n}; E_n^{(0)})$, which is associated with the first order in the $\alpha$ expansion, and the series converges. However, if $V$ is strong, this expansion fails.

In the problems we are dealing with, it is possible to write the Hamiltonian as $H = H_0 + \Delta H$, where $\Delta H$ is the scattering potential (photoemission) or the potential plus coupling to the atomic levels (atom-surface collision). As seen in the Chapters 3 and 4, however, the interaction with the conduction band renormalizes the potential by a factor proportional to the number of participating energy scales. In the continuum limit, that number grows to infinity, and even the first-order term diverges. The coupling to the atomic levels is likewise renormalized.


\section{Adiabatic Approximation}

Understanding adiabaticity is crucial in nearly all areas of physics as it serves as a powerful tool for dealing with time-dependent properties. Quantum adiabatic processes, in particular, have raised much interest due to their relevance in quantum computing, quantum thermodynamics, and various nonequilibrium condensed-matter phenomena.

The pioneers in the subject were Born and Fock, in 1928 \cite{Born}. Their paper establishes the Adiabatic Theorem, which states that if a quantum system is initially one of its eigenstates and the external parameters of the system change slowly enough, the system will remain in its instantaneous eigenstate throughout the process. At zero temperature, the Quantum Adiabatic Criterion (QAC) is expressed by the inequality
\begin{eqnarray}\label{QAC}
\frac{\hbar|\bra{0} \Dot{H}(t) \ket{n}|}{(E_n - E_0)^2} \ll 1.
\end{eqnarray}
Here, $\dot{H}$ represents the time derivative of the Hamiltonian, while $\ket{0}$ and $\ket{n}$ denote the ground state and an excited state of $H$ with corresponding eigenenergies $E_0$ and $E_n$. This expression provides an estimate of the occupation of the excited state $\ket{n}$. If it tends to zero, the occupation of the ground state tends to one, indicating that the evolution is adiabatic. 

The QAC was initially derived under the assumption of a discrete and non-degenerate spectrum \cite{Born}. The theorem has been generalized in various ways. Nevertheless, even after nearly a century of work, far from trivial questions concerning the applicability of the QAC to many-body Hamiltonians remain unanswered. Moreover, the QAC is applicable only at zero temperature, and breaks down in the continuum limit.

The adiabatic approximation takes form when, upon examining Eq. \eqref{QAC}, we conclude at first look that only the eigenstates with energies close to the ground state are significant. If this proposition is true, then there exists an energy cut-off $E_{\mathrm{cut-off}}$, beyond which only eigenstates with energies below it are considered important $\{ \ket{\varphi_n^*} \in \{ \ket{\varphi_n} \} / E_n \leq E_{\mathrm{cut-off}} \}$. However, as discussed in the introduction and in the initial chapters, high-energy excited states can also play a role in the problems we are dealing with here. This is a consequence some characteristics of the system, like the presence of the Anderson orthogonality and the rapid changes in the Hamiltonian, indicating that the adiabatic approximation may not be suitable for these type of problems.

\newpage
\section{Born-Oppenheimer approximation}

The Born-Oppenheimer approximation (BO) is traditionally employed in problems involving atomic nuclei and electrons and successfully describes many systems, including the metals. This approximation decouples the electronic wave functions from the atomic motion under the assumption that the two subsystems are governed by distinct characteristic time scales.  Typically, the time scale of electronic processes is much shorter than that of atomic processes. However, in hydrogen-surface collision, this approximation is not warranted. This is because the characteristic electronic times are defined by the uncertainty principle $E_{\mathrm{elec.}}.\Delta t \sim \hbar$ and can hence be as long as the characteristic times for the atomic motion. 

To investigate further, let us start out with the Hamiltonian \eqref{H_Colision}, which can be written more simply as 
\begin{equation}
H_c(z) = \frac{P^2_z}{2M} + H_e(z),
\end{equation}
where $H_e(z)$ represents the electronic contribution, while the atomic contribution is the kinetic term $\frac{P^2_z}{2M}$. Let us then consider $\{ \ket{\varphi_n(z)} \}$ as the set of eigenstates and $\{ {E_n(z)} \}$ as the eigenenergies of $H_c(z)$. Then, we can expand the time-dependent wave function as
\begin{equation}\label{AUX: LN1}
\ket{\Psi(z,t)} = \sum_n \chi_n(z,t) \ket{\varphi_n(z)},
\end{equation} 
where $\chi_n(z,t)$ is the nuclear wave function for each possible electronic state.

Although the time-dependent Schrödinger equation 
\begin{equation}
    i \hbar \partial_t \ket{\Psi(z,t)}= \left[  \frac{P^2_z}{2M} + H_e(z) \right]  \ket{\Psi(z,t)}
\end{equation}
cannot be solved analytically, the expansion \eqref{AUX: LN1} leads to an analogous differential equation for the nuclear wave functions $\chi_n(z,t)$
\begin{equation}\label{ATOMIC_SEQ}
\begin{aligned}
i \hbar {\partial_t \chi_n(z,t) } &= E^{e}_n (z) \chi_n(z,t) - \frac{\hbar^2}{2M} \partial_z^2 \chi_n(z,t)
 \\  
-&\frac{\hbar^2}{M} \sum_{n'} \partial_z \chi_{n'} (z,t) \langle \varphi_n(z) \lvert \partial_z \lvert \varphi_{n'}(z) \rangle \\
-& \frac{\hbar^2}{2M} \sum_{n'} \chi_{n'} (z,t) \langle \varphi_n(z) \lvert \partial_z^2 \lvert \varphi_{n'}(z) \rangle.
\end{aligned}
\end{equation}

On the right-hand side of Eq. \eqref{ATOMIC_SEQ}, there are four terms. The energy in the first term, $E^{e}_n (z)$, is known as the Born-Oppenheimer Potential Energy Surface. The Potential Energy Surface is analogous to the potential in the Schrodinger equation. The second term is analogous to the kinetic energy $\frac{\hbar^2 k_n^2}{2M}$, while the remaining two terms couple the atomic wave functions for different Potential Energy Surfaces. 
 
\newpage

Applying the differential operator $\partial_z$ to the  expression $H_e(z) \ket{ \varphi_n(z)} = E_n(z) \ket{ \varphi_n(z)}$, it is easy to show that 
\begin{equation}
\bra{\varphi_m} \partial_z \ket{\varphi_n} = \frac{-\bra{\varphi_m} \partial_z H_e(z) \ket{\varphi_n}}{E_m(z) - E_n(z) }
\end{equation}
and 
\begin{equation}
\bra{\varphi_m} \partial_z^2 \ket{\varphi_n} \approx \frac{-\bra{\varphi_m} \partial_z^2 H_e(z) \ket{\varphi_n}}{E_m(z) - E_n(z)}+ \frac{\bra{\varphi_m} (\partial_z H_e(z) )^2 \ket{\varphi_n}}{(E_m(z) - E_n(z))^2} 
\end{equation}
for states on different surface energies $E_m(z) \neq E_n(z)$. For typical problems such as isolated atoms or collisions between two atoms, these energy differences are typically on the scale of eV, corresponding to time scales of femtoseconds (fs).

Let us now return to Eq. \eqref{ATOMIC_SEQ}. We can use the discussion from the last paragraph to estimate the energy scale of each term. The energy scale of the atom is on the order of $\frac{\hbar^2 k_0^2}{2M}$, where the off-diagonal electronic contributions in the Hamiltonian are on the energy scale of 
\begin{equation}
\frac{\hbar^2 k_0}{M}\frac{|\bra{\varphi_m} \partial_z H_e(z) \ket{\varphi_n}|}{|E_m(z)-E_n(z)|}, \\
\frac{\hbar^2}{2M}\frac{|\bra{\varphi_m} \partial_z^2 H_e(z) \ket{\varphi_n}|}{|E_m(z) - E_n(z)|}, \\ ~\mathrm{and}~
\frac{\hbar^2}{2M}\frac{\bra{\varphi_m} (\partial_z H_e(z) )^2 \ket{\varphi_n}}{(E_m(z) - E_n(z))^2}.
\end{equation}
Since the energy differences are typically $ |E_m(z) - E_n(z)| \sim 1 $ eV, and considering the lightest atom, hydrogen, with a mass of approximately $ M \sim 938 \, \mathrm{MeV/c^2} $, these energy scales are usually very small, and the corresponding time scales very big, by the Heisenberg uncertainty principle.

In this exact situation, the Born-Oppenheimer approximation is valid. The differences in time scales effectively decouple the atomic motion from the electronic motion. Therefore, the time-dependent Schrödinger equation simplifies to:
\begin{equation}\label{BO-APROX}
\begin{aligned}
&i \hbar {\partial_t \chi_n^{(BO)}(z,t) } \approx E^{e}_n (z) \chi_n^{(BO)}(z,t) - \frac{\hbar^2}{2M} \partial_z^2 \chi_n^{(BO)}(z,t). \\
\end{aligned}
\end{equation}
Which concludes our simple derivation of the BO approximation for atoms.

However, when electronic time scales are comparable to the atomic time scale, as in metallic bands, these terms cannot be ignored. The last two terms in Eq. \eqref{ATOMIC_SEQ}, which are dropped in the Born-Oppenheimer approximation, explain the non-adiabaticities and the nuclear energy losses, allowing the atom to become trapped at the surface. In the BO approximation, a single Potential Energy Surface controls the dynamics of the collision. The atom is simply reflected by the surface and never gets trapped on it, as Eq. \eqref{BO-APROX} shows. The BO approximation is therefore unsuitable for our proposes.

\newpage

\section{Fermi's Golden Rule}

Given a initial Hamiltonian $H_0$ with known eigenstates ${ \ket{\phi_n} }$ and eigenenergies ${ {E^{(0)}_n} }$, and introducing a small time-dependent harmonic perturbation $V(t) = Ve^{+i\omega t} + \hc$, we have a time-dependent Hamiltonian $H(t) = H_0 + V(t)$. Considering the system initially in a pure eigenstate $\ket{\phi_i}$, so that $\ket{\Psi(0)} = \ket{\phi_i}$, it is not difficult to show that the transition probability to a final state $\ket{\phi_f}$ is
\begin{equation}\label{TP_FGR}
\begin{aligned}
|\braket{\phi_f}{\Psi(t)}|^2 \approx 4|\bra{\phi_f} V \ket{\phi_i}|^2 \frac{\sin^2\left( \frac{\left(E_f^{(0)} - E_i^{(0)} - \hbar\omega\right)t}{2\hbar}   \right)}{ \hspace{0.5cm} \left(E_f^{(0)} - E_i^{(0)} - \hbar\omega\right)^2}
\end{aligned}.
\end{equation}

The transition rate is the ratio $R_{if} = \frac{|\braket{\phi_f}{\Psi(t)}|^2}{t}$. It follows from Eq. \eqref{TP_FGR} that, unless the denominator vanish, this ratio approaches zero at long times. Since
\begin{equation}
    \lim_{t\rightarrow \infty} \frac{\sin^2(\alpha t)}{(\alpha)^2 t} \equiv \pi \delta(\alpha),
\end{equation}
we associate the long-time ratio with the delta function: 
\begin{equation}\label{FGR}
\begin{aligned}
R_{if} = \lim_{t \rightarrow \infty} \frac{|\braket{\phi_f}{\Psi(t)}|^2}{t}   \approx \frac{2 \pi}{\hbar} |\bra{\phi_f} V \ket{\phi_i}|^2 \delta\left(E_f^{(0)} - E_i^{(0)} - \hbar\omega\right)
\end{aligned}.
\end{equation}
This expression is known as the Fermi's golden rule.

Fermi's golden rule \cite{Sakurai_Napolitano_2020} states that if a system starts in an initial eigenstate $ \ket{\phi_i} $ of $ H_0 $, then, for long times after a harmonic perturbation $ V(t) = Ve^{+i\omega t} + \text{h.c.} $ is applied, the transition rate $ R_{if} $ is constant, if the final state satisfies the energy conservation condition $ E_f^{(0)} = E_i^{(0)} + \hbar\omega $. These principles can be used to estimate and explain the transition from the initial state $ \ket{\phi_i} $ to the final state $ \ket{\phi_f} $. However, this rule is quantitatively accurate only when $ V $ is a small perturbation, which is not the case for the problems considered here. Nonetheless, it can still be applied qualitatively to explain transitions between two quantum states; Chapter 2 has already offered an example.

In summary, the approximations often used in treatments of non-equilibrium problems cannot answer the questions we are interested in. This understood, we treat the atom-surface collision problem numerically. Specifically, our approach comprises two numerical methods, the NRG and the Crank-Nicolson procedure, which are discussed in Chapter 6.


%% file: CAP/Cap6.tex
\chapter{Mathematical tools}

Before proceeding further, let us provide a brief overview of the mathematical tools required to numerically solve the atom-surface model. The Crank-Nicolson procedure describes the temporal evolution of a physical system. The Numerical Renormalization Group (NRG) allows the numerical diagonalization of the electronic Hamiltonian with no additional approximations other than the discretization and truncation. Projection of one basis, numerically obtained via the NRG, onto another basis provide the information needed to describe the temporal evolution. Since high-energy many-body states can significantly influence time-dependent properties, such as those discussed in Chapter 3 on photoemission, we developed a specialized truncation method for the NRG.

\section{Crank-Nicolson method}

The Crank-Nicolson method \cite{crank_nicolson_1947} is a remarkably stable finite-differences technique, first developed to describe heat conduction. We will use it to solve the time-dependent Schrödinger equation, for a time-dependent Hamiltonian $H(t)$. We can define the temporal evolution operator $U(t,t_0)$, which takes a physical system from an initial time $t_0$ to a time $t$, i.e., $\ket{\Psi(t)} = U(t,t_0) \ket{\Psi(t_0)}$. The temporal evolution operator obeys the partial differential equation $i \hbar \frac{\partial U(t,t_0)}{\partial t} = H(t) U(t,t_0)$. For an infinitesimal time variation, the equation become $U(t+\delta t, t) = e^{-iH(t) \delta t / \hbar}$. Next, discretize the time axis by defining \( t_j = \Delta t \, j \), where \( \Delta t \) is a small time interval. The discrete temporal evolution operator can now be expressed approximately as
\begin{align}\label{CN:1}
U(t_j +\Delta t ; t_j) \approx e^{-iH(t_j + \Delta t/2) \Delta t /\hbar} = \frac{1- \frac{i\Delta t }{2\hbar} H(t_j + \Delta t /2)}{1 + \frac{i\Delta t }{2\hbar} H(t_j + \Delta t /2)} + \mathcal{O} ( \Delta t^3). 
\end{align}
This equality, instead of the expansion $U(t_j +\Delta t ; t_j) \approx 1 - i\frac{\Delta t }{\hbar}  H(t_j) $ constitutes the core of Crank-Nicolson approach.

This expansion preserves the normalization of the wave function; choosing the median time between $t_j$ and $t_{j+1}$ reduces the error associated with discretization. Thus, using Eq. \eqref{CN:1}, as $\ket{\Psi(t)} = U(t,t_0) \ket{\Psi(t_0)}$, we can numerically find the temporal evolution of a physical system through the CN procedure, described by:
\begin{align}\label{CN:2}
\left(1 + \frac{i\Delta t }{2\hbar} H(t_j + \tau/2)\right)\ket{\Psi(t_{j+1})} = \left(1- \frac{i\Delta t }{2\hbar} H(t_j + \Delta t /2)\right) \ket{\Psi (t_j)}.
\end{align}

\newpage
\section{The Numerical Renormalization Group}

The NRG method is a powerful numerical technique. In essence, it discretizes model Hamiltonians that typically describe correlated impurity orbitals coupled to states in the conduction band structure. Although the orbitals in the problems we are interested in may not necessarily describe impurities, the procedure applies as well. Diagonalizing a Hamiltonian using the NRG method involves two main steps. First, the conduction band is discretized logarithmically, which preserves the scale invariance of the conduction band. This step is parameterized by dimensionless parameter $\Lambda > 1$. The resulting discrete states are projected onto the NRG basis using the Lanczos transformation. These two steps are extensively discussed in detail in \cite{RevModPhys.80.395}. Ultimately, we obtain a scaled and truncated Hamiltonian of the form:
\begin{align}\label{NRG:Eq1}
H_{{\tilde N}}^\mathrm{NRG}= \frac{1}{D_{\tilde N}}\left(H_0({d,f_0 }) + \sum_{n=0}^{\tilde N-1}\tau_n(\fd{n}\fn{n+1}+\hc)\right),
\end{align}
where $\{f_n\}$ represents the  Lanczos transformed operators of the band. $\tilde N$ is the number of links in the Wilson chain. The $H_0({d, f_0})$ term contains only operators $d$ and $f_0$ and describe the impurity and how its coupling to the band. For instance, $H_0({d, f_0}) = \varepsilon_d d^\dagger d + V(d^\dagger f_0 + f_0^\dagger c_d)$ is the SIAM. The coupling $\tau_n \sim \Lambda^{-n/2}$ between the NRG operators decays exponentially with $n$, $D_{\tilde N} = D\frac{1-\Lambda^{-1}}{\log \Lambda}\Lambda^{-({\tilde N}-1)/2}$ denotes the scaled bandwidth, and $D$ is the initial bandwidth.

It can be observed that $H_{\tilde N}^{NRG}$ couples the impurity states $d$ and the band operator $f_0$ through $V(d^{\dagger}f_0 + \text{h.c.})$ terms in $H_0({d, f_0})$, while the operators $f_n$ couple to each other through the hopping term
\begin{align}\label{NRG:Acoplamento}
\tau_n = \left(D\frac{1-\Lambda^{-1}}{\log \Lambda}\right) \frac{\left(1-\Lambda^{-n-1}\right)\Lambda^{-n/2}}{\sqrt{1-\Lambda^{-2n-1}}\sqrt{1-\Lambda^{-2n-3}}}= \left(D\frac{1-\Lambda^{-1}}{\log \Lambda}\right) \Lambda^{-n/2} + \mathcal{O} (\Lambda^{-n}).
\end{align}
Here, $\Lambda^{-n}$ decays rapidly for typical $\Lambda$s.

\begin{figure}[htb!]
\centering
\begin{tabular}{ll}
\hspace{-0.5 cm}
\includegraphics[scale=0.4]{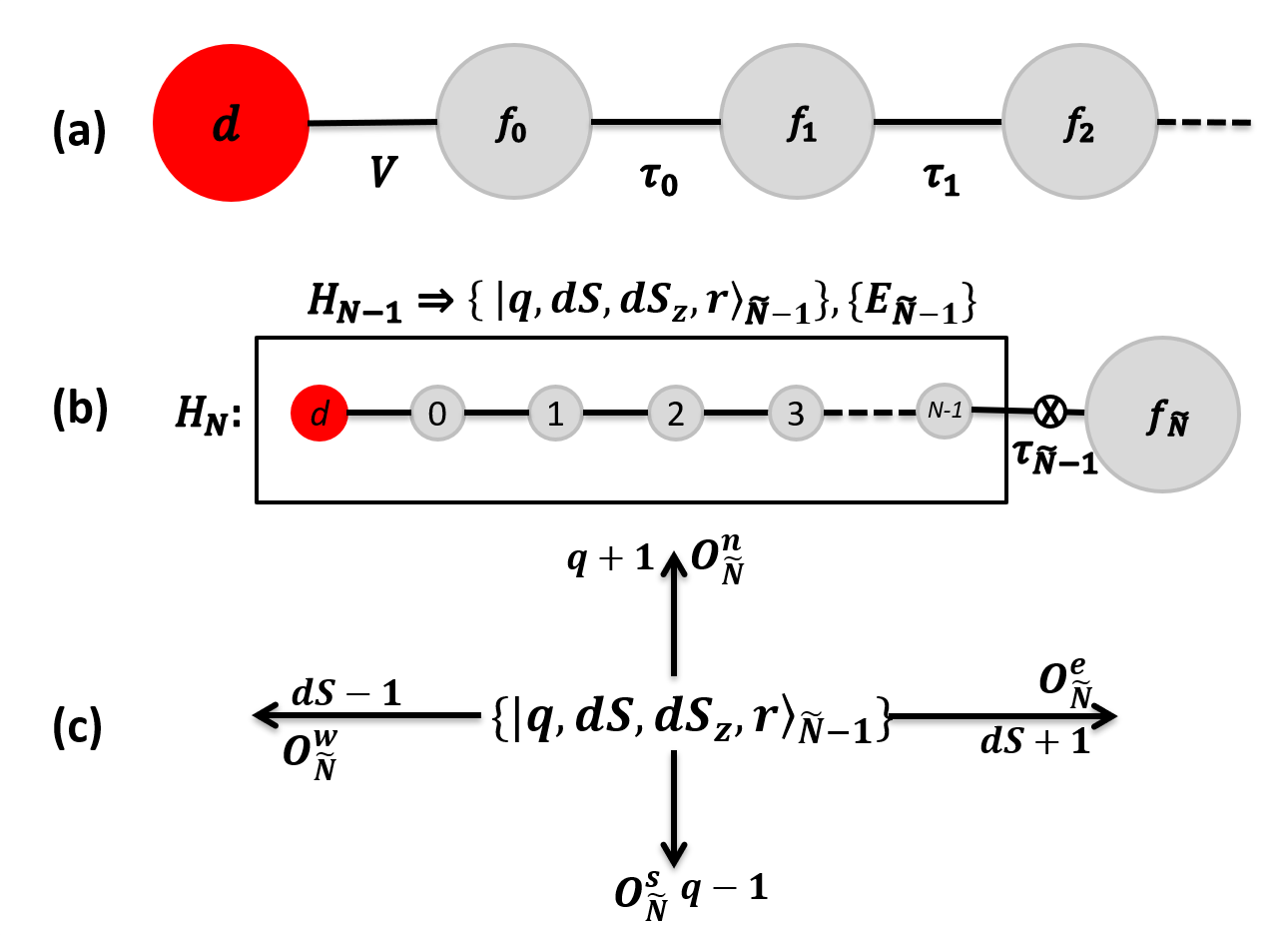}
& \hspace{-0.2cm}
\includegraphics[scale=0.40]{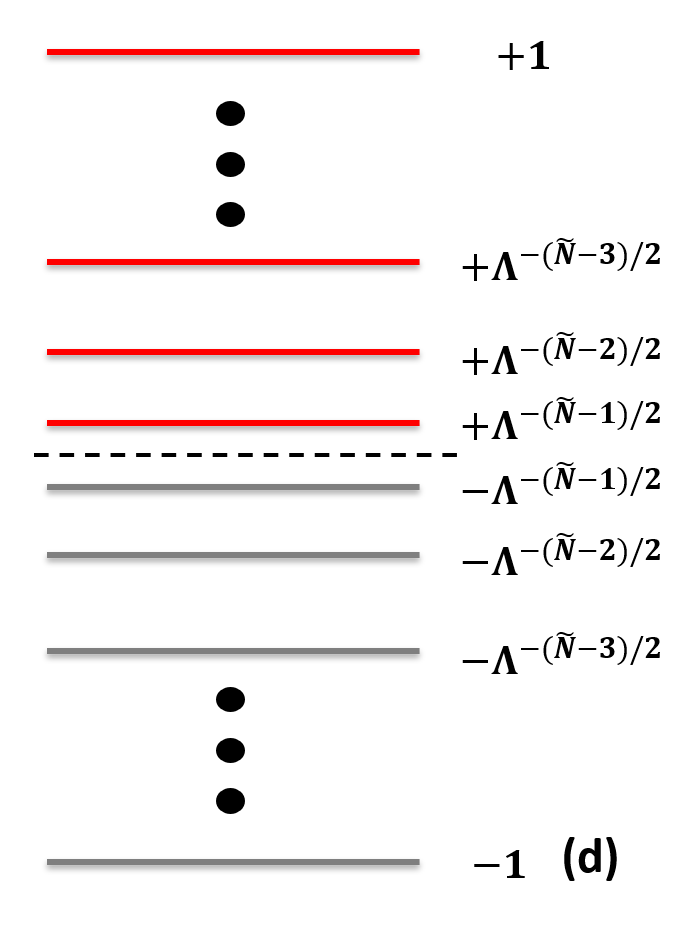}
\end{tabular}
\vspace{-0.4cm}
\caption{\footnotesize (a) Interpretation of H \eqref{NRG:Eq1} as a chain, coupling the impurity orbital $d$ to the first conduction orbital $f_0$ by the term $V$ and the conduction sites to each other by the coupling $\tau_n$. (b) Assemblage of the Hamiltonian in iteration $\tilde N$ from the eigenstates and eigenenergies of iteration $\tilde N-1$; Procedure that numerically enables the diagonalization of $H_{\tilde N}$. (c) Assembly of the primitive base of interaction $\tilde N$ from the proper base of interaction $\tilde N-1$. We define a \textit{gender} for the new base, north, south, east, or west according to the change of charge and spin associated with the operation in each direction. (d) After the NRG procedure, the energy spectrum of the band, which was initially continuously distributed from $-1$ to $+1$ (in units of $D$), is now logarithmically distributed in powers of $\Lambda^{-1/2}$, from $1$ to $\Lambda^{-(\tilde N - 1)/2}$ and $-1$ to $-\Lambda^{-(\tilde N - 1)/2}$. \Sou}
\label{NRG}
\end{figure}

Figure \ref{NRG} (a) depicts the Hamiltonian \eqref{NRG:Eq1} schematically. As a consequence of the discretization, the energy spectrum of the band, which was initially continuously distributed from $-1$ to $+1$ (in units of $D$), is now logarithmically distributed in powers of $\Lambda^{-1/2}$, from $1$ to $\Lambda^{-(\tilde N - 1)/2}$, as shown in Fig. (\ref{NRG}) (d)). This property allows truncation of the Hamiltonian at a maximum iteration $\tilde{N}$, assuming that energy differences below $\Lambda^{-\tilde{N}/2}$ are not important. This truncation is necessary since the original NRG basis, despite being discrete, is infinite.

In Eq. \eqref{NRG:Eq1}, separating the terms that depend on $f_N$ from the others, we can rewrite this Hamiltonian in the form
\begin{align}\label{NRG:Eq2}
H_{\tilde N} = \sqrt\Lambda H_{\tilde N-1} + \left(\frac{\tau_{\tilde N-1}}{D_{\tilde N}}\right)(\fd{\tilde N-1}\fn{\tilde N}+\hc),
\end{align}
expressing the Hamiltonian at iteration $\tilde N$ in terms of the Hamiltonian at iteration $\tilde N-1$ plus a term that couples the operators $f_{\tilde N-1}$ and $f_{\tilde N}$. Equation \eqref{NRG:Eq2} defines an iterative procedure, as illustrated in Fig.~\ref{NRG}(b).

Following this approach, we start from iteration 0 with \( H_0({ c_d, f_0 }) \), diagonalize it, and use the resulting eigenstates and eigenvalues to construct iteration 1. This process is repeated until iteration \( N \), allowing us to determine the eigenenergies of \( H_{\tilde{N}} \) and the relevant matrix elements. This strategy will be employed to numerically diagonalize the Hamiltonian \eqref{NRG:Eq1}.
\vspace{-0.25cm}

A crucial step of the NRG procedure is the construction of the base of iteration $\tilde N$ from that of $\tilde N-1$. For this, we combine the eigenstates of the previous iteration $\tilde N-1$ with the operators $f_{\tilde N}$. One way to do this is to use the operator $O_{g}^{\tilde N}$, which takes each eigenstate of the previous iteration and combines it with the operators $f_{\tilde N}$ in order to obtain 4 new states, adding $\pm 1$ in the charge $q$ or twice the spin $dS$. These new states, each with its gender $g$, can be associated with directions analogous to the compass rose (see Fig. (\ref{NRG})(c)), north $n$ ($q' = q + 1$), east $e$ ($dS'= dS + 1$), south $s$ ($q' = q - 1$), and west $w$ ($dS'= dS - 1$). Thus, we can define this transformation as $\lvert q; 2S; 2S_z; p; g \rangle_{\tilde N} = O^{\tilde N}_{g}\lvert q'; 2S'; 2S'z; r \rangle_{\tilde N-1}$, where for each direction or gender, this transformation results in:
\begin{equation}\label{NRG:Base}
  \begin{aligned}  
&\lvert q; dS; dS; p \rangle_{\tilde N}^{n} = f_{\tilde N\upA}^{\dagger}f_{\tilde N\downA}^{\dagger} \lvert q-1; dS; dS; r \rangle_{\tilde N-1},\\
&\lvert q; dS; dS; p \rangle_{\tilde N}^{e} =  f_{\tilde N\upA}^{\dagger}\lvert q; dS-1; dS-1; r \rangle_{\tilde N-1},\\
&\lvert q; dS; dS; p \rangle_{\tilde N}^{s} = \lvert q+1; dS; dS; r \rangle_{\tilde N-1},\\
&\lvert q; dS; dS; p \rangle_{\tilde N}^{w} = \sqrt{\frac{2S+1}{2S+2}} f_{\tilde N\downA}^{\dagger}\lvert q; dS+1; dS+1; r \rangle_{\tilde N-1} \\& \hspace{2.5cm} -  \sqrt{\frac{1}{2S+2}} f_{\tilde N\upA}^{\dagger}\lvert q; dS+1; dS-1; r \rangle_{\tilde N-1}.
\end{aligned}
\end{equation}
Where we choose $S_z = S$, since neither problem we are dealing with depends on the spin component $S_z$, or in other words $[H,S_z]=0$.

Now, using this new primitive base ${\lvert q; 2S; 2S_z; p; g \rangle_{\tilde N}}$, for each $(q,dS)$ sector, we can rewrite the Hamiltonian \eqref{NRG:Eq2} as:
\begin{align}\label{NRG:Itera}
H_{\tilde N,(p',p | g', g)} = \sqrt \Lambda E_{\tilde N-1,r} \delta\PC{r'_{p',g'}; r_{p,g}} + \PC{\frac{\tau_{\tilde N-1}}{D_{\tilde N}}} \PC{\mathcal{M}_{(p',p | g', g)}+\mathcal{M}^*_{(p,p' | g, g')}},
\end{align}
where $p$ refers to a state in the primitive base, $r$ to the eigenstate of the previous iteration that generated the state $p$, and $g$ is the gender of the state $p$. The matrix $\mathcal{M}$ is obtained by sandwiching, using this new primitive base, the operator $\fd{\tilde N-1}\fn{\tilde N}$ and depends on the genders of these bases. From the definition of the primitive base in Eq. \eqref{NRG:Base}, it is not difficult to show that the non-zero elements of the matrix $\mathcal{M}$, for the pair of genders $(g',g)$, take the form:
\begin{equation}\label{NRG:Matrix}
\begin{aligned}
&\langle q; dS; p'; g' \lvert f_{\tilde N-1}^{\dagger} f_{\tilde N} \lvert q; dS; p ; g \rangle_{\tilde N}= \hspace{6cm} \\
&\hspace{2cm} (s,e): \langle q+1; dS; r' \lvert f_{\tilde N-1}^{\dagger} \lvert q; dS-1; r \rangle_{\tilde N-1},\\
&\hspace{2cm} (e,n): \langle q; dS-1; r' \lvert f_{\tilde N-1}^{\dagger} \lvert q-1; dS; r \rangle_{\tilde N-1},\\
&\hspace{2cm} (s,w): \langle q+1; dS; r' \lvert f_{\tilde N-1}^{\dagger} \lvert q; dS+1; r \rangle_{\tilde N-1},\\
&\hspace{2cm} (w,n): -\sqrt{\frac{2S+2}{2S+1}} \langle q; dS+1; r' \lvert f_{\tilde N-1}^{\dagger} \lvert q-1; dS; r \rangle_{\tilde N-1}.\\
\end{aligned}
\end{equation}

Therefore, to find the elements of the matrix $\mathcal M$ and, consequently, assemble the Hamiltonian of interaction $\tilde N$, we need the matrix elements of the operator $f_{\tilde N-1}^{\dagger}$ calculated at iteration $\tilde N-1$. With Eq. \eqref{NRG:Base} and a few manipulations, we can calculate the non-zero elements of this operator via
\begin{equation}\label{NRG:Matrix_2}
\begin{aligned}
\langle q+1; dS+d\sigma; p'; g' \lvert f_{N'}^{\dagger} \lvert q; dS; p; g \rangle_{N'} &= \\
(\mathrm{for}~d\sigma = +1): ~~~& \delta_{(r',r)}\PR{{\delta_{(g',e)}\delta_{(g,s)} + \sqrt{\frac{2S+1}{2S+2} }\delta_{(g',n )}\delta_{(g,w )} }}, \\
(\mathrm{for}~d\sigma = -1): ~~~& \delta_{(r',r)}\PR{{\delta_{(g',w)}\delta_{(g,s)} - \sqrt{\frac{2S+1}{2S} }\delta_{(g',n )}\delta_{(g,e)} }}. \\
\end{aligned}
\end{equation}
To calculate the matrix elements of $f_{N'}^\dagger$ on the proper base ${\lvert q; dS; r \rangle_{N'}}$, simply multiply the matrix elements on the primitive by the rotation matrix $\mathcal{D}^{(N')}$ resulting from the diagonalization.
\begin{equation}
    \lvert q; dS; r \rangle_{N'} = \sum_{p} \mathcal{D}_{r,p}^{(N')}(q,dS)\lvert q; dS; p \rangle_{N'}.
\end{equation}

These are all the ingredients needed to numerically diagonalize a Hamiltonian using the NRG procedure. In summary, at each iteration $n$, starting from $n=1$, we assemble the primitive base $\lvert q; dS; r \rangle_{n}$ from the proper base of iteration $n-1$, and then the Hamiltonian $H_n$ of the iteration is numerically diagonalized and the eigenenergies $E_{n,r}$ and non-zero elements of the operator $f_n^{\dagger}$ are saved for the next iteration, where they will be used to assemble $H_{n+1}$ (Eq. \ref{NRG:Eq2}). This sequence of steps is applied until the maximum iteration $n = N$, which meets the truncation criterion adopted for each problem.

The concept of fixed points plays capital role in renormalization-group theory \cite{RevModPhys.47.773}. From Eq. \eqref{NRG:Eq2} we can regard the creation of the new iteration $H_{N+1}$ from $H_N$ as a renormalization group transformation $\mathcal{T}(H_{N'}) = H_{N'+1}$. A fixed point Hamiltonian $H^*$ satisfies $\mathcal{T}( H^*) = H^*$. In the NRG transformation, the charge-spin sectors into which the spectrum of the Hamiltonian splits change when the transformation changes from odd to even iterations or vice-versa. For this reason, to work with the same charge-spin sector, we must apply the NRG transformation in Eq. \eqref{NRG:Eq2} twice: $\mathcal{T}^2( H_{\tilde N-1}^*) = H_{\tilde N+1}^*$. If the energy spectra of $H_{\tilde N-1}^*$ and $H_{\tilde N+1}^*$ are the same, then $H^*$ is a fixed point of the NRG transformation.


The SIAM model exhibits four well-known fixed points depending on the energy scale \cite{RevModPhys.47.773,RevModPhys.80.395}. First is the free orbital regime, where the quantum state is a superposition of the impurity orbital with $n_d=1, 0$, or $2$. Second is the mixed-valence fixed point. Third is the local moment regime, where the quantum state comprises only the impurity states with $n_d=1$ with spin up or down. Finally, for small energy scales, comes the strong coupling regime, where the impurity forms a ground state singlet with the $f_0$ orbital. The transition form the local moment to the strong coupling is known as the Kondo temperature $T_K$ \cite{Haldane_1978}:
\begin{equation}\label{Kondo}
\begin{aligned}
T_K = \sqrt{\frac{\pi \rho V^2 U }{4}} \exp\left( - \frac{|\varepsilon_d||\varepsilon_d + U|}{\rho V^2 U} \right)
\end{aligned}
\end{equation}

We have briefly explained how we will apply the NRG method to diagonalize a family of Hamiltonian of orbitals coupled to a conduction band, and independent of the $S_z$ component of the spin. Certain theoretical aspects of NRG have not been addressed, and mathematical passages have been hidden to avoid cluttering the text. The NRG code we will use was written in C++ and simultaneously diagonalizes up to two Hamiltonians, which will be necessary to calculate the projection between two bases, which in turn is indispensable to describe the temporal evolution.

\section{Projections between two different NRG basis}

Given two distinct Hamiltonian, formed by the same orbitals coupled to the same conduction band, but with different coupling terms and/or energy levels, we want to discover here what is the projection of the basis that diagonalizes the first $(1)$ onto the basis that diagonalizes the second $(2)$. One way to do this is by writing the proper basis as
\begin{align}\label{proj:Eq1}
\lvert q; dS; r; (i) \rangle_{N'} = \sum_{p} \mathcal{D}_{r,p}^{(N')(i)}(q,dS) O_{g(p)}^{N'} \lvert q'(p); dS'(p); l(p); (i) \rangle_{N'-1},
\end{align}
where the primitive basis can be written as $\lvert q; dS; p; (i) \rangle_{N'} = O_{g(p)}^{N'} \lvert q'(p); dS'(p); l(p); (i) \rangle_{N'-1}$ (transformation defined in Eq. \eqref{NRG:Base}), in terms of the proper basis of the previous iteration. Note that the inner product between two proper states from different bases $\langle q; dS; p'; (1) \lvert q; dS; p; (2) \rangle_{N'}$ is only nonzero if the genders of the primitive states are the same, since the operator $O_g^{N'}$ has only combinations of the operators $f_{N'}$. Therefore, we can write the projection as
\begin{equation}\label{proj:Eq2}
\begin{aligned}
\langle q; dS; r';& (1) \lvert q; dS; r; (2) \rangle_{N'} = \\
&\sum_{p,p'} \PC{\mathcal{D}_{r,p}^{(N')(1)}(q,dS)}^* \mathcal{D}_{r,p}^{(N')(2)}(q,dS) \delta_{g'(p'),g(p)} \langle q'; dS'; l'; (1) \lvert q'; dS'; l; (2) \rangle_{N'-1},
\end{aligned}
\end{equation}
from iteration $N'$ in terms of the projection of the previous iteration. This projection can be obtained by repeating this procedure until reaching iteration $0$, where the Hamiltonian depends only on the same operators $f_0$ and $d$, and therefore the same basis, i.e. 
\begin{eqnarray}
\langle q; dS; p'; (1) \lvert q; dS; p; (2) \rangle_{0} = \delta_{p,p'}.
\end{eqnarray}

\begin{figure}[hbt!]
		\centering
        \includegraphics[scale=0.45]{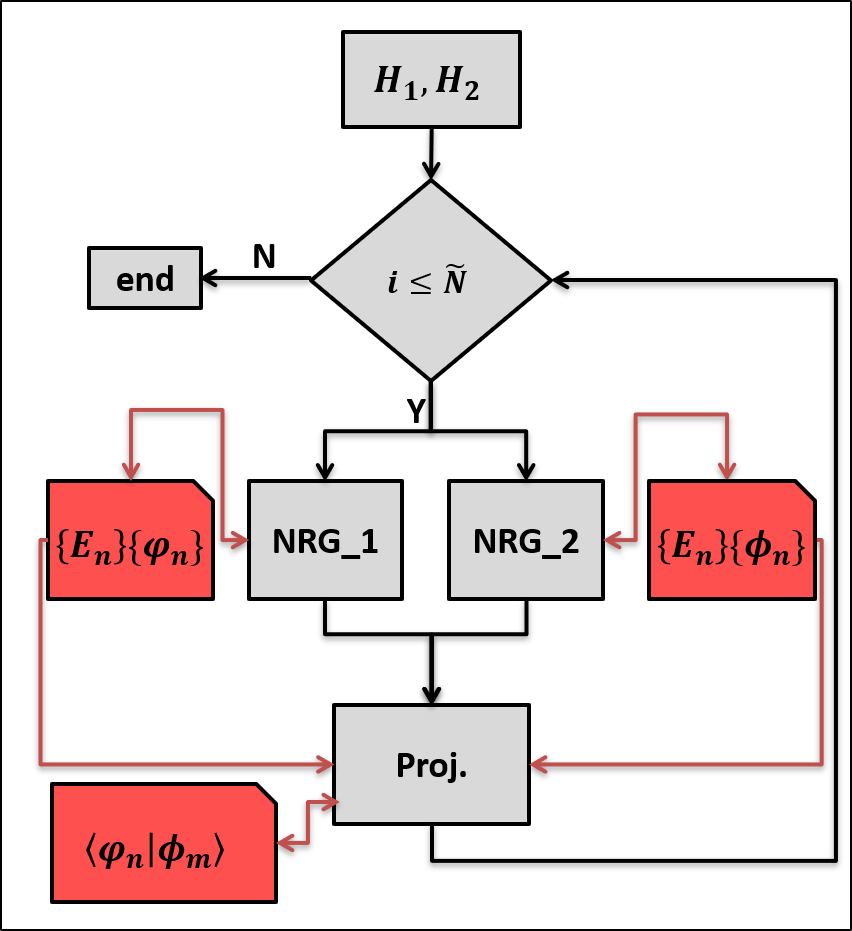}
        \vspace{-0.02cm}
	\caption{\footnotesize Procedure that computes the projections between the bases of two different Hamiltonians $H_1$ and $H_2$. At each iteration $i$, the NRG procedure (Eq. \eqref{NRG:Itera}) is used to compute and save the new energies $\{E_n\}$, eigenstates ($\{\phi_n\}$, $\{\varphi_n\}$), and the new matrix $\mathcal{\hat{M}}$ (Eq. \eqref{NRG:Matrix}). From the eigenstates of each Hamiltonian, we compute the projections using Eq. \eqref{proj:Eq2} and save the projections for the next iteration. We repeat this procedure until $i = \tilde{N}$. \Sou}
    \label{NRG_procedure}
\end{figure}

Thus, starting from iteration $0$ where we know the projections by the above equation, we can calculate the projections of iteration $\tilde{N}$ by repetitively using Eq. \eqref{proj:Eq2}, which requires the eigenvector matrices obtained through NRG for each of the Hamiltonians $(1)$ and $(2)$, as shown schematically by the algorithm flowchart in Fig. \ref{NRG_procedure}. Since these matrices are already calculated during the diagonalization process, the only additional computational cost is the calculation of Eq. \eqref{proj:Eq2}, which grows with the square of the number of states in each charge and spin sector $(q, dS)$. 

Another way to perform this calculation and find the projections would be to calculate and save the matrix elements of each operator $f_{n \leq \tilde N}$ for each of the Hamiltonians, which would require a lot more computational resources and time.


\newpage

\section{Quasi-Block-Diagonal Matrix Approximation}

To achieve accurate results using NRG, it is essential to iteratively solve the Hamiltonian until it converges to a fixed point, over numerous interactions. However, the NRG basis grows rapidly, making it unfeasible to solve the eigenvalue problem for the entire basis. Usually, this challenge is addressed by truncating the basis above an ultraviolet energy cut-off. However, there are certain problems, such as the photoemission problem discussed in Chapter 3, in which is necessary to consider high-energy excited states. One approach that attempted to keep information about high-energy states was the TDNRG method proposed by Anders and Schiller \cite{PhysRevLett.95.196801}. 

Here, we discuss a different, simple idea exploring the nature of the NRG technique. As shown by Eqs. \eqref{NRG:Acoplamento} and \eqref{NRG:Eq1}, in both the NRG and eNRG methods the hopping term $\tau_n$ in the Hamiltonian decreases exponentially. Each new interaction $n+1$ introduces a new level $f_{n+1}^\dagger$, correcting the energy by $\Delta\varepsilon_{n+1} \sim \tau_{n}$ and generating new levels through the tensorial product of the previous basis with this new orbital. Consequently, if two levels are separated by energy differences much greater than $\tau_n$, the coupling between these levels is very small, as Eq. \eqref{NRG:Itera} shows.

\newpage
In such a situation, one might consider whether it is possible to break the Hamiltonian into small pieces and solve each one individually, given that their coupling decreases very rapidly with $\Lambda$ for typical values of the discretization parameter. To explore this idea, let us begin with a Hamiltonian divided into two sectors, as follows:
\begin{eqnarray}
\hat H = \begin{pmatrix}
\hat H_1 & \mu \hat T \\
\mu \hat T^\dagger &\hat H_2 \\
\end{pmatrix}. 
\end{eqnarray}

Let us assume that we initially know the eigenvalues and eigenvectors of each diagonal block,
given by the eigenvalue equation:
\begin{eqnarray}
\hat H_1 \ket{ \Vec{v}_1^{(0)} } = \epsilon^{(0)}_1  \ket{\Vec{v}_1^{(0)}} \mathrm{~and~} \hat H_2 \ket{\Vec{v}_2^{(0)}} = \epsilon^{(0)}_2 \ket{\Vec{v}_2^{(0)}}. 
\end{eqnarray}

Our goal is to solve the eigenvalue equation $\hat H \ket{\mathbf{v} }= \epsilon \ket{\mathbf{v}}$ using the known solutions for each block. Consider an energy $\epsilon$ inside the first block. We can easily demonstrate that $\Vec{v}_2 = - \mu (\hat H_2 - \epsilon \hat I )^{-1} \hat T^\dagger \Vec{v}_1$, if $\mathrm{det}(\hat H_2 - \epsilon \hat I ) \neq 0$, which is not an issue since we are seeking energies within the first block. Thus, we obtain a new eigenvalue equation as follows:
\begin{eqnarray}
\left[ \hat H_1 - \mu^2 \hat T (\hat H_2 - \epsilon \hat I )^{-1} \hat T^\dagger \right] \ket{\Vec{v}_1} = \epsilon \ket{\Vec{v}_1} .
\end{eqnarray}
\begin{eqnarray}
\mathrm{det}\left[\hat H_1 - \epsilon \hat I - \mu^2 \hat T (\hat H_2 - \epsilon \hat I )^{-1} \hat T^\dagger\right] = 0.
\end{eqnarray}

The subsequent steps are too long to elaborate on in the main text, so a full derivation is provided in Appendix \ref{QBDM Aproximation}. The main points to consider during the derivation are that the coupling $\mu$ is very small (not an issue since $\mu = \tau_n$) and that the block is sufficiently large to ensure accurate results. Using this consideration, the Appendix proves that 
\begin{eqnarray}\label{error_energy_}
\epsilon_l =  \epsilon_l^{(0)} + \mathcal{O}({\mu^2}),
\end{eqnarray}
\begin{eqnarray}\label{error_vector_}
\ket{\Vec{v}_l} = \ket{\Vec{v}_{l}^{(0)}} + \mathcal{O}({\mu}).
\end{eqnarray}
Clearly, the same proof demonstrates that Eqs. \eqref{error_energy} and \eqref{error_vector_} are also valid for block 2, simply by selecting an energy within the spectrum of $E_2$.

\begin{figure}[hbt!]
		\centering
        \includegraphics[scale=0.45]{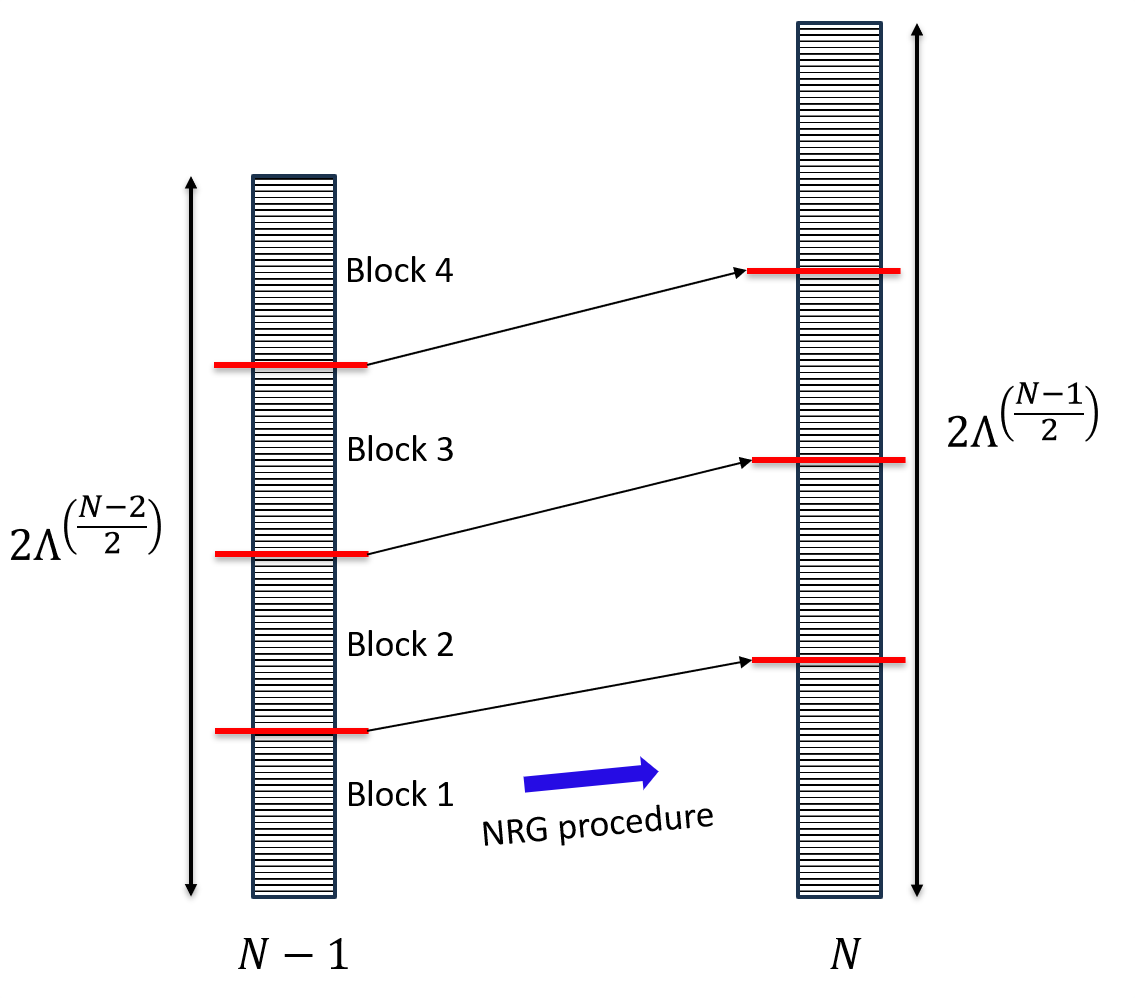}
        \vspace{-0.2cm}
	\caption{\footnotesize Block diagonal procedure. We solve the NRG Hamiltonian up to iteration $N-1$ without the traditional ultraviolet cutoff, which is typically used to reduce the number of eigenstates in the basis. We save all the energies $E_{N-1}$, eigenvectors $\mathcal{D}_{N-1}$ (if needed), and the matrix $\mathcal{M}_{N-1}$ of iteration $N-1$. Then, we break the energies into $\kappa$ pieces, as well as the matrix $\mathcal{M_{N-1}}$ and other relevant matrices (here we also kept $\mathcal{D}_{N-1}$). Next, we use the energy $E^{(j)}_{N-1}$ and the matrices $\mathcal{M}^{(j)}_{N-1}$ and $\mathcal{D}^{(j)}_{N-1}$ of each block $j=1, \ldots, \kappa$ to build the NRG Hamiltonian of this energy sector in iteration $N$ following the NRG procedure in Eq. \eqref{NRG:Itera_BDA}. Then, we diagonalize each block, save the energies and the important matrices, and after that, repeat this procedure to build iteration $N+1$. \Sou}
       \label{Block_Diagonal_procedure}
\end{figure}

To implement this idea within the NRG method, schematically represented in the Fig. \ref{Block_Diagonal_procedure}, we start by choosing an iteration $N$ that is large enough so that the truncation error $\mu = \tau_{N-1} = D\Lambda^{-(N-1)/2}$ becomes sufficiently small. Then we solve the NRG Hamiltonian up to iteration $N-1$ without the traditional ultraviolet cutoff, which is typically used to reduce the number of eigenstates in the basis. We save all the energies $E$, eigenvectors $\mathcal{D}$ (if needed), and the matrix $\mathcal{M}$ of iteration $N-1$. Then, we break the energies into $\kappa$ pieces, as well as the matrix $\mathcal{M}$ and other relevant matrices (here we also keep $\mathcal{D}$).

Next, we use the energy $E^{(j)}_{N-1}$ and the matrices $\mathcal{M}^{(j)}_{N-1}$ and $\mathcal{D}^{(j)}_{N-1}$ of each block $j=1, \ldots, \kappa$ of the iteration $N-1$ to build the NRG Hamiltonian of this energy block $j$, for each $(q,dS)$ sector, in iteration $N$ following the NRG procedure 
\begin{align}\label{NRG:Itera_BDA}
H_{N,(p',p | g', g)}^{(j)} = \sqrt \Lambda E_{N-1,r}^{(j)} \delta\PC{r'_{p',g'}; r_{p,g}} + \PC{\frac{\tau_{N-1}}{D_{N}}} \PC{\mathcal{M}_{(p',p | g', g)}^{(j)}+\left(\mathcal{M}^{(j)}_{(p,p' | g, g')}\right)^*}.
\end{align}
Then, we diagonalize Hamiltonian $H_{ N,(p',p | g', g)}^{(j)}$ for each block $j$, save the energies and the important matrices in a new $E_N$, $\mathcal{D}_N$ and $\mathcal{M}_N$, and after that, we can repeat this procedure to build iteration $N+1$. Furthermore, since $\tau_{N} < \tau_{N-1}$, this procedure can be applied for the subsequent iterations $N+2$ without significantly increasing the error.

The number of eigenvalues and states for each block grows by a factor of 4 from iteration $N-1$ to $N$ . This information is important for choosing the appropriate $\kappa$ necessary to generate a Hamiltonian for each block $H_{N,(p',p | g', g)}^{(j)}$ with a numerically treatable dimension. The condition $\kappa \ll \mathrm{dim}_N$ is necessary condition to guarantee accuracy. In practice, we obtained good results even when each block has around 20 energies, i.e., $\mathrm{dim}_N / \kappa \sim 20$, where $\mathrm{dim}_N$ is the dimension of a $(Q,dS)$ sector of the iteration $N$ with the largest number of eigenstates.
 
The numerical cost of the NRG procedure, without the ultraviolet cutoff can be approximated as proportional to $\mathrm{max}(\mathrm{dim}_N)^3$, where $\mathrm{max}(\mathrm{dim}_N)$ represents the dimension of the largest charge-spin sector. If we break the spectrum into $\kappa$ pieces, the cost becomes $\kappa \times \mathrm{max}(\mathrm{dim}_N/\kappa)^3 = \mathrm{max}(\mathrm{dim}_N)^3/\kappa^2$, a significant decrease in computational cost with $\kappa$, with controllable error.

To split the energy spectrum into $\kappa$ blocks, one can either set the maximum number of eigenstates for each block or divide the spectrum by energy intervals. However, it is important to note that the scaled eigenvalues grow by a factor of $\Lambda^{1/2}$ from iteration $N-1$ to $N$ , and the number of states by a factor of 4. Thus, to keep the matrices with approximately the same length throughout this procedure, instead of splitting the spectrum into $\kappa$ pieces, we must to divide it into $4 \kappa$ pieces. Clearly, the number of blocks will grow by a factor of 4 at each new iteration.

This exponential growth of the number of blocks by $4^{(N+\Delta N)} \kappa$, where $\Delta N$ is the number of extra iterations after the block diagonal procedure starts (at iteration $N$), is an inconvenience. However, it allows us to solve the NRG procedure without an ultraviolet cutoff, up to a Wilson chain size of $\tilde{N} = 9$, or even bigger if one has access to high-performance computers. Without ultraviolet cutoff the traditional procedure would involve diagonalization of a matrix of size $\sim 2^{18} = 2.6 \times 10^5$, a very challenging task even for the best computers in the world. Instead, we can break that matrix into 1024 independent pieces (which can be perfectly parallelized) and diagonalize matrices with dimensions $\sim 256$, finding the energies and the important matrix elements with small error.

Even more efficiently, if one knows the important energy scales of the system, this method can be used as an accurate magnifying glass to zoom in on the properties of such crucial energy scales. The significant energy scales of the problem can often be roughly estimated from smaller iterations, such as $N=4$ or 5, by examining the elements of the matrix that one wants to compute accurately. This is always possible because, up to $N = 4$ or 5, or even $N = 6$, the Hamiltonian can be diagonalized without the ultraviolet truncation. Then, one can examine the spectrum to identify the important energy scales and apply this strategy to find the energies and matrix elements for these specific energy scales. As an example, in the photoemission problem we found that there are two important energy scales: one around the ground state energy and another around the bound state energy.

To sum up, the block-diagonal procedure, combined with the NRG, give access to states with high energy differences from the initial state while maintaining controllable computational cost and controllable error. This procedure, combined with projections between two different bases and the Crank-Nicolson method, offers an accurate and efficient approach to describe the time evolution of strongly correlated systems, even when high-energy states play a important role in accurately describing the system's behavior. The following section we discuss the accuracy of our method.

\section{Preliminary results}

In this section, we present the illustrative results obtained with the mathematical techniques discussed in the previous sections. We begin with the NRG iterative procedure. To ensure that we are correctly solving the Hamiltonian and obtaining accurate results, we first apply the NRG method to the non-interacting case $U=0$, for a metal-impurity system with a localized scattering potential, and compare it with results obtained by direct diagonalization of this quadratic Hamiltonian. The quantitative results of this comparison are provided in Appendix \ref{NRGxDirectDiagonalization}, where we compare the energies and projections computed with various cutoff energies, with those resulting from the direct diagonalization.  Our NRG code is functional and obtaining good results. As expected, the accuracy of the results increases with the cutoff energy.

Next, to test the block-diagonal approximation, we consider the electronic part of the Hamiltonian in Eq. \eqref{H_Colision}, which describes the atomic-surface collision. After the NRG transformation, this Hamiltonian can be written as:
\begin{align}\label{H:NResults}
H_{\tilde N}(z) =& \left[\varepsilon_d n_d + U n_{d\upA} n_{d\downA} \right] +\sum_{n=0}^{\tilde N-1}\tau_n(\fd{n}\fn{n+1}+\hc)
 \nonumber \\ &+  V_0 e^{-z/r} \left(\cd{d}\fn{0}+\hc \right)+ \frac{W_0}{4(z+z_{\text{im}})}(n_{d}-1)^2\fd{0}\fn{0}.
\end{align}
Here, we are considering the set of parameters: $\varepsilon_d = -13.6$ eV, $U=12.8$ eV, $V_0 = 5.0 $ eV, $W_0 = -27.211 $ eV, $z_{\text{im}} =  1\mathfrak{a}_0$  and $r = 1.667 \mathfrak{a}_0$, where $\mathfrak{a}_0$ is the Bohr radius. The energies are normalized by the bandwidth $D = 4.3$ eV. The NRG parameters are $\tilde N=7$ and $\Lambda=6$.

\begin{figure}[hbt!]
		\centering
        \includegraphics[scale=0.85]{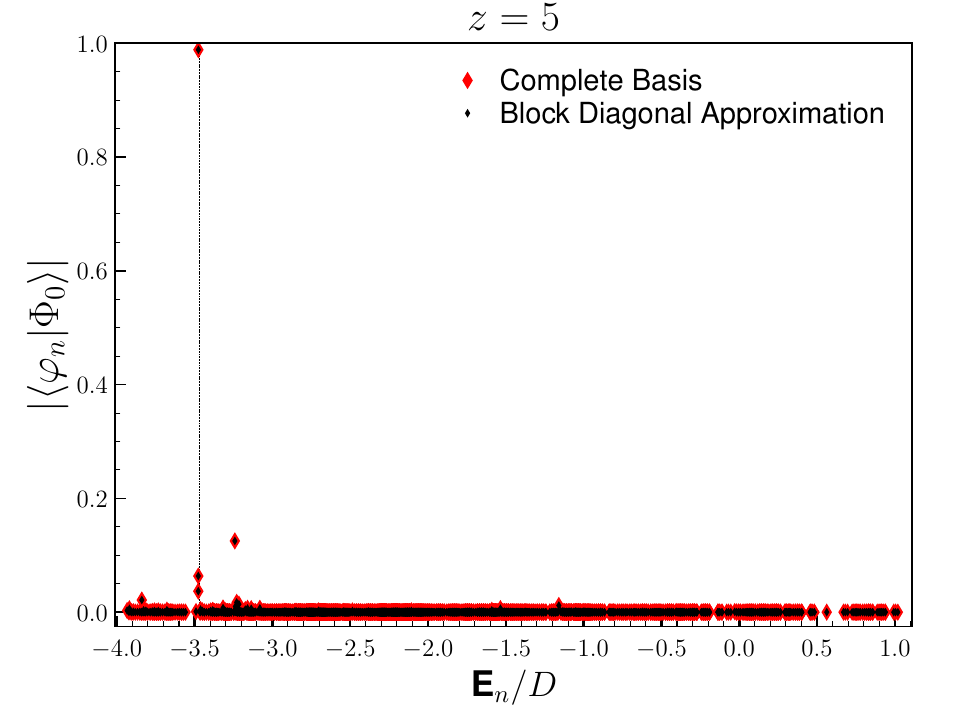}
        \includegraphics[scale=0.85]{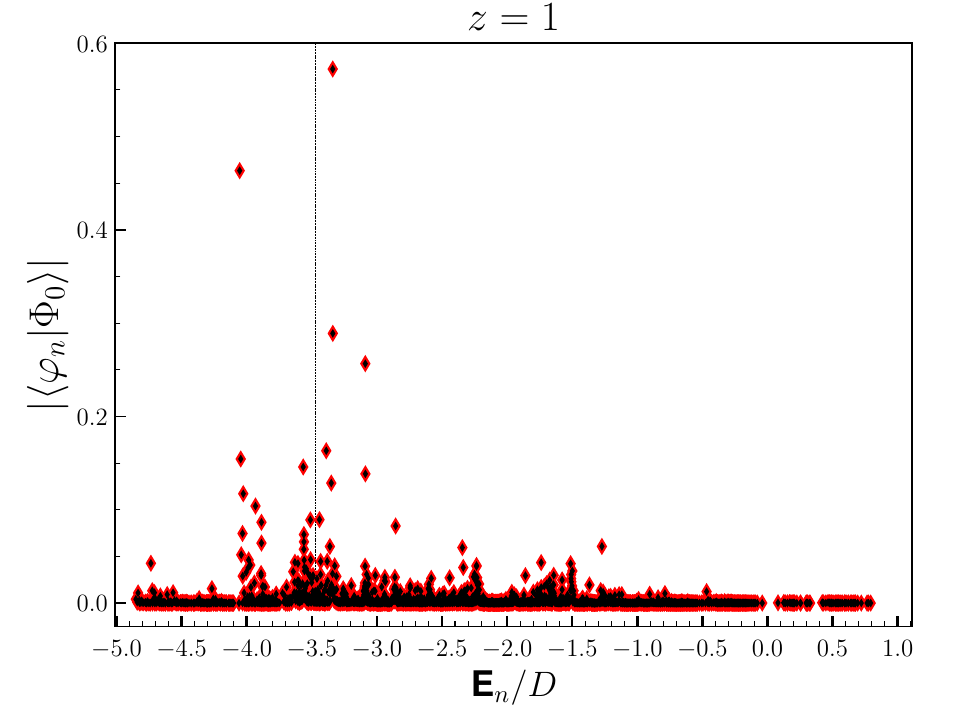}
        \vspace{-0.3cm}
	\caption{\footnotesize Projections $\braket{\varphi_n(z)}{\Phi_0}$ versus Energy $E_n(z)$ resulting from diagonalization of the the Hamiltonian \eqref{H:NResults}, with the complete basis (red triangles) and with the block-diagonal approximation (black triangles), at $z=5 \mathfrak{a}_0$ (Top panel) and $z=1 \mathfrak{a}_0$ (bottom panel). In both panels, the dashed vertical line indicates the initial energy. \Sou}
       \label{Block_Diagonal_Approximation_Results}
\end{figure}

The initial many-body electronic state $\ket{\Phi_0}$ represents the neutral atom far from the surface. Far from the surface the hydrogen atom and the band are decoupled, the electronic configuration $\ket{\Phi_0}$ being a product of $n_d = 1$ atomic state with the ground state of the band. We have computed the projection $\braket{\varphi_n(z)}{\Phi_0}$, where $\{\varphi_n(z) \}$ is the complete set of many-body eigenstates of $H_N(z)$. Figure \ref{Block_Diagonal_Approximation_Results} shows the results of the projections $\braket{\varphi_n(z)}{\Phi_0}$ versus energy $E_n(z)$ from the Hamiltonian \eqref{H:NResults}. The red triangles were computed without ultraviolet truncation (complete basis), while the black triangles were computed by the block-diagonal procedure with $\kappa = 5$, initializing at $N=5$, considering two distances from the surface $z = 5 \mathfrak{a}_0 \mathrm{~and~}1 \mathfrak{a}_0$ . The errors can  be estimated using Eq. \eqref{error_energy_} and Eq. \eqref{error_vector_}, with $\mu = D_7/D = 6^{-(7-1)/2} = 0.005$. The error in the energy is below $10^{-5}$ for all $E_n(z)$ computed and the maximum value of the error in the  computed projections was $0.5 ~D_7/D$, consistent with the expected error.

~
\newpage
From Fig. \ref{Block_Diagonal_Approximation_Results}, two important observations can be made. First, as $z$ decreases both the hybridization and the scattering potential increase, causing the initial state to spread out in the energy spectrum and hybridize with many others many-body states. States with energy far way from the initial energy have significant projection onto the initial state. This advises against ultraviolet truncation, such strategies will fail for this type of problem. Secondly, the block-diagonal approximation accurately reproduces the results of the complete basis calculation, but with smaller computational cost. For $\tilde N=7$,  the complete set of the many-body basis requires 4 Gb of RAM and takes 9 hour on a computer with an 8-core i7 processor with 8 Gb of RAM. The block-diagonal procedure with only 5 blocks reduces the memory requirement to 250 Mb and the time to just 28 minutes.

In the top panel of  Fig. \ref{Block_Diagonal_Approximation_Results}, the atom is far from the surface, in its neutral configuration, and the hybridization $V(z)$ is nearly zero. In this scenario, the initial neutral many-body electronic configuration is one of the eigenstates, making it orthogonal to any other many-body state in the ionized sector. However, as the atom approaches the surface, the hybridization gradually increases, and the initial many-body configuration is no longer an eigenstate and becomes a composition of the eigenstates at each $z$.

As the atom approaches the surface, the scattering potential grows in absolute value, causing the energies of ionized many-body levels to decrease. Some of these levels may cross the initial neutral level, as schematically illustrated in Fig. \ref{Draw_Collision}. When such energy level crossover occurs, the levels can hybridize with the initial state, significantly increasing the projection between them. This effect is evident in Fig. \ref{Block_Diagonal_Approximation_Results} for \( z = 5 \) (top plot). In this plot, the primary projection peak near 1 (at \( -3.5 D \)) corresponds to the electronic configuration that remains neutral. However, smaller peaks, observed at \( -3.5 D \), correspond to ionized states crossing the initial energy and hybridizing with the initial neutral configuration. As a result, their projections on the initial state become considerable. As $z$ decreases, many other ionized many-body energy levels drop and cross the initial energy, hybridizing with the neutral configuration. For $z=1$, as shown in Fig. \ref{Block_Diagonal_Approximation_Results} (bottom plot), the ionized states have already hybridized with the initially neutral configuration. At this point, the initial neutral state is a superposition of many ionized and neutral electronic configurations, spreading out in the energy spectrum.

To fully understand the results in Fig. \ref{Block_Diagonal_Approximation_Results}, we must examine the energy spectrum of the electronic Hamiltonian in Eq. \eqref{H:NResults} at each position \(z\). This will be one of the focuses of the following chapter. With the numerical tools developed to diagonalize the interacting electronic Hamiltonian and a strategy to compute the time evolution based on the Crank-Nicolson method, the next chapter will concentrate on discussing the electronic behavior, numerically solving the atomic-surface collision, and analyzing the behaviors that can be extracted from these solutions.

%% file: CAP/Cap7.tex
\chapter{Sticking Coefficient}

Finally, in this section, we discuss the physics of the collision between an initially neutral hydrogen atom and a metallic surface. Qualitatively, the problem was discussed in the introduction, and we proposed a Hamiltonian \eqref{H_Colision} to describe the phenomena in Chapter 2. However, the proposed model considers a large number of electrons in the metallic band, making the model Hamiltonian intractable. To reduce the number of electronic degrees of freedom while still accurately representing the metal, and thus making the problem solvable, we used the NRG method to transform the electronic Hamiltonian in Eq. \eqref{H_Colision} into the Hamiltonian in Eq. \eqref{H:NResults}.

Since the NRG basis, although discrete, is infinite, in practice it needs to be truncated at energy $\Delta \varepsilon = 2 D\Lambda^{-\frac{\tilde N}{2}}$ (where $D$ is half the bandwidth of the conduction band). One way to define this infrared cutoff energy is to explore the Heisenberg uncertainty principle, expressed by the condition $\Delta\varepsilon T = \hbar$, where $T$ is the collision time.  The Hamiltonian then takes the numerically tractable form:
\vspace{-0.4cm}
\begin{align}\label{eq:4}
\mathcal{H}_{\tilde N}(z) =& \left(\varepsilon_d n_d + U n_{d\upA} n_{d\downA} \right) +\sum_{n=0}^{\tilde N-1}\tau_n(\fd{n}\fn{n+1}+\hc)+  V_0 e^{-z/r} \left(\cd{d}\fn{0}+\hc \right)
 \nonumber \\ & +\frac{W_0}{4(z+z_{\text{im}})}(n_{d}-1)^2\fd{0}\fn{0} + \dfrac{P_z^2}{2M} + \mathcal{V}_z.
\end{align}
Here, the $\fn{n}$ ($n=0,1,\ldots$,$\tilde N$) are the Fermi operators of the NRG basis, coupled to each other by coefficients $\tau_n$ that decay exponentially with $n$, $H_d = \varepsilon_d n_d + U n_{d\upA}n_{d\downA}$ represents the atomic levels, and the atomic potential $\mathcal{V}_z$ comprises two contributions: an infinite barrier for $z \le 0$ (representing the repulsion from the atoms on the surface), and the absorbing potential at the end of the "box"~ to simulate the atom moving away from the surface after the collision.

Now, we need realistic values for the parameters in the Hamiltonian \eqref{eq:4} to describe the atomic hydrogen impinging upon a Cu surface. The atomic orbital of the hydrogen has an energy of $\varepsilon_d = -13.6$ eV in reference to the vacuum, and the Coulomb repulsion $U = 12.8$ eV, so the doubly occupied state has energy $-14.4$ eV. Then, the $\text{H}^{-}$ ion is more stable than the $\text{H}^{+}$ ion, and slightly more stable than neutral H, so the hydrogen atom tends to become negatively ionized near the surface. The constant in the image-charge potential is $W_0 = -27.21$ eV.$\mathfrak{a}_0$ and the scattering potential decays with $1/({4(z+z_{\text{im}})})$.

The position of the image-charge $z_{\text{im}} = 1.0  \mathfrak{a}_0$ and $V_0 = 6.803 \, \text{eV}$ can be obtained from a DFT computation adjusted to reproduce the result for the binding energy (2.39 eV) of H adsorption on Cu surfaces \cite{SupCu,CHULKOV1999330,WEI20154102}. Also from DFT \cite{CHULKOV1999330}, the Cu surface can be approximated by a flat band with $D = 4.3 \, \text{eV}$ and a Fermi energy of $-4.00 \, \text{eV}$ relative to the vacuum. The hybridization decays exponentially as the atom moves away from the surface. Since the neutral H atom has a radius of 1\Rb~and the ion $\mathrm{H}^{-}$, of 2.6\Rb, we expect that $1 \le r \le 2.6$\Rb. From Ref \cite{WEI20154102}, $r \approx 1.67$\Rb.

In the dynamics of the atom-surface collision, two relevant time scales deserve special attention. The first is artificial: the time intervals $\Delta t = 10^{-3}$ fs, into which we divide the time axis, which must be sufficiently small to describe the fastest electronic transitions. The second time scale is nuclear. The hydrogen atom ($M = 938.27$ [MeV/c²]) has a typical initial kinetic energy of a few meV, but it accelerates to energies of the order of a few eV when close to the surface, due to the image charge potential. The time during which the atom interacts with the surface corresponds to the time it remains in the region $[0, 2r]$. The nuclear time scale is numerically determined by simulating a non-interacting hydrogen atom on the Born-Oppenheimer potential surfaces. From the simulation, for an initial kinetic energy \(K_0 = 0.3\) eV, a time on the order of $T = 200 \frac{\hbar}{D}$ is needed to properly account for the nucleus-surface collision.

Slow electrons, with time scales larger than the collision time, and consequently with low energies, less than $0.005D$, will not have enough time to participate and therefore will not contribute significantly to the collision dynamics. This defines the criterion for truncating the NRG Hamiltonian, which has already been discussed earlier. Using $\Lambda = 8$, the condition $\Lambda^{-\tilde N/2} = 0.005$ results in $\tilde N = 5$, which correspond to a conduction band with six spin-degenerate single-particle levels. The half-filled band will accommodate six electrons. With the additional electron from the neutral atom, the many-body state whose time evolution we must follow encompasses seven electrons. Account taker of spin ($S^2$ and $S_z$) conservation, the dimension of the pertinent sector of the electronic spectrum is  784.

Therefore, to simulate the collision and find the sticking coefficient $\mathcal{S}$, we first need to compute the temporal evolution of the system over $z$ coordinates in the interval $0 \le z \le 9$\Rb, which we divide uniformly into 800 intervals, and over times in the interval $0 \le t \le 500$ fs, divided into  $\Delta t = 0.001$ fs. To achieve this, we need to handle matrix operations, as multiplications and inversions, with a square matrix dimension of $800 \times 784$, which makes the problem numerically intractable, even with high performance cluster resources. Using only $50$ states, the Hamiltonian is describe by a square matrix with dimensions $40000\times40000$, which typically requires 100 Gb of memory RAM to perform the calculations with this matrix.

As the scattering potential becomes strong near the surface, the presence of states like the bound state discussed in the photoemission problem cannot be neglected. Consequently, high-energy excited states play a crucial role in describing the phenomenon, necessitating the abandonment of the ultraviolet cut-off energy. At this stage, we can apply the insights gained from the photoemission analysis to identify the important many-body states and extract the relevant physics from the Hamiltonian \eqref{eq:4}. With realistic parameter values in Eq. \eqref{eq:4} for the atomic hydrogen and copper surface interactions, we can now discuss how to numerically determine the time-evolution operator for this process, extract the physics of the collision, and ultimately compute the sticking coefficient.


\section{Atomic motion and non-adiabaticities}

The wave function of the system, governed by the Hamiltonian in Eq. \eqref{eq:4}, includes both the nuclear coordinate $z$ and the electronic states $ \{\ket{\mathfrak{e}}\} $. At $t=0$, the atomic wave function is a Gaussian, centered at  $z_0$, moving with kinetic energy $K_0 = \frac{\hbar^2 k_0^2}{2M}$ towards the surface. Additionally, the initial electronic wave function $\ket{\phi_0}$ is a product of the neutral hydrogen state and the metallic band in its ground state, as the atom is initially far from the surface. Thus, keeping this in mind, the total wave function at $ t=0$ can be written as:
\begin{equation}\label{wave_function_t_0}
    \ket{\Psi(z,0)} = Be^{-\frac{{(z-z_0)^2}}{2 \eta}} e^{-ik_0 z} \ket{\phi_0}.
\end{equation}

Since $V(z)$ decays rapidly with  $z$, $\ket{\phi_0}$ is an electronic eigenstate when the atom is far from the surface. However, as the atom approaches the surface, $V(z)$ becomes significant, leading to a hybridization of the neutral and ionized sectors. As a result, for each coordinate $z$ near the surface, the state $\ket{\phi_0}$ decomposes into a linear combination $ \ket{\phi_0} = \sum_m c_m(z) \ket{\varphi_m(z)}$ of the many-body eigenstates of the electronic Hamiltonian $H_e(z)$.

As discussed in Section 5.3, when atomic and electronic states are coupled for each coordinate $z$, the non-zero terms $\langle \varphi_n | \partial_z | \varphi_m \rangle$ and $\langle \varphi_n | \partial_z^2 | \varphi_m \rangle$ introduce non-adiabatic effects, enabling transitions between electronic eigenstates. If these non-adiabatic processes are minimal (i.e., the atom moves slowly within the interaction region), transitions  $\ket{\varphi_m(z)} \rightarrow \ket{\varphi_{n \ne m}(z)}$ are unlikely. In this case, the atom collides with the surface and is reflected, and as it moves away from the surface, it returns to its initial electronic configuration $\ket{\phi_0}$. Since the atom does not lose energy in this process, it simply leaves the box, and the sticking coefficient is zero.

On the other hand, if these non-adiabatic terms are significant, after the decomposition  $\ket{\phi_0} = \sum_m c_m(z) \ket{\varphi_m(z)}$, pairs of particle-hole excitations can be created from each electronic configuration $\varphi_m(z)$. In this case, the atom collides with the surface, interacts with it, and is reflected. However, as it moves away from the surface, it has now a significant probability of transitioning to an ionized state. One ionized state feels an attractive potential toward the surface. The creation of particle-hole excitations during this interaction can steal enough kinetic energy from the atom, potentially leading to the atom becoming bound to the surface.
\newpage

These discussions are crucial for understanding the results of our simulations. However, we will carry out the calculations in the basis \( \{ \varphi_n(L) \} = \{ \phi_n \} \). Since \( V(L) \approx 0 \), the electronic states are a product of the atomic electronic configuration and the band configuration, implying that the atom is either ionized or neutral. In this basis \( \{ \phi_n \} \), we can expand the wave function as
\begin{equation}
    \lvert \bm{\Psi}(z,t) \rangle = \sum_n \chi_n(z,t) \lvert \phi_n \rangle
\end{equation}
and the time-dependent Schrödinger equation results in
\begin{align}\label{SEQ.Collision_}
i\hbar \partial_t  \vec{\chi}(z,t) = \left[\hat P(z) \hat E_e(z) \hat P^\dagger(z) +\hat{\mathcal{V}}(z) - \frac{\hbar^2\partial_z^2}{2M} \right]\vec{\chi}(z,t).
\end{align}
Here, the change of basis matrix $\hat{P}(z)$ and the eigenenergy matrix $\hat{E}_e(z)$ are found by the NRG procedure for each $z$ described in Chapter 6.

\section{Born-Oppenheimer potentials and the important states}

Now, to compute the electronic part of the Hamiltonian in Eq. \eqref{eq:4} for each position $z$ and find the energies $\hat E_e(z)$ and the projection matrix $\hat P(z)$, necessary to compute the time-evolution, we combined the values of the parameters that we mentioned before in the Table \ref{Paramenters_1}. The relative energies are in reference to the vacuum. To avoid confusion, the electronic part of the Hamiltonian \eqref{eq:4} is written as:
\begin{align}\label{H_e(z)}
H_e(z) =& \left( \varepsilon_d n_d + U n_{d\upA}n_{d\downA}\right)+\sum_{n=0}^{\tilde N-1}\tau_n(\fd{n}\fn{n+1}+\hc) + V_0 e^{-z/r} \left(\cd{d}\fn{0}+\hc \right) \nonumber \\ &+ \frac{W_0}{4(z+z_{\text{im}})}(n_{d}-1)^2\fd{0}\fn{0}.
\end{align} \vspace{-0.5cm}
\begin{table}[h!]
\caption{\footnotesize  Values of the parameters to simulate the electronic contribution for the hydrogen-Cu surface collision. The energies are in reference to the vacuum of the isolated hydrogen atom, and they can be refined in relation of the Cu surface Fermi energy.}
\vspace{-0.5cm}
\begin{center}
\begin{tabular}{| c | c | c | c | c |}
\hline
Parameter & Value [unities]& ~ & Parameter & Value [unities] \\
\hline
$\varepsilon_d$ & -13.6 [eV] & ~ & $D$ & 4.3 [eV] \\
$U$ & 12.8 [eV] & ~ & $\epsilon_F$ & -4.0 [eV] \\
$V_0$ & 6.803 [eV] & ~ & $r$ & 1.67 [\Rb] \\
$W_0$ & -27.21 [eV.\Rb] & ~ & $z_{\text{im}}$ & 1.00 [\Rb] \\
$\tilde N$ & 5 & ~ & $\Lambda$ & 8.0 \\
\hline
\end{tabular}
\end{center}
\begin{flushleft}
		\footnotesize Source: By the author. \
\end{flushleft}
\label{Paramenters_1}
\end{table}

The next step is to compute numerically the energies and projection matrices of the Hamiltonian in \eqref{H_e(z)} using the parameters in Table \ref{Paramenters_1}. We can observe from Eq. \eqref{H_e(z)} that if $V(z) = 0$ ($z\gg r$) the metallic band and the atom are decoupled. In this case, if $n_d = 1$, the scattering potential is zero and the electronic levels in the band are the same as the unperturbed metal $H_B^0$. However, if $n_d \neq 1$, the atom is ionized and the scattering potential appears. In the latter case, the Hamiltonian of the band becomes $H_B = H_B^0 + W f_0^\dagger f_0$, which is exactly the same Hamiltonian for the photoemission problem (see Eq. \eqref{H_photo_eNRG}).

\begin{figure}[hbt!]
		\centering
        \hspace{0.25cm}
        \includegraphics[scale=0.37]{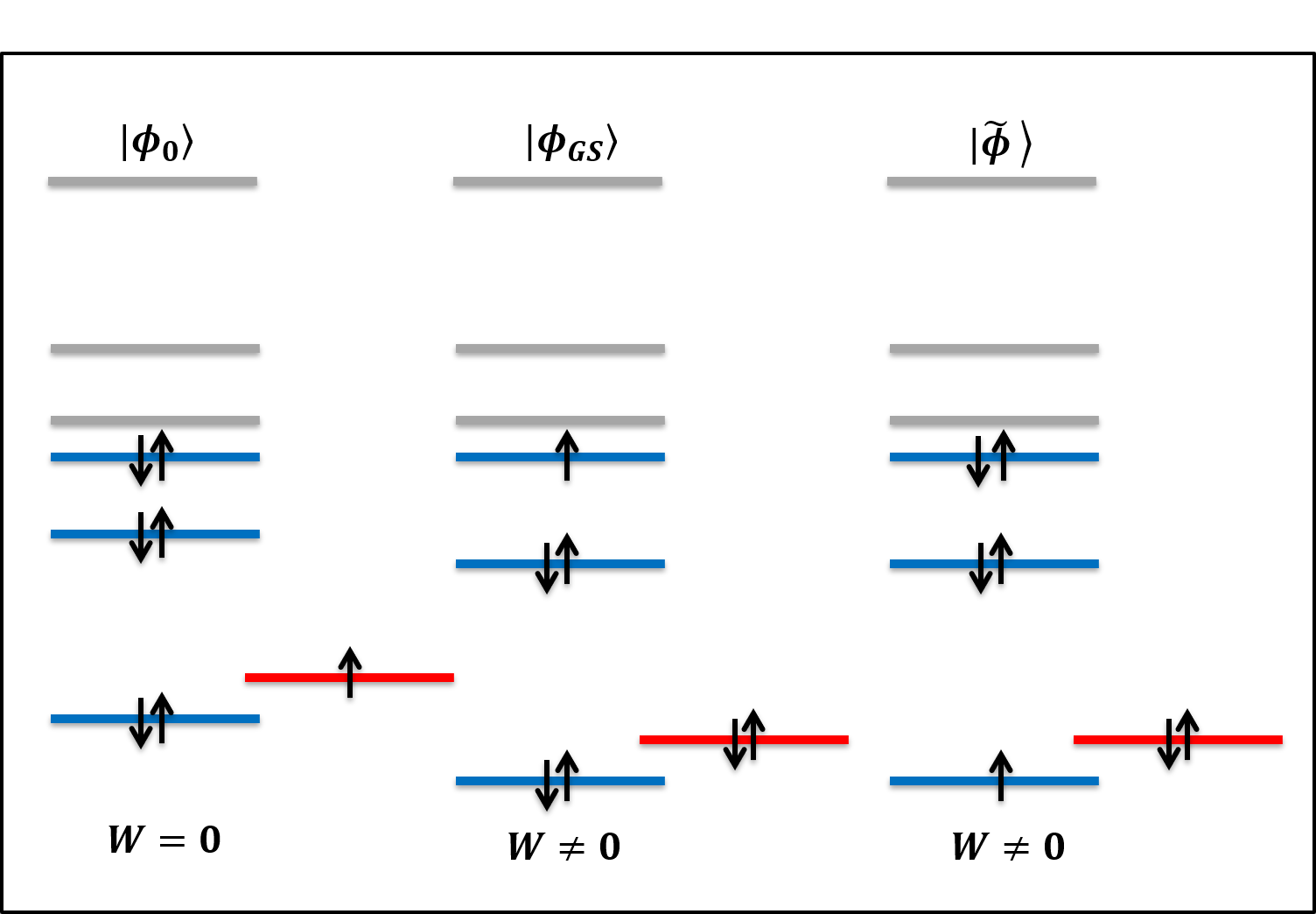}
        \vspace{-0.15cm}
	    \caption{\footnotesize Schematic representation of the lowest energy electronic configuration with the atom neutral ($\ket{\phi_0}$), the lowest energy electronic configuration ($\ket{\phi_{GS}}$), and the smallest energy excitation from the bound state $\ket{\tilde\phi} = g_{-\Delta\varepsilon}^\dagger g_{B} \ket{\phi_{GS}}$. The red level represent the electronic configuration of the atom.  \Sou}
        \label{Eletronic_Levels_Collision}
\end{figure}

In this situation, in Fig. \ref{Eletronic_Levels_Collision} we schematically represent the lowest energy electronic configuration with the atom neutral, i.e $\ket{\phi_0}$, and the lowest energy electronic configuration, keeping the same number of electrons, when the atom is ionized positively, i.e $\ket{\phi_{GS}}$. Additionally, the smallest energy excitation from the bound state is represented as $\ket{\tilde\phi} = g_{-\Delta\varepsilon}^\dagger g_{B} \ket{\phi_{GS}}$. We can observe that $\ket{\phi_0}$ behaves like the initial ground state in the photoemission problem, and $\ket{\phi_{GS}}$ and $\ket{\tilde{\phi}}$ are analogous to the final ground state and the excitations from the bound state, respectively. This indicates that the photoemission problem not only has a similar Hamiltonian but also exhibits similar behaviors when considering only the electronic contribution, which means that we need pay attention to high energy excited states, especially when evolves excitations from this bound state level.

Before proceeding, it is important to note that the many-body spectrum of the Hamiltonian consists of 784 states. While the case with \( \tilde{N} = 5 \) takes only 6 minutes to compute without applying the state cutoff, repeating this calculation 800 times would be a significant task. However, by using our block approximation procedure described in Section 6.3, the computation time is reduced to just a few seconds, allowing us to complete the calculations in a few hours rather than days.

\newpage
\begin{figure}[hbt!]
		\centering
        \includegraphics[scale=0.83]{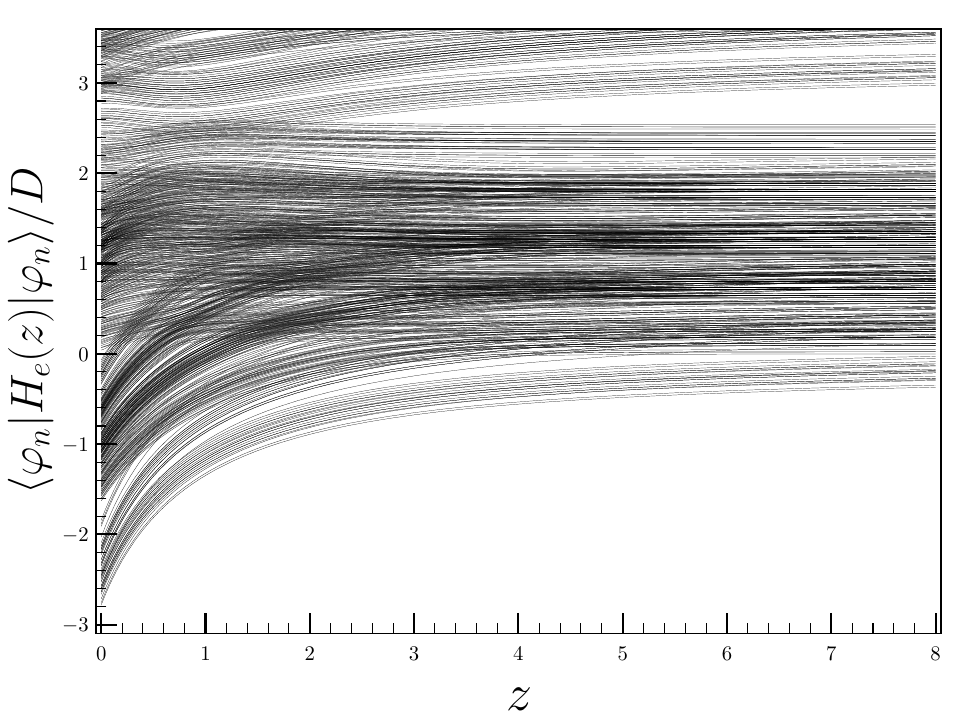}
	\includegraphics[scale=0.83]{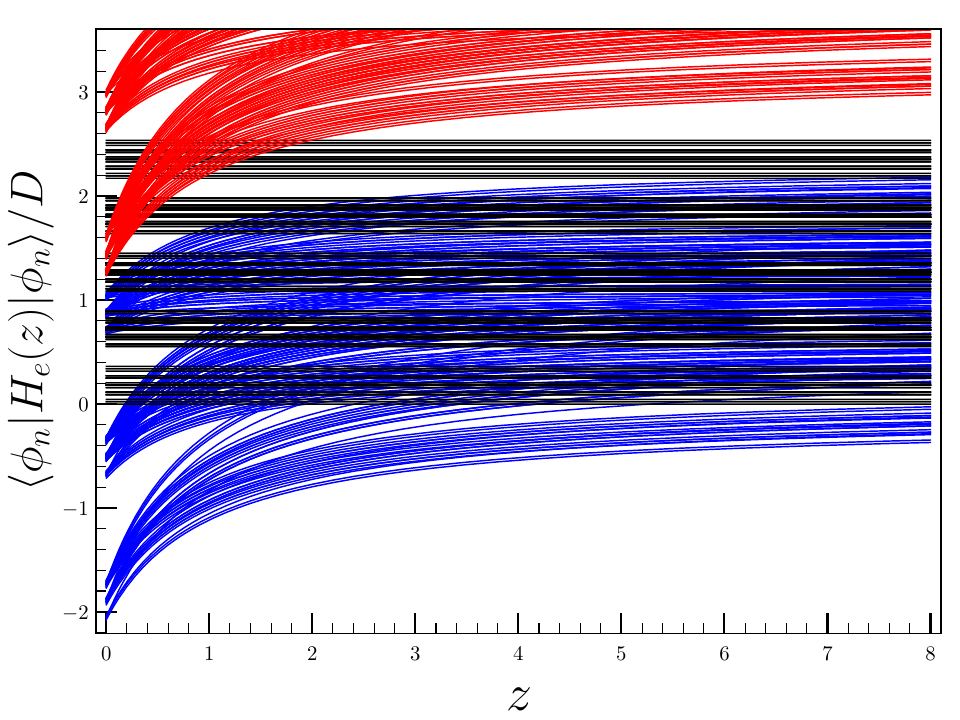}
        \vspace{-0.35cm}
	    \caption{\footnotesize (Top panel) BO potentials, or the many-body energy spectrum of the \eqref{H_e(z)} for each $z$. The energies are in reference to the initial state energy, where the hydrogen atom is neutral and the band is in the ground state and the atom is far away from the surface. (Bottom panel) Expected value of  $H_e(z)$ \eqref{H_e(z)} for each electronic state and $z$ using the initial basis at $z=L$. The energies are referenced to $\phi_0$ energy. The color indicates how many electrons are in the hydrogen atom: red for 0, black for 1, and blue for 2.  \Sou }
        \label{Eletronic_Potentials_}
\end{figure}

\newpage
Fig.  \ref{Eletronic_Potentials_} (Top panel) shows the many-body energy spectrum of the $H_e(z)$ \eqref{H_e(z)}, for each position $z$. The energies are referenced to the initial state energy, where the hydrogen atom is neutral far away from the surface and the band is in the ground state (with $V=0$ and $W=0$). Far from the surface, $V(z) \approx 0$, the atom is decoupled from the metal, and the electronic states are products of the atomic levels with the band configuration. As $z$ decreases, the ionized states drop in energy due to the scattering potential $W(z)$, as shown in the Fig. \ref{Eletronic_Potentials_} (Bottom panel), where the black states are neutral, blue are negatively ionized, and red are positively ionized. For small $z$, the ionized states and the neutral states hybridize and it becomes difficult to distinguish each state by the atomic charge.

Since $2\varepsilon_d + U < \varepsilon_d$, the ground state at $z = L$ occurs when the atom has 2 electrons, becoming negatively ionized. The ground state and small excitations from it decrease in energy as $z$ decreases, given that $|W(z)|$ grows. Then, the energy difference between these levels and the initial state also grows. Additionally, particle-hole excitations from the ground state, which are also ionized and have energy greater than the initial neutral level at $z = L$, also decrease in energy as $|W(z)|$ increases and can cross the initial neutral level. When it happens, these levels can hybridize with the initial level. This behavior is evident from Fig. \ref{Eletronic_Potentials_}.

If the atom becomes ionized, the metallic band experiences the scattering potential term and exhibits a bound state similar to the one in the photoemission problem. From Figs. \ref{Eletronic_Potentials_}, we observe that around $z = 5$, there is a crossover between an excited ionized level and the initial electronic configuration. Upon investigating this electronic level, we discover that it corresponds to a particle-hole excitation from the bound state level to a level near the Fermi energy, described by $\ket{\tilde{\phi}} = g_{-\Delta\varepsilon}^\dagger g_B \ket{\phi_{GS}}$. In other words, this level behaves like the smaller energy unplugged states identified in the photoemission problem and can exhibit a high projection with the initial neutral level.

\begin{figure}[hbt!]
		\centering
        \begin{tabular}{ll}
        \hspace{-0.65cm}
        \includegraphics[scale=0.515]{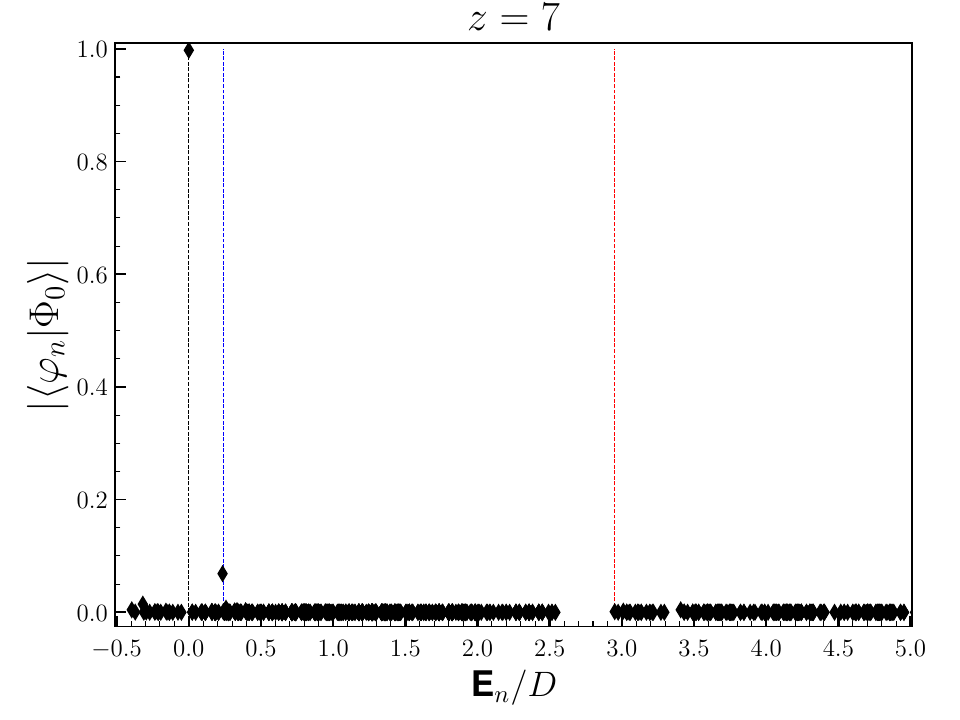}
        & \hspace{-0.9cm}
        \includegraphics[scale=0.515]{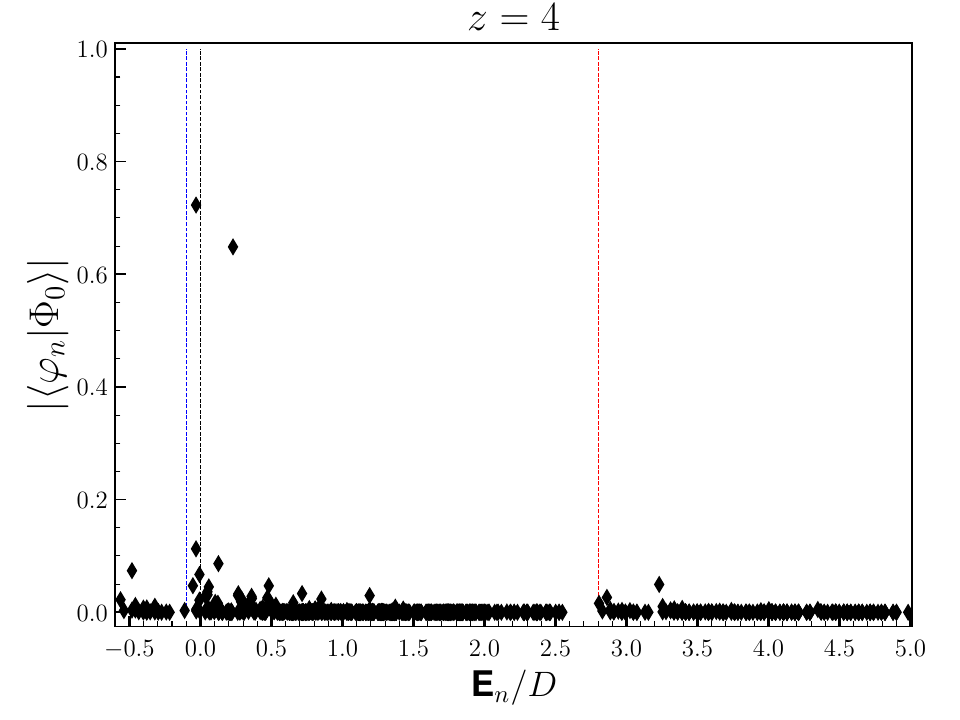}
        \end{tabular}
        \begin{tabular}{ll}
        \hspace{-0.65cm}
        \includegraphics[scale=0.515]{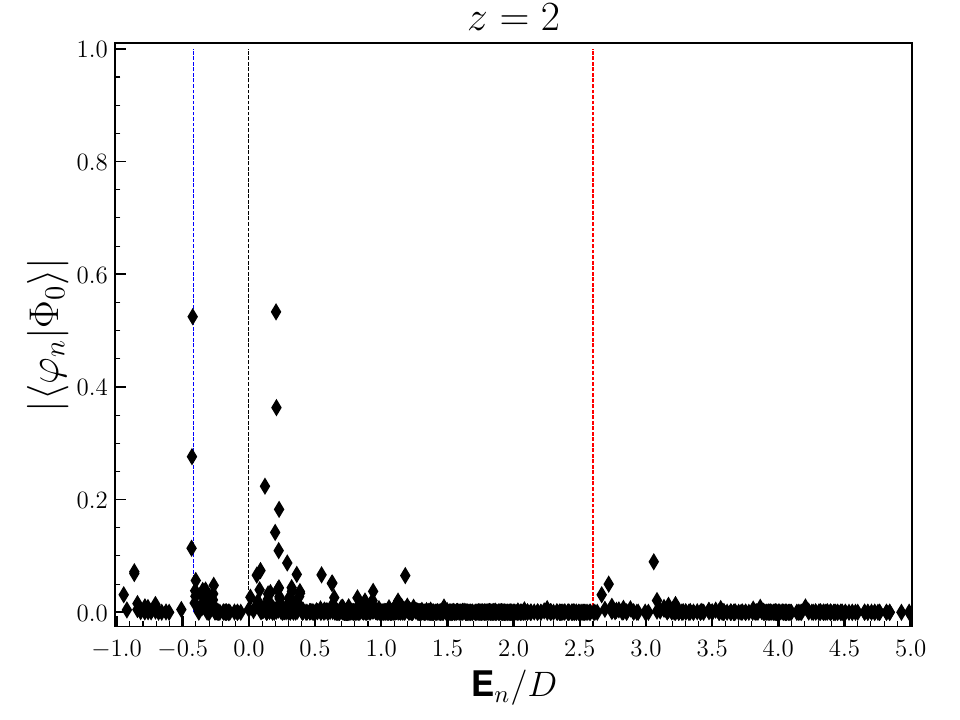}
        & \hspace{-0.9cm}
        \includegraphics[scale=0.515]{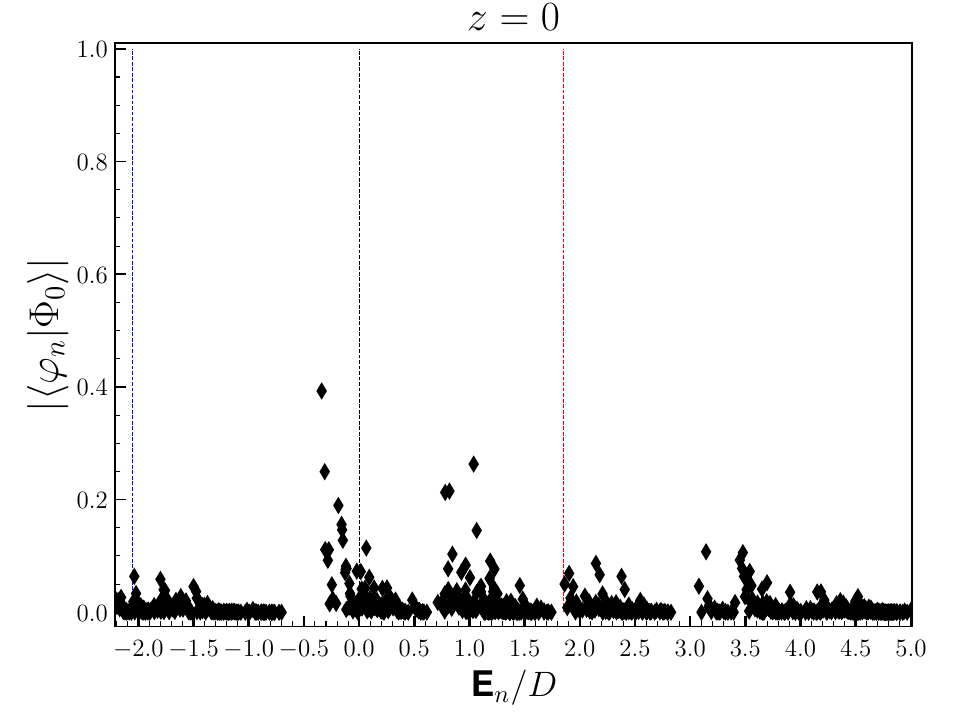}
        \end{tabular}
    \vspace{-0.6cm}
	\caption{\footnotesize Projections $\braket{\varphi_n(z)}{\Phi_0}$ versus Energy $E_n(z)$ from the $H_e(z)$ \eqref{H_e(z)}, considering $z = 7, 4, 2 \mathrm{~and~}0$ \Rb. The initial energy are represented with the dashed black line in $E_n = 0$. The red line represents the smaller energy with the atom positively ionized. The blue line represent the smaller energy of the one particle-hole excitation from the bound state level $\ket{\tilde\phi}$. \Sou }
       \label{Projections_Results}
\end{figure}

Fig. \ref{Projections_Results} shows the projection of the initial state  $\ket{\Phi_0}$, where the hydrogen atom is neutral and the band is in its ground state, for $z = 0, 2, 4,$ and  $7$ \Rb. For  $z \gg 1$, $V(z) $ and $ W(z)$ are small, and the band states remain decoupled from the atomic state. In this regime,  $\ket{\Phi_0}$ is a many-body eigenstate of the $H_e$ and the difference between the ground state energy and the initial energy is $|\varepsilon_d + U| = 0.16D$. However, as  $z$ decreases, two effects occur: first, the ground state energy shifts due to the increasingly negative $W(z)$, and the energy of the ionized states drop. Second, the many-body state  $\ket{\Phi_0}$ is no longer an eigenstate, becoming a linear combination of the states  $\{ \ket{\varphi_n(z) } \}$. 

In these results, we do not observe a significant projection from either the positively ionized sector or the ground state and its small energy excitations. In contrast, the unplugged states have significant projection  for all $z < 7$. This indicates that the unplugged states, similarly to the states we identified in the photoemission, are crucial for capturing the dynamics of the atom-surface interaction within this region.

\newpage

After this observation, we can remove the positively ionized sector, the low-energy negatively ionized levels, and the high-energy neutral levels from the total basis without significantly affecting the physical behavior of the collision. This reduction dramatically decreases the basis from the initial 784 states to just 180 states. However, this number still results in matrices that are too large to compute. With this in mind, we will propose an effective spinless Hamiltonian that focuses solely on the neutral and negatively ionized electronic configurations of interest in the following section, and we will demonstrate that this effective Hamiltonian can describes well the important electronic states.

\newpage 

\section{Electronic spinless effective  Hamiltonian}

To extract the effective spinless Hamiltonian that captures the physics from the Hamiltonian in Eq. \eqref{H_e(z)} we start out with the many-body states
\begin{equation}
    \ket{\phi_0} = d_{\downA}^\dagger \prod_{\varepsilon_k < \epsilon_F} \tilde a_{k\upA}^\dagger \tilde a_{k\downA}^\dagger \ket{0}
\end{equation}
and 
\begin{equation}
\ket{\tilde\phi} = d_{\upA}^\dagger d_{\downA}^\dagger g_{B \upA } \prod_{\epsilon_k < \epsilon_F} g_{k\upA}^\dagger g_{k\downA}^\dagger \ket{0}    
\end{equation}
where $\tilde a_k$ and $g_k$ are the one-particle levels for the neutral sector and the ionized sector, respectively.

To be more specific, we can look at the Hamiltonian \eqref{H_e(z)} in the first iteration, given by
\begin{align}
    H_{(\tilde N=0)}(z) = \varepsilon_d c_d^\dagger c_d + U n_{d\upA}n_{d\downA} + W(z) f_0^\dagger f_0 (n_d - 1)^2 + V(z)(f_0^\dagger c_d + c_d^\dagger f_0).
\end{align}
Eq. \eqref{Base_H0} shows the basis and their energies, at iteration $\tilde N = 0$.
\begin{align}\label{Base_H0}
(N_e = 0, dS = 0):
~&\lvert 0, 0, 0 \rangle = \lvert 0 \rangle \nonumber ~ &E&(0,0,0) = 0, \\
(N_e = 1,dS = 1):~
&\lvert 1, 1, 0 \rangle = d_{\upA}^\dagger \lvert 0 \rangle \nonumber ~ &E&(1,1,0) = \varepsilon_d, \\
&\lvert 1, 1, 1 \rangle = f_{0\upA}^\dagger \lvert 0 \rangle \nonumber ~ &E&(1,1,1) = W(z), \\
(N_e = 2,dS = 0):~
&\lvert 2, 0, 0 \rangle = (d_{\upA}^\dagger f_{0\downA}^\dagger)_{\mathrm{s}}\lvert 0\rangle\nonumber ~ &E&(2,0,0) = \varepsilon_d, \\
&\lvert 2, 0, 1 \rangle = d_{\upA}^\dagger d_{\downA}^\dagger \lvert 0 \rangle \nonumber ~ &E&(2,0,1) = 2\varepsilon_d + U, \\
&\lvert 2, 0, 2 \rangle = f_{0\upA}^\dagger f_{0\downA}^\dagger \lvert 0 \rangle \nonumber ~ &E&(2,0,2) = 2W(z), \\
(N_e = 2,dS = 2):~
&\lvert 2, 2, 0 \rangle = (d_{\upA}^\dagger f_{0\downA}^\dagger)_{\mathrm{t}} \lvert 0 \rangle \nonumber ~ &E&(2,2,0) = \varepsilon_d, \\
(N_e = 3,dS = 1):~
&\lvert 3, 1, 0\rangle = d_{\upA}^\dagger f_{0\upA}^\dagger f_{0\downA}^\dagger\lvert 0 \rangle \nonumber ~ &E&(3,1,0) =  \varepsilon_d, \\
&\lvert 3, 1, 1\rangle = f_{0\upA}^\dagger d_{\upA}^\dagger d_{\downA}^\dagger \lvert 0 \rangle \nonumber ~ &E&(3,1,1) = 2\varepsilon_d + U + W(z), \\
(N_e = 4,dS = 0):~
&\lvert 1, 1, 1 \rangle = d_{\upA}^\dagger d_{\downA}^\dagger f_{0\upA}^\dagger f_{0\downA}^\dagger \lvert 0 \rangle ~ &E&(4,0,0) = 2\varepsilon_d + U + 2W(z).
\end{align}

Observe that, the $\tilde N = 0$ Hamiltonian contains no conduction-band energies. But, the $\tilde N = 1$ introduces the highest single-particle level of the band via the term 
\begin{equation}
    H^\mathrm{Band}_{\tilde N=1} = \tau_0(f_0^\dagger f_1 + f_1^\dagger f_0), \nonumber
\end{equation} 
which has eigenenergies $-\tau_0$ and $+\tau_0$ with respectively eigenstates 
\begin{equation}
    \ket{-} = \frac{1}{\sqrt{2}} f_0^\dagger \ket{0} - \frac{1}{\sqrt{2}} f_1^\dagger \ket{0} ~ \mathrm{and}~\ket{+} = \frac{1}{\sqrt{2}} f_0^\dagger \ket{0} + \frac{1}{\sqrt{2}} f_1^\dagger \ket{0}. \nonumber
\end{equation}

Clearly, the states in \eqref{Base_H0} that do not include the  $d^\dagger$ operator in the  $\tilde{N} = 0$ iteration belong to the $n_d = 0$ sector and will produce descendants with energies $E > |\varepsilon_d|$ relative to the $ n_d = 1$ sector. These states do not contribute significantly.  Additionally, the states in \eqref{Base_H0} that do not include the  $f_0^\dagger$ operator will generate descendants with energies at least $E > \tau_0$ and  their contribution is also insignificant. After this consideration, only the states that are descendants from $(d_{\upA}^\dagger f_{0\downA}^\dagger)_{\mathrm{s/t}}\lvert 0\rangle$, $d_{\upA}^\dagger f_{0\upA}^\dagger f_{0\downA}^\dagger\lvert 0 \rangle$, $f_{0\upA}^\dagger d_{\upA}^\dagger d_{\downA}^\dagger \lvert 0 \rangle$, and $ d_{\upA}^\dagger d_{\downA}^\dagger f_{0\upA}^\dagger f_{0\downA}^\dagger \lvert 0 \rangle$ will be relevant.

Defining the effective vacuum state as
\begin{equation}
\ket{\tilde 0} \equiv (d^\dagger f_0^\dagger)_{\mathrm{s}}\lvert 0\rangle,
\end{equation} 
observe that we can construct all the important states by using only the spinless operators $d^\dagger$ and $f_0^\dagger$:
\begin{align}\label{Base_H0_eff}
(\tilde{N}_e = 0):~
&\lvert 0, 0 \rangle = \lvert \tilde{0} \rangle = (d^\dagger f_0^\dagger)_{\mathrm{s}}\lvert 0\rangle\nonumber ~ &\tilde E&(0,0) = \varepsilon_d , \\
(\tilde N_e = 1):~
&\lvert 1, 0\rangle = f_0^\dagger \lvert \tilde{0} \rangle =   d_{\upA}^\dagger f_{0\upA}^\dagger f_{0\downA}^\dagger\lvert 0 \rangle \nonumber ~ &\tilde E&(1,0) =  \varepsilon_d, \\
&\lvert 1, 1\rangle = d^\dagger \lvert \tilde{0} \rangle = f_{0\upA}^\dagger d_{\upA}^\dagger d_{\downA}^\dagger \lvert 0 \rangle \nonumber ~ &\tilde E&(1,1) = 2\varepsilon_d + U + W(z) + \Delta, \\
(\tilde N_e = 2):~
&\lvert 2, 0 \rangle = d^\dagger f_0^\dagger \lvert \tilde{0} \rangle  = d_{\upA}^\dagger d_{\downA}^\dagger f_{0\upA}^\dagger f_{0\downA}^\dagger \lvert 0 \rangle ~ &\tilde E&(2,0) = 2\varepsilon_d + U + 2W(z) + \Delta. 
\end{align}

Electronic eigenstates belonging to the $n_d=2$ sector at large $z$ with energies that are smaller than the ground state of the $n_d=1$ sector contribute little to the atomic-surface collision because the image potential pushes down their energies as $z$ becomes smaller and they not hybridize significant with the neutral sector (see Figs. \ref{Eletronic_Potentials_} and \ref{Projections_Results}). To eliminate such eigenvalues from the spectrum of the spinless Hamiltonian,  we add a positive energy 
\begin{equation}
    \Delta \equiv \tau_0 \coth \left(\frac{1}{W(L)} \right)
\end{equation}
to the $n_d = 2$ sector, equal to the energy necessary to excite a particle from the bound state to the Fermi level at $z=L$. In practice, this expedient eliminates the plugged states in the $n_d=2$ sector and gives prominence  to the unplugged ones.

After these considerations and re-normalizing the effective vacuum energy as $\varepsilon_d$, we can express the the effective spinless Hamiltonian as: 
\begin{align}\label{H_eff}
    H_{N=0}^{\mathrm{eff}} (z) = \tilde\varepsilon_d(z) \tilde d^\dagger \tilde d  + W(z) f_0^\dagger f_0 \tilde n_d + V(z)(f_0^\dagger \tilde d + \tilde d^\dagger f_0).
\end{align}
Here, the operator $\tilde d^\dagger$ changes the atom from neutral to negatively ionized, the latter configuration having energy
\begin{equation}
    \tilde\varepsilon_d(z) \equiv \varepsilon_d + U + W(z) + \Delta.
\end{equation}
\newpage
~
~
\begin{figure}[hbt!]
	\centering
         \includegraphics[scale=0.815]{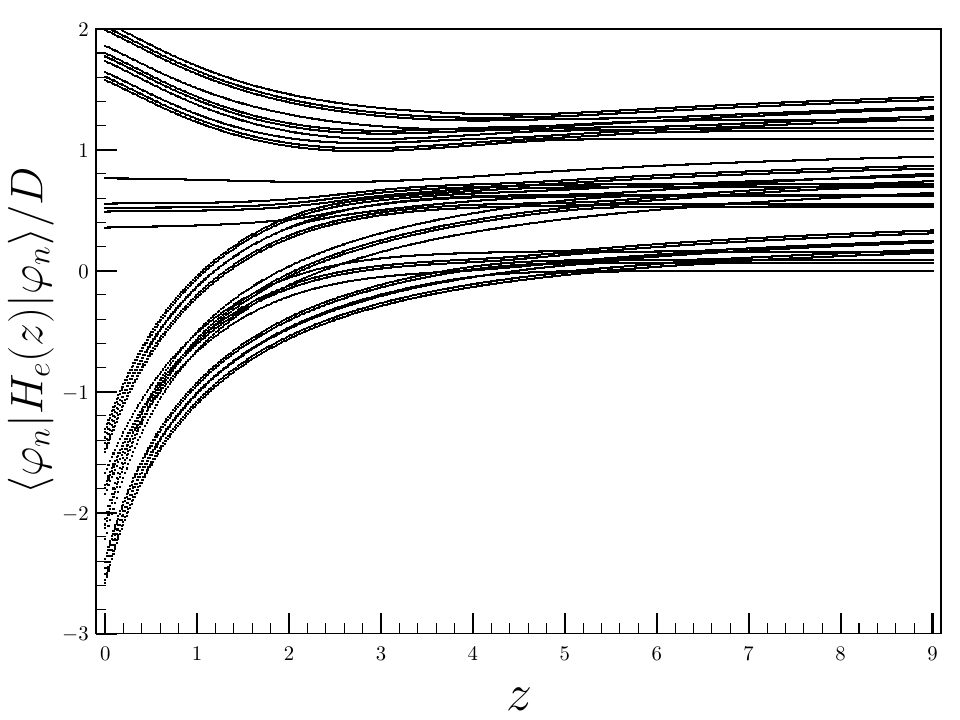}
        \hspace{-0.3cm}
	\includegraphics[scale=0.815]{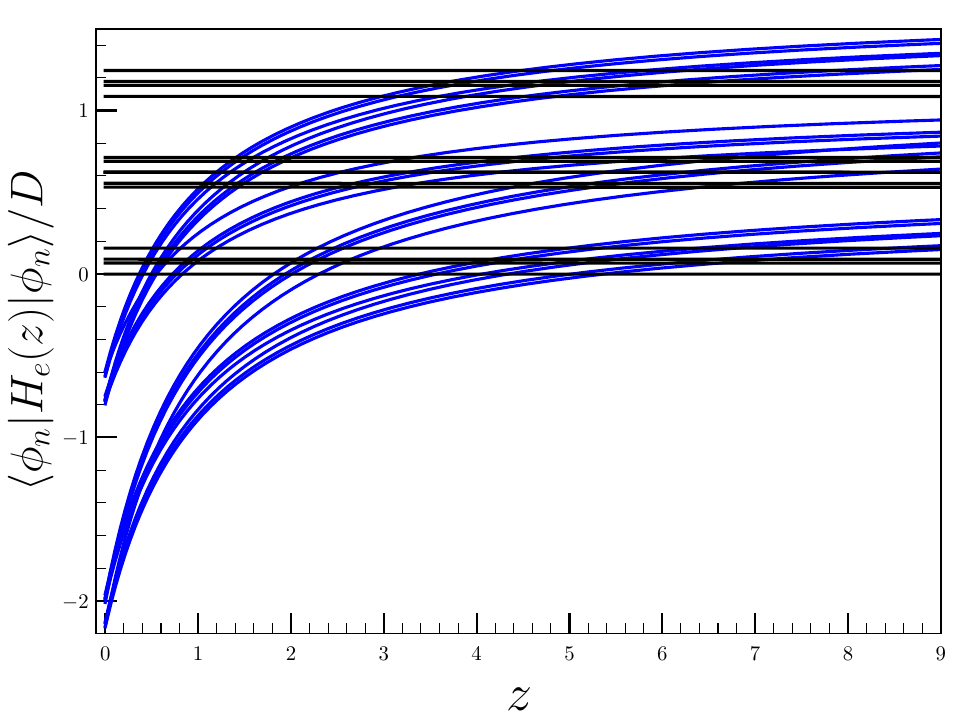}
        \vspace{-0.3cm}
	\caption{\footnotesize (Top panel) Expected value of $ H^{\mathrm{eff}} (z) $ \eqref{H_eff} using the basis at $z=L$ $\{ \ket{\phi_n} \}$. The color counts the electrons in the H orbital: black for 1, and blue for 2. (Bottom panel) Many-body energy spectrum of the $ H^{\mathrm{eff}} (z) $ \eqref{H_eff} as a function of $z$. \Sou}
        \label{Spinless_Potentials}
\end{figure} 
~
~
\newpage
The energy spectrum and the expected values of $ H^{\mathrm{eff}}(z)$ with $V(z) = 0$, so that $n_d$ is conserved are shown in Fig. \ref{Spinless_Potentials}. The results are similar to those found in Fig. \ref{Eletronic_Potentials_}, except for the $n_d= 0$ sector and the low energy $n_d=2$ states. The crossing between the energies for the initial neutral level $\ket{\phi_0}$ and for $\ket{\tilde \phi}$ occurs at approximately the same position (around $5$\Rb) where the crossing occurs for the original Hamiltonian, and the ionized levels drop similarly. Moreover, from Fig. \ref{Spinless_Potentials} (Top panel), we can observe that $\ket{\phi_0}$ and $\ket{\tilde \phi}$ hybridize at $z= 5$ and continue to hybridize with other ionized levels that cross the initial levels. Thus, the effective spinless Hamiltonian captures the  important levels of $H_e(z)$, except for the degeneracies originating from multiple choices of possible excitations with the same energy for different spin combinations.


Once we have the energies and the projections, we can build the Hamiltonian of each position $z$ by 
\begin{equation}
    \hat H_e(z) = \hat P(z) \hat E_e(z) P^\dagger (z).
\end{equation}
Here, $\hat E_e(z)$ is a diagonal matrix with the eigenvalues computed by the NRG. Now the total number of many-body states drops to only 35, which allows us to build the Hamiltonian that combines the atomic motion with the electronic behavior.

\section{Matrix representation of the Hamiltonian}

We are now ready to build the matrix representation of the total Hamiltonian $\mathcal{\hat{H}}$, which contains the electronic $\hat H_e$ Hamiltonian, the atomic kinetic energy $\hat K$ term and the atomic potential $\mathcal{\hat V}$ for each $z_q = \Delta z \cdot q$ ($q=0,1,...,800$), where $\Delta z = \frac{9.00}{800}$\Rb. Let us start with the atomic kinetic energy matrix $\hat{K}$, which  is proportional to the second derivative operator $\partial_z$, which in the central finite-difference approximation can be written as
\begin{equation}
[\hat \partial_z^2]_\mathrm{FD} \approx \frac{1}{\Delta z^2} 
\begin{pmatrix}
-5/2 & 4/3 &-1/12&  0  & 0   & ...\\
 4/3 &-5/2 & 4/3 &-1/12& 0   & ...\\
-1/12& 4/3 &-5/2 & 4/3 &-1/12& ...\\
\vdots&\vdots&\vdots&\vdots & \vdots& ...\\
 ... & 0   &  0  &-1/12& 4/3 &-5/2\\
\end{pmatrix}.
\end{equation}
The error from the discretization becomes smaller as $\Delta z$ decreases. By using the analytical solution for the free particle in a box, we verified that dividing the box into 800 pieces was sufficient to simulate 
the free hydrogen atom accurately.

Now, since the atomic part of the Hamiltonian is coupled with the electronic part, the atom behaves differently for each electronic configuration, and we must consider the electronic configuration in addition to the atomic position for all atomic operators. Then, the correct second-order derivative operator for our problem is
\begin{equation}
   \hat \partial ^2 \equiv [\hat \partial_z^2]_\mathrm{FD} \otimes \hat I_e.
\end{equation}
Here, $\otimes$ represents the tensor product and $\hat I_e$ is the identity operator for the electronic states.

The atomic kinetic energy operator is, therefore,
\begin{equation}\label{Kinetic_part}
\hat K  \equiv \frac{-1}{2M \Delta z^2} 
\begin{pmatrix}
-\frac{5}{2}\hat I_e & \frac{4}{3}\hat I_e&-\frac{1}{12}\hat I_e&  0           & 0   & ...\\
 \frac{4}{3}\hat I_e &-\frac{5}{2}\hat I_e& \frac{4}{3} \hat I_e&-\frac{1}{12} \hat I_e& 0   & ...\\
-\frac{1}{12}\hat I_e& \frac{4}{3}\hat I_e&-\frac{5}{2} \hat I_e& \frac{4}{3} \hat I_e &-\frac{1}{12}\hat I_e & ...\\
\vdots&\vdots&\vdots&\vdots & \vdots& ...\\
 ... & 0   &  0  &-\frac{1}{12}\hat I_e & \frac{4}{3}\hat I_e &-\frac{5}{2}\hat I_e \\
\end{pmatrix}.
\end{equation}

The surface reflection potential at $z = 0$ is automatically incorporated by the discretized derivative $\hat{\partial}_z^2$. However, the box has a finite size, which presents a challenge since the atom must have the freedom to collide with the surface, be reflected, and move away. To address this, we introduced an absorbing potential at the end of the box, in the region from  $z = L - z_c$ to $z = L$, where $z_c$ controls the size of this region as
\begin{eqnarray}
\mathcal{\hat V} \equiv -i\mathcal{V}
\begin{pmatrix}
0 & ... & 0 & ... & 0 \\
\vdots & \vdots & \vdots &... & \vdots \\
0 & ... & \hat I_e(L - z_c) & ... & 0 \\
\vdots  & \vdots & \vdots &... & \vdots \\
0 & ... & 0 & ... &  \hat I_e(L) \\
\end{pmatrix},
\end{eqnarray}
where  $\mathcal{V}$ controls the intensity of this absorbing potential. We determined this amplitude by numerically computing the quasi-free atom case ($\mathcal{\hat H} = \hat K + \mathcal{\hat V}$) and choosing $\mathcal{V}$ to make the probability of atomic reflection at the end of the box less than $1\%$.

Likewise, the electronic part of the Hamiltonian acquires the matricial form 
\begin{equation}
   \hat H_e \equiv \hat I_z \otimes \hat H_e(z_q),
\end{equation}
that is
\begin{eqnarray}
\hat H_e \equiv 
\begin{pmatrix}
\hat H_e(0) & 0 & ... & 0 \\
0 &  \hat H_e(\Delta z) & ... & 0 \\
\vdots & \vdots &... & \vdots \\
0 & 0 & ... &  \hat H_e(L) \\
\end{pmatrix}.
\end{eqnarray}

In this matrix representation, the atomic wavefunction can be simplified by grouping the atomic distribution for each electronic configuration and position $z$ into a vector as follows
\begin{equation}\label{wave_function_vector}
\vec{\chi}(t)  \equiv  
\begin{pmatrix}
& \vec{\chi}(0,t) \\
& \vec{\chi}(\Delta z,t) \\
& \vec{\chi}(2\Delta z,t) \\
& \vdots \\
&  \vec{\chi}(L,t) \\
\end{pmatrix},
\end{equation}

After that, the Schrödinger equation for the collision becomes simply
\begin{align}\label{SEQ.Collision_Matrix}
i\hbar \partial_t  \vec{\chi}(t) = \left[ \hat H_e + \hat{\mathcal{V}} + K \right]\vec{\chi}(t),
\end{align}
and it is not difficult to show that the solution is
\begin{align}
\vec{\chi}(t) = \exp\left[-i\frac{t}{\hbar} \left(\hat H_e + \hat{\mathcal{V}} + K \right) \right] \vec{\chi}(0).
\end{align}

However, the numerical complexity of the diagonalization, necessary to compute the exponential on the right-hand sde, is $\mathcal{O}(\mathrm{size}^3)$. In contrast, the Crank-Nicholson procedure \eqref{CN:2}, which depends on matrix inversions and multiplications, has complexity of $\mathcal{O}(\mathrm{size}^{2.4})$, which for large matrices results in a substantial difference in the computational resources required. Explicitly, by the Crank-Nicholson method, the evolution of the system can be found by computing
\begin{align}\label{wave_function_CN}
\vec{\chi}(m. \Delta t) \approx \left[   \left( 1 + i\frac{\Delta t}{2\hbar} \mathcal {\hat H} \right)^{-1} \left( 1 - i\frac{\Delta t}{2\hbar} \mathcal{ \hat H} \right)   \right]^m  \vec{\chi}(0), 
\end{align}
at time $t_m = m. \Delta t $.

\section{Atomic-surface collisions}

Finally, we are now ready to compute the atomic wavefunction and find the sticking coefficient.Table \ref{Paramenters_2} collects the parameters defining our computation.
\begin{table}[h!]
\caption{\footnotesize Parameters in or simulation of the H-Cu surface collision. }
\begin{center}
\begin{tabular}{| c | c | c | c | c |}
\hline
Parameter & Value [unities]& ~ & Parameter & Value [unities] \\
\hline
$M$ & 938.27 [MeV/c²] & ~ & $K_0$ & $\in$ [0,1] [eV] \\
$z_0$ & 5.75 [Bohr] & ~ & $\Delta t$ & 0.00001 [fs] \\
$L$ & 9.0 [Bohr] & ~ & $\Delta z$ & 0.01125 [Bohr] \\
$\eta$ & 0.6 [$\mathrm{Bohr}^2$] & ~ &  $z_c $&  $1.5$ [Bohr] \\
\hline
\end{tabular}
\end{center}
\begin{flushleft}
		\footnotesize Source: By the author. \
\end{flushleft}
\label{Paramenters_2}
\end{table}

The initial wave function vector $\vec{\chi}(0)$ in the matrix representation can be found by Eqs. \eqref{wave_function_t_0} and \eqref{wave_function_vector}.  Equation \eqref{wave_function_CN} then yields the atomic wavefunction along the box as function of the time, for each electronic configuration. After that, in our notation, the probability of finding the atom in the region $[z_A, z_B]$, in the electronic configuration $\ket{\phi_n}$, at the instant of time $t$ is found by
\begin{equation}
    P_n[t;z_A,z_B] = \int_{z_A}^{z_B} dz  \left[\vec{\chi}(z,t)\right]_n^* \left[\vec{\chi}(z,t)\right]_n, 
\end{equation}
or if one is interested only in the probability to find the hydrogen atom in this region 
\begin{equation}
    P[t;z_A,z_B] = \int_{z_A}^{z_B} dz  \vec{\chi}^\dagger (z,t) \vec{\chi}(z,t).
\end{equation}

\begin{figure}[hbt!]
		\centering
        \begin{tabular}{ll}
        \hspace{-0.5cm}
        \includegraphics[scale=0.5]{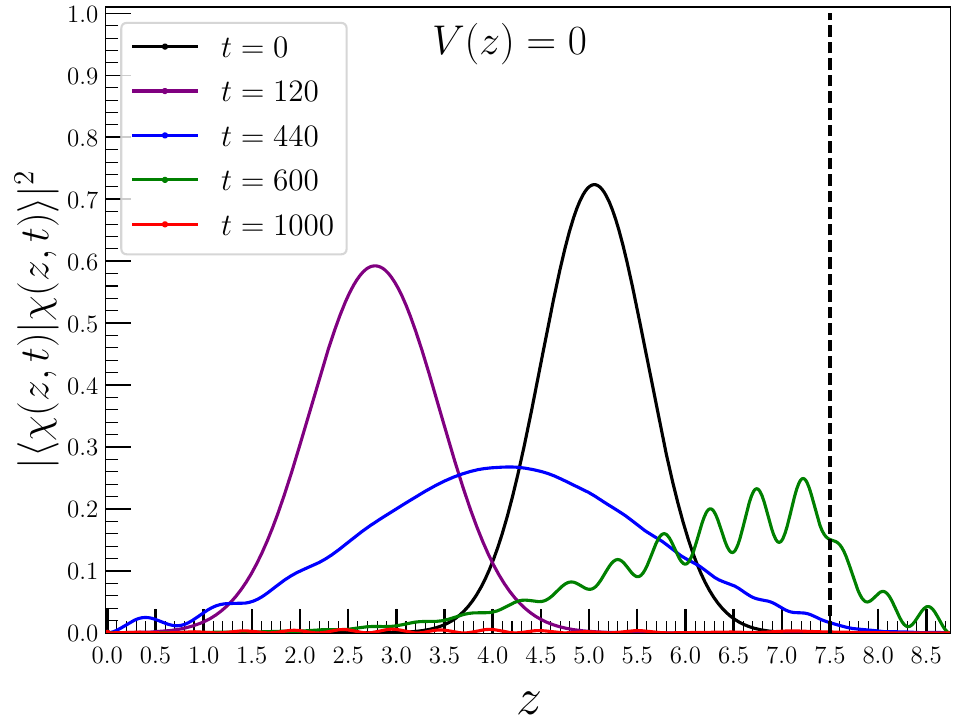}
        & \hspace{-0.63cm}
        \includegraphics[scale=0.5]{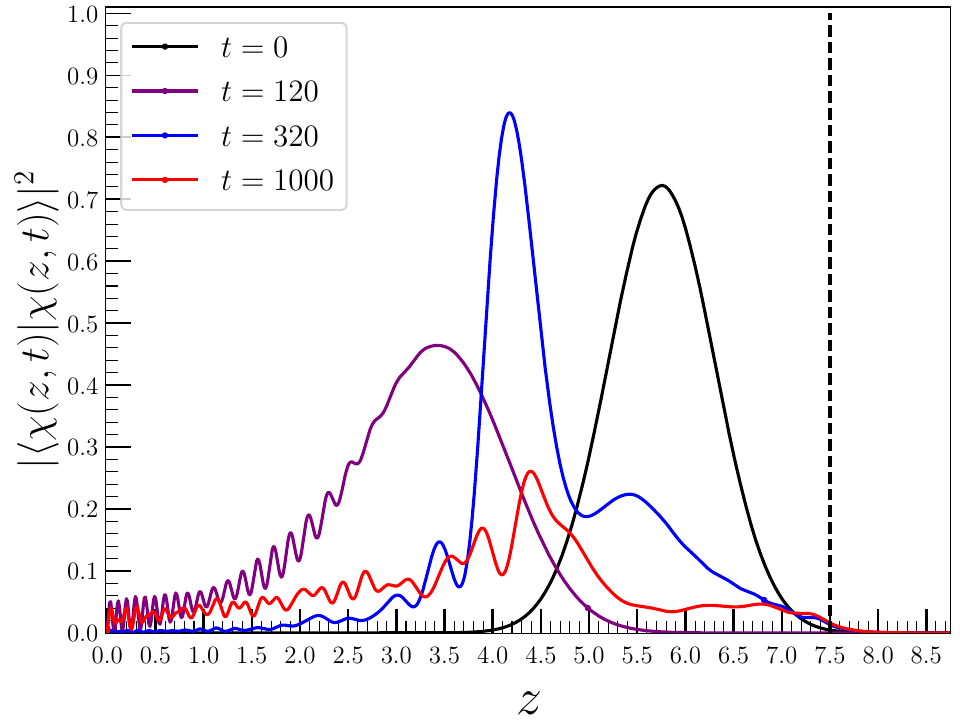}
        \end{tabular}
        \vspace{-0.5cm}
	    \caption{\footnotesize Atomic probability distribution as a function of position at indicated time with $K_0=0.3$ eV. (Left panel) Atomic orbital decoupled from the metal $V(z) = 0$. (Right panel) Atomic orbital coupled with the surface by $V(z)=V_0 \exp(-z/r)$. The dashed vertical black line delimits the absorbing region. Here, $z$ is in atomic units, and the time in units of $\frac{\hbar}{D}$. \Sou}
        \label{Atomic_Collision}
\end{figure}

Figure \ref{Atomic_Collision} shows the squared atomic wavefunction $ |\braket{\chi(z,t)}{\chi(z,t)}|^2 = \vec{\chi}^\dagger (z,t) \vec{\chi}(z,t)$, for a hydrogen atom with initial kinetic energy $0.3$ eV as a function of $z$ at different times, obtained from the parameters defined in Tables \ref{Paramenters_1} and \ref{Paramenters_2}. The left panel represents the case $V(z) = 0$, which bars electronic transfer between the surface and the atom, while the right panel represents the case $V(z) = V_0 \exp(-z/r)$ (electrons can be transferred near the surface).

Consider first the left panel ($V(z) = 0$). At $t=0$, the probability distribution is a Gaussian centered at $z_0$ (black curve). The atom moves towards the surface, collides with it, is reflected, and moves away from the surface (blue curve). After the atom enters the absorbing region, at $ t = 1000 \frac{\hbar}{D} $, the probability of finding the atom inside the box  is less than 1\% (red curve) and drops  to less than 0.1\%. 

Let us now focus on the right panel of Fig. \ref{Atomic_Collision} where $V(z)=V_0 \exp(-z/r)$. At time $ t=0 $, the probability distribution is the same as the previously case (black curve), the atom is neutral and the surface is in its initial ground state. The hybridization of the atomic orbital adds structures to the probability destiny. As time passes, the atom approaches the surface, the hybridization grows, allowing electronic transitions between the atom and the surface. The image-charge potential is switched one by the ionization and accelerates the atom and the atom towards the surface. Notice that, at time $ t = 120 \frac{\hbar}{D} $ (purple curve), the leading tail of the distribution stretches over the interval $0 \le z \le 1.5$\Rb, which is virtually empty, at the same time, in the left-panel - clear evidence of acceleration. 

Then, the accelerated atom collides with the surface, is reflected and moves away from the surface. At this point, the behavior is similar to the non-coupled case, but the atomic center of mass reaches  $ z = 4.5 $\Rb~ at $t = 320 \frac{\hbar}{D}$ (blue curve), whereas in the non-coupled case, it takes more time to reach the same position, another evidence of the acceleration by the image-charge potential. Additionally, at  $ t = 320 \frac{\hbar}{D} $ (blue curve), the atomic wavefunction is nearly split, around $z=5$\Rb, into two packets. The right-hand packet has sufficient kinetic energy to leave the box, while the left-hand packet has been trapped by the image potential and is at a turning point.

For long times, at $t=1000 \frac{\hbar}{D}$, the red curve shows that a significant fraction of the wavefunction remais in the box. The region $0 \le z \le 5$\Rb contains more than 30\% of the probability to find the atom, an indication that the probability to binding to the surface is substantial. In contrast to the elastic collision shown in the left panel, the hybridization with the metallic orbitals leads to non-adiabatic effects that have significantly robbed energy from the nuclear motion.

One of these non-adiabatic effects is the creation of pairs of particle-hole excitations in the metallic band. A less obvious mechanism is differentiated acceleration, where instead of the atom losing all the energy gained from the image-charge potential, the right-hand packet escapes with part of it. The kinetic energy of the right-hand packet in the blue plot, calculated as the average kinetic energy in the range $5$\Rb $\le z \le 9$\Rb, is larger than the original 0.3 eV. This packet leaves the box with added energy, a gain achieved at the expense of the left-hand packet.

\begin{figure}[hbt!]
	\centering
        \includegraphics[scale=0.75]{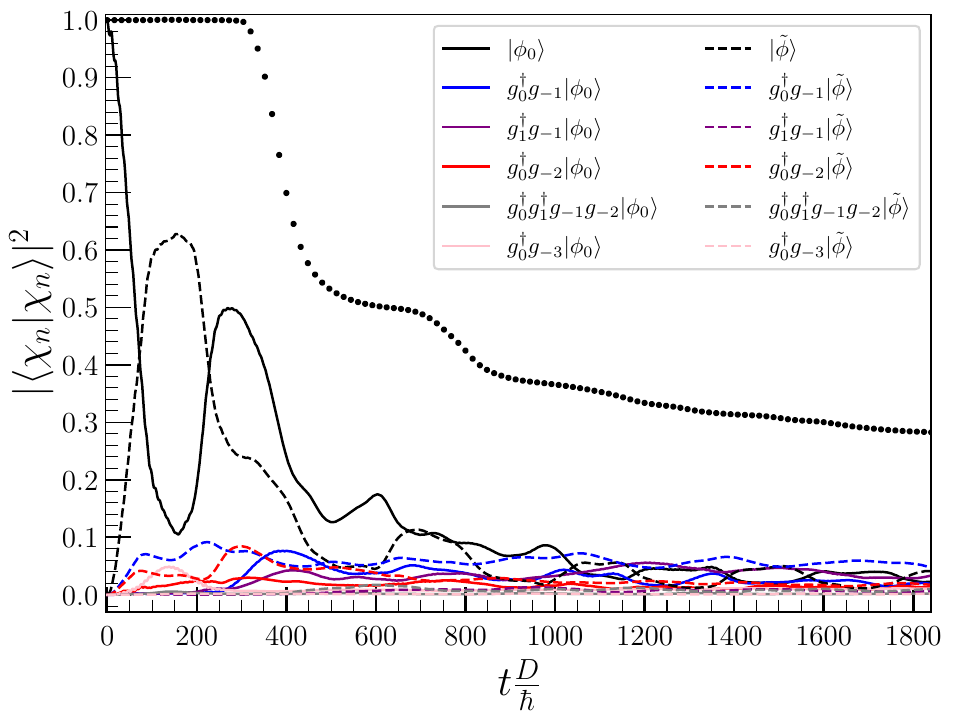}
        \vspace{-0.5cm}
	\caption{\footnotesize Probability to find the electrons in the indicated electronic configuration, as function of time. The solid lines represent electronic configurations where the atom is neutral and the dashed line represents the ionized atom. The operator $ g_k^\dagger g_{k'} $ indicates particle-hole excitations. Since $ \tilde{N} = 5 $, $ k = \{-3, -2, -1, 0, 1, 2\} $ represents the six levels of the conduction band. The total probability of finding the atom inside the box is shown by circular black dots. \Sou }
        \label{Atomic_Sticking_0.3}
\end{figure}

Figure \ref{Atomic_Sticking_0.3} shows the time-dependent probability to find the electrons in twelve configurations, computed by 
\begin{equation}\label{TP_C}
    |\braket{\chi_n(t) }{\chi_n(t)}|^2 \equiv  P_n[t;0,L] = \int_{0}^{L} dz  \left[\vec{\chi}(z,t)\right]_n^* \left[\vec{\chi}(z,t)\right]_n. 
\end{equation}
For convenient reference, each electronic configuration  $\ket{\phi_n}$ is described by the electronic eigenstates at $z=L$, where $V(z) \approx 0$ and the occupation of the atomic orbital is conserved. The solid lines represents electronic configurations with $\tilde n_d = 0$ (neutral atom), while the dashed ones correspond to $\tilde n_d = 1$ (the ionized electronic configurations).

As discussed in Section 7.2, two important things happen with the initial neutral electronic configuration. First, since the electronic Hamiltonian depends on \(z\), as the hydrogen moves towards the surface, the initial neutral state is no longer an eigenstate and decomposes into the eigenstates at each position \(z\). As the initial state hybridizes with the neutral state, near the surface these eigenstates become a combination of neutral and ionized states. This decomposition alone does not steal significant energy from the hydrogen; after interacting with the surface, the atom leaves the box. However, if the atom is moving quickly along this trajectory, in addition to the decomposition of the neutral state into eigenstates, high-energy particle-hole excitations appear, and these excitations steal a significant part of the kinetic energy.

Far from the surface, the atom-surface coupling is close to zero, transitions being improbable. As the atom approaches the surface and the coupling gradually increases, electronic transitions between the atom and the surface become possible. Near $z= 5$ \Rb, Fig. \ref{Spinless_Potentials} (bottom panel) shows the first crossing, between $\ket{\tilde{\phi}}$ and the initial state $\ket{\phi_0}$. A consequence of this crossing, in Fig. \ref{Atomic_Sticking_0.3} the solid black line decays rapidly at short times while the dashed black line rises steeply. Since $\ket{\phi_0}$ and $\ket{\tilde{\phi}}$ are the most probable states at short times, the two curves oscillate in opposition until other electronic configurations acquire significant  probabilities. The first half-cycle of this oscillation ends at $t = 180 \frac{\hbar}{D}$, when the atom is fully squashed against the surface and starts to move away from the metal. At this point, the probability of finding the electrons in the many-body state \(\ket{\tilde{\phi}}\) reaches its maximum.

Due to the image-charge potential, which increases in magnitude as the atom approaches the surface, the attractive potential lowers the energies of the ionized states, allowing particle-hole excitations from \(\ket{\tilde{\phi}}\) to cross the energy of the initial state and hybridize with it, as shown in Fig. \ref{Eletronic_Levels_Collision}. We can observe from Fig. \ref{Atomic_Sticking_0.3} that the probability to find the electrons in these ionized excited states (dashed blue and red curves) also increase for $t \in \left[0,180  \frac{\hbar}{D} \right]$, and these ionized levels also have significant probabilities.

After that, $t \in \left[180  \frac{\hbar}{D}, 300  \frac{\hbar}{D} \right]$, the atom is reflected and moves away from the surface, as shown in Fig. \ref{Atomic_Collision}. As the atom returns to its initial position, the state $\ket{\tilde{\phi}}$ transitions back to the initial state $\ket{\tilde{\phi}} \rightarrow \ket{\phi_0}$, completing the oscillation observed in Fig. \ref{Atomic_Sticking_0.3} at $t = 300 \frac{\hbar}{D}$, where the black solid curve reaches its maximum again. But along this return trajectory to the initial position, the state $\ket{\tilde{\phi}}$ also redistributes the probability to particle-hole excited states from $\ket{\phi_0}$ and $\ket{\tilde{\phi}}$, as indicated by the dashed blue and red curves, as well as the solid blue and red curves. 

For $t=350 \frac{\hbar}{D}$, the total black circles, representing the total probability initiate the sharp descent associated with the departure of the high-kinetic energy packet.  The remaining package has less energy than initially and becomes bound to the surface, as can be clearly observed in both Figs. \ref{Atomic_Sticking_0.3} and \ref{Atomic_Collision} for longer times.

\begin{figure}[hbt!]
		\centering
        \includegraphics[scale=0.6]{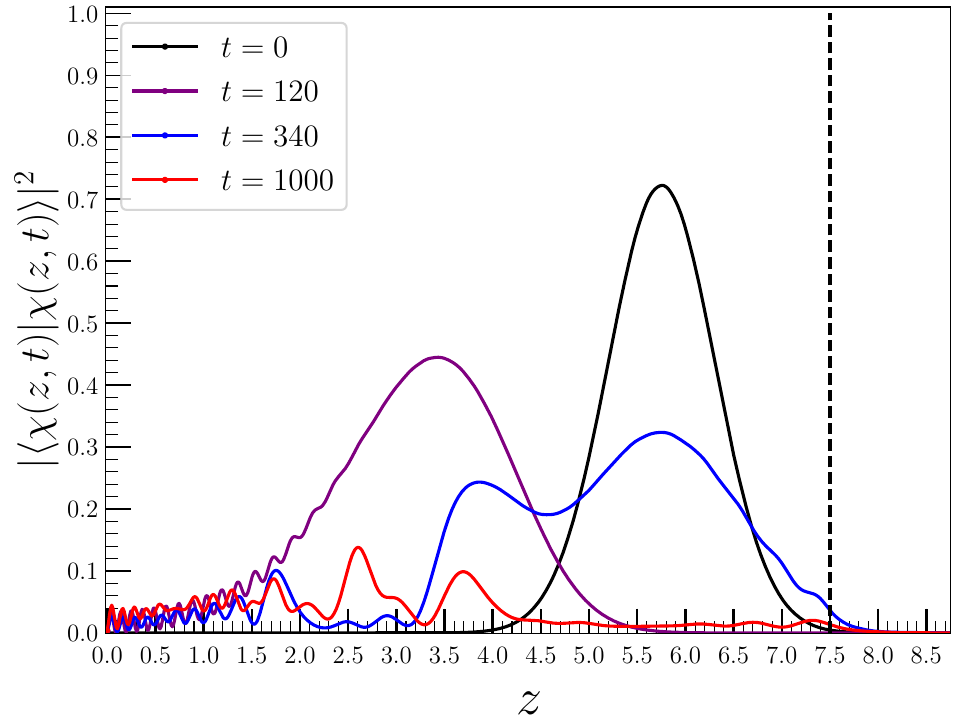}
        \includegraphics[scale=0.7]{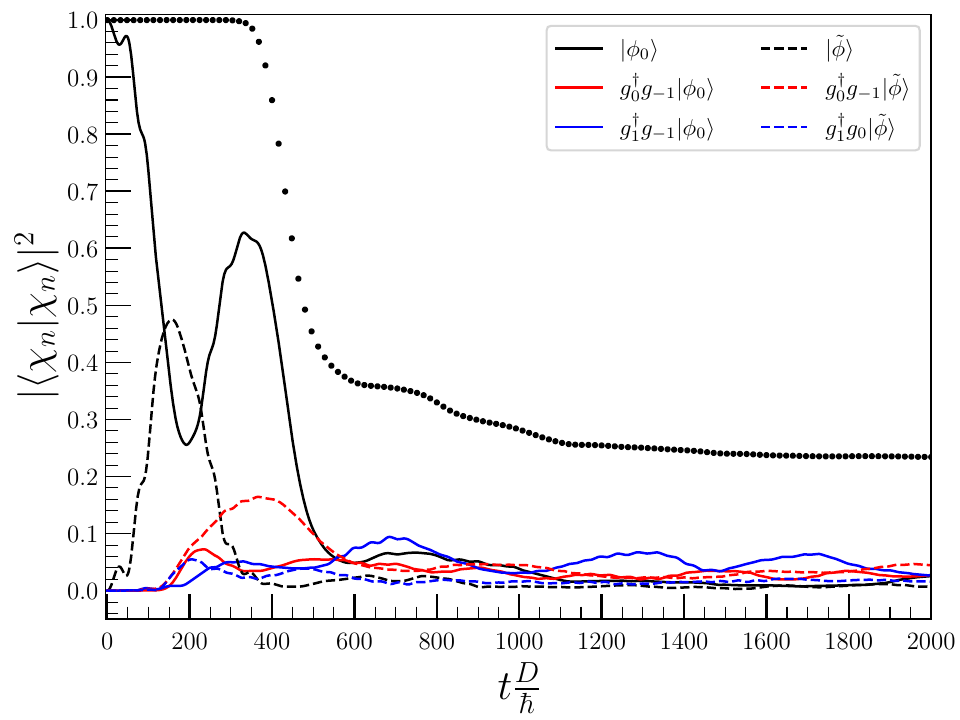}
        \vspace{-0.4cm}
	    \caption{\footnotesize (Top panel) Atomic probability distribution as a function of position at indicated time with $K_0=0.3$ eV. (Bottom panel) Probability to find the electrons in the indicated electronic configuration, as function of time. For this plot we used the NRG Hamiltonian \eqref{eq:4} with $\Lambda = 12$ and $\tilde N = 3$. \Sou}
        \label{Atomic_Sticking_0.3_WSpin}
\end{figure}

\newpage
Figure \ref{Atomic_Sticking_0.3_WSpin} verify that the spinless model preserves the physical content of the original Hamiltonian. The top panel is analogous to Fig. \ref{Atomic_Collision}, and the bottom one, to Fig. \ref{Atomic_Sticking_0.3}. The probability densities in both sets of plots resulted from the diagonalization of the NRG Hamiltonian \eqref{eq:4} with $\Lambda = 12$ and $\tilde N = 3$. This coarse description of the conduction band was necessary to keep the number of electronic states within practical limits. 

While quantitatively distinct, the plots in the corresponding sets of figures, Fig. \ref{Atomic_Collision} compared to Fig. \ref{Atomic_Sticking_0.3_WSpin} (top) and Fig. \ref{Atomic_Sticking_0.3} to Fig. \ref{Atomic_Sticking_0.3_WSpin} (bottom), are semi-quantitatively identical. All features identified in our discussions of the spinless model - the trapping mechanism, the evaporation of part of the wave function, the oscillatory time dependence of the electronic probabilities - all reappear in Fig. \ref{Atomic_Sticking_0.3_WSpin}.

\begin{figure}[hbt!]
	\centering
        \includegraphics[scale=0.48]{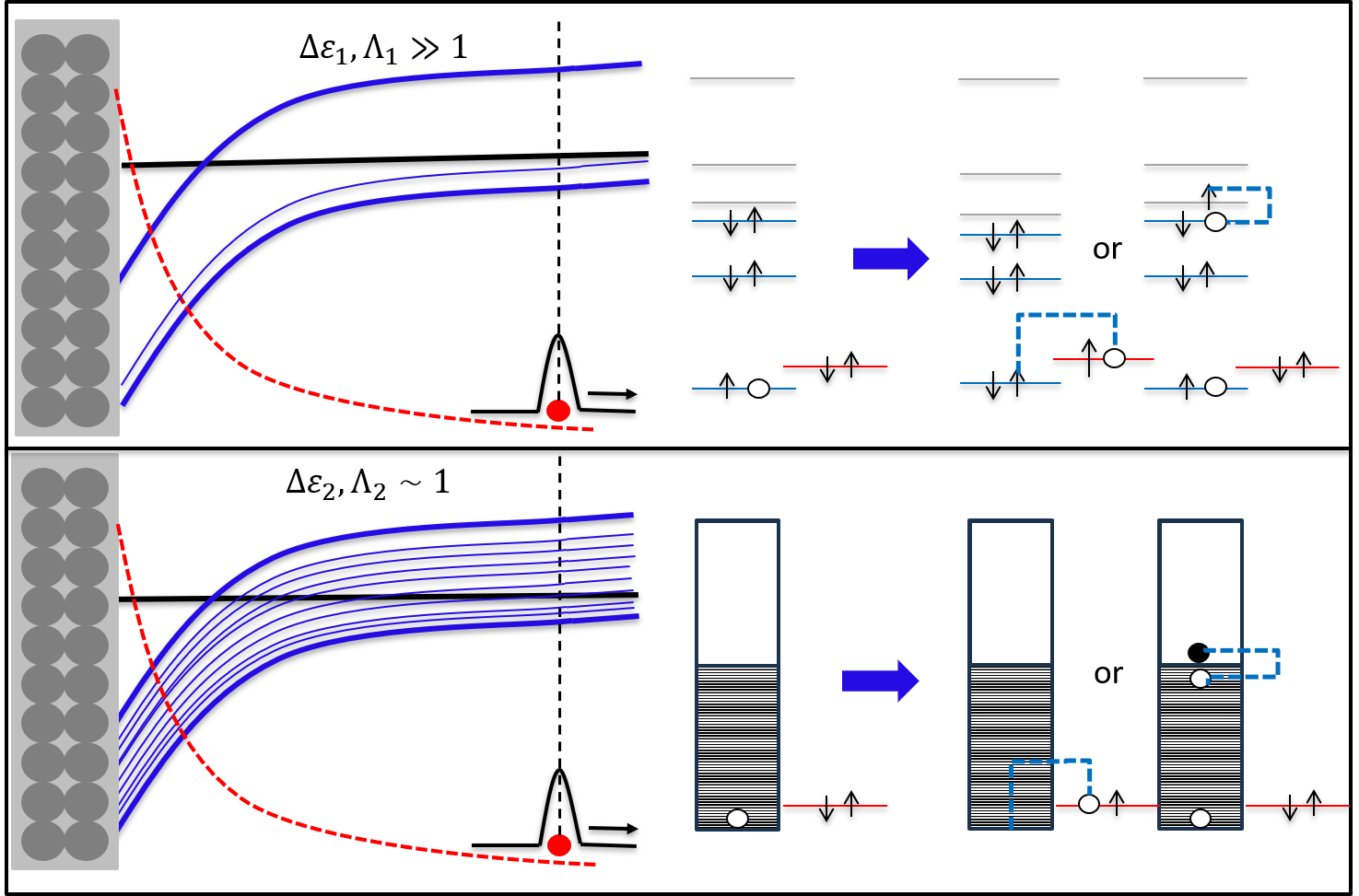}
	\caption{\footnotesize Schematic representation of the energy of the electronic configuration near the turning point around $z=5$\Rb. The black line represents the neutral atom, while the blue levels represent ionized states that cross the initial energy. The red dashed line represents the atom-surface coupling, which decreases exponentially with the distance, the white circles represent holes, and the blue dashed line represent electronic transitions. The top panel shows the NRG logarithmic discretization, only a few levels in the band, so the atom can only stay in this ionized state, transition to the neutral state, or transition to a small excited ionized state. The bottom panel shows the band levels are uniformly distributed, with many possible levels between the ionized and the neutral states (black line). \Sou }
        \label{Atomic_lifetime}
\end{figure}

Unfortunately, the lifetime of the atomic-surface binding in our computations is not infinite. Neither Eq. \eqref{eq:4} nor the spinless Hamiltonian are realistic for times much longer than the simulated time. In this situation, smaller energy scales need to be considered in the calculation. Since there is no energy dissipation term in the Hamiltonian, once trapped in the image-charge potential, the H atom will maintain an oscillatory motion, repeatedly moving from the surface to the turning point around $z=5$\Rb. As schematically shown in Fig. \ref{Atomic_lifetime} (top panel), when the trapped atom reaches the turning point, the small energy difference between the ionized state $\ket{\tilde{\phi}}$ and the lowest neutral state $\ket{\phi_0}$, the Heisenberg uncertainty principle allows transitions to this neutral state, losing kinetic energy. In the neutral state at $z \ge 5$\Rb, the H atom does not feel the image-charge potential, and is free to move away from the surface, and leave the box. The probability of finding the hydrogen inside the box gradually decrease.

However, as we discussed in the last paragraph, this finite lifetime effect should only appear if the surface band is discrete, as shown in Fig. \ref{Atomic_lifetime} (top panel). As the trapped ionized atom moves in the direction of the turning point at position $z=5$\Rb, it has a finite probability of transitioning to the neutral state, represented by the black horizontal line, because there are only a few levels in between. 

In the real situation, there are many ionized levels with energy uniformly distributed between these two levels, as shown in Fig. \ref{Atomic_lifetime} (bottom panel). These ionized levels behave like the Fermi gas system in the presence of a continuously changing scattering potential, as discussed in Chapter 4. This makes it more likely that, over time, more small-energy particle-hole pairs will be created, which continue to steal the atomic kinetic energy, trapping the atom in the image-charge potential and extending the lifetime to infinity.

Additionally, dissipative effects that naturally arise in this type of collision, such as the creation of phonons on the surface and infrared radiation produced by the ion's acceleration, would gradually drain the hydrogen's kinetic energy, pulling the atom closer to the surface and extending its lifetime indefinitely. These factors would lead to a stable hydrogen-surface binding. However, due to limited computational resources, we were unable to account for these effects in the hydrogen-surface collision simulations.

Keeping that in mind, to avoid the finite lifetime that results from computational limitations, we define the sticking coefficient as the remaining probability of finding the hydrogen atom inside the box after it has collided with the surface and returned to its initial position five times. At this time scale, our simulations for the non-interacting case with \(V(z) = 0\) show that the probability of finding the atom inside the box is less than 0.1\%. For the cases discussed in this section, we used $K_0 = 0.3$ eV, and based on this criterion, we found a sticking coefficient of 28\% using the spinless effective Hamiltonian and 25\% using the original electronic Hamiltonian with $\tilde N = 3$. In the following section, we will apply this criterion to compute the sticking coefficient for hydrogen-Cu surface collisions at different initial atomic kinetic energies $K_0$.

\newpage

\section{Computing the Sticking Coefficient}

\begin{figure}[hbt!]
		\centering
        \includegraphics[scale=0.75]{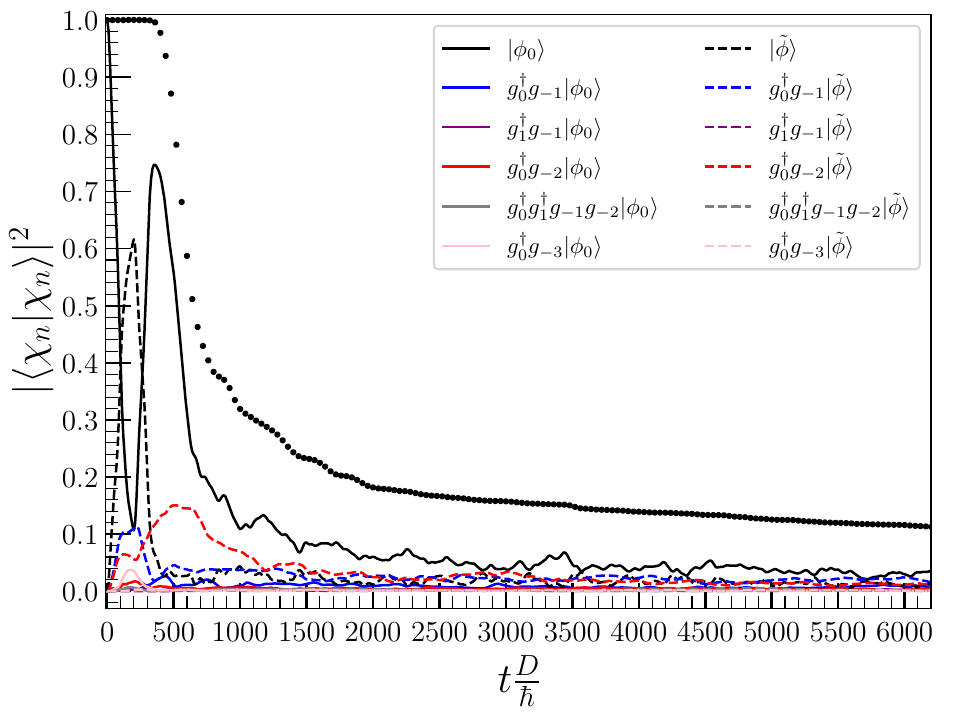}
        \includegraphics[scale=0.75]{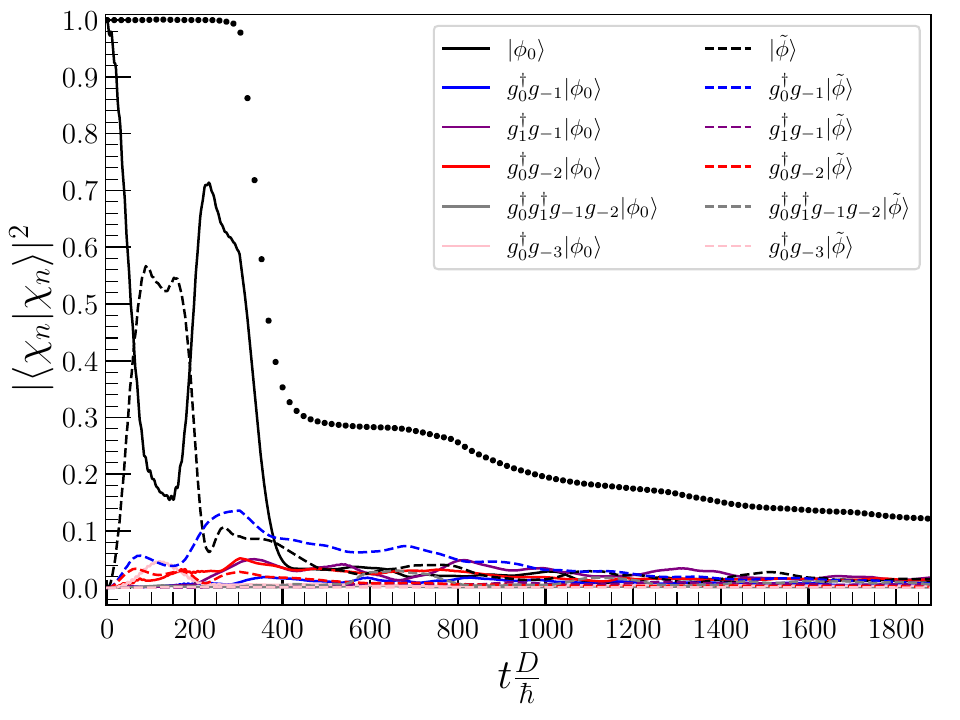}
        \vspace{-0.5cm}
	\caption{\footnotesize The total probability to find the atom inside the box over time (circular black dots ) for $K_0 =0.1$ eV (top panel) and for $K_0 = 0.5$ eV (bottom panel). The probability to find the H for different electronic configurations: Solid lines represent electronic states where the atom is neutral, and the dashed line the ionized states.  \Sou}
        \label{Atomic_Sticking}
\end{figure}

By our criterion for the sticking, the required simulation time as a function of the initial kinetic energy is shown in Table \ref{time_table}. For \(K_0 = 0.3\) eV, the example we just discussed, the sticking coefficient is \(\mathcal{S} = 0.28\%\). Figure \ref{Atomic_Sticking} shows the results for \(K_0 = 0.1\) eV and \(K_0 = 0.5\) eV, with the sticking coefficients found to be \(\mathcal{S} = 11\%\) for \(K_0 = 0.1\) eV and \(\mathcal{S} = 18\%\) for \(K_0 = 0.5\) eV.

~
\begin{table}[ht!]
\caption{\footnotesize The time required for the probability of finding a quasi-free hydrogen atom (\(V(z) = 0\)) inside the box to drop below 0.1\% is computed from our simulations.}
\begin{center}
\begin{tabular}{| c | c | c | c | c | c | c |c | c |}
\hline
$K_0[eV]$              &0.1& 0.2& 0.3 & 0.4 & 0.5 &  0.6 & 0.7 & 0.8  \\
\hline
$t_\mathrm{max} [\hbar/D]$ &5500& 2800 & 2000 & 1800 & 1600 & 1500 & 1400 & 1250 \\
\hline
\end{tabular}
\end{center}
\begin{flushleft}
		\footnotesize Source: By the author. \
\end{flushleft}
\label{time_table}
\end{table}

Qualitatively, the behavior of the hydrogen motion for different values of $K_0$ are very similar. The atom starts at the initial position $z_0$, moves towards the surface, interacts with the Cu surface, and is accelerated due to the image-charge potential. On its trajectory back to the initial position, the atomic wave function divides into two parts: one part goes away from the box (absorbed by the imaginary potential at the end of the box), and the other part reaches the returning point and becomes trapped by the surface, forming the hydrogen-surface bond.

On the other hand, the chance of finding the hydrogen in the box, clearly depends on the initial kinetic energy, as observed by comparing Fig. \ref{Atomic_Sticking} and Fig. \ref{Atomic_Sticking_0.3}. To understand the quantitative differences between the results for $K_0 = 0.3$ eV in Fig. \ref{Atomic_Sticking_0.3} and the results for $K_0 = 0.1$ eV and $K_0 = 0.5$ eV in Fig. \ref{Atomic_Sticking}, we analyze the electronic configuration of the hydrogen over time for both situations. 

Focusing on the top panel of Fig. \ref{Atomic_Sticking} ($K_0 = 0.1$ eV), we observe that after the hydrogen interacts with the surface, transitioning from the neutral to the ionized state $\ket{\phi_0} \rightarrow \ket{\tilde \phi}$ and being reflected, the probability of the hydrogen transitioning back to the electronic neutral state is higher than that observed for the $K_0 = 0.3$ eV case. When the atom interacts with the surface, the initial neutral state (black solid curve) transitions to mostly ionized states, indicated by the dashed curves with different colors—first to the black dashed curve, then to the blue, and followed by the red dashed curve, in the energy difference sequence that cross the initial neutral energy and hybridize with it.

\newpage

This behavior is a result of the crossing energy levels, indicating that what we observe is mostly the decomposition of the initial state into the electronic eigenstates at each position $z$, which are a combination of the neutral and ionized states, caused by the hybridization. This decomposition reflects the changes in the electronic Hamiltonian as the atom moves in the box, and if there are no extra non-adiabatic transitions, the atom does not lose significant kinetic energy.

This occurs because the atom with lower initial kinetic energy moves more slowly, causing the atom to experience the hybridization and the image-charge potential more gradually over each time interval. As a result, only small-energy particle-hole excitations become significant, analogous to the discussions in Chapter 4. In this case, the hydrogen loses less kinetic energy and tends to recover its initial electronic neutral configuration when it reaches the returning point at $z = 5$\Rb. Consequently, most of the atomic wave function does not feel the image-charge potential and leaves the box, resulting in a small sticking coefficient $\mathcal{S} = 11\%$.

In contrast, for  $K_0 = 0.5$ eV (bottom panel of Fig. \ref{Atomic_Sticking}), the atom moves faster, causing the electronic Hamiltonian to change significantly at each instant of time, which tends  to create higher-energy pairs of particle-hole excitations. These non-adiabaticities is evidenced by the neutral excitations (solid blue, red, and pink curves), nearly absent for the $K_0 = 0.1$ eV. Additionally, there is a significant difference between the electronic transitions observed for the cases $K_0 = 0.1$, $0.3$, and $0.5$ eV.

However, when $K_0 = 0.5$ eV, the hydrogen spends less time interacting with the surface. According to the Heisenberg uncertainty principle, small energy excitations do not have sufficient time to contribute. Consequently, after the hydrogen interacts with the surface, transitioning from the initial neutral state to the ionized state $\ket{\phi_0} \rightarrow \ket{\tilde{\phi}}$ and being reflected, the probability of the hydrogen transitioning back to the electronic neutral state is higher than in the $K_0 = 0.3$ eV case. The reason for this behavior is that the time the hydrogen remains in the region  $z \in [0, 2r]$ is not long enough for all transitions to occur as in the $K_0 = 0.3$ eV case, resulting in fewer particle-hole pairs being created and a smaller sticking coefficient.

\begin{figure}[hbt!]
		\centering
        \includegraphics[scale=0.84]{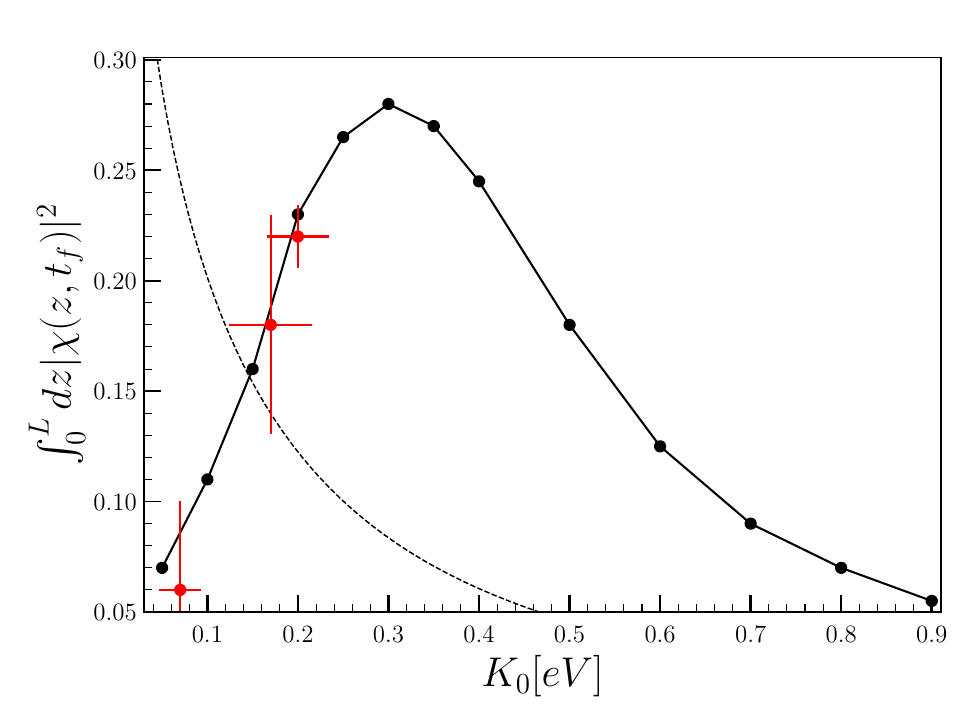}
        \vspace{-0.5cm}
	\caption{\footnotesize Sticking coefficient for different initial kinetic energies $K_0$. The black dots represent the contribution from the electronic processes to the sticking coefficient, using the parameters in Tables \ref{Paramenters_1} and \ref{Paramenters_2}. The dashed black line represents the phonon contribution computed by Bischler et al., Phys. Rev. Lett., 1993. The red dots represent experimental values from the literature (see text for the references). \Sou }
        \label{Sticking}
\end{figure}

Finally, the black dots in Fig. \ref{Sticking} shows the calculated sticking coefficient for various kinetic energies of incidence, along with experimental data for hydrogen Cu surface sticking (red dots). In addition, the contribution from phonons for the sticking  are shown by the dashed black line, adapted from Bischler et al., Phys. Rev. Lett., 1993 \cite{PhysRevLett.70.3603}. The experimental data obtained by Bischler \cite{PhysRevLett.70.3603} suggest a sticking coefficient value of approximately 0.18 for a quasithermal beam of hydrogen atoms with a mean kinetic energy of 0.16 eV on a Cu(110) surface. Another experimental study indicates that the sticking coefficient an incident kinetic energy of 0.07 eV falls between 0.02 and 0.1 on a Cu(110) surface \cite{10.1063/1.471006}. Experiments using a beam of neutral hydrogen atoms with an initial kinetic energy of 0.2 eV on Cu(111) surface resulted in a sticking coefficient of 0.22 \cite{HOFMAN2018153,10.1063/1.480145}.

The results in Fig. \ref{Sticking} are in line with our discussions of the collision. The atomic motion is coupled with the electronic configurations. Since the image-charge potential $W(z)$ and the coupling $V(z)$ depend on the atomic position as shown in Fig. \ref{Eletronic_Levels_Collision}, fast movement of the atom close to the surface causes a sudden change in the electronic Hamiltonian. These changes can create pairs of particle-hole excitations, which absorb a considerable part of the atomic kinetic energy, resulting in the atomic-surface bounding.

However, the number of particle-hole pairs created depends on the time the atom remains inside the region  $z \in [0, 2r]$, which decreases as the kinetic energy $K_0$ increases. If the initial hydrogen kinetic energy is too high, the atom passes too quickly through the region where the hybridization is significant. Under these circumstances, the atom does not have enough time to transition to the ionized excitations, as shown by Fig. \ref{Atomic_Sticking},  and does not lose enough kinetic energy to become trapped, resulting in minimal sticking.

Otherwise, if the atom moves slowly, the Hamiltonian changes gradually, only small energy particle-hole excitations will be created. In the absence of dissipation, the atom will merely accelerate, collide with the surface, and rebound. This is a reversible process, which leads to very little sticking.

Our findings clearly demonstrate this, as shown in Fig. \ref{Sticking}. Only at intermediate energy scales, where there is a compromise between the conditions needed to create non-adiabatic effects and sufficient time for these effects to manifest, that we observe a maximum probability of binding to the surface. Specifically, around $ K_0 = 0.3 $ eV, the atom has the highest probability of binding to the surface, and this probability drops as $ K_0 $ deviates from this value. Fig. \ref{Sticking} also shows that phonons, by themselves, cannot account for the experimental data \cite{PhysRevLett.70.3603,10.1063/1.471006,HOFMAN2018153,10.1063/1.480145}. Our results, which focus on the non-adiabatic contributions from the electronic degrees of freedom explain the experimental findings well. This indicates that the primary sticking mechanism in these interactions is more likely due to the complex interplay between the atomic motion and the electronic excitations rather than solely phonon interactions.



The non-adiabatic effects that arise from atom motion primarily depend on the atomic-surface coupling. While most parameters of the Cu surface-hydrogen collision model are determined by experiment or DFT calculations, the range parameter $ r $  is difficult to estimate. Table \ref{Stiking_table} shows the results of varying $ r $, from $r=1$ (neutral hydrogen atomic radius) to $r=2.6$ (negatively ionized atomic radius),  for different values of $ K_0 $.

\begin{table}[ht!]
\caption{\footnotesize Sticking coefficient for different values of $r$ and $K_0$.}
\begin{center}
\begin{tabular}{| c | c | c | c | c | c | c |}
\hline
- -       &$K_0[eV]$& 0.15& 0.2 & 0.3 & 0.4 &  0.5 \\
\hline
$r$ [Bohr]& - -     & - - & - - & - - & - - &  - - \\
\hline
1.00      &- -      & 10\% & 11\%& 10\%& 7\% &  5\% \\
\hline
1.67      &- -      & 16\% &23\% &  28\% & 24.5\% & 18\% \\
\hline
2.60      &- -      & 20\% &24\% & 30\% & 32.5\% & 27\% \\
\hline
\end{tabular}
\end{center}
\begin{flushleft}
		\footnotesize Source: By the author. \
\end{flushleft}
\label{Stiking_table}
\end{table}

As $ r $ increases, the initial atomic kinetic energy that maximizes the sticking shifts to higher values. Likewise, the maximum sticking coefficient also increases with $ r $. This behavior occurs because, as previously discussed, the sticking is maximized when there is a balance between the conditions needed to create non-adiabatic effects and the time required for these effects to manifest. For example, considering the reference case at $ r = 1.67 $\Rb , the maximum sticking occurs for $ K_0 = 0.3 $ eV. For smaller $ r $, when $ K_0 = 0.3 $ eV, the atom spends less time inside the interaction region. Consequently, the balance between kinetic energy and time tends to happen at a lower kinetic energy, resulting in less energy loss and a lower maximum sticking coefficient ($ \mathcal{S} $). 

In contrast, for larger $ r $, this balance tends to occur at a higher initial kinetic energy. Additionally, as $ r $ increases, the interaction between the hydrogen atom and the surface starts earlier, leading to a higher maximum $ \mathcal{S} $.

In summary, our spinless model describes the hydrogen-Cu surface collision well. Future work with this model may yield more results for the hydrogen-Cu surface collision, such as the electronic friction, the production of hydride (ion $\mathrm{H}^-$) from neutral hydrogen, and the survival probability for the ions.

%% file: CAP/Cap8.tex
\chapter{Conclusion}

This work, has delved into the physics of time-dependent many-body systems. Specifically, we have studied (1) x-ray photoemission from a simple metal, (2) a Fermi gas subject to a growing scattering potential, and (3) the collision of a H atom with a metallic surface. The three Hamiltonians being similar, we have built our treatment of each problem upon the experience gained with the previous ones, that is, the problem (2) upon (1), and problem (3) upon (1) and (2). 

The photoemission problem has a long history. Experimental physicists prefer the frequency domain, and so the theoretical attention has been focused on the energy dependence of the photoemission current. In this context, the most important advance came in 1970, when Doniach and Sunjic described this dependence in the vicinity of the threshold energy for photoexcitation. The remarkable success of their results turned attention away from higher energies. 

In our time-domain analysis, by contrast, the contributions to the photocurrent amplitude from a class of high-energy excitations gained prominence, namely the amplitude due to excitations from the bound state created by the attractive scattering potential to single-particle states above the Fermi level. Such excitations behave as x-ray absorption problem study by Nozières-De Dominicis. The contribution , from the unplugged final many-body states, interfere with the amplitude due to the plugged states, final many-body states in which the bound state is filled, giving rise to eye-catching oscillations in the plots of the photocurrent as a function of time.

Our results have shown that, while the amplitude from the plugged states follow the Doniach-Sunjic law, the amplitude from the unplugged ones decay following the Nozières-De Dominicis law. The weaker the potential, the faster is the decay of the latter, relative to the former. For weak potentials the contribution from the unplugged states decays so fast that the oscillations are barely visible. For strong $W$s, the oscillations are long lived.  These oscillations  give rise to a higher-energy threshold in the energy domain, which corresponds to the broad high-energy peak often seen in experimental spectra. The broadening, due to the sort lifetime of the vacant bound state, makes the asymmetry associated with the threshold behavior difficult to distinguish; experimental work focused on the high-energy behavior will be necessary before the Nozières-De Dominicis exponent can be identified. The physical interpretation of the damped oscillations in the time dependence of the photoemission current is one of the most important results in this thesis.

Another important conclusion was derived from our study of the Fermi gas subject to an up-ramping localized potential.  The issue, here, is whether the gas remains in the instantaneous ground state up to end of the ramp. On the basis of a simple approximation, in which only the ground state and single-particle-hole excitations are allowed in final states, we have been able to derive an expression defining the conditions under which the probability of transition to an excited state remains bellow a specified threshold value. Numerical computation of the same probability has confirmed the accuracy of our expression.

The results shows striking disagreement with the quantum adiabatic criterion, according to which the threshold probability should be proportional to the square of the slope of the ramp-up. Instead, we have found the probability to obey a power law, in which the argument is the number of electrons participating in the process - limited by the uncertainty principle - and the exponent depends only on the phase shift introduced by the maximal scattering potential. Thus, even trough the threshold probability depends on the maximal potential and on the time necessary to reach it, it is by no means a function of the ratio.

Finally, using our experience with the Fermi-gas and the photoemission, we have been able to describe the electronic transitions in the hydrogen-Cu surface collision. The electronic transitions mimic the evolution of the Fermi gas under the ramp-up potential when the atom is far from the surface and mimic the evolution under the sudden turning on of the potential in the photoemission problem when the atom is close to the surface. However, the motion of the atom being coupled to the evolution of the electrons, this qualitative depiction is too simplistic, in the same way that the Born-Oppenheimer approximation is inadequate.

Trustworthy analysis of the problem calls for full treatment of the atomic and electronic degrees of freedom. The NRG method takes care of the latter; even with this simplification, however, we have found the computational cost of following the evolution of the wave function to be prohibitive. This led to the substitution of the more realistic spin-degenerate Hamiltonian by the simplest spinless one. The derivation of the spinless Hamiltonian combined inferences drawn from our previous analyses and the Fermi gas with the results from the diagonalization of the spin-degenerate electronic Hamiltonian.

The substitution has allowed us to simulate the collision and compute the sticking coefficient. We have computed the sticking coefficient as a function of the initial kinetic energy, far from the surface, and interpreted the results physically. The coefficient peak is at intermediate energies, around $0.3$  eV. At low energies, the probability of adsorption is small because the slow movement only allows low-energy particle-hole excitations, close to the Fermi level, to participate, which makes the collision quasi-adiabatic.  At high kinetic energies, at the opposite extreme, there is little time for transitions between the initial, neutral atomic state and the ionized state. Since the sticking calls for ionization, the corresponding sticking coefficient is minute. Only at intermediate kinetic energies is there sufficient nonadiabaticity and enough time for the electrons transitions to rob energy and trap the nucleus. Only at intermediate energies is there balance between the conditions for non-adiabatic effects and the time for these effects to take place. Only then is the probability of the atomic-surface binding substantive.

Additionally, the sticking coefficients we have computed are in better agreement with the available experimental data than the phonon mechanism, reproducing well the experimental data when the initial kinetic energy is in the region of $0.1 - 0.25$ eV. Our approach can be extended to simulate collisions and features of the model can be modified to make it more realistic. 

A more important aspect nevertheless deserves attention is this paragraph. All the models we have studied suffer from a shortcoming that obstructs the computation of time-dependent properties: the lack of mechanism to dampen low-energy excitations. Even the simplest mode, the one discussed in Chapter 3, allows cascading of energy from high- to low-energy modes. The transient contribution from the unplugged states to the photoemission current is the most striking example. Nonetheless, the low-energy electron-hole excitations cannot decay. In the real systems, the coupling to the electromagnetic fields and to vibration modes damp high- and low-frequency osculations and broaden peaks and thresholds in experimental spectra. The extension of the simple model to encompass terms that represent couplings of this sort is, therefore, a natural extension of our work.


In closing, we highlight the importance of high-energy states in the computation of time-dependent properties. In the photoemission problem, the particle-hole excitations from the bound state to vacant conduction levels shape the oscillations characteristics of the calculated photocurrents. In the atom-surface collision problem, our preliminary results showed that truncation of the NRG spectrum, even eliminating of a few states at the high-energy tail of the spectrum, is sufficient to introduces significant deviations in the calculated sticking coefficients. The development of the block-diagonal approach described in Chapter 6, which makes the computation of excited states in the iterative diagonalization of the NRG Hamiltonians more efficient, was motivated by these findings.

\section{Technical results}

Throughout this thesis, new technical results have been presented. Here, we highlight these contributions, which we have not found documented in the literature.

From a more mathematical perspective, we derived accurate expressions based on the Cauchy determinant formula. These include the projections between the initial ground state and the final ground state in Eq. \eqref{AOC EQ II}, as well as the projection between the initial ground state and the lowest-energy unplugged state in Eq. \eqref{Bound_State_MB_proj}. Additionally, for the x-ray photoemission problem, we provided analytical expressions for the Green's function in the frequency domain (see Eq. \eqref{ImGw_}) and for the fidelity in the time domain (see Eq. \eqref{Fidelity_time}), along with the relative contribution of excitations from the bound state, quantified by the ratio defined in Eq. \eqref{Relative_Contribution_BS}. 

For the Fermi gas problem, we provided an analytical equation that effectively describes the time evolution of the system (see Eq. \eqref{SEFM_}), the probability of finding the system in the ground state (see Eq. \eqref{AS}), and the adiabatic diagram (see Fig. \ref{Fig: Phase_Diagram}).

In Chapter 6, we proposed a new procedure, based on the block-diagonal approximation demonstrated in this thesis, to account for the contributions of high-energy excited states in the desired physical properties of NRG-like Hamiltonians. We showed that this procedure not only works with high accuracy but can also reduce the computational cost and increase the efficiency of the classical NRG approach.

Additionally, in Chapter 7, we introduced an effective spinless Hamiltonian that accurately describes the atomic-surface collision. Using this effective Hamiltonian, we calculated the sticking coefficient for a hydrogen atom, initially neutral, colliding with a Cu surface as a function of its initial kinetic energy.

%% file: CAP/Apendices.tex

\begin{apendicesenv}
\partapendices

\chapter{Analytical Diagonalization}\label{Appendix_Analytical_Diagonalization}

Let us start with the Hamiltonian in Eq. \eqref{H_photo_MI}. In order to simplify it, we can consider a flat band and write the localized scattering potential term in the momentum-space, which transforms
$W a_0^\dagger a_0 \rightarrow \frac{W}{N} \sum_{k,q} \tilde{a}_k^\dagger\tilde{a}_q $. Then, the Hamiltonian is now:
\begin{align}\label{H_photo_ap}
H = \sum_{k}\varepsilon_{k}\tilde{a}^\dagger_{k}\tilde{a}_{k} + \frac{W}{N} \sum_{k,q} \tilde{a}_k^\dagger\tilde{a}_q.
\end{align}
Where $\varepsilon_{k} = k.\Delta\varepsilon$, $\Delta\varepsilon$ is the energy gap and $k$ an integer such that $-N/2 \leq k \leq N/2$.

Furthermore, we can diagonalize this Hamiltonian if we can write it in the form $H = \sum_{m} \epsilon_{m} g_{m}^\dagger g_m$, where the operator $g_m$ can be written as a linear combination of the operators $\{\tilde a_k\}$ as
\begin{eqnarray}\label{gm}
g_m = \sum_k \alpha_{k,m} \tilde a_{k}.
\end{eqnarray}

Before proceeding, we can calculate the main commutators of $H$ for this problem. After some manipulations, we arrive at the following expressions:
\begin{eqnarray}\label{[H,Ck]}
[H,\tilde a_k] = (-1) \left[\varepsilon_k \tilde a_k + \frac{W}{N} \sum_q \tilde a_q    \right],
\end{eqnarray}
\begin{eqnarray}\label{[H,gm]}
[H,g_m] = -\epsilon_{m}g_m.
\end{eqnarray}

Using the equality defined in equation \eqref{[H,gm]} by substituting $g_m$ with the linear combination defined in \eqref{gm} and after some algebraic manipulations, we arrive at:
\begin{eqnarray*}
\sum_k \left[ \varepsilon_k \alpha_{k,m} +\frac{W}{N} \sum_q \alpha_{q,m} \right] \tilde a_k = \sum_k \epsilon_m \alpha_{k,m}\tilde a_k.
\end{eqnarray*}

As the operators $\tilde a_k$ are independent for each $k$, we can separate this equation into
\begin{eqnarray}\label{E__I}
\left( \epsilon_m-\varepsilon_k \right) \alpha_{k,m} - \frac{W}{N} \sum_q \alpha_{q,m} = 0.
\end{eqnarray}

Thus,  by applying a sum over $k$ to both sides of Eq. \eqref{E__I}, we obtain
\begin{eqnarray}
\sum_k \alpha_{k,m} = W \left( \frac{1}{N} \sum_k \frac{1}{\left( \epsilon_m-\varepsilon_k \right)} \right) \sum_q \alpha_{q,m},
\end{eqnarray}
where the sums $\sum_q \alpha_{q,m}$ and $\sum_k \alpha_{k,m}$ are equal and non null, resulting in 
\begin{eqnarray}\label{Auxiliar_I}
1 = W \left( \frac{1}{N} \sum_k \frac{1}{\left( \epsilon_m-\varepsilon_k \right)} \right).
\end{eqnarray}

\newpage
Before proceeding, it is necessary to solve the sum that appears in the previous equation. To do so, and to simplify the calculations, we will consider a flat density of states, $\rho= \frac{1}{\Delta\varepsilon N}$, where $\Delta\varepsilon$ is the difference between two consecutive energy levels. Furthermore, the energy levels have an approximate energy given by $\varepsilon_q = q \Delta\varepsilon$. Considering that the $W$ in the Hamiltonian modify the initial energy levels by a quantity $\delta$, known as phase-shift, we can write
\begin{eqnarray}
\epsilon_m = \varepsilon_m - \frac{\Delta\varepsilon}{\pi} \delta_m. 
\end{eqnarray}
The sum in question appears again at several points during the calculations, so it is convenient to define a variable $\xi_m$ such that:
\begin{eqnarray}
\xi_m = \frac{1}{N} \sum_q \frac{1}{\epsilon_m-\varepsilon_q - \frac{\Delta\varepsilon}{\pi} \delta_m} = -\rho \sum_q \frac{1}{q - m + \frac{\delta_m}{\pi}}.
\end{eqnarray}

To solve this sum, we can use the Sommerfeld-Watson transformation \cite{sommerfeld1949partial}. Considering a sum of the form $\sum_n f(n)$, where $f(z)$ is a function with non-integer poles, we can define a function $F(z) = \pi f(z) \cot (\pi z)$. The function $F(z)$ has simple integer poles $\{n\}$ coming from the term $\sin (\pi z)$ and non-integer poles $\{z_{n_i}\}$ coming from the term $f(z)$. The integral around a closed curve of $F(z)$ in a complex domain satisfies the equality:
\begin{eqnarray}
\frac{1}{2\pi i} \oint F(z) dz = \sum_n \text{Res}(F(z),z=n) + \sum_{n_i} \text{Res}(F(z),z=z_{n_i}).
\end{eqnarray}

The residue $\text{Res}(F(z),z=n)$ is calculated by:
\begin{eqnarray}
\text{Res}(F(z),z=n)= \lim_{z \rightarrow n} F(z)(z-n) =\lim_{z \rightarrow n} \pi f(z)\cos(\pi z) \frac{(z-n)}{\sin(\pi z)} = f(n).
\end{eqnarray}

Thus, the integral can be rewritten as:
\begin{eqnarray}
\frac{1}{2\pi i} \oint F(z) dz = \sum_n f(n) + \sum_{n_i} \pi \cot (\pi z_{n_i}) \text{Res}(f(z),z=z_{n_i}).
\end{eqnarray}

Or, since we are interested in finding the sum:
\begin{eqnarray}
\sum_{n= -\infty}^{\infty} f(n) = \frac{1}{2 i} \oint f(z) \mathrm{cot}(\pi z) dz - \pi \sum_{n_i} \mathrm{cot}(\pi z_{n_i}) \text{Res}(f(z),z=z_{n_i}).
\end{eqnarray}

The discussion conducted in the previous paragraphs can be applied to find the value of $\xi_m$, using $f(n)=\frac{-\rho}{z-\left(m-\frac{\delta_m}{\pi}\right)}$ with a single simple pole $z_1= m-\frac{\delta_m}{\pi}$, where the sum becomes:
\begin{eqnarray}\label{ComplexIntegral}
 \xi_m = -\rho \frac{1}{2i} \oint \frac{\mathrm{cot} (\pi z)}{z-\left(m-\frac{\delta_m}{\pi}\right)}  dz + \rho \pi \mathrm{cot}  (\pi m - \delta_m).
\end{eqnarray}

The last term can be simplified to $\cot (\pi m - \delta_m) = -\cot (\delta_m)$. Now we only need to find the integral. For this purpose, it is necessary to define a closed path in the domain of the function $F(z)=f(z) \mathrm{cot}  (z)$ as shown in Fig.\ref{Sommerfeld-Watson}. Choosing this path and taking the limit where the box size tends to infinity, once $\lim_{(\mathrm{Im} z \rightarrow \pm \infty)}\cot(\mathrm{Re}(z) +i\mathrm{Im} z) = \pm -i$, it is not difficult to show that the contribution to $\xi_m$ is $\rho \mathcal{P} \int_{-D}^{+D} \frac{d\varepsilon}{\epsilon_m - \varepsilon - \Delta\varepsilon\frac{\delta_m}{\pi} }$. Then, we obtain:
\begin{eqnarray}\label{soma}
\xi_m =  - \rho \pi \mathrm{cot}(\delta_m) + \rho \mathcal{P}~\int_{-D}^{+D} \frac{d\varepsilon}{\epsilon_m - \varepsilon - \Delta\varepsilon\frac{\delta_m}{\pi} }. 
\end{eqnarray}
Here, $\mathcal{P}$ represents the Cauchy principal part, $D$ is the width of the band, the contribution $-\pi\rho \cot(\delta_l)$ comes from the pole inside the range $[-D,+D]$, and $\rho = \frac{1}{D}$.

\begin{figure}[hbt!]
		\centering
        \begin{tabular}{ll}
        \hspace{-0.5cm}
        \includegraphics[scale=0.44]{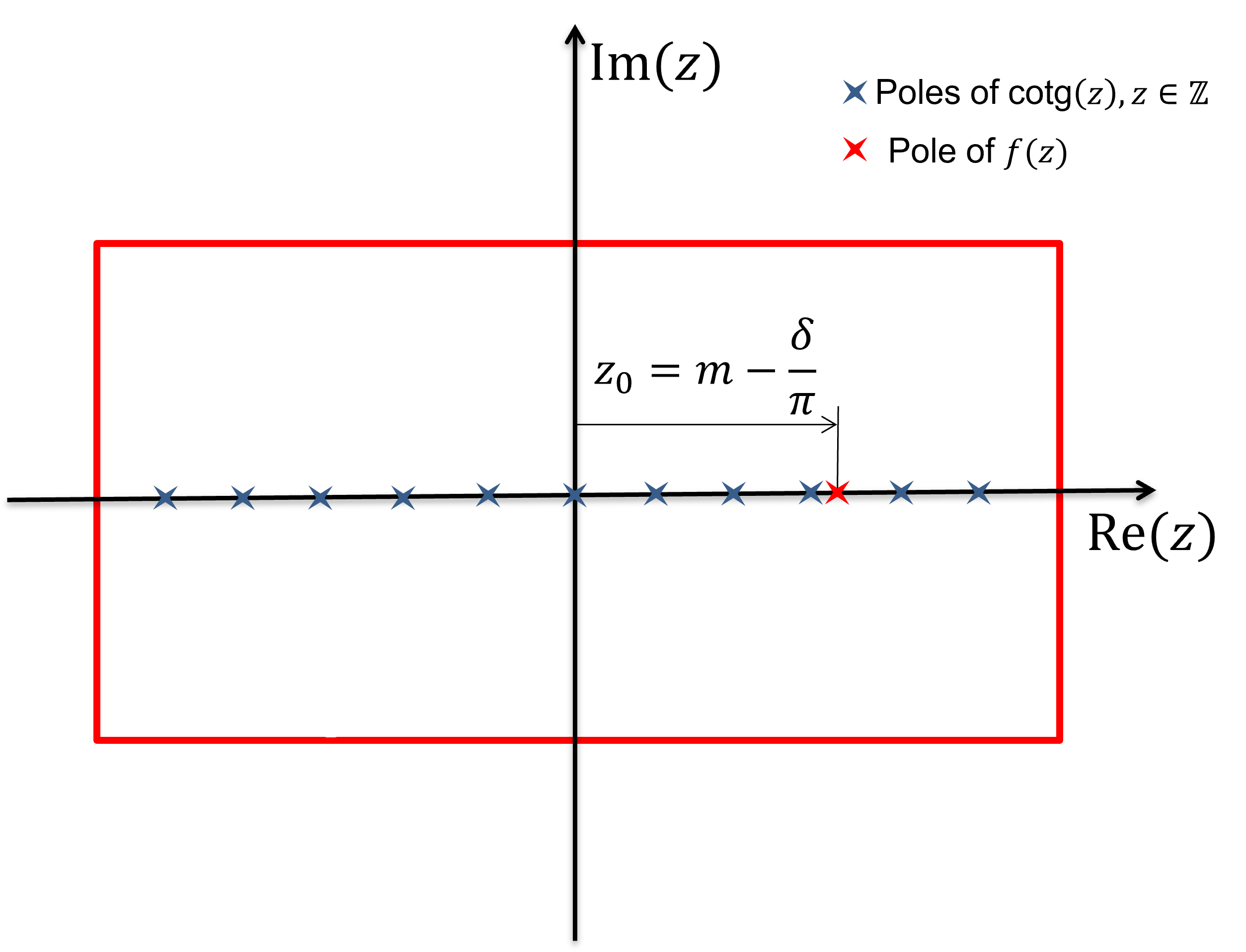}
        & \hspace{-0.7cm}
        \includegraphics[scale=0.4]{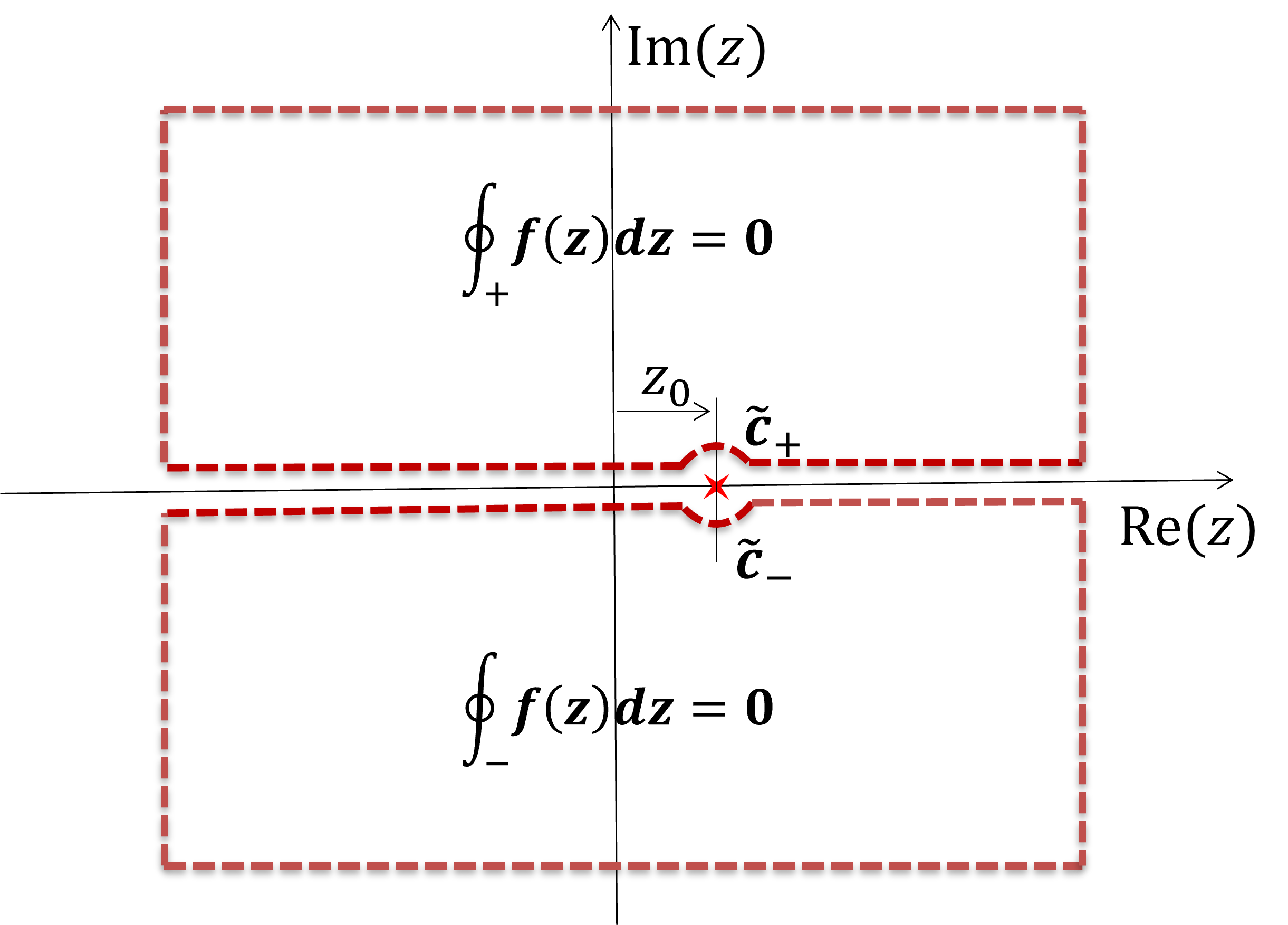}
        \end{tabular}
    \vspace{-0.4cm}
	\caption{\footnotesize (Left) Integration path of \eqref{ComplexIntegral}. (Right) In the Upper and Bottom boundaries of the integration path $\lim_{(\mathrm{Im} z \rightarrow \pm \infty)}\cot(\mathrm{Re}(z) +i\mathrm{Im} z) = \pm -i$ and the contour can be deformed resulting in the expression in Eq. \eqref{soma}. \Sou}
       \label{Sommerfeld-Watson}
\end{figure} 

Finally, the sum $\xi_m$ has been found, and we can proceed. The eigenvalues $\epsilon_m$ can be found through the implicitly expression $\epsilon_m = \varepsilon_m - \frac{\Delta\varepsilon}{\pi} \delta_m $ and after this procedure, the phase shift can be obtained from Eq. \eqref{Auxiliar_I} using $\xi_l = \frac{1}{W}$, where
\begin{eqnarray}\label{xi_ap}
\xi_l = \frac{1}{N} \sum_{q} \frac{1}{\varepsilon_l -\varepsilon_q - \frac{\Delta \e}{\pi} \delta_l} = -\pi\rho \cot\delta_l + \rho~\mathcal{P} \int_{-D}^{+D} \frac{1}{\epsilon_l -\varepsilon} d\varepsilon.
\end{eqnarray}
Inverting this expression to find the phase shift, once $\log\left(\frac{|\varepsilon_l - D|}{|\varepsilon_l + D|}\right) \approx 0$,  we obtain:
\begin{eqnarray}\label{phase_shift_ap}
&\delta_l = \mathrm{atan} \left(\frac{-\pi\rho W}{1 - \rho W \log\left(\frac{D+\varepsilon_l}{D  - \varepsilon_l}\right) } \right)  \approx \mathrm{atan} \left(-\pi\rho W\right) .
\end{eqnarray}

At this point we already found the eigenenergies, now we need the eigenvector. To this aim, we start by isolating the Eq. \eqref{E__I} and squaring both sides and then applying the sum over $k$, we obtain
\begin{eqnarray}
\sum_k |\alpha_{k,m}|^2 = \frac{W^2}{N} \left( \frac{1}{N} \sum_k \frac{1}{\left( \epsilon_m-\epsilon_k \right) ^2} \right) \left( \sum_q \alpha_{q,m} \right)^2.\end{eqnarray}
Where $\frac{1}{N} \sum_k \frac{1}{\left( \epsilon_m-\epsilon_k \right) ^2}=- \frac{d\xi_m}{d\epsilon_m}= \left( \frac{\rho}{\Delta\varepsilon}|\frac{\pi}{\sin \delta_m}|^2  \right)$ and $\sum_k |\alpha_{k,m}|^2 = 1$, resulting in
\begin{eqnarray}\label{Coef.alpha}
1 = \frac{W^2}{N} \left( - \frac{d\xi_m}{d\epsilon_m} \right) \left( \sum_q \alpha_{q,m} \right)^2.
\end{eqnarray}

Finally, the coefficients $\alpha_{l,k}$ of the eigenstates $g_l^\dagger = \sum_k \alpha_{l,k} \tilde a_k^\dagger$ of the Hamiltonian \eqref{H_photo_ap} with energy $\epsilon_l$ can be obtained by:
\begin{eqnarray}\label{Band_proj_}
\alpha_{l,k} = \pm \frac{1}{\epsilon_l - \varepsilon_k} \sqrt{\frac{1}{N}\left(\frac{-\partial \xi_l }{\partial \epsilon_l}\right)^{-1} } = \pm \frac{\sin \delta_l }{\pi} \frac{\Delta\varepsilon}{\varepsilon_l - \varepsilon_k - \frac{\delta_l}{\pi}\Delta\varepsilon}.
\end{eqnarray}

If we consider $W \rightarrow 0$, then $\alpha_{k,m} \rightarrow 1$, since the eigenvectors become exactly the initial basis. In this limit, $\epsilon_m \rightarrow \varepsilon_m$ and therefore $\delta_m \rightarrow 0$
\begin{eqnarray}
\lim_{W \rightarrow 0} \alpha_{k,m} = \lim_{\delta_m \rightarrow 0} \pm  \frac{\Delta\varepsilon}{\pi} \frac{\sin \delta_m}{\left( -\frac{\Delta\varepsilon}{\pi} \delta_m \right)} = \pm (-1).
\end{eqnarray}
Therefore, the negative solution is the only possible one! We have found the correct coefficients $\alpha_{k,m}$
\begin{eqnarray}
 \alpha_{k,m} = -\frac{\Delta \varepsilon}{\left( \epsilon_m-\epsilon_k \right)} \frac{\sin \delta_m}{\pi} .
\end{eqnarray}
together with the energy eigenvalues $\epsilon_m$ found earlier.

\begin{figure}[hbt!]
		\centering
        \begin{tabular}{ll}
        \hspace{-0.5cm}
        \includegraphics[scale=0.50]{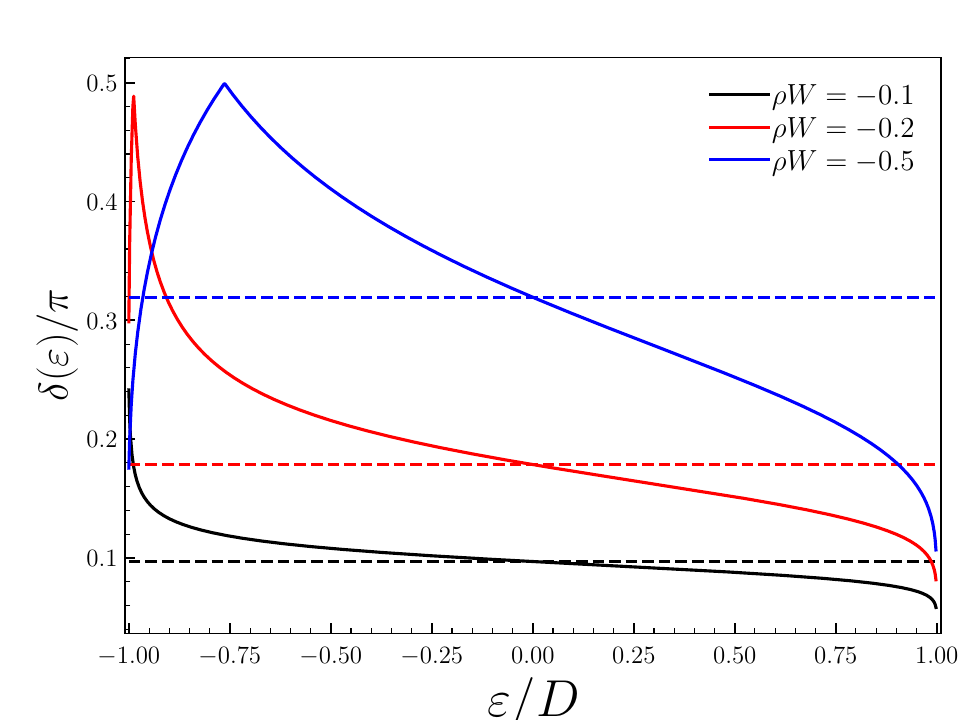}
        & \hspace{-0.65cm}
        \includegraphics[scale=0.50]{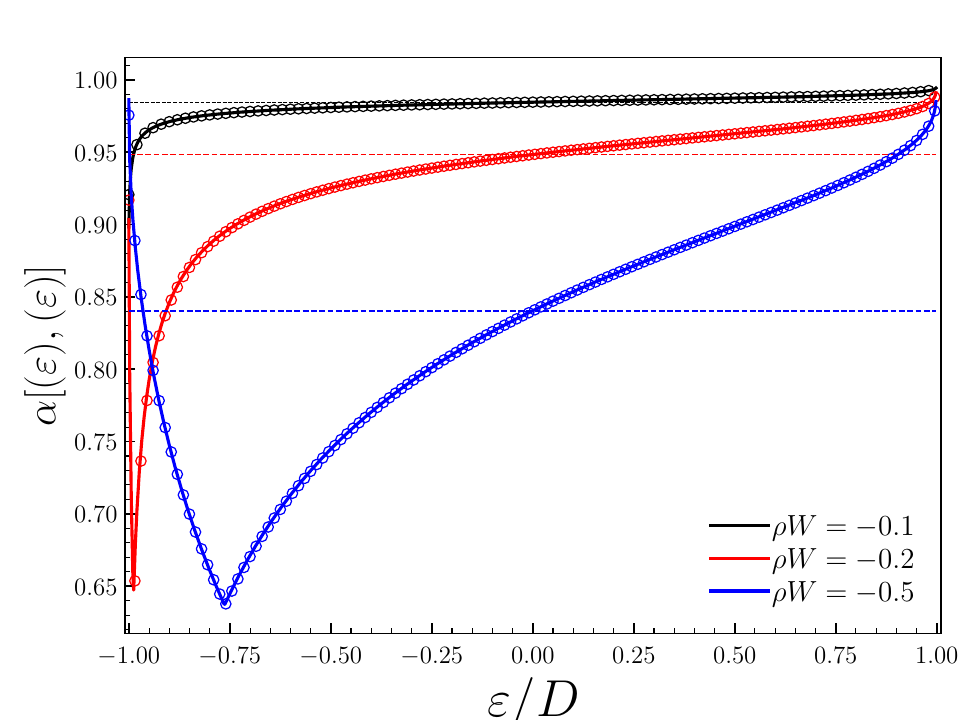}
        \end{tabular}
	\caption{\footnotesize (RIGHT) we show the values of $\delta(\varepsilon)$ computed by Eq. \eqref{phase_shift} (solid line), and  the value computed by Eq. \eqref{phase_shift} (dashed line), for different values of $\rho W$. Once $l$ represents the energy levels $\varepsilon_l$, then $\delta_l=\delta(\varepsilon_l)$, this notation is more convenient for comparing the data.(LEFT) we show the values of the diagonal coefficients computed by Eq. \eqref{Band_proj} considering the $\delta(\varepsilon)$ in Eq. \eqref{phase_shift} (solid line), the values computed considering the $\delta$ constant (dashed line), and the numerical values computed by direct diagonalization (circular dots). \Sou}
        \label{Analytical_Coefficients}
\end{figure} 

Before proceeding, we verified the validity of the Eq. \eqref{phase_shift_ap} by comparing the diagonal coefficients obtained by the phase shift formula and by direct diagonalization. In Fig. \ref{Analytical_Coefficients}, the right panel displays the phase shift values: the solid line represents those calculated using Eq. \eqref{phase_shift_ap}, while the dashed line shows values obtained from Eq. \eqref{phase_shift} for various $\rho W$ values (black is $ -0.1$, red is $-0.2$ and blue is $-0.5$). In the left panel, we depict the values of the diagonal coefficients: the solid line represents those computed by Eq. \eqref{Band_proj_} using the correct phase shift, the dashed line shows values computed considering a constant phase shift, and the numerical values computed by direct diagonalization are denoted by circular dots. We notice a strong agreement between the expression derived from Eq. \eqref{Band_proj_} and the numerical outcomes. It is notable that for $\rho W \leq 0.1$, indeed the approximation for the phase shift hold well. However, deviations from the numerical results become significant when $\rho W \geq 0.2$. Additionally, as $\rho W$ increases, a distortion of the phase shift near the bottom of the band is noticeable. This distortion arises due to the emergence of a new level at the bottom of the band with energy $\epsilon_B < - D$.

These expressions for the phase shift in the Eq. \eqref{phase_shift_ap} and for the coefficients in Eq. \eqref{Band_proj_} have been known for more than fifty years \cite{Anderson_1967} and are the basis of many important works \cite{Mahan,Nozieres,Doniach_1970,PhysRevB.24.4863}. However, Eq. \eqref{xi_ap} is only applicable when $ |\epsilon_l| \leq D$. It breaks down when the level is the bottom of the band, $\varepsilon_B = -D$, as $W<0$ implies $\epsilon_B < -D$, meaning there is no pole within the range $[-D,+D]$ and the $\cot$ term needs to be removed from  Eq. \eqref{xi_ap}. In this situation 
\begin{eqnarray}
\xi_B =\rho \mathcal{P} \int_{-D}^{+D} \frac{1}{\epsilon_{B} -\varepsilon} d\varepsilon = \frac{1}{W}.
\end{eqnarray} 
Solving this equation, we find the new energy as:
\begin{eqnarray}\label{Bond_State_energy_ap}
\epsilon_{B} = -D.\mathrm{coth}\left(  \frac{1}{-2\rho W} \right).
\end{eqnarray}

Furthermore, the eigenoperator with energy $\epsilon_B$ is $g_B^\dagger = \sum_k \alpha_{B,k} \tilde a_k^\dagger$, where the coefficients $\alpha_{B,k}$ with the initial levels $\varepsilon_k$ of the band can be obtained using a strategy similar to that used for Eq. \eqref{Band_proj_}, with $\xi_B = \rho \mathcal{P} \int_{-D}^{+D} \frac{1}{\epsilon_{B} -\varepsilon} d\varepsilon$, for $\epsilon_B < -D$, as:
\begin{eqnarray}\label{Bond_State_proj_}
\alpha_{B,k} = \sqrt{\frac{\Delta\varepsilon}{2D}} \frac{\sqrt{\epsilon_B^2 - D^2}}{-\epsilon_B + \varepsilon_k} \rightarrow \sqrt{\frac{\Delta\varepsilon}{D}} \frac{\sqrt{(-\epsilon_B+\Delta\varepsilon)^2 - D^2}}{(-\epsilon_B+\Delta\varepsilon) + \varepsilon_k}.
\end{eqnarray}
Including a slight modification $-\epsilon_B \rightarrow - \epsilon_B + \Delta\varepsilon$ to ensure that $\lim_{\epsilon_B \rightarrow - D} (\alpha_{B,-{N}/{2}}) = 1$.

\chapter{Anderson Orthogonality Catastrophe}\label{AOC proof}

The Anderson orthogonality catastrophe is well understood by the Eq. \eqref{AOC EQ}. However, the expression in Eq. \eqref{AOC EQ} is an approximation for the determinant in Eq. \eqref{Slater}. A more accurate approximation can be found using the determinant formula for a Cauchy matrix \cite{cauchy1841exercices}, once the matrix coefficients can be expressed as $m_{ij} = \frac{1}{u_i - v_j}$. We can start with the Slater determinant: 
\begin{eqnarray}
\braket{\phi_0}{\varphi_0} = \mathrm{det}
\begin{pmatrix}
\{\tilde a_{-1},g_{-1}^\dagger\} & \{\tilde a_{-1},g_{-2}^\dagger\} & ... & \{\tilde a_{-1},g_{-N_e}^\dagger\} \\
\{\tilde a_{-2},g_{-1}^\dagger\} & \{\tilde a_{-2},g_{-2}^\dagger\} & ... & \{\tilde a_{-2},g_{-N_e}^\dagger\} \\
\vdots & \vdots & ... & \vdots \\
\{\tilde a_{-N_e},g_{-1}^\dagger\} & \{\tilde a_{-N_2},g_{-2}^\dagger\} & ... & \{\tilde a_{-N_e},g_{-N_e}^\dagger\} \\
\end{pmatrix}. 
\end{eqnarray}

The coefficients of the matrix $\alpha_{m,k} = \{\tilde a_{k},g_{m}^\dagger\} $ can be obtained by the expression
\begin{eqnarray}\label{Coef_M}
\alpha_{m,k} = -\frac{\Delta \varepsilon}{\left( \epsilon_m-\varepsilon_k \right)} \frac{\sin \delta_m}{\pi } \approx -\frac{\Delta \varepsilon}{\left( \varepsilon_m - \varepsilon_k  - \frac{\delta}{\pi}\Delta\varepsilon \right)} \frac{\sin \delta}{\pi } =  \frac{\sin \delta}{\pi } \frac{1}{m-k-\frac{\delta}{\pi}},
\end{eqnarray}
where the phase shift can be obtained by $\tan(\delta_m) \approx -\pi\rho W$. This expression using the phase shift constant is a good approximation only for small values of the scattering potential $W$. But for now, let us assume that $W$ is small.

Then, using the Eq. \eqref{Coef_M} we can approximate the projections between the two ground states as
\begin{eqnarray}\label{Proj. MI}
\braket{\phi_0}{\varphi_0}^* = \left( -\frac{\sin\delta}{\pi} \right)^{N_e} \mathrm{det} \left( \left\{ \frac{1}{i-j-\frac{\delta}{\pi}} \right\} \right)  . 
\end{eqnarray}
Here $\mathrm{det} \left( \left\{ \frac{1}{i-j-\frac{\delta}{\pi}} \right\} \right)$ is the determinant of the matrix composed by elements $\frac{1}{i-j-\frac{\delta}{\pi}}$.

One matrix that can be written as $ \left( \left\{ \frac{1}{x_i-y_j} \right\} \right)$ is known as Cauchy matrix and its determinant can be expressed by the Cauchy determinant formula: 
\begin{eqnarray}
\mathrm{det} \left( \left\{ \frac{1}{x_i-y_j} \right\} \right) = \frac{\prod_{i=2}^{N_e} \prod_{j < i} (x_i - x_j)(y_j - y_i)}{\prod_{i=1}^{N_e} \prod_{j=1}^{N_e} (x_i - y_j) }
\end{eqnarray}

In Eq. \eqref{Proj. MI}, the only quantity whose value is unknown is $\mathrm{det} \left( \left\{ \frac{1}{i-j-\frac{\delta}{\pi}} \right\} \right)$. Therefore, we only need to apply the Cauchy determinant formula to find the projection. Applying the Cauchy formula to this quantity, we obtain:
\begin{eqnarray}\label{DetM_I}
\mathrm{det} \left( \left\{ \frac{1}{i-j-\frac{\delta}{\pi}} \right\} \right) = \frac{\prod_{i=2}^{N_e} \prod_{j < i} (i - j)(j - i)}{\prod_{i=1}^{N_e} \prod_{j=1}^{N_e} (i - j - \frac{\delta}{\pi}) }.
\end{eqnarray}

Before proceeding, we can simplify some terms in the Eq. \eqref{DetM_I}. The first term is: 
\begin{eqnarray}
\prod_{j=1}^{N_e} \left(i - j - \frac{\delta}{\pi}\right) &=& (-1)^{N_e} \frac{\gamma\left(N_e + 1 - i + \frac{\delta}{\pi}\right)}{\gamma\left(- i + \frac{\delta}{\pi}\right)} \\ &=& (-1)^{N_e} \frac{\gamma\left(N_e + 1 - i + \frac{\delta}{\pi}\right)\gamma\left( i - \frac{\delta}{\pi}\right)}{(-1)^{i+1} \left( \frac{\sin\delta}{\pi} \right)},
\end{eqnarray}
where we used the well known gamma function $\gamma(x)$ to simplify the term. Here we used the gamma function identity $\gamma(1-z)\gamma(z) = \frac{\pi}{\sin(\pi z)}$.

Another term that can be simplified is:
\begin{eqnarray}
\prod_{j < i} (i - j)(j - i) = (-1)^{i-1}[(i-1)!]^2
\end{eqnarray}

Now, returning the simplified terms to the original expression in Eq. \eqref{DetM_I} we obtain:
\begin{eqnarray}
\mathrm{det} \left( \left\{ \frac{1}{i-j-\frac{\delta}{\pi}} \right\} \right) &=& \left(-\frac{\sin\delta}{\pi} \right)^{-N_e} \frac{\prod_{i=2}^{N_e} [(i-1)!]^2 }{\prod_{i=1}^{N_e} \gamma\left(N_e + 1 - i + \frac{\delta}{\pi}\right)\gamma\left( i - \frac{\delta}{\pi}\right) } \\ &=& \left(-\frac{\sin\delta}{\pi} \right)^{-N_e} \frac{\prod_{i=2}^{N_e} [(i-1)!]^2 }{\prod_{i=1}^{N_e} \gamma\left(i + \frac{\delta}{\pi}\right)\gamma\left( i - \frac{\delta}{\pi}\right) }.
\end{eqnarray}
In the last step, we utilized the fact that the product over all possible $i$ and the product over all possible values of $N_e+1-i$ yield exactly the same result.

After some manipulations, we can write the determinant as:
\begin{eqnarray}
\mathrm{det} \left(\left\{ \frac{1}{i-j-\frac{\delta}{\pi}} \right\}\right) &=& \left(-\frac{\sin\delta}{\pi} \right)^{-N_e}  \frac{\sin\delta}{\delta} \prod_{i=1}^{N_e-1} \frac{[(i)!]^2 }{\gamma\left(1+i + \frac{\delta}{\pi}\right)\gamma\left(1+ i - \frac{\delta}{\pi}\right)}.
\end{eqnarray}

Now, since we are dealing with the gamma function and factorials, we can employ an expression derived from the well-known Stirling approximation:
\begin{eqnarray}\label{starling}
\gamma(1+n+x) \approx n! \left(1 + \frac{x}{n} \right)^{n+0.5} (n+x)^x e^{-x}.
\end{eqnarray}

Using the Stirling approximation in the determinant expression we obtain:
\begin{eqnarray}\label{Det_MII}
\mathrm{det} \left( \left\{ \frac{1}{i-j-\frac{\delta}{\pi}} \right\} \right) &\approx& \left(-\frac{\sin\delta}{\pi} \right)^{-N_e}  \frac{\sin\delta}{\delta} \prod_{i=1}^{N_e-1} \frac{1}{\left( 1 - \left(\frac{\delta}{\pi}\right)^2 \frac{1}{i^2}\right)^{i+0.5}} \frac{\left(i - \frac{\delta}{\pi} \right)^\frac{\delta}{\pi} }{\left( i + \frac{\delta}{\pi} \right)^\frac{\delta}{\pi} }.
\end{eqnarray}

Before proceeding, we can simplify the terms in expression \eqref{Det_MII}. The first term can be approximated by considering that $P = \prod_i a_i \rightarrow \ln P = \sum_i \ln(a_i)$ followed by the series expansion of the logarithmic function until second order. 
\begin{eqnarray}
\prod_{i=1}^{N_e-1}  \left( 1 - \left(\frac{\delta}{\pi}\right)^2 \frac{1}{i^2}\right)^{i+0.5} \approx N_e^{\left(\frac{\delta}{\pi}\right)^2} e^{\left(\frac{\delta}{\pi}\right)^2 \left( C_{EM} + \frac{\pi^2}{12} \right)} .
\end{eqnarray}

The second term can be simplified by using the proprieties of the product and the gamma function.
\begin{eqnarray}
\prod_{i=1}^{N_e-1}  \frac{\left(i - \frac{\delta}{\pi} \right)^\frac{\delta}{\pi} }{\left( i + \frac{\delta}{\pi} \right)^\frac{\delta}{\pi}} =  \left( \frac{\gamma\left(1+\frac{\delta}{\pi}\right)\gamma\left(N_e-\frac{\delta}{\pi}\right)}{\gamma\left(1-\frac{\delta}{\pi}\right)\gamma\left(N_e+\frac{\delta}{\pi}\right)} \right)^\frac{\delta}{\pi}  \approx \left( \frac{\gamma\left(1+\frac{\delta}{\pi}\right)}{\gamma\left(1-\frac{\delta}{\pi}\right)} \right)^\frac{\delta}{\pi} N_e^{-2\left(\frac{\delta}{\pi}\right)^2}.
\end{eqnarray}

Finally, returning the obtained values after the simplifications in the determinant we can approximate the projection between two determinant, before and after the scattering potential as:
\begin{eqnarray}\label{Proj__II}
\braket{\phi_0}{\varphi_0}^* \approx \left(\frac{\sin\delta}{\delta} \right)^{1+\frac{\delta}{\pi}} \left( \gamma\left(1+\frac{\delta}{\pi}\right)\right)^{2\frac{\delta}{\pi}} e^{\left(C_{\mathrm{EM}} + \frac{\pi^2}{12}\right)\left(\frac{\delta}{\pi}\right)^2}N_e^{-\left(\frac{\delta}{\pi}\right)^2}.
\end{eqnarray}
where $\gamma(x)$ is the well known Gamma function and $C_{\mathrm{EM}}$ is the Euler-Mascheroni Constant.

However the expression \eqref{Proj__II} is only valid for small values of $W$. Correctly computing the projections between the two ground states for any value of $W$ is challenging. For $-W/D > 0.5$, the $\delta_l \approx \delta$ approximation is no longer applicable, and the Cauchy formula does not provide much help in this case. But we could consider that $\braket{\phi_0}{\varphi_0}^*$ is taking account the main contribution and the projection could be expressed as $\braket{\phi_0}{\varphi_0} = \mathcal{F}(\delta,N_e). \braket{\phi_0}{\varphi_0}^*$. 

To find the function $\mathcal{F}(\delta,N_e) = \frac{\braket{\phi_0}{\varphi_0}(\delta,N_e)}{\braket{\phi_0}{\varphi_0}^*(\delta,N_e)}$, we started by fixing the value of $N_e$ and searching for a fitting function, using the bounding conditions that $\mathcal{F}(0,N_e) = 1$ and the asymptotic behavior of the function as $\delta$ approaches $\frac{\pi}{2}$. By setting $N_e = 600$, we estimated that $\mathcal{F}(\delta,N_e=600) =  \left(\frac{\sin\delta}{\delta} \right)^{-2+1.25\left( \frac{\delta}{\pi} \right)^4 } \left( \gamma\left(1+\frac{\delta}{\pi}\right)\right)^{-2\frac{\delta}{\pi}}$. The results of this correction in the projections is shown in the Fig. \ref{MB Proj.} (left panel). After that, we compute numerically the ratio $ \frac{\braket{\phi_0}{\varphi_0}(\delta,2N_e)}{\braket{\phi_0}{\varphi_0}(\delta,N_e)}$ for different values of $N_e$ and we observed that exist a little deviation from the $N_e^{-\left(\frac{\delta}{\pi}\right)^2}$ factor, leading to the more accurately factor $\left(\frac{N_e}{600}\right)^{\frac{1}{2} \left(\frac{\delta}{\pi}\right)^2 } N_e^{-\left(\frac{\delta}{\pi}\right)^2}$, as shown in Fig. \ref{Fitting_function}. Summarizing, by computing numerically $\mathcal{F}(\delta,N_e) = \frac{\braket{\phi_0}{\varphi_0}(\delta,N_e)}{\braket{\phi_0}{\varphi_0}^*(\delta,N_e)}$, we approximated the determinant by the following expression:
\begin{eqnarray}
\braket{\phi_0}{\varphi_0} \approx \left(\frac{\sin\delta}{\delta} \right)^{-1+\frac{\delta}{\pi} + 1.25\left( \frac{\delta}{\pi} \right)^4 } e^{\left(C_{\mathrm{EM}} + \frac{\pi^2}{12}\right)\left(\frac{\delta}{\pi}\right)^2}\left(\frac{N_e}{600}\right)^{\frac{1}{2} \left(\frac{\delta}{\pi}\right)^2 } N_e^{-\left(\frac{\delta}{\pi}\right)^2}.
\end{eqnarray}

In Fig. \ref{MB Proj.}, we compare the numerical results with Eqs. \eqref{AOC EQ II} and \eqref{Bound_State_MB_proj} for different values of $W$ (left panel) and for different values of $N$ (right panel). In this figure, the quantity $\braket{\phi_0}{\Phi}$ is used generically to represent both $\braket{\phi_0}{\varphi_0}$ and $\braket{\phi_0}{\bar\varphi}$. The black circles and the solid black line represent the numerical and analytical results for $\braket{\phi_0}{\varphi_0}$, respectively, while the red circles and the solid red line represent the numerical and analytical results for $\braket{\phi_0}{\bar\varphi}$, respectively. From this picture, we observe that the expressions from Eqs. \eqref{AOC EQ II} and \eqref{Bound_State_MB_proj} are good approximations for the quantities we discussed.
\newpage

\begin{figure}[hbt!]
		\centering
        \includegraphics[scale=0.65]{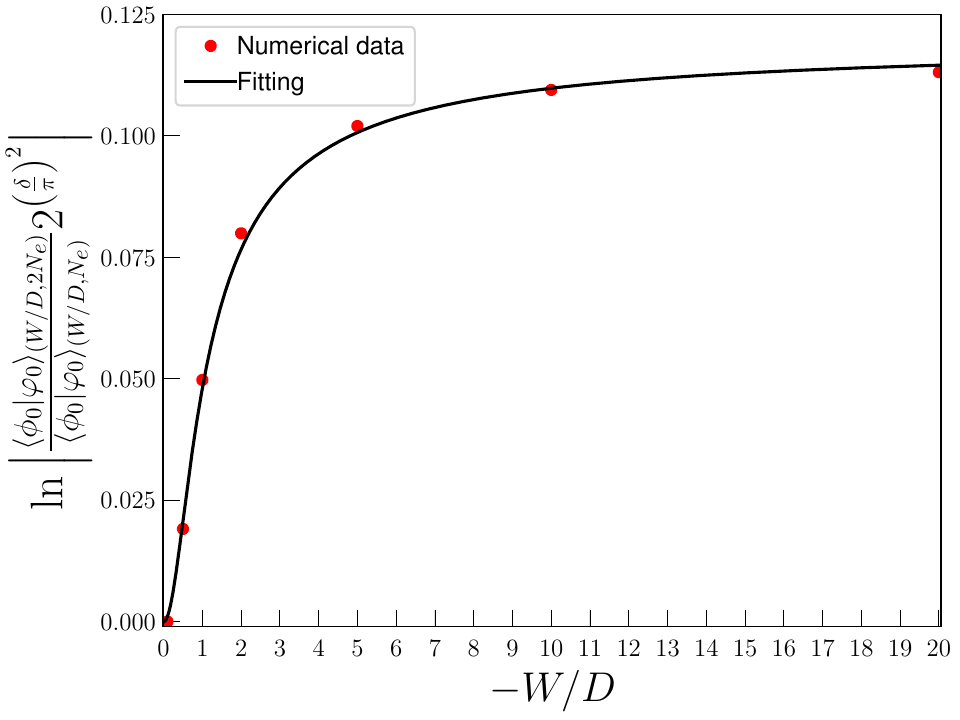}
	  \caption{\footnotesize Finding the factor $\left(\frac{N_e}{600}\right)^{\frac{1}{2} \left(\frac{\delta}{\pi}\right)^2 } N_e^{-\left(\frac{\delta}{\pi}\right)^2}$ from the numerical data obtained by direct diagonalization. \Sou}
        \label{Fitting_function}
\end{figure} 

The overlap $\braket{\phi_0}{\bar\varphi} = \bra{\phi_0} g_{0}^\dagger g_{B} \ket{\varphi_0}$ can be determined using a Slater determinant similar to Eq. \eqref{Slater}, but with the last column replaced by the coefficients $\alpha_{B,h} \rightarrow \alpha_{p,h}$. However, solving this analytically becomes challenging when $W$ is large. Therefore, approximating the determinant by considering that the dominant term is on the diagonal, times a fitting function, we can approximate $|\bra{\phi_0} g_{0}^\dagger g_{B} \ket{\varphi_0}|$ as:
\begin{eqnarray}
 {|\bra{\phi_0} g_{0}^\dagger g_{B} \ket{\varphi_0}|}  \approx \frac{2}{\pi} \left| \frac{{-e_B - D}}{-e_B} \right| \left(\frac{\sin\delta}{\delta}\right)^{- 2\pi\left(\frac{\delta}{\pi}\right) - \left(\frac{\delta}{\pi}\right)^4 }  \left(\frac{N_e}{600}\right)^{-\frac{1}{2} \left( 1-\frac{2\delta}{\pi} \right) } \braket{\phi_0}{\varphi_0}.
\end{eqnarray} Here, we derive this fitting function by comparing the results with the numerical values obtained for different values of $\delta$, a similar method as we did to approximate the $\braket{\phi_0}{\varphi_0}$. The results of this approximation are shown in Fig. \ref{MB Proj.}.

Here, we found that the critical exponent in the Anderson orthogonality has a different factor than what was found by Anderson. Specifically, this factor is $\left({N_e}\right)^{ -\frac{1}{2} \left(\frac{\delta}{\pi}\right)^2 }$, instead of $\left({N_e}\right)^{-\left(\frac{\delta}{\pi}\right)^2 }$. This correction by the factor of $1/2$ has been found in more recent and rigorous works on the Anderson orthogonality catastrophe \cite{Gebert2014-fu}.

\chapter{Doniach - Sunjic power law}\label{DN law}
\section{Frequency Domain}
Here, we aim to compute $G(\omega)$ for a Fermi gas system in response to a sudden change in the scattering potential. These calculations were already conducted by Nozeries and Dominicis in 1969. Let us consider the imaginary part of the Green function $G(\omega)$, denoted as $\mathrm{Im}G(\hbar\omega)$:
\begin{eqnarray}\label{G1}
\mathrm{Im}G(\hbar\omega)=\pi\sum_{n=0} |\braket{\varphi_n}{\phi_0}|^2 \delta(\hbar\omega - E_n)
\end{eqnarray}
where $E_n$ represents the energy of the associated many-body state $\ket{\varphi_n}$.

We will utilize the commonly accepted approximation that diagonal elements of the Slater determinant $\bra{\phi_0}\prod_{j_n} g_{k_{j_n}}^\dagger g_{h_{j_n}} \ket{\varphi_0}$ are dominant. This allows us to express the many-body projections with the initial ground state as:
\begin{eqnarray}\label{P00}
{\braket{\phi_0}{\varphi_n}} \approx \left[\prod_{j_n} \left(\frac{\delta}{\pi} \right) \frac{\Delta\varepsilon}{\varepsilon_{k_{j_n}} - \varepsilon_{h_{j_n}} - \frac{\delta}{\pi} \Delta\varepsilon} \right] {\braket{\phi_0}{\varphi_0}}.
\end{eqnarray}
Here, $\varepsilon_{k_{j_n}}$ and $\varepsilon_{h_{j_n}}$ represent the energies of the particle and hole level, respectively, where this parameter ${j_n}$ depends on the many-body state $\ket{\varphi_n}$ and represents the pairs of particle-hole excitations which compose this state.

Given that $E_0 = 0$ and the energy $E_n = \sum_{{j_n}} (\varepsilon_{k_{j_n}} - \varepsilon_{h_{j_n}})$ represents the energy of many-body excitations from the ground state, and utilizing the Eq. \eqref{P00} $\Omega_n^2 = \prod_{j_n} \left(\frac{\delta}{\pi} \right)^2 \frac{{\Delta\varepsilon}^2}{\left(\varepsilon_{k_{j_n}} - \varepsilon_{h_{j_n}} - \frac{\delta}{\pi} \Delta\varepsilon\right)^2}$ in the Eq. \eqref{G1}, we can derive:
\begin{eqnarray}\label{G2}
\mathrm{Im}G(\hbar\omega) &=& \pi |\braket{\varphi_0}{\phi_0}|^2 \sum_{n=0} \Omega_n^2 \delta\left(\hbar\omega - \sum_{{j_n}} (\varepsilon_{k_{j_n}} - \varepsilon_{h_{j_n}}) \right).
\end{eqnarray}

Now, let us transform the sum over all many-body states into a sum over all possible particle-hole excitations, with $p$ representing the number of excitations:
\begin{eqnarray}\label{Change_sum}
\sum_{n=0} \rightarrow  \sum_{p=0}\left(\frac{1}{p!} \sum_{h_1}  \sum_{k_1} \sum_{{h_2} \neq \{h_1\}} \sum_{{k_2} \neq \{k_1 \}} ... \sum_{{h_p} \neq \{h_1,..,h_{p-1}\}} \sum_{{k_p} \neq \{k_1,..,k_{p-1}\}} \right) 
\end{eqnarray}
Here, the $\neq$ constraint ensures the fermionic occupation rule (Pauli's exclusion principle).

Substituting this new sum in Eq. \eqref{Change_sum} into the expression for the imaginary part of the Green function in Eq. \eqref{G2}, we get:
\begin{align}\label{G3}
\frac{\mathrm{Im}G(\hbar\omega)}{ \pi |\braket{\varphi_0}{\phi_0}|^2} =  \mathlarger{\sum}_{p=0}& \frac{1}{p!} \sum_{h_1}  \sum_{k_1} ... \sum_{{h_p} \neq \{h_1,..,h_{p-1}\}} \sum_{{k_p} \neq \{k_1,..,k_{p-1}\}} \nonumber \\
& \left[\prod_{j=1}^p \left(\frac{\delta}{\pi} \right) \frac{\Delta\varepsilon}{\varepsilon_{k_{j}} - \varepsilon_{h_{j}} - \frac{\delta}{\pi} \Delta\varepsilon} \right]^2 \delta\left(\hbar\omega - \sum_{{j=1}}^p (\varepsilon_{k_{j}} - \varepsilon_{h_{j}}) \right).
\end{align}

We can then demonstrate that:
\begin{eqnarray}\label{Sum1}
\sum_{k_1} \sum_{{k_2} \neq \{k_1\}}  \Delta\varepsilon^2 = \sum_{k_1}  \sum_{{k_2}} \Delta\varepsilon^2 - \Delta\varepsilon \sum_{k_1} \sum_{k_2} \Delta\varepsilon~\delta^K_{k_1,k_2}. 
\end{eqnarray}

Substituting the Eq. \eqref{Sum1} into Eq. \eqref{Change_sum} we find the Eq. \eqref{Sum2}, which implies that for a small $\Delta\varepsilon$, the number of possible states in the sum becomes very large, making the rigorous fermionic rule in the sum a minor correction.
\begin{eqnarray}\label{Sum2}
 \sum_{h_1}  \sum_{k_1} ... \sum_{{h_p} \neq \{h_1,..,h_{p-1}\}} \sum_{{k_p} \neq \{k_1,..,k_{p-1}\}} \Delta\varepsilon^{2p}  = \sum_{h_1}  \sum_{k_1} ... \sum_{{h_p}} \sum_{{k_p} } \Delta\varepsilon^{2p} + \mathcal{O}(\Delta\varepsilon).
\end{eqnarray}

Additionally, for $\varepsilon_{k_{j}} - \varepsilon_{h_{j}} \ge \Delta\varepsilon > \frac{\delta}{\pi} \Delta\varepsilon$, we can show that:
\begin{align}\label{Re1}
\left[\left(\frac{\delta}{\pi} \right) \frac{\Delta\varepsilon}{\varepsilon_{k_{j}} - \varepsilon_{h_{j}} - \frac{\delta}{\pi} \Delta\varepsilon} \right]^2 = \left[\left(\frac{\delta}{\pi} \right) \frac{\Delta\varepsilon}{\varepsilon_{k_{j}} - \varepsilon_{h_{j}}} \right]^2 + \mathcal{O}\left( \Delta\varepsilon^3 \right).
\end{align}

By incorporating these approximations in the Eq. \eqref{Sum2} and Eq.\eqref{Re1} into the expression for the imaginary part of the Green function, we arrive at:
\begin{align}\label{G4}
\frac{\mathrm{Im}G(\hbar\omega)}{ \pi |\braket{\varphi_0}{\phi_0}|^2} =  \mathlarger{\sum}_{p=0}&\frac{1}{p!}  \left(\frac{\delta}{\pi} \right)^{2p} \sum_{h_1}  \sum_{k_1} ... \sum_{{h_p}} \sum_{{k_p}} \nonumber \\
& \left[\prod_{j=1}^p \frac{\Delta\varepsilon}{\varepsilon_{k_{j}} - \varepsilon_{h_{j}}} \right]^2 \delta\left(\hbar\omega - \sum_{{j=1}}^p (\varepsilon_{k_{j}} - \varepsilon_{h_{j}}) \right) + \mathcal{O}(\Delta\varepsilon).
\end{align}

Since the energy difference $ \Delta\varepsilon \le \epsilon = \epsilon_k - \epsilon_h \le D$ represents the particle-hole excitation energy, and $\Delta\varepsilon \le -\epsilon_h \le \epsilon$, we can modify the sum domain into a sum over the excitation energy $\epsilon$ as:
\begin{eqnarray}\label{Change_Domain}
\sum_{\epsilon_h} \sum_{\epsilon_{k} }  \Delta\varepsilon^2 f(\epsilon_k - \epsilon_h) = \sum_{\epsilon} \Delta\varepsilon f(\epsilon)  \sum_{-\epsilon_h<\epsilon} \Delta\varepsilon = \sum_{\epsilon} \Delta\varepsilon~ \epsilon. f(\epsilon) + \mathcal{O}(\Delta\varepsilon).
\end{eqnarray}

By applying the previously equation in Eq. \eqref{Change_Domain} to the imaginary part of the Green function in Eq. \eqref{G4}, we get:
\begin{align}
\frac{\mathrm{Im}G(\hbar\omega)}{\pi |\braket{\varphi_0}{\phi_0}|^2} =  \mathlarger{\sum}_{p=0}&\frac{1}{p!}  \left(\frac{\delta}{\pi} \right)^{2p} \sum_{\epsilon_1} ... \sum_{\epsilon_p} \left[\prod_{j=1}^p \frac{\Delta\varepsilon}{\epsilon_j} \right] \delta\left(\hbar\omega - \sum_{{j=1}}^p \epsilon_j \right) + \mathcal{O}(\Delta\varepsilon).
\end{align}

Considering the sum can be approximated by an integral, and this approximation lead to a error in the other of $\Delta\varepsilon$ once there is no pole inside the sum since $D \ge \epsilon_j \ge \Delta\varepsilon$, them we can rewrite the imaginary part of the Green's function in Eq. \eqref{G4} as:
\begin{align}
\frac{\mathrm{Im}G(\hbar\omega)}{\pi |\braket{\varphi_0}{\phi_0}|^2} =  \mathlarger{\sum}_{p=0}&\frac{1}{p!}  \left(\frac{\delta}{\pi} \right)^{2p} \int_{\Delta\varepsilon}^{D}  \frac{{d\epsilon_1}}{\epsilon_1} ... \int_{\Delta\varepsilon}^{D} \frac{d{\epsilon_p}}{\epsilon_p} \delta\left(\hbar\omega - \sum_{{j=1}}^p \epsilon_j \right) + \mathcal{O}(\Delta\varepsilon).
\end{align}

Now, by using the Dirac delta function $\hbar\omega = \sum_{{j=1}}^p \epsilon_j$, which means that $\epsilon_j < \hbar\omega $ for any $j$ and $\hbar\omega > 0$. Then, the imaginary part of the Green's function can be written as: 
\begin{align}\label{G5}
\frac{\mathrm{Im}G(\hbar\omega)}{\pi |\braket{\varphi_0}{\phi_0}|^2} = \mathlarger{\sum}_{p=0}&\frac{1}{p!}  \left(\frac{\delta}{\pi} \right)^{2p} \int_{\Delta\varepsilon}^{\hbar\omega - \Delta\varepsilon} \frac{{d\epsilon_1}}{\epsilon_1} ... \int_{\Delta\varepsilon}^{\hbar\omega - \Delta\varepsilon} \frac{ d{\epsilon_{p-1}}}{\epsilon_{p-1}}\frac{\Theta(\hbar\omega)}{\left(\hbar\omega - \sum_{{j=1}}^{p-1} \epsilon_j \right)} + \mathcal{O}(\Delta\varepsilon).
\end{align}

Now we need to solve the integral inside the sum over $p$ in the $\frac{\mathrm{Im}G(\hbar\omega)}{\pi |\braket{\varphi_0}{\phi_0}|^2}$. To do that we define a new quantity:
\begin{align}\label{g1}
 \mathcal{G}_p (\hbar\omega) = \frac{1}{p!}  \left(\frac{\delta}{\pi} \right)^{2p} \int_{\Delta\varepsilon}^{\hbar\omega - \Delta\varepsilon} \frac{{d\epsilon_1}}{\epsilon_1} ... \int_{\Delta\varepsilon}^{\hbar\omega - \Delta\varepsilon} \frac{ d{\epsilon_{p-1}}}{\epsilon_{p-1}}\frac{1}{\left(\hbar\omega - \sum_{{j=1}}^{p-1} \epsilon_j \right)}.
\end{align}

We can easily show that in the Eq. \eqref{g1} $\mathcal{G}_p$ obeys the recursive equation as: 
\begin{align}\label{recursive}
 \mathcal{G}_p (\hbar\omega) = \frac{1}{p}  \left(\frac{\delta}{\pi} \right)^{2} \int_{\Delta\varepsilon}^{\hbar\omega - \Delta\varepsilon} \frac{{d\epsilon}}{\epsilon}\mathcal{G}_{p-1} (\hbar\omega - \epsilon)
\end{align}

Once the $p=0$ represents no excitations, clearly the $ \mathcal{G}_0 (\hbar\omega)$ is: 
\begin{align}
 \mathcal{G}_0 (\hbar\omega) = \delta(\hbar\omega).
\end{align}

By using the recursive equation in Eq.\eqref{recursive} we can easily show that:
\begin{align}
 \mathcal{G}_1 (\hbar\omega) = \left(\frac{\delta}{\pi} \right)^{2} \frac{1}{\hbar\omega},
\end{align}
and
\begin{align}
 \mathcal{G}_2 (\hbar\omega) = \left(\frac{\delta}{\pi} \right)^{4} \frac{\ln{\left(\hbar\omega/\Delta\varepsilon \right)}}{\hbar\omega}. 
\end{align}

Once the structure of the $ \mathcal{G}_1 (\hbar\omega)$ and $ \mathcal{G}_2 (\hbar\omega)$ follow a pattern, let us now consider that the main part of this quantity $\mathcal{G}_p$ is:
\begin{align}\label{g_ind}
 \mathcal{G}_p (\hbar\omega) \approx \frac{1}{(p-1)!}\left(\frac{\delta}{\pi} \right)^{2p} \frac{\ln^{p-1}{\left(\hbar\omega/\Delta\varepsilon \right)}}{\hbar\omega}.
\end{align}

Then, by using the recursive equation in Eq. \eqref{recursive} we find:
\begin{align}\label{g2}
 \mathcal{G}_{p+1} (\hbar\omega) \approx  \frac{p}{(p+1)!}  \frac{ \left(\frac{\delta}{\pi} \right)^{2(p+1)}}{\hbar\omega} \int_{\Delta\varepsilon}^{\hbar\omega - \Delta\varepsilon} \frac{{d\epsilon}}{\epsilon} \left\{ 1 + \left[ \ln\left(\frac{\hbar\omega-\epsilon}{\hbar\omega}\right) + \ln\left(\frac{\hbar\omega}{\Delta\varepsilon}\right) \right]^{p-1} \right\}.
\end{align}

To solve the expression Eq. \eqref{g2} let us remember the binomial expansion: 
\begin{align}\label{Binomial}
\left[\ln\left(\frac{\hbar\omega-\epsilon}{\hbar\omega}\right) + \ln\left(\frac{\hbar\omega}{\Delta\varepsilon}\right) \right]^{p-1} = \sum_{q=0}^{p-1} \frac{(p-1)!}{q!(p-1-q)!}  \ln^{p-1-q}\left(\frac{\hbar\omega}{\Delta\varepsilon}\right) \ln^q\left(\frac{\hbar\omega-\epsilon}{\hbar\omega}\right).
\end{align}

Also, we use the substitution rule in the integral to simplify the equations as:
\begin{align}\label{Substitution}
u = \ln\left(\frac{\hbar\omega-\epsilon}{\hbar\omega}\right) \rightarrow \frac{d\epsilon}{\epsilon} = \frac{e^u}{1-e^u}(-du).
\end{align}

Now using the Eqs. \eqref{Substitution} and \eqref{Binomial} into the Eq. \eqref{g2}, after some manipulations and taking off the first term in the sum over $q$ we find that:
\begin{align}\label{g3}
 \mathcal{G}_{p+1} (\hbar\omega) =  \frac{p}{(p+1)!}  \frac{ \left(\frac{\delta}{\pi} \right)^{2(p+1)}}{\hbar\omega} \left\{ \frac{\ln^{p}{\left(\hbar\omega/\Delta\varepsilon \right)}}{p} +\ln^{p}{\left(\hbar\omega/\Delta\varepsilon \right)} + I_{p+1} \right\},
\end{align}
where the last term contains the remains terms of the sum and can be written as: 
\begin{align}\label{Int1}
I_{p+1} = \sum_{q=1}^{p-1} \frac{(p-1)!(-1)^q}{q!(p-1-q)!} \ln^{p-1-q}\left(\frac{\hbar\omega}{\Delta\varepsilon}\right) \int_0^{\ln(\hbar\omega/\Delta\varepsilon)} du \frac{e^{-u} u^q}{1-e^{-u}}.
\end{align}

Now, once the $ 0 < e^{-u} < 1$, then we can use the geometric series formula: 
\begin{align}
\frac{e^{-u}}{1-e^{-u}} = \sum_{j=1}^{\infty} e^{-u} .
\end{align}

Then we can substitute the geometric series formula into the integral in Eq. \eqref{Int1} as:
\begin{align}
\int_0^{\ln(\hbar\omega/\Delta\varepsilon)} du \frac{e^{-u} u^q}{1-e^{-u}} = \sum_{j=1}^{\infty} \frac{1}{j^{q+1}}\int_0^{j\ln(\hbar\omega/\Delta\varepsilon)} du~ {e^{-u} u^q}.
\end{align}

The exact solution for this integral is not hard to find. One only needs to use integration by parts and identify patterns. Then, the exact solution is: 
\begin{align}\label{Int2}
\sum_{j=1}^{\infty} \frac{1}{j^{q+1}}\int_0^{j\ln(\hbar\omega/\Delta\varepsilon)} du~ {e^{-u} u^q} = q!\left(\zeta^{\mathrm{R}}(q+1) - \sum_{j=1}^\infty \frac{1}{j^{q+1}}\left(\frac{\Delta\varepsilon}{\hbar\omega}\right)^j \sum_{l=0}^q \frac{\left(j \ln(\hbar\omega/\Delta\varepsilon) \right)^l}{l!} \right).
\end{align}
Here $\zeta^{\mathrm{R}}(q+1)$ denotes the famous Riemann Zeta function. However, this value is still very complicated to deal with and does not provide much assistance. Nevertheless, using this result, and remembering that $\sum_n \frac{1}{n!} x^n =  e^x$, we can write:
\begin{align}
0 \le \sum_{j=1}^{\infty} \frac{1}{j^{q+1}}\int_0^{j\ln(\hbar\omega/\Delta\varepsilon)} du~ {e^{-u} u^q}  \le q!\zeta^{\mathrm{R}}(q+1).
\end{align}

Substituting the above result in the Eq.\eqref{Int1} we find the upper boundaries: 
\begin{align}
0 \le I_{p+1} \le \left| \sum_{q=1}^{p-1} \frac{(p-1)!(-1)^q \zeta^R(q+1)}{(p-1-q)!} \ln^{p-1-q}\left(\frac{\hbar\omega}{\Delta\varepsilon}\right) \right| .
\end{align}

Now, changing the sum order and considering that $\frac{\zeta^R(p-q)}{q!} \approx \frac{1}{q!}$: 
\begin{align}
I_{p+1} \le \left| (p-1)! \sum_{q=0}^{p-1} \frac{(-1)^q \zeta^R(p-q)}{(q)!} \ln^{q}\left(\frac{\hbar\omega}{\Delta\varepsilon}\right) \right| \approx  \left| (p-1)! \sum_{q=0}^{p-1} \frac{(-1)^q}{(q)!}\ln^{q}\left(\frac{\hbar\omega}{\Delta\varepsilon}\right) \right|.
\end{align}

If $\hbar\omega > \Delta\varepsilon$, then Eq. \eqref{Int2} is zero. Thus, we can consider $\hbar\omega > \Delta\varepsilon$ and $\ln\left(\frac{\hbar\omega}{\Delta\varepsilon}\right) > 1$. Using the well-known truncation error for the exponential function, $\left|e^{-x} - \left|\sum_{j=0}^{n-1} \frac{(-x)^j}{j!} \right| \right| \le \frac{x^n}{n!}$ for $x > 0$, we can write: 
\begin{align}
I_{p+1} \le (p-1)!~e^{-\ln\left(\frac{\hbar\omega}{\Delta\varepsilon}\right)} + \frac{(p-1)!}{p!} \ln^{p}\left(\frac{\hbar\omega}{\Delta\varepsilon}\right) .
\end{align}
This equation is also valid for $\hbar\omega = \Delta\varepsilon$ since $I_{p+1} = 0$ in this case, as previously discussed. Therefore, we find:
\begin{align}
 \frac{p+1}{p} \ln^{p}{\left(\hbar\omega/\Delta\varepsilon \right)} \le \frac{p+1}{p} \ln^{p}{\left(\hbar\omega/\Delta\varepsilon \right)} + I_{p+1} \le \frac{p+2}{p} \ln^{p}{\left(\hbar\omega/\Delta\varepsilon \right)} + \mathcal{O}(\Delta\varepsilon).
\end{align}

So, the principal terms in $\mathcal{G}_{p+1}$ are indeed:
\begin{align}
 \mathcal{G}_{p+1} (\hbar\omega) = \frac{1}{p!} \left(\frac{\delta}{\pi} \right)^{2(p+1)} \frac{\ln^{p}{\left(\hbar\omega/\Delta\varepsilon \right)}}{\hbar\omega} \left[ 1  + \mathcal{O}\left(\frac{1}{(p+1)}\right) + \mathcal{O}(\Delta\varepsilon)  \right].
\end{align}
Where $p \geq 2$, then, we have shown by finite induction that indeed the expression in Eq. \eqref{g_ind} represents the main part of the quantity $\mathcal{G}_{p}$.

Now, substituting these results in the imaginary part of the Green's function in the Eq. \eqref{G5} we get:
\begin{align}
\frac{\mathrm{Im}G(\hbar\omega)}{\pi |\braket{\varphi_0}{\phi_0}|^2} \approx \delta(\hbar\omega) + \mathlarger{\sum}_{p=1}&\frac{1}{(p-1)!}  \left(\frac{\delta}{\pi} \right)^{2p} \frac{\ln^{p-1}{\left(\hbar\omega/\Delta\varepsilon \right)}}{\hbar\omega} {\Theta(\hbar\omega)}.
\end{align}

\begin{align}
\frac{\mathrm{Im}G(\hbar\omega)}{\pi |\braket{\varphi_0}{\phi_0}|^2} \approx \delta(\hbar\omega) + \left(\frac{\delta}{\pi} \right)^{2} \frac{1}{{\hbar\omega}}\left( {\sum}_{p=0}  \frac{1}{p!} \left(\frac{\delta}{\pi} \right)^{2p}  \ln^{p}\left(\hbar\omega/\Delta\varepsilon \right)  \right)  {\Theta(\hbar\omega)}  .
\end{align}

\begin{align}
\frac{\mathrm{Im}G(\hbar\omega)}{\pi |\braket{\varphi_0}{\phi_0}|^2} \approx \delta(\hbar\omega) + \left(\frac{\delta}{\pi} \right)^{2} \frac{1}{{\hbar\omega}} \left(\frac{\Delta\varepsilon}{\hbar\omega}\right)^{-\left(\frac{\delta}{\pi} \right)^{2}}  {\Theta(\hbar\omega)}  .
\end{align}

Finally,  we can write $\mathrm{Im}G(\hbar\omega) $ as
\begin{eqnarray}
\mathrm{Im}G(\hbar\omega) &\approx& \pi |\braket{\varphi_0}{\phi_0}|^2 \delta(\hbar\omega) + \mathcal{K} \left( {\hbar\omega}\right)^{-1+\left(\frac{\delta}{\pi}\right)^2}\Theta(\hbar\omega).
\end{eqnarray}
Here $\mathcal{K}= \pi \left(\frac{\delta}{\pi}\right)^2 |\braket{\varphi_0}{\phi_0}|^2 (\Delta\varepsilon)^{-\left(\frac{\delta}{\pi}\right)^2}$. This power law in the frequency domain is known as the Doniach-Sunjic law.

\newpage
\section{Time Domain}

Our main goal here is to compute the Fidelity $ \mathcal{F} = |\braket{\phi_0}{\Psi(t)}|^2$, but to do it we need first compute:
\begin{eqnarray}\label{Ft}
\braket{\phi_0}{\Psi(t)} = \sum_{n=0} |\braket{\varphi_n}{\phi_0}|^2 \exp\left(-i \frac{t}{\hbar} E_n\right).
\end{eqnarray}

Before computing $\braket{\phi_0}{\Psi(t)} $, we will partition the many-body basis to isolate the usual contributions to $\braket{\phi_0}{\Psi(t)} $ from those arising from high-energy excitations of the bound state level. To accomplish this, we will define the set $\{\ket{\varphi_n'}\}$ as the many-body states where the bound state level remains occupied, and the set $\{\ket{\bar\varphi_n}\}$ as the many-body excitations that include one particle-hole excitation from the bound state level. Clearly, the total many-body basis is the union of these two sets
\begin{equation}
    \{\ket{\varphi_n}\} = \{\ket{\varphi_n '}\} \cup \{\ket{\bar\varphi_n}\}.
\end{equation}

The reason for this partitioning is that we can now express $\ket{\varphi_n'}$ as a product of pairs of particle-hole excitations from the ground state, 
\begin{equation}
    \ket{\varphi_n'} = \prod_{j_n} g_{p_{j_n}}^\dagger g_{h_{j_n}} \ket{\varphi_0},
\end{equation}
where $\varepsilon_{p_{j_n}}$ and $\varepsilon_{h_{j_n}}$ represent the energies of the particle and the hole level, respectively, and the set of possible $j_n$ depends on the many-body configuration and represents what is the pairs of particle-hole excitations which compose this state. These pairs of particle-hole excitations do not involve the bound level by construction. Similarly, we can express $\ket{\bar\varphi_n}$ as a product of pairs of particle-hole excitations from the special state $\ket{\bar\varphi} = g_0^\dagger g_B \ket{\varphi_0}$ as 
\begin{equation}
    \ket{\bar\varphi_n} = \prod_{j_n} g_{p_{j_n}}^\dagger g_{h_{j_n}} \ket{\bar\varphi}.
\end{equation}

It is important to note that the energy of the state that is a particle-hole excitations from the ground state $\ket{\varphi_n'}$ is 
\begin{equation}
    E_n' = \sum_{j_n} (\varepsilon_{p_{j_n}} - \varepsilon_{h_{j_n}})
\end{equation}
and the energy of $\ket{\bar\varphi_n}$ is
\begin{equation}
    \bar E_n = -\epsilon_B + \sum_{j_n} (\varepsilon_{p_{j_n}} - \varepsilon_{h_{j_n}}),
\end{equation}
as there is already one particle-hole excitation from the bound level to the first level above the Fermi level.

Let us rewrite Eq. \eqref{Ft} after partitioning the energy spectrum into non-excitations from the bound level ${\ket{\varphi_n'}}$ and with one-particle-hole excitations from the bound level ${\ket{\bar\varphi_n}}$ as:
\begin{equation}\label{FtII}
 \braket{\phi_0}{\Psi(t)} = \sum_{n=0} |\braket{\varphi_n}{\phi_0}|^2 \exp(-i \omega_n t )
 +\sum_{n=0} |\braket{\bar\varphi_n}{\phi_0}|^2  \exp(-i {t} (\omega_B + \omega_n))
\end{equation}
where $\hbar\omega_n = \sum_{{j_n}} (\varepsilon_{p_{j_n}} - \varepsilon_{h_{j_n}})$ represents the energy of the many-body excitations from the ground state, except from the bound level, and $\hbar \omega_B = |\epsilon_B|$ represents the energy of the bound level. 

We will utilize the commonly accepted approximation that diagonal elements of the Slater determinant $\bra{\phi_0}\prod_{j_n} g_{k_{j_n}}^\dagger g_{h_{j_n}} \ket{\varphi_0}$ are dominant. Then, we can use the Eqs. \eqref{Proj_II} and \eqref{Proj_III} to rewrite the Eq. \eqref{FtII} as
\begin{align}\label{FtIII}
\braket{\phi_0}{\Psi(t)} &= |\braket{\varphi_0}{\phi_0}|^2 \sum_{n=0} \Omega_n^2(\delta) \exp(-i \omega_n t ) \nonumber \\
&+ |\braket{\bar\varphi}{\phi_0}|^2 e^{-i \omega_B t } \sum_{n=0} \Omega_n^2(\pi-\delta) \exp(-i \omega_n t ), 
\end{align}
where
\begin{equation}\label{P002}
    \Omega_n^2(\delta) \approx \prod_{j_n} \left(\frac{\delta}{\pi} \right)^2 \frac{{\Delta\varepsilon}^2}{\left(\varepsilon_{p_{j_n}} - \varepsilon_{h_{j_n}} - \frac{\delta}{\pi} \Delta\varepsilon\right)^2}.
\end{equation}
Here, $\varepsilon_{k_{j_n}}$ and $\varepsilon_{h_{j_n}}$ represent the energies of the particle and hole level, respectively, where this parameter ${j_n}$ depends on the many-body state $\ket{\varphi_n}$ and represents the pairs of particle-hole excitations which compose this state.

\subsection*{An approximated formula for the projection}

Before procedure, let us define a new function 
\begin{eqnarray}\label{f}
f(\delta; t) = \sum_{n=0} \Omega_n^2(\delta) \exp(-i E_n t/ \hbar ).
\end{eqnarray}

Now, let us transform the sum over all many-body states into a sum over all possible particle-hole excitations, with $p$ representing the number of excitations:
\begin{eqnarray}\label{Change_sum2}
\sum_{n=0} \rightarrow  \sum_{p=0}\left(\frac{1}{p!} \sum_{h_1}  \sum_{k_1} \sum_{{h_2} \neq \{h_1\}} \sum_{{k_2} \neq \{k_1 \}} ... \sum_{{h_p} \neq \{h_1,..,h_{p-1}\}} \sum_{{k_p} \neq \{k_1,..,k_{p-1}\}} \right) 
\end{eqnarray}
Here, the $\neq$ constraint ensures the fermionic occupation rule (Pauli's exclusion principle).

We can then demonstrate that:
\begin{eqnarray}\label{Sum12}
\sum_{k_1} \sum_{{k_2} \neq \{k_1\}}  \Delta\varepsilon^2 = \sum_{k_1}  \sum_{{k_2}} \Delta\varepsilon^2 - \Delta\varepsilon \sum_{k_1} \sum_{k_2} \Delta\varepsilon~\delta^K_{k_1,k_2}. 
\end{eqnarray}

Substituting the Eq. \eqref{Sum12} into Eq. \eqref{Change_sum2} we find the Eq. \eqref{Sum22}, which implies that for a small $\Delta\varepsilon$, the number of possible states in the sum becomes very large, making the rigorous fermionic rule in the sum a minor correction.
\begin{eqnarray}\label{Sum22}
 \sum_{h_1}  \sum_{k_1} ... \sum_{{h_p} \neq \{h_1,..,h_{p-1}\}} \sum_{{k_p} \neq \{k_1,..,k_{p-1}\}} \Delta\varepsilon^{2p}  = \sum_{h_1}  \sum_{k_1} ... \sum_{{h_p}} \sum_{{k_p} } \Delta\varepsilon^{2p} + \mathcal{O}(\Delta\varepsilon).
\end{eqnarray}

Additionally, for $\varepsilon_{k_{j}} - \varepsilon_{h_{j}} \ge \Delta\varepsilon > \frac{\delta}{\pi} \Delta\varepsilon$, we can show that:
\begin{align}\label{Re12}
\left[\left(\frac{\delta}{\pi} \right) \frac{\Delta\varepsilon}{\varepsilon_{k_{j}} - \varepsilon_{h_{j}} - \frac{\delta}{\pi} \Delta\varepsilon} \right]^2 = \left[\left(\frac{\delta}{\pi} \right) \frac{\Delta\varepsilon}{\varepsilon_{k_{j}} - \varepsilon_{h_{j}}} \right]^2 + \mathcal{O}\left( \Delta\varepsilon^3 \right).
\end{align}

Substituting this new sum in Eq. \eqref{Change_sum2} and Eq. \eqref{P002} into $f(t)$, we get:
\begin{align}\label{G4_2}
 f(\delta;t) \approx {\sum}_{p=0}&\frac{1}{p!}  \left(\frac{\delta}{\pi} \right)^{2p} \sum_{h_1}  \sum_{k_1} ... \sum_{{h_p}} \sum_{{k_p}} \left[\prod_{j=1}^p \frac{\Delta\varepsilon}{\varepsilon_{k_{j}} - \varepsilon_{h_{j}}} \right]^2 e^{-i\frac{t}{\hbar}\sum_{j=1}^p  \left(  \varepsilon_{k_j} - \varepsilon_{h_j}\right)} + \mathcal{O}(\Delta\varepsilon).
\end{align}

Since the energy difference $ \Delta\varepsilon \le \epsilon = \varepsilon_k - \varepsilon_h \le D$ represents the particle-hole excitation energy, and $\Delta\varepsilon \le -\epsilon_h \le \epsilon$, we can modify the sum domain into a sum over $\epsilon$ as:
\begin{eqnarray}\label{Change_Domain22}
\sum_{\epsilon_h} \sum_{\epsilon_{k} }  \Delta\varepsilon^2 \mathcal{Q}(\varepsilon_k - \varepsilon_h) = \sum_{\epsilon} \Delta\varepsilon \mathcal{Q}(\epsilon)  \sum_{-\epsilon_h<\epsilon} \Delta\varepsilon = \sum_{\epsilon} \Delta\varepsilon~ \epsilon. \mathcal{Q}(\epsilon) + \mathcal{O}(\Delta\varepsilon).
\end{eqnarray}
\begin{align}\label{G42}
 f(\delta;t) \approx {\sum}_{p=0}&\frac{1}{p!}  \left(\frac{\delta}{\pi} \right)^{2p} \sum_{q_1} \Delta\varepsilon \frac{ e^{-i\frac{t}{\hbar}\epsilon_{q_1}} }{\epsilon_{q_1} }  ... \sum_{{q_p}} \Delta\varepsilon \frac{ e^{-i\frac{t}{\hbar}\epsilon_{q_p}} }{\epsilon_{q_p} }  + \mathcal{O}(\Delta\varepsilon).
\end{align}

Considering that the sum can be approximated by an integral, and this approximation leads to an error in the other of $\Delta\varepsilon$ once there is no pole inside the sum over $D \ge \epsilon_j \ge \Delta\varepsilon$, then we can rewrite:
\begin{align}\label{G4_II2}
 f(\delta;t) \approx {\sum}_{p=0}&\frac{1}{p!}  \left(\frac{\delta}{\pi} \right)^{2p} \int_{\Delta\varepsilon}^{D} d\epsilon_{q_1} \frac{ e^{-i\frac{t}{\hbar}\epsilon_{q_1}} }{\epsilon_{q_1} }  ... \int_{\Delta\varepsilon}^{D} d\epsilon_{q_p} \frac{ e^{-i\frac{t}{\hbar}\epsilon_{q_p}} }{\epsilon_{q_p} }  + \mathcal{O}(\Delta\varepsilon).
\end{align}

The integrals over $\epsilon_{q_j}$ with $j=1,...,p$ are independent, then we can write: 
\begin{align}\label{G4_III2}
 f(\delta;t) \approx {\sum}_{p=0}&\frac{1}{p!}  \left(\frac{\delta}{\pi} \right)^{2p}  \left[ \int_{\Delta\varepsilon}^{D} d\epsilon \frac{ e^{-i\frac{t}{\hbar}{\epsilon }}}{\epsilon }  \right]^p  \rightarrow  \exp \left(\left(\frac{\delta}{\pi}\right)^2 \int_{\Delta \varepsilon}^{D}  d\epsilon \frac{\exp(-i \epsilon t / \hbar )}{\epsilon} \right), 
\end{align}
and
\begin{eqnarray}\label{Unk}
 \sum_{n=0} \Omega_n^2(\delta) \exp(-i \omega_n t ) \approx \exp \left(\left(\frac{\delta}{\pi}\right)^2 \int_{\Delta \varepsilon}^{D}  d\epsilon \frac{\exp(-i \epsilon t / \hbar )}{\epsilon} \right)\left(1 + \mathcal{O} \left(\frac{\delta}{\pi} \right)^4 \right) + \mathcal{O}(\Delta\varepsilon).~~
\end{eqnarray}

Once we found an approximation for the function $f(\delta;t)$, we can use it to write the projection as
\begin{align}\label{AUXX_III}
\braket{\phi_0}{\Psi(t)} = |\braket{\varphi_0}{\phi_0}|^2 f(\delta;t) + |\braket{\bar\varphi}{\phi_0}|^2 f(\pi-\delta;t) e^{-i |\epsilon_B| \frac{t}{\hbar} }, 
\end{align}
with the boundary condition $\braket{\phi_0}{\Psi(0)} = 1$.

At $t=0$, it is not difficult to show that $f(\delta;0) = N_e^{\left(\frac{\delta}{\pi}\right)^2}$, and the projection
\begin{align}
\braket{\phi_0}{\Psi(0)} = 1 =  |\braket{\varphi_0}{\phi_0}|^2 N_e^{\left(\frac{\delta}{\pi}\right)^2}  + |\braket{\bar\varphi}{\phi_0}|^2 N_e^{\left(\frac{\delta}{\pi}\right)^2}, 
\end{align}
and to keep the above statement true for any circumstances we can set
\begin{align}\label{AUXX_I}
|\braket{\varphi_0}{\phi_0}|^2 = \frac{1}{1+r} N_e^{-\left(\frac{\delta}{\pi}\right)^2} , 
\end{align}
and 
\begin{align}\label{AUXX_II}
|\braket{\bar\varphi}{\phi_0}|^2 = \frac{r}{1+r} N_e^{-\left(\frac{\delta}{\pi}\right)^2}.
\end{align}
The parameter $r$ is defined by
\begin{eqnarray}\label{ratio}
r = \frac{|\braket{\bar\varphi}{\phi_0}|^2}{|\braket{\varphi_0}{\phi_0}|^2}.
\end{eqnarray}

Now, let us note that:
\begin{eqnarray}\label{AUXX_IV}
\exp \left(-\left(\frac{\delta}{\pi}\right)^2 \int_{\Delta \varepsilon}^{D}  d\epsilon \frac{1}{\epsilon} \right) = N_e^{- \left(\frac{\delta}{\pi} \right)^2}.
\end{eqnarray}

By substituting the Eq. \eqref{AUXX_IV}, \eqref{AUXX_I} and \eqref{AUXX_II} into the Eq. \eqref{AUXX_III}, we found that the projection can be written as
\begin{align}\label{AUXX_V}
\braket{\phi_0}{\Psi(t)} &\approx \frac{1}{1+r} \exp \left(-\left(\frac{\delta}{\pi}\right)^2 \int_{\Delta \varepsilon}^{D}  d\epsilon \frac{1-\exp(-i \epsilon t / \hbar )}{\epsilon} \right) \nonumber  \\ &+ 
\frac{r .\exp\left( -i \frac{|\epsilon_B| t}{\hbar} \right) }{1+r} \exp \left(-\left(1-\frac{\delta}{\pi}\right)^2 \int_{\Delta \varepsilon}^{D}  d\epsilon \frac{1-\exp(-i \epsilon t / \hbar )}{\epsilon} \right).
\end{align}
Here, if one compute the Fourier transformation of the projection, the obtained result will be the same as we obtained for the green function in the last section in the frequency domain.

\subsection*{Computing the fidelity}

Finally, we can used the Eq. \eqref{AUXX_V} to compute the fidelity, only is necessary to square the modulus of the projection. But before doing so, we can define two new functions
\begin{align}\label{AUXX_VI}
\mathcal{R}(t) = r \times \exp \left(-\left(1-\frac{2\delta}{\pi}\right) \int_{\Delta \varepsilon}^{D}  d\epsilon \frac{1-\cos( \epsilon t / \hbar )}{\epsilon} \right),
\end{align}
\begin{align}\label{AUXX_VII}
\Upsilon(t) = \left(1-\frac{2\delta}{\pi}\right) \int_{\Delta \varepsilon}^{D}  d\epsilon \frac{\sin( \epsilon t / \hbar )}{\epsilon},
\end{align}
and after some mathematical manipulations the fidelity become 
\begin{align}\label{Fidelity}
\mathcal{F}(t) &= |\braket{\phi_0}{\Psi(t)}|^2  \nonumber \\ 
&\approx \frac{\left(1 + \mathcal{R}(t)^2 + 2\mathcal{R}(t) \cos\left(\frac{\epsilon_B t}{\hbar} + \Upsilon(t) \right) \right)}{(1+r)^2}  \exp \left( -2\left(\frac{\delta}{\pi}\right)^2 \int_{\Delta \varepsilon}^{D} d\epsilon \frac{1 - \cos( \epsilon t / \hbar )}{\epsilon} \right).~~
\end{align}
Here the function $\mathcal{R}(0) = r$ and slowly decay with time and the function $\Upsilon(t)$ slowly increase.

To rewrite the fidelity in terms of known functions, let us start by writing the fidelity as 
\begin{eqnarray}\label{Fidelity_G}
\mathcal{F}(t) \approx \frac{\left(1 + \mathcal{R}(t)^2 + 2\mathcal{R}(t) \cos\left(\frac{\epsilon_B t}{\hbar}  + \Upsilon(t) \right) \right)}{(1+r)^2} e^{-2\left(\frac{\delta}{\pi}\right)^2 \left(\ln \left(\frac{D}{\Delta\varepsilon}\right) - \int_{\Delta \varepsilon}^{D} d\epsilon \frac{\cos(\epsilon t /\hbar )}{\epsilon} \right)},
\end{eqnarray}
and by using substitution on the integral exponential term, we found
\begin{eqnarray}
\mathcal{F}(t) \approx \frac{\left(1 + \mathcal{R}(t)^2 + 2\mathcal{R}(t) \cos\left(\frac{\epsilon_B t }{\hbar} + \Upsilon(t) \right) \right)}{(1+r)^2} e^{-2\left(\frac{\delta}{\pi}\right)^2 \left(\ln \left(\frac{D}{\Delta\varepsilon}\right) - \int_{\frac{\Delta\varepsilon t}{\hbar} }^{\frac{D t}{\hbar}} du \frac{\cos(u)}{u} \right)}.
\end{eqnarray}

Now, we can modify the integral in the exponent in terms on the well known cosine integral function $\mathrm{CI}(x)$ as
\begin{eqnarray}
\mathcal{F}(t) \approx \frac{\left(1 + \mathcal{R}(t)^2 + 2\mathcal{R}(t) \cos\left(\frac{\epsilon_B t }{\hbar} + \Upsilon(t) \right) \right)}{(1+r)^2} e^{ -2\left(\frac{\delta}{\pi}\right)^2 \left(\ln \left(\frac{D}{\Delta\varepsilon}\right) + \mathrm{CI}\left(\frac{\Delta\varepsilon  t}{\hbar} \right) - \mathrm{CI}\left( \frac{D t}{\hbar} \right) \right)}.
\end{eqnarray}

\subsection*{Doniach-Sunjic Behavior of the fidelity}

\begin{eqnarray}\label{Fidelity_G3}
\mathcal{F}(t) \approx \frac{\left(1 + \mathcal{R}(t)^2 + 2\mathcal{R}(t) \cos\left(\frac{\epsilon_B t + \Upsilon(t)}{\hbar}\right) \right)}{(1+r)^2} \exp \left( -2\left(\frac{\delta}{\pi}\right)^2 \int_{\Delta \varepsilon}^{D} d\epsilon \frac{1 - \cos( \epsilon t / \hbar )}{\epsilon} \right).
\end{eqnarray}

To try to estimate the exponent behavior, let us now define the new function:
\begin{eqnarray}
\mathcal{Q}(t) =\int_{\Delta \varepsilon}^{D} d\epsilon \frac{1 - \cos( \epsilon t / \hbar )}{\epsilon}
\end{eqnarray}

It is not difficult to show that: 
\begin{eqnarray}
&\frac{d\mathcal{Q}(t)}{dt}& = \frac{1}{t} \left(\cos(\Delta\varepsilon t/\hbar)-\cos(Dt/\hbar)\right) \nonumber \\
&\frac{d^2\mathcal{Q}(t)}{dt^2}& = \frac{-1}{t^2} \left(\cos(\Delta\varepsilon t/\hbar)-\cos(Dt/\hbar)\right) + ... \nonumber \\ 
&\frac{d^3\mathcal{Q}(t)}{dt^3}& = \frac{2}{t^3} \left(\cos(\Delta\varepsilon t/\hbar)-\cos(Dt/\hbar)\right) + ... \nonumber
\end{eqnarray}
Also, we can observe the asymptotic behavior of the function $\mathcal{Q}(t)$ by computing the limit: 
\begin{eqnarray}
\lim_{t \rightarrow \infty} \frac{\int_{\Delta \varepsilon}^{D} d\epsilon \frac{1 - \cos( \epsilon t / \hbar )}{\epsilon}}{\ln t} = \lim_{t \rightarrow \infty} \left(\cos(\Delta\varepsilon t/\hbar)-\cos(Dt/\hbar)\right).
\end{eqnarray}

Then, using the asymptotic behavior for $t \gg 1$:
\begin{eqnarray}
\lim_{t \gg 1} {\int_{\Delta \varepsilon}^{D} d\epsilon \frac{1 - \cos( \epsilon t / \hbar )}{\epsilon}} \sim  {\ln t}.
\end{eqnarray}

\begin{figure}[htb!]
		\centering
        \includegraphics[scale=0.6]{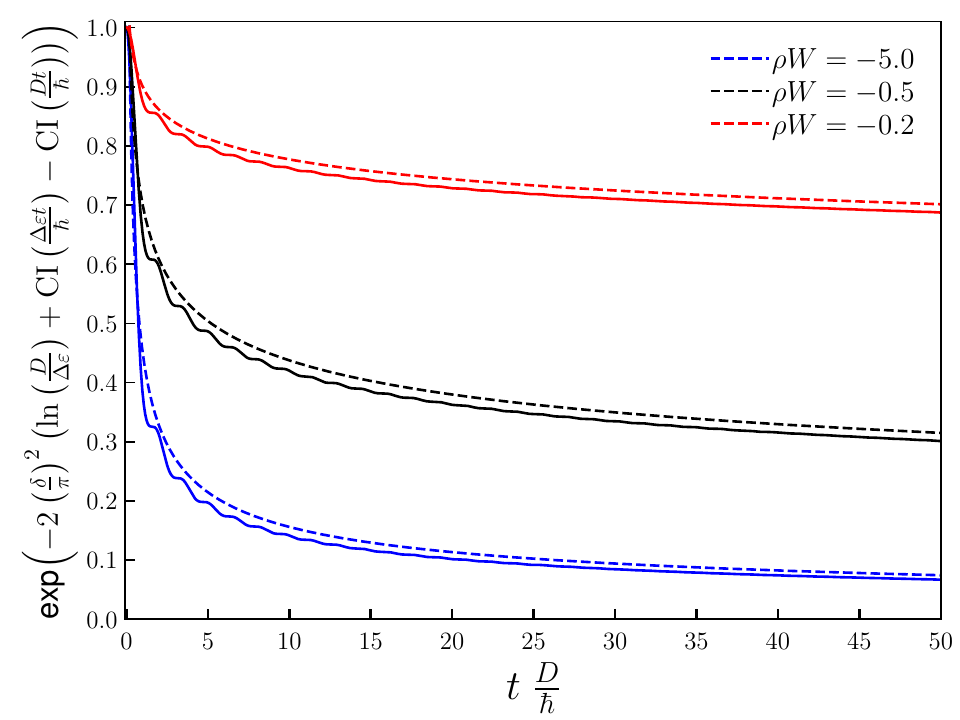}
        \vspace{-0.2cm}
	    \caption{\footnotesize Here we plot $e^{ -2\left(\frac{\delta}{\pi}\right)^2 \left(\ln \left(\frac{D}{\Delta\varepsilon}\right) + \mathrm{CI}\left(\frac{\Delta\varepsilon  t}{\hbar} \right) - \mathrm{CI}\left( \frac{D t}{\hbar} \right) \right)}$ (solid line) and the Doniach Sunjic law $t^{-2 \left(\frac{\delta}{\pi}\right)^2}$ (dashed line) as function of time for different values of $\rho W$. \Sou}
	   \label{Asymptotic}
\end{figure}

Using this asymptotic behavior for $t \gg 1$ in the Fidelity, after some manipulations, we can show that:
\begin{eqnarray}\label{Fidelity_G4}
\mathcal{F}(t) \sim \frac{1}{(1+r)^2} \left(1 + r^2~t^{-2\left(1+2\frac{\delta}{\pi}\right)} + 2r~t^{-\left(1+2\frac{\delta}{\pi}\right)} \cos\left(\frac{\epsilon_B t}{\hbar}\right) \right) t^{-2\left(\frac{\delta}{\pi}\right)^2} .
\end{eqnarray}

\subsection*{Numerical x analytical fidelity}

To derive the analytical equation for the fidelity, we used some approximations, such as the phase shift constant and the diagonal dominance in the Slater determinant. These approximations are valid near the Fermi levels but do not work very well for the high energy levels. As we observe from Fig. \ref{Fidelity_NumexAnaly}, the analytical equation inherently follows the Doniach-Sunjic law and shows good results when compared with direct diagonalization, explaining the fidelity behavior for all time scales. We can also observe that the analytical equation converges to the exact values of the fidelity for long times and the error drops, once this time scales is associated with the small energies excitations around the Fermi level, where the approximations we considered are accurate.

\begin{figure}[htb!]
		\centering
        \includegraphics[scale=0.6]{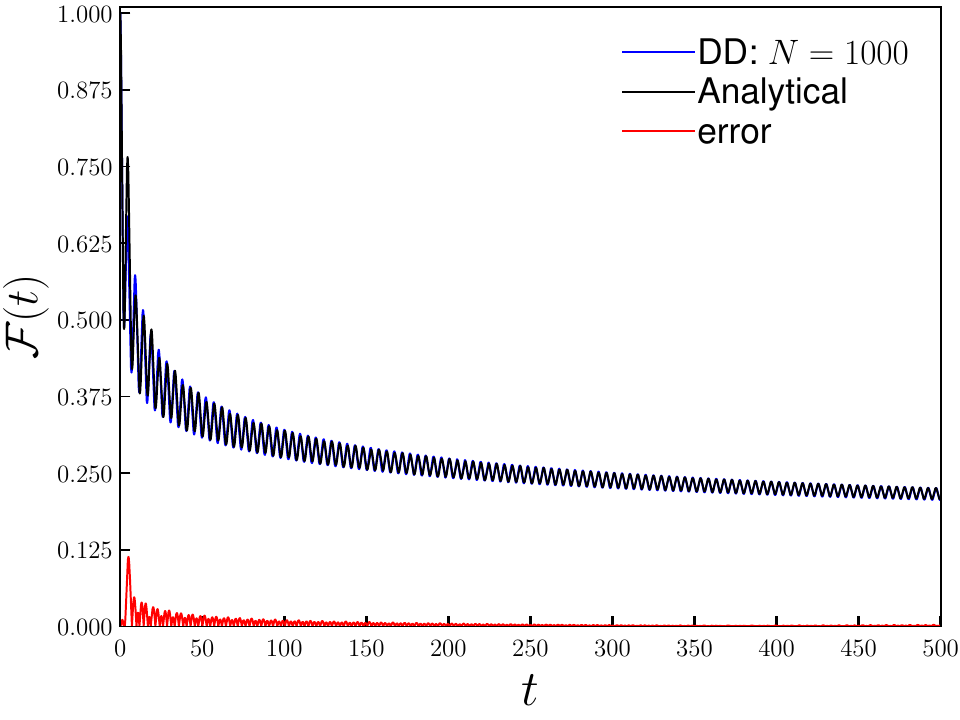}
        \includegraphics[scale=0.6]{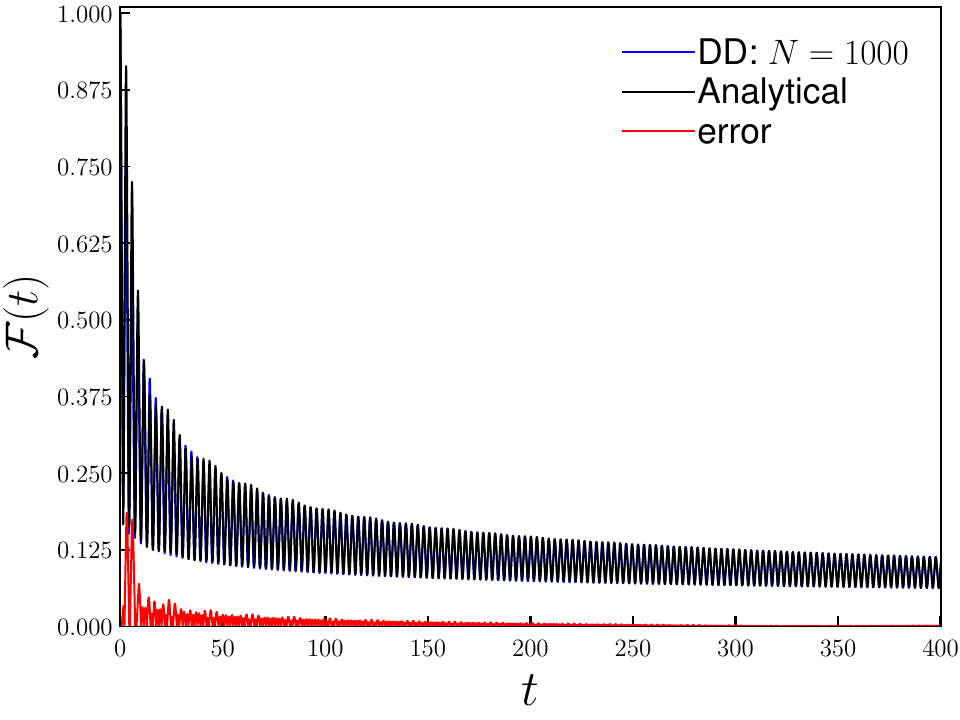}
        \vspace{-0.5cm}
	    \caption{\footnotesize  Comparing the analytical results (dashed black line) with results from direct diagonalization (solid blue line) using $W=-1D$ and $N=1000$ (Top plot)  and  $W=-2D$ and $N=1000$ (Bottom plot). The red line shows the absolute value of the error in the analytical expression by computing $|\mathcal{F}_{DD} - \mathcal{F}_{\mathrm{Analytical}}|(t)$. \Sou}
	   \label{Fidelity_NumexAnaly}
\end{figure}

\chapter{Continuous scattering potential}\label{Model_Continuous_W}

The quadratic Hamiltonian in Eq. \eqref{H_photo} is well-known for a fermionic gas with a localized scattering potential and can be diagonalized analytically, as we showed in the last section. The eigenstates of a single particle and the eigenenergies of \eqref{H_photo} are given by the creation operator $g_l^\dagger(t) = \sum_q \alpha_{l,q}(t) \tilde{a}_q^\dagger$ and the energy $\epsilon_l(t)= {\varepsilon}_l -\left(\frac{\delta(t)}{\pi}\right)\Delta\varepsilon$. Here, the change of basis coefficient is $\alpha_{l,q}(t) = \frac{\sin{\delta}(t)}{\pi} \frac{\Delta\varepsilon}{{\varepsilon}_q - {\varepsilon}_l + \frac{\delta(t)}{\pi}\Delta\varepsilon }$. Note that the differences in energy levels $\epsilon_p(t) - \epsilon_{h}(t) = \Delta\varepsilon (p - h)$ do not depend on phase shift or time.

Once the Hamiltonian is time dependent, we need to connect the instantaneous basis at different times. To do it we can start re-writing ${H}(t>t')$ as 
\begin{equation}\label{H_tt'}
\begin{aligned}
{H}(t) &= {H}(t') + \frac{W(t)-W(t')}{N} \sum_{k,k'} \tilde{a}_k^\dagger\tilde{a}_{k'} \\ &= H(t') + \frac{\Delta W\cos^2\delta(t')}{N} \sum_{q,q'} g_q^\dagger(t')g_{q'}(t').
\end{aligned}
\end{equation}
Here $\Delta W = W(t) - W(t')$, and we use the fact that the term $\sum_{q,q'} \tilde{a}_q^\dagger\tilde{a}_{q'}$ can be written in terms of $\{g_q^\dagger (t')\}$ as $\sum_{k,k',q,q'} \alpha_{q,k}(t') \alpha_{q',k'}(t') g_q^\dagger(t') {g}{q'}(t')$, and $\sum_{k,k'} \alpha_{q,k}(t') \alpha^{q',k'}(t') = \cos^2\delta(t') $, since $\sum_k \alpha_{q,k}(t') = \frac{1}{W(t')} \frac{\sin(\delta(t'))}{\rho\pi}$ and $\tan \delta(t') = -\pi \rho W(t')$ as shown in Appendix \ref{Appendix_Analytical_Diagonalization}.

Then, we can write the operators $g_l^\dagger(t) = \sum_q {\beta}_{l,q} {g}_q^\dagger(t')$ and the energies $\epsilon_l(t) = \epsilon_l(t') -\left(\frac{\Delta\delta}{\pi}\right)\Delta\varepsilon$, with $\beta_{l,q} = -\frac{\sin\Delta\delta}{\pi} \frac{\Delta\varepsilon}{\epsilon_l(t') - \epsilon_q(t') - \frac{\Delta\delta}{\pi}\Delta\varepsilon}$ and $\Delta\delta = \delta(t) - \delta(t') + \mathcal{O}(\Delta\delta^2) $. This shows that the change of basis transformation from $t'$ to $t$ kept the same structure as $\alpha_{k,q}$.

To track the time evolution of the many-body system, we need to solve the Schrödinger equation, which, in the instantaneous basis, give to us the time propagation of the coefficients by 
\begin{eqnarray}\label{SE}
\frac{d\tilde{c}_n(t)}{dt}=-\sum_m \tilde{c}_m(t)\bra{\varphi_n(t)} {\partial _t} \ket{\varphi_m(t)} e^{-\frac{i}{\hbar}\int_0^t (E_m(t') - E_n(t')) dt' }.
\end{eqnarray}
Here, the coefficient $\tilde{c}_n(t) = c_{n}(t) e^{+\frac{i}{\hbar}\int_0^t dt' E_n(t')}$ contains information about the instantaneous occupation of the many-body eigenstate $\ket{\varphi_n(t)}$, since $|\tilde c_n|^2 = | c_n|^2$.

The time derivative term $\bra{\varphi_n} {\partial _t} \ket{\varphi_m}$ in the differential equation \eqref{SE} can be written explicit in the limit definition of the derivative by
\begin{eqnarray}\label{lim}
\bra{\varphi_n}\partial_t\ket{\varphi_m}=\lim_{\Delta t \rightarrow 0} \left(\frac{\delta_{n,m}^K - \braket{\varphi_n(t)}{\varphi_m(t-\Delta t)} }{\Delta t} \right).
\end{eqnarray}
Here $\delta^K_{n,m}$ is the Kronecker delta.

However, even knowing the general analytical single-particle solution, to build the many-body wave function, we need to deal with many-body projections $\braket{\varphi_n(t)}{\varphi_m(t')}$, which are obtained by Slater determinants of matrices with elements $\alpha_{n,q}$. For this aim we can think about the $\ket{\varphi_n(t)}$ as a product of pair particle-hole excitations $\{ g_{p_j}^\dagger g_{h_j} \}$ of the configuration $\ket{\varphi_m(t)}$. Then, for small $\Delta\delta$, the determinant can be approximate by a diagonal matrix determinant as
\begin{eqnarray}\label{Proj}
\frac{\braket{\varphi_n(t)}{\varphi_m(t')}}{\braket{\varphi_m(t)}{\varphi_m(t')}} \approx \left[ \prod_j \left(\frac{\Delta\delta}{\pi} \right) \frac{\Delta\varepsilon}{\varepsilon_{p_j} - \varepsilon_{h_j} + \frac{\Delta\delta}{\pi} \Delta\varepsilon} \right]. \hspace{0.2cm}
\end{eqnarray}

Before proceeding, let us use Eq. \eqref{AOC EQ} to determine that the projection $\braket{\varphi_m(t)}{\varphi_m(t')} \sim N_e^{-\left(\frac{\Delta\delta}{\pi}\right)^2}$. Now, let us explicitly compute some cases by combining Eq. \eqref{Proj} with Eq. \eqref{lim} and Eq. \eqref{AOC EQ}, but without taking the limit $\Delta t \rightarrow 0$.
\begin{eqnarray}
&\bra{\varphi_n}\partial_t\ket{\varphi_n} \approx  \frac{1 - N_e^{-\left(\frac{\Delta\delta}{\pi}\right)^2}  }{\Delta t} = 0 + \mathcal{O}\left( \frac{\Delta\delta^2}{\Delta t} \right) \nonumber, \\
&\bra{\varphi_n}\partial_t ~ g_p^\dagger g_h \ket{\varphi_n} \approx  -\frac{1}{\pi}\frac{\Delta \delta}{\Delta t} \frac{\Delta\varepsilon}{\epsilon_p - \varepsilon_h}N_e^{-\left(\frac{\Delta\delta}{\pi}\right)^2} =  -\frac{1}{\pi}\frac{\Delta \delta}{\Delta t} \frac{\Delta\varepsilon}{\varepsilon_p - \varepsilon_h}  +  \mathcal{O}\left( \frac{\Delta\delta^2}{\Delta t} \right)  \nonumber, \\
&\bra{\varphi_n}\partial_t ~ g_p^\dagger g_h  g_{p'}^\dagger g_{h'} \ket{\varphi_n} \approx -\frac{1}{\pi^2}\frac{\Delta \delta^2}{\Delta t} \frac{\Delta\varepsilon}{\epsilon_p - \varepsilon_h} \frac{\Delta\varepsilon}{\epsilon_{p'} - \varepsilon_{h'}} N_e^{-\left(\frac{\Delta\delta}{\pi}\right)^2} = 0 +  \mathcal{O}\left( \frac{\Delta\delta^2}{\Delta t} \right) . 
\end{eqnarray}

Then, combining the Eq.\eqref{Proj} with the Eq. \eqref{lim}, the terms $\mathcal{O}(\Delta t)$ and  $\mathcal{O}\left(\Delta \delta = \frac{d \delta}{dt} \Delta t\right)$ vanish in the limit when $\Delta t \rightarrow 0$, then we obtain
\begin{eqnarray}
\bra{\varphi_n}\partial_t\ket{\varphi_m} = \delta^{K}_{(\{ph\},1)} . \frac{1}{\pi}\frac{d\delta(t)}{dt} \frac{\Delta\varepsilon}{\varepsilon_p - \varepsilon_h}.
\end{eqnarray}
Here $\delta^K_{(\{ph\},1)}$ just means that this expression is different from zero only if $\ket{\varphi_n} = g_p^\dagger g_h \ket{\varphi_m}$, for $h \neq p$, or, in other words, two many-body instantaneous eigenstates are coupled only when it is possible to write one as a particle-hole excitation from the other one.

Finally, we can re-write the Eq \eqref{SE} as
\begin{equation}\label{SEFM}
\begin{aligned}
\frac{d\tilde{c}_n(t)}{dt} = -\frac{1}{\pi} \frac{d\delta(t)}{dt}\sum_p\sum_{h\neq p} \tilde{c}_{n,p,h}(t) \frac{\Delta\varepsilon}{\varepsilon_p - \varepsilon_h} e^{-i(\varepsilon_p - \varepsilon_h)\frac{t}{\hbar}},
\end{aligned}
\end{equation}
where the coefficients $\tilde{c}_{n,p,h}$ are the occupation of the eigenstate $\ket{\varphi_{n,p,h}} = g_p^\dagger g_h \ket{\varphi_n }$.

This result demonstrates that for this type of system, when the localized scattering potential changes continuously, a coupling exists between two many-body instantaneous eigenstates only when it is possible to represent one many-body eigenstate as a particle-hole excitation from the other one. Furthermore, the Eq. \eqref{SEFM} is exact, with no approximation or perturbation theory employed in the deduction. Even the high-order terms in $\Delta\delta$ that are not explicitly shown here vanish after taking the limit of $\Delta t \rightarrow 0$, once $\Delta\delta \rightarrow 0$ in this case.

\chapter{Diagrammatic Dyson's series}\label{diagrams_dayson}

In this thesis, more than one time, we needed to deal with differential equations that represents the time evolution of the many-body system like
\begin{eqnarray}
\frac{d c_n(t) }{dt } = \sum_{m\neq n} V_{nm}(t) c_m(t) ~~~ \mathrm{or}~~~ \frac{d \vec{c}_t }{dt } = \hat V_t \vec{c}_t,
\end{eqnarray}
where $c_n(t) = \braket{\varphi_n(t)}{\Psi(t)}$, $|c_n(t)|^2$ represents the probabilities to find the system in the many-body state $\ket{\varphi_m (t) }$ and $V_{nm}(t)$ represents the coupling between two states.

One approach to solve the above differential equations, is the well-known Dyson series solution
\begin{eqnarray}\label{Dayson}
\vec{c}_t = \mathcal{\hat T} \left\{ \exp \left( \int_0^t dt' \hat V_{t'} \right) \right\} \vec{c}_0.
\end{eqnarray}
Here, $\mathcal{\hat T}$ is the time order operator.

Even if we will solve numerically the differential equation, the Dyson series solution allow us to explore and understand more easily some aspects of the system. Now, let us consider that the system starts totally in one initial configuration, so $c_i(0) = 1$, in this situation the Eq. \eqref{Dayson} can be expanded as
\begin{eqnarray}\label{Dayson_}
c_f(t) = \delta_{f,i} + \int_0^t dt_1 V_{f,i}(t_1) + \frac{1}{2} \int_0^t dt_1 \int_0^t dt_2  \mathcal{\hat T} \sum_k V_{f,k}(t_2) V_{k,i}(t_1) +~\cdots ~,
\end{eqnarray}
and the diagrammatic representation of the Eq. \eqref{Dayson_} are shown in the Fig. \ref{fig:Feynman_Dayson} up to the second order in the potential $V$ expansion.

In Fig. \ref{fig:Feynman_Dayson}, the dashed horizontal line represents the ground state ($\ket{\tilde 0} \equiv \ket{\varphi_0}$) and is analogous to the vacuum in the traditional Feynman diagram seen in particle physics. The other electronic states are represented by particle-hole excitations from the ground state $\ket{\varphi_n} =  \prod_j g^\dagger_{p_j} g_{h_j} \ket{\varphi_0}$, depicted by lines with arrows indicating the creation of electrons and anti-electrons (holes) (as shown on the right-hand side of Fig. \ref{fig:Feynman} (middle) and (right)). Thus, after the initial electronic configuration (analogous to the vacuum) interacts with the potential, it may create one (or more) pairs of particle-hole excitations from this configuration. As shown in \ref{fig:Feynman_Dayson_}, on the left-hand side of the diagram is the initial state, which interacts with the potential. The possible electronic configurations over time can be observed in each diagram by making a vertical cut and examining the electronic configuration expressed as particle-holes created from this initial state.

Let us consider that the system at $t=0$ is in the initial configuration $\ket{\varphi_i}$ ($c_i(0)=1$). The initial state interacts with the potential. We want to know the possible final configurations $\ket{\varphi_f}$. In Figures \ref{fig:Feynman_Dayson}, we illustrate the principal transitions, arranged by the expansion in the potential. (Left) In the zero-order expansion, the system remains in the same initial configuration $\braket{\varphi_f}{\Psi}^{(0)} \sim 1$. (Middle) The first-order expansion of the potential, with diagrams illustrating the transition of the initial state to the final many-body state with one (or more) particle-hole excitation from the initial configuration. (Right) The second-order expansion of the potential, with diagrams illustrating the transition of the initial state to the final many-body state through an intermediate state.

\begin{figure}[hbt!]
    \centering
    \includegraphics[scale=0.4]{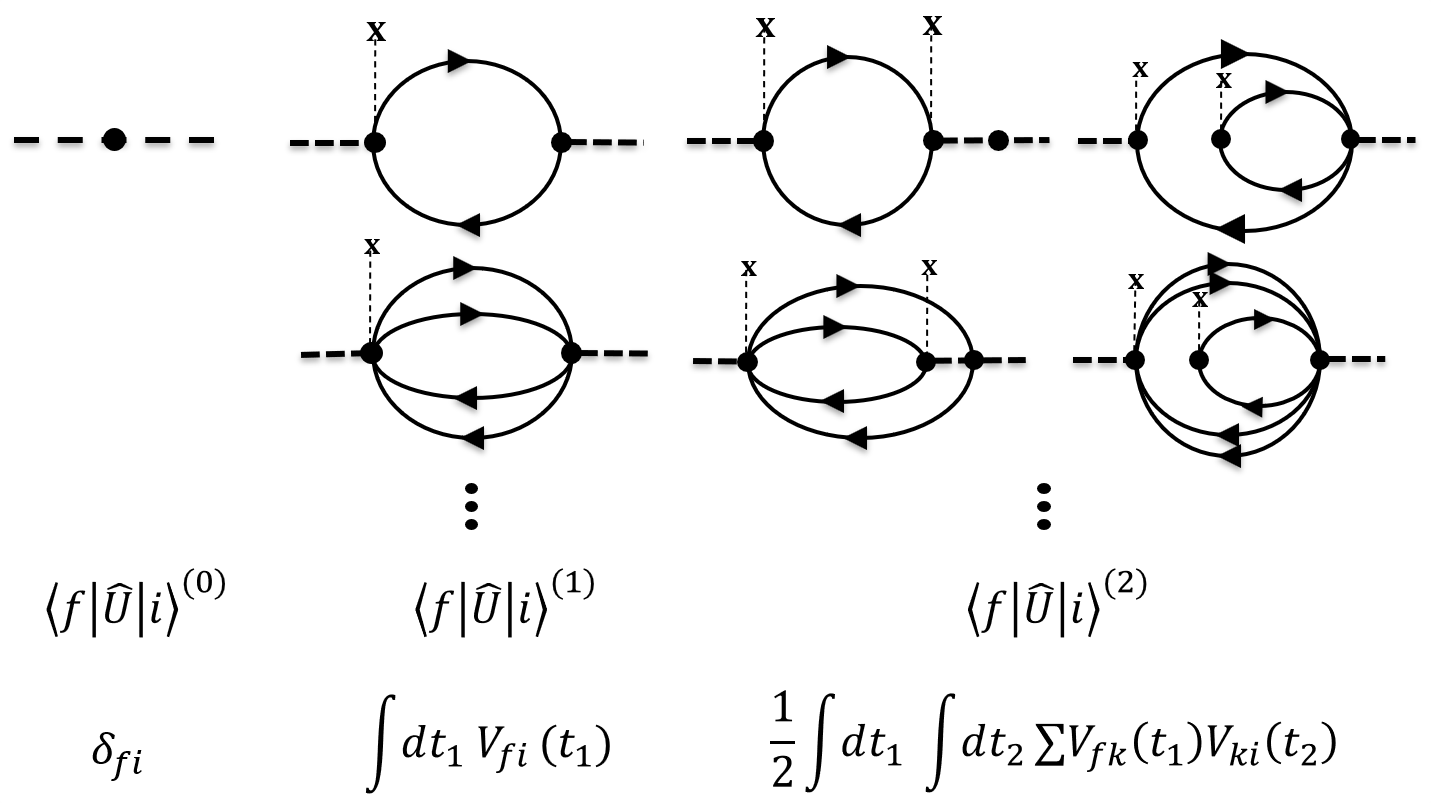}
    \vspace{-0.25cm}
    \caption{\footnotesize (Left) The zeroth-order diagram illustrates the transition of a many-body state with energy $ E_i $ to the final many-body state with the same configuration. (Middle) The first-order expansion of the potential, with diagrams illustrating the transition of the initial state to the final many-body state with one (or more) particle-hole excitation from the initial configuration. (Right) The second-order expansion of the potential, with diagrams illustrating the transition of the initial state to the final many-body state through an intermediate state. \Sou}
    \label{fig:Feynman_Dayson}
\end{figure}

\begin{figure}[hbt!]
    \centering
    \includegraphics[scale=0.38]{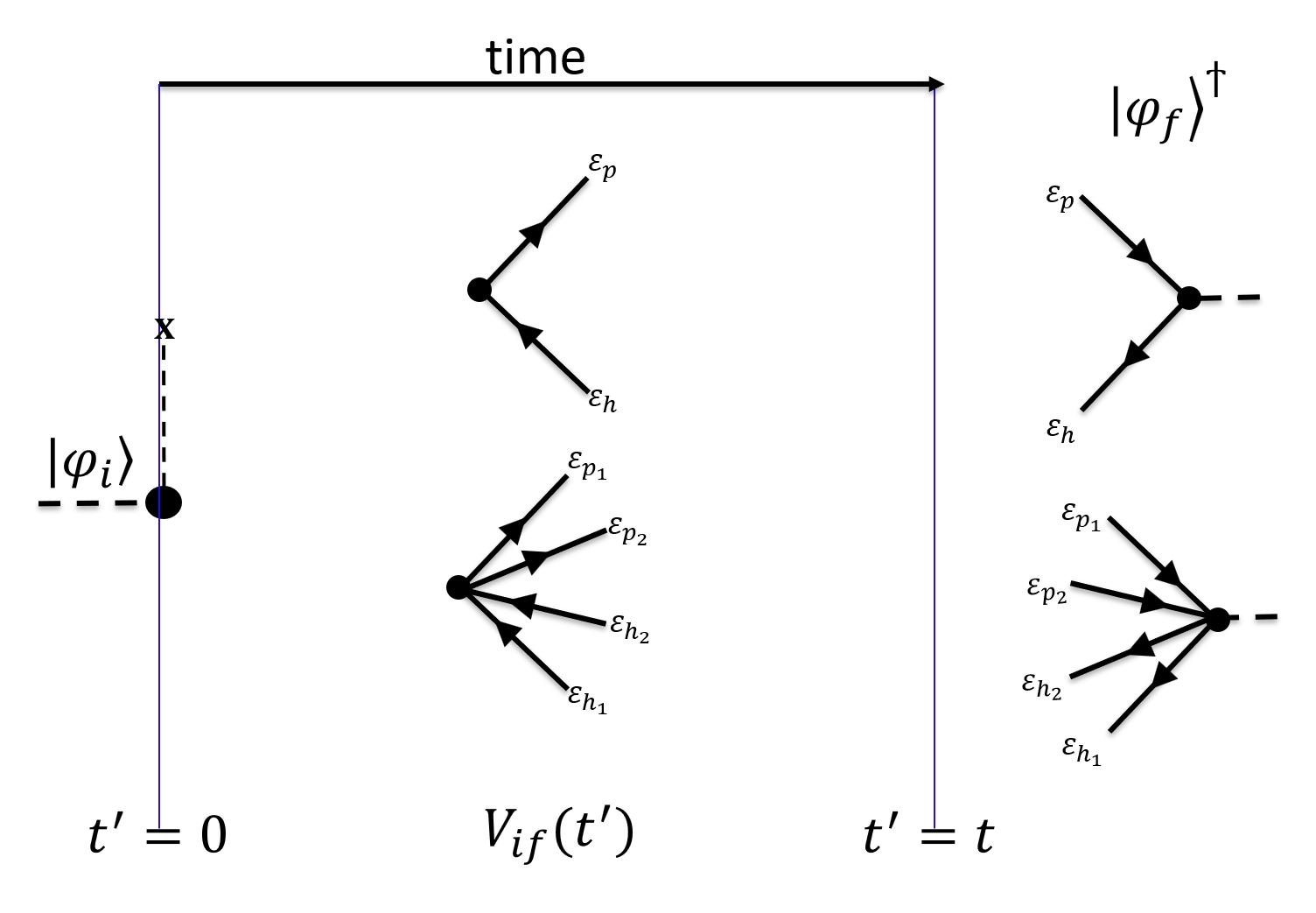}
    \vspace{-0.25cm}
    \caption{\footnotesize  Details of the construction of the diagrams shown in Fig. \ref{fig:Feynman_Dayson}. On the left-hand side of the diagram is the initial state, which interacts with the potential, represented by the X. The possible electronic configurations over time can be observed in each diagram by making a vertical cut and examining the electronic configuration expressed as particle-holes created from this initial state. \Sou}
    \label{fig:Feynman_Dayson_}
\end{figure}

\chapter{Estimating the Coefficients}\label{|C_0|}

Let us start from the Eq. \eqref{SEFM_}, considering $\frac{d\delta}{dt}$ constant, we can estimate the contribution of the single-particle-hole excitations by
\begin{equation}
\begin{aligned}
\tilde{c}_{p,h}(t) \approx   \frac{1}{\pi} \frac{d\delta}{dt}  \frac{1}{p + h}  \int_0^t dt' \tilde{c}_{0}(t') e^{+i(p +h)\frac{\Delta\varepsilon t'}{\hbar}}.
\end{aligned}
\end{equation}

Now, assuming $ c_0(t') \approx 1 $, we can solve the integral and obtain an approximation for the coefficient of a single-particle-hole excitation as follows:
\begin{equation}\label{1PH}
\begin{aligned}
\tilde{c}_{p,h}(t) \approx \frac{1}{\pi} \frac{d\delta}{dt}  \frac{e^{+i\frac{\Delta\varepsilon(p +h)  T}{2\hbar}} }{p + h} \frac{\sin\left( \frac{\Delta\varepsilon (p+h) T}{2\hbar} \right) }{\left( \frac{\Delta\varepsilon(p+h)}{2\hbar} \right)}.
\end{aligned}
\end{equation}

By the normalization, the probability to found the system in the ground state can be obtained by
\begin{equation}
\begin{aligned}
|\tilde{c}_{0}(T)|^2  \approx 1 - \sum_{p,h} |\tilde{c}_{p,h}(t) |^2 + ...,
\end{aligned}
\end{equation}
considering that the probability of finding the system in a state with two or more particle-hole pairs is very small. By using the Eq. \eqref{1PH}, we find that
\begin{equation}
\begin{aligned}
|\tilde{c}_{0}(T)|^2  \approx 1 - \left(\frac{1}{\pi} \frac{d\delta}{dt} \right)^2 \sum_{p,h}   \frac{1}{(p + h)^2} \frac{\sin^2\left( \frac{\Delta\varepsilon (p+h) T}{2\hbar} \right) }{\left( \frac{\Delta\varepsilon(p+h)}{2\hbar} \right)^2}  + ... ~.
\end{aligned}
\end{equation}

By approximating the sum $ \sum_{p>0 ,h \ge 1} $ with the integral $ \int_0^{N_e} \int_1^{N_e} dp \, dh $, once there are no poles within the summation range, we find, after some manipulations, that:
\begin{equation}
\begin{aligned}
|\tilde{c}_{0}(T)|^2  \approx 1 - \left(\frac{1}{\pi} \frac{d\delta}{dt} T \right)^2 \int_{\frac{\Delta\varepsilon T}{2\hbar}}^{\frac{DT}{2\hbar}} du \frac{\sin^2 u}{u^3} + ... ~.
\end{aligned}
\end{equation}

The function $ f(x) = \sin^2(x)/x^3 $ behaves as $ 1/x $ for $ x < 1 $ and as $ 1/x^3 $ for $ x > 1 $. Therefore, if $ \frac{D T}{2\hbar} > 1 $, we can approximately split the integral into two parts as follows:
\begin{equation}
\begin{aligned}
|\tilde{c}_{0}(t)|^2  \approx 1 - \left(\frac{1}{\pi} \frac{d\delta}{dt} T \right)^2 \left(\int_{\frac{\Delta\varepsilon T}{2\hbar}}^{1} du \frac{1}{u} + \int_{1}^{\frac{DT}{2\hbar}} du \frac{1}{u^3}  \right) + ... ~.
\end{aligned}
\end{equation}

Since \( \ln \left( \frac{1}{T} \right) > \frac{1}{T^2} \), we consider only the first interval in the integral, leading to the following result:
\begin{equation}
\begin{aligned}
|\tilde{c}_{0}(T)|^2  \approx 1 - \left(\frac{1}{\pi} \frac{d\delta}{dt} T \right)^2 \ln\left(\frac{2\hbar}{\Delta\varepsilon T}\right) + ...
\rightarrow \left(\frac{2\hbar}{\Delta\varepsilon T}\right)^{-\left(\frac{1}{\pi} \frac{d\delta}{dt} T \right)^2 } .
\end{aligned}
\end{equation}




\chapter{Quasi-Block-Diagonal Matrix}\label{QBDM Aproximation}

Here, our aim is to demonstrate that for certain Hamiltonians, it's feasible to decompose them into pieces and solve the eigenvalue problem for each piece without sacrificing accuracy significantly. We'll begin with a Hamiltonian consisting of two sectors:
\begin{eqnarray}
\hat H = \begin{pmatrix}
\hat H_1 & \mu \hat T \\
\mu \hat T^\dagger &\hat H_2 \\
\end{pmatrix}. 
\end{eqnarray}

Let us assume that we initially know the eigenvalues and eigenvectors of each sector, given by the eigenvalue equation:
\begin{eqnarray}
\hat H_1 \Vec{v}_1^{(0)} = \epsilon^{(0)}_1 \Vec{v}_1^{(0)} \mathrm{~and~} \hat H_2 \Vec{v}_2^{(0)} = \epsilon^{(0)}_2 \Vec{v}_2^{(0)}. 
\end{eqnarray}

The total eigenvalues equation for the Hamiltonian is:
\begin{eqnarray}
\hat H \Vec{v} = \epsilon \Vec{v} \rightarrow 
\begin{pmatrix}
\hat H_1 & \mu \hat T \\
\mu \hat T^\dagger & \hat H_2 \\
\end{pmatrix}
\begin{pmatrix}
\Vec{v}_1\\
\Vec{v}_2\\
\end{pmatrix}
= \epsilon
\begin{pmatrix}
\Vec{v}_1\\
\Vec{v}_2\\
\end{pmatrix}
. 
\end{eqnarray}

Let us focus on the first sector, we can easily show that $\Vec{v}_2 = - \mu (\hat H_2 - \epsilon \hat I )^{-1} \hat T^\dagger$. If $\mathrm{det}(\hat H_2 - \epsilon \hat I ) \neq 0$, then we find a new eigenvalue equation as follows:
\begin{eqnarray}\label{Eigen_Eq}
\left[ \hat H_1 - \mu^2 \hat T (\hat H_2 - \epsilon \hat I )^{-1} \hat T^\dagger \right] \Vec{v}_1 = \epsilon \Vec{v}_1 .
\end{eqnarray}
\begin{eqnarray}
\mathrm{det}\left[\hat H_1 - \epsilon \hat I - \mu^2 \hat T (\hat H_2 - \epsilon \hat I )^{-1} \hat T^\dagger\right] = 0.
\end{eqnarray}

Using a combination of Sylvester's determinant theorem and the well-known relation ${\mathrm{det}(I+\mu X)=1+\mu \mathrm{tr}(X) +\mathcal{O}\left(\mu^{2}\right)}$, the determinant can be approximated by:
\begin{eqnarray}\label{Determinat}
&\mathrm{det}\left[\hat H_1 - \epsilon \hat I - \mu^2 \hat T (\hat H_2 - \epsilon \hat I )^{-1} \hat T^\dagger\right] =
\nonumber \\ &\mathrm{det}\left[\hat H_1 - \epsilon \hat I \right]  - \mu^2 \mathrm{det}\left[\hat H_1 - \epsilon \hat I \right] . \mathrm{tr}\left[(\hat H_1 - \epsilon \hat I)^{-1} \hat T (\hat H_2 - \epsilon \hat I )^{-1} \hat T^\dagger \right] + \mathcal{O}(\mu^4).
\end{eqnarray}

Once we choose the basis of the Hamiltonian as $\{ \Vec{v}_1^{(0)} \} \bigcup \{ \Vec{v}_2^{(0)} \}$, then $H_1$ and $H_2$ are diagonal in this basis and we can write: 
\begin{eqnarray}\label{Diagonal}
\left[(\hat H_i - \epsilon \hat I) \right] = \sum_k (\epsilon_{i,k}^{(0)} - \epsilon )\ket{k}\bra{k},\\
\left[(\hat H_i - \epsilon \hat I) \right]^{-1} = \sum_k \frac{1}{(\epsilon_{i,k}^{(0)} - \epsilon )}\ket{k}\bra{k}, \\
\mathrm{det}\left[(\hat H_i - \epsilon \hat I) \right] = \prod_k (\epsilon_{i,k}^{(0)} - \epsilon ).
\end{eqnarray}

By the eigenvalue equation in Eq. \eqref{Eigen_Eq} using Eq. \eqref{Determinat} we can show that:
\begin{eqnarray}\label{EqII}
\mathrm{det}\left[\hat H_1 - \epsilon \hat I \right]\left(1 - \mu^2. \mathrm{tr}\left[(\hat H_1 - \epsilon \hat I)^{-1} \hat T (\hat H_2 - \epsilon \hat I )^{-1} \hat T^\dagger \right] \right) + \mathcal{O}(\mu^4) = 0.
\end{eqnarray}

Now, to solve this determinant, we need to compute this trace in Eq. \eqref{EqII}. Then, by defing
\begin{eqnarray}
\hat T  = \sum_{k,\bar k} \beta_{k,\bar k} \ket{k}\bra{\bar k}.
\end{eqnarray}
where $\ket{k} \in \{ \Vec{v}_1^{(0)} \} $ and $\ket{\bar k} \in \{ \Vec{v}_2^{(0)} \} $, then by using the Eq. \eqref{Diagonal} we can show that:
\begin{eqnarray}
\left[(\hat H_1 - \epsilon \hat I)^{-1} \hat T (\hat H_2 - \epsilon \hat I )^{-1} \hat T^\dagger \right] = \sum_{k,k'} \sum_{\bar k} \frac{\beta_{k',\bar k}\beta_{k,\bar k}^*}{ (\epsilon_{1,k'}^{(0)} - \epsilon )(\epsilon_{2,\bar k}^{(0)} - \epsilon )}  \ket{k'}\bra{k}.
\end{eqnarray}

Let us consider solutions with energy close to $\epsilon_{1,l}^{(0)}$. We can set $\epsilon = \epsilon_{1,k}^{(0)} -\Delta\varepsilon_l$. Moreover, considering that this energy can be degenerate with multiplicity $r_l$, then $\mathrm{det}\left[\hat H_1 - \epsilon \hat I \right] = (\Delta\varepsilon_l)^{r_l} \prod_{k \neq \{l\}} (\epsilon_{1,k}^{(0)} - \epsilon)$, where $k \neq \{l\}$ indicates that we are excluding all energy levels with energy $\epsilon_{1,k}^{(0)}$ from the product. Then, we can write the Eq.\eqref{EqII} as:
\begin{eqnarray}\label{EqIII}
(\Delta\varepsilon_l)^{r_l}\left(1 - \mu^2. \sum_{k, \bar k} \frac{|\beta_{k,\bar k}|^2}{ (\epsilon_{1,k}^{(0)} - \epsilon )(\epsilon_{2,\bar k}^{(0)} - \epsilon )}  \right) = + \mathcal{O}\left( \frac{\mu^4}{\prod_{k \neq \{l\}} (\epsilon_{1,k}^{(0)} - \epsilon )} \right).
\end{eqnarray}

Now we will carefully examine the sum to isolate the significant terms. By splitting the sum as $\sum_k = \sum_{\epsilon_{1,k}^{(0)} = \epsilon_{1,l}^{(0)}} + \sum_{\epsilon_{1,k}^{(0)} \neq \epsilon_{1,l}^{(0)}}$, we obtain:
\begin{eqnarray}\label{inside_sum}
\sum_{k, \bar k} \frac{|\beta_{k,\bar k}|^2}{ (\epsilon_{1,k}^{(0)} - \epsilon )(\epsilon_{2,\bar k}^{(0)} - \epsilon )}  \approx  \frac{r_l}{\Delta\varepsilon_l} \sum_{\bar k}  \frac{|\beta_{l,\bar k}|^2}{(\epsilon_{2,\bar k}^{(0)} - \epsilon_{1,l}^{(0)} )} + \sum_{k \neq \{ l\}, \bar k} \frac{|\beta_{k,\bar k}|^2}{(\epsilon_{1,k}^{(0)} - \epsilon_{1,l}^{(0)} )(\epsilon_{2,\bar k}^{(0)} - \epsilon_{1,l}^{(0)})}.
\end{eqnarray}

Substituting the Eq. \eqref{inside_sum} into Eq.\eqref{EqIII}, we find:
\begin{eqnarray}\label{EqIV}
(\Delta\varepsilon_l)^{r_l} \left( 1 - \mu^2  \sum_{k \neq \{ l\}, \bar k} \frac{|\beta_{k,\bar k}|^2}{(\epsilon_{1,k}^{(0)} - \epsilon_{1,l}^{(0)} )(\epsilon_{2,\bar k}^{(0)} - \epsilon_{1,l}^{(0)})} \right) - r_l\mu^2.(\Delta\varepsilon_l)^{r_l-1} \sum_{\bar k}  \frac{|\beta_{l,\bar k}|^2}{(\epsilon_{2,\bar k}^{(0)} - \epsilon_{1,l}^{(0)} )} \nonumber
\end{eqnarray}
\begin{eqnarray}
= \mathcal{O}\left( \frac{\mu^4}{\prod_{k \neq \{l\}} (\epsilon_{1,k}^{(0)} - \epsilon_{1,l}^{(0)} )} \right).
\end{eqnarray}

Considering that the distribution of energies in the interval $E_a < \epsilon_{1,l}^{(0)} < E_b$ is approximately uniform, with $\mathrm{max}\left(|\epsilon_{1,k}^{(0)} - E_a|,|\epsilon_{1,k}^{(0)} - E_b|\right) \gg \Delta\varepsilon$ and $|\epsilon_{1,k}^{(0)} - \epsilon_{1,l}^{(0)}| > \Delta\varepsilon$, we can demonstrate that:
\begin{eqnarray}
  P &=& {\prod_{k \neq \{l\}} (\epsilon_{1,k}^{(0)} - \epsilon_{1,l}^{(0)})} = \pm {\prod_{k \neq \{l\}} |\epsilon_{1,k}^{(0)} - \epsilon_{1,l}^{(0)}|}.\nonumber \\
 \log |P| &=& {\sum_{k \neq \{l\}} \log|\epsilon_{1,k}^{(0)} - \epsilon_{1,l}^{(0)}| }  \nonumber \\ 
 &\approx&\frac{1}{\Delta\varepsilon} \int_{E_a}^{\epsilon_{1,l}^{(0)} - \Delta\varepsilon} d\epsilon \log(\epsilon_{1,l}^{(0)} - \epsilon) + \frac{1}{\Delta\varepsilon} \int_{\epsilon_{1,l}^{(0)} + \Delta\varepsilon}^{E_b} d\epsilon \log(\epsilon - \epsilon_{1,l}^{(0)}) \nonumber \\
&\approx&\frac{(\epsilon_{1,l}^{(0)} - E_a)\log(\epsilon_{1,l}^{(0)} - E_a) + (E_b - \epsilon_{1,l}^{(0)})\log(E_b - \epsilon_{1,l}^{(0)})}{\Delta\varepsilon} > 1.
\end{eqnarray}

Using the Eq. \eqref{EqIV}, after shown that the product ${\prod_{k \neq \{l\}} |\epsilon_{1,k}^{(0)} - \epsilon_{1,l}^{(0)}|} > 1$, we find that: 
\begin{eqnarray}
(\Delta\varepsilon_l)^{r_l-1} \left(\Delta\varepsilon_l - r_l\mu^2.\sum_{\bar k}  \frac{|\beta_{l,\bar k}|^2}{(\epsilon_{2,\bar k}^{(0)} - \epsilon_{1,l}^{(0)} )} \right) = \mathcal{O}(\mu^4). 
\end{eqnarray}

Finally, we can approximate the error in the energy by considering $\epsilon \approx \epsilon_{1,\bar k}^{(0)}$ by: 
\begin{eqnarray}\label{error_energy}
\Delta\varepsilon_l \sim r_l\mu^2.\sum_{\bar k}  \frac{|\beta_{l,\bar k}|^2}{(\epsilon_{2,\bar k}^{(0)} - \epsilon_{1,l}^{(0)} )}.
\end{eqnarray}
Clearly, this approximation is better when $\mu$ is very small or the two energy sector are separated by high energy differences. 

Now to find the correction for the eigenvector, we can start by considering $\Vec{v}_{l} = \Vec{v}_{l}^{(0)} + \delta \Vec{v}_{l}$ in Eq. \eqref{Eigen_Eq} as: 
\begin{eqnarray}
\left[ \hat H_1 - \mu^2 \hat T (\hat H_2 - \epsilon \hat I )^{-1} \hat T^\dagger \right] (\Vec{v}_l^{(0)} + \delta \Vec{v}_{l}) = (\epsilon_{1,\bar l}^{(0)}  - \Delta\varepsilon_l ) (\Vec{v}_{l}^{(0)} + \delta \Vec{v}_{l}) .
\end{eqnarray}

The following algebraic steps are straightforward, so we will present them without additional explanation.
\begin{eqnarray}
\left[-\mu^2 \hat T (\hat H_2 - \epsilon \hat I )^{-1} \hat T^\dagger + \Delta\varepsilon_l \right] \Vec{v}_l^{(0)} = 
\left[- \hat H_1 + \epsilon_{1,\bar l}^{(0)} +\mu^2 \hat T (\hat H_2 - \epsilon \hat I )^{-1} \hat T^\dagger - \Delta\varepsilon_l \right]\delta\Vec{v}_l .
\end{eqnarray}

\begin{eqnarray}
\delta \Vec{v}_l =  \sum_{k} u_k \Vec{v}_k^{(0)} .
\end{eqnarray}

\begin{eqnarray}
\left[-\mu^2 \hat T (\hat H_2 - \epsilon \hat I )^{-1} \hat T^\dagger + \Delta\varepsilon_l \right] \Vec{v}_l^{(0)} = \sum_k  u_k \left[-\epsilon_{1,\bar k}^{(0)}  + \epsilon_{1,\bar l}^{(0)} +\mu^2 \hat T (\hat H_2 - \epsilon \hat I )^{-1} \hat T^\dagger - \Delta\varepsilon_l \right]\Vec{v}_k^{(0)} .\nonumber
\end{eqnarray}

\begin{eqnarray}
\sum_{k \neq l}  u_k \left[-\epsilon_{1,\bar k}^{(0)}  + \epsilon_{1,\bar l}^{(0)} +\mu^2 \hat T (\hat H_2 - \epsilon \hat I )^{-1} \hat T^\dagger - \Delta\varepsilon_l \right]\Vec{v}_k^{(0)} = 0.
\end{eqnarray}

\begin{eqnarray}
\sum_{k \neq l}  u_k \left[\epsilon_{1,\bar k}^{(0)}  - \epsilon_{1,\bar l}^{(0)} +\Delta\varepsilon_l\right]\Vec{v}_k^{(0)} = \mu^2  \sum_{k \neq l} u_k \left[\hat T (\hat H_2 + (\epsilon_{1,\bar l}^{(0)} -\Delta\varepsilon_l) \hat I )^{-1} \hat T^\dagger \right]\Vec{v}_k^{(0)}.
\end{eqnarray}

The above expression represents a system of linear equations, which, for an eigenstate with non-degenerate energy, once $\left|\epsilon_{1,\bar k}^{(0)}  - \epsilon_{1,\bar l}^{(0)}\right| > \Delta\varepsilon$ clearly $u_{k \neq l} \sim \mathcal{O}({\mu^2/\Delta\varepsilon})$.  Once $|\Vec{v}_l^{(0)} + \delta\Vec{v}_l|=1$, then $(1-u_l)^2 + \sum_{k \neq l} u_l^2 = 1$ $\rightarrow (u_l) \sim \mathcal{O}({\mu^2}/\Delta\varepsilon)$. But if the energy $\epsilon_{1,\bar l}^{(0)}$ is degenerated with $r_l$ level with the same energy, then, in this situation, we need a little bit more manipulations, starting with: 
\begin{eqnarray}
\sum_{k \neq \{l\}}  u_k \left[\epsilon_{1,\bar k}^{(0)}  - \epsilon_{1,\bar l}^{(0)} +\Delta\varepsilon_l\right]\Vec{v}_k^{(0)} + \Delta\varepsilon_l \sum_{j=1}^{r_l-1}  u_{l,j}  \Vec{v}_{l,j}^{(0)} = \nonumber \\ \mu^2  \sum_{k \neq l} u_k \left[\hat T (\hat H_2 + (\epsilon_{1,\bar l}^{(0)} -\Delta\varepsilon_l) \hat I )^{-1} \hat T^\dagger \right]\Vec{v}_k^{(0)}.
\end{eqnarray}

\begin{eqnarray}
\left[\hat T (\hat H_2 - \epsilon \hat I )^{-1} \hat T^\dagger \right] = \sum_{k,k'} \sum_{\bar k} \frac{\beta_{k',\bar k}\beta_{k,\bar k}^*}{(\epsilon_{2,\bar k}^{(0)} - \epsilon )}  \ket{k'}\bra{k}.
\end{eqnarray}

\begin{eqnarray}
\sum_{k \neq \{l\}}  u_k \left[\epsilon_{1,\bar k}^{(0)}  - \epsilon_{1,\bar l}^{(0)} +\Delta\varepsilon_l\right]\Vec{v}_k^{(0)} + \Delta\varepsilon_l \sum_{j=1}^{r_l-1}  u_{l,j}  \Vec{v}_{l,j}^{(0)} = \nonumber \\ \mu^2  \sum_{k'} \left( \sum_{\bar k, k \neq l} u_k \frac{\beta_{k',\bar k}\beta_{k,\bar k}^*}{(\epsilon_{2,\bar k}^{(0)} - \epsilon )} \right) \Vec{v}_{k'}^{(0)}.
\end{eqnarray}

\begin{eqnarray}
{k \neq \{l\}} :   u_k \left[\epsilon_{1,\bar k}^{(0)}  - \epsilon \right]  =  \mu^2 \left( \sum_{\bar k', k \neq l} u_{k'} \frac{\beta_{k,\bar k}\beta_{k',\bar k}^*}{(\epsilon_{2,\bar k}^{(0)} - \epsilon )} \right).
\end{eqnarray}

Once again, we can observe that the above expression represents a system of linear equations. When $\left|\epsilon_{1, k}^{(0)} - \epsilon_{1, l}^{(0)}\right| > \Delta\varepsilon $, it becomes clear that $u_{k \neq \{l\}} \sim \mathcal{O}({\mu^2}/\Delta\varepsilon)$. Since $|\Vec{v}_l|=1$, we have $(1-u_l)^2 + \sum_{j=1}^{r_l-1} u_{j,l}^2 + \sum_{k \neq {l}} u_l^2 = 1$. This implies $-2(u_l) + \sum_{j=1}^{r_l} u_{j,l}^2 = -\mathcal{O}({\mu^4}/\Delta\varepsilon^2)$. The equations $\sum_{j=1}^{r-1} x_{j}^2 + (1-x_r)^2 = (1-\mathcal{O}({\mu^4/\Delta\varepsilon^2}) ) $ represent a sphere with dimension $r$ and radius $\sqrt{1-\mathcal{O}({\mu^4}/\Delta\varepsilon^2)}$, indicating that $\Vec{v}_l = \sum_j {u_{j,l}} \Vec{v}_{j,l}^{(0)} + \mathcal{O}({\mu^2}/\Delta\varepsilon)$, which is a linear combination of the degenerated sector. In such situations, we can set $\Vec{v}_l = \Vec{v}_{l}^{(0)}$, completing the proof that indeed $\Vec{v}_l = (1-\mathcal{O}({\mu^2/\Delta\varepsilon}) ) \Vec{v}_{l}^{(0)} + \sum_{k\neq l} \mathcal{O}({\mu^2/\Delta\varepsilon}) \Vec{v}_{l}^{(0)} $.

\chapter{NRG results x Direct Diagonalization}\label{NRGxDirectDiagonalization}

We want to calculate and verify the results of the projections of an NRG solution basis for a certain $W_1$ onto another NRG solution basis for a certain $W_2$. Both problems we want to address, whether the electronic states of the adsorption coefficient problem or x-ray absorption, have a similar Hamiltonian, when $U=0$, in the form
\begin{align} \label{TP:1}
 H_N = \e_d \cd{d}\cn{d}+\sum_{n=0}^{N-1}\tau_n(\fd{n}\fn{n+1}+\hc) +\sqrt{2}V \left(\cd{d}\fn{0}+\hc \right)+2W_i\fd{0}\fn{0},
\end{align}
where $i=1,2$. This Hamiltonian can be diagonalized in the basis of operators of a single particle $c_d$, $f_0$, $f_1$, ..., $f_N$, generating a set of $N+2$ single-particle operators $\{ g_l \}$ with energies from $\epsilon_1$ to $\epsilon_{N+2}$ (depicted in Fig.~\ref{E_levels}(a)). The ground state of many-body systems is the combination of two electrons in each of the $\frac{N}{2}+1$ lowest energy levels. The construction of excited states is done by removing one or more electrons from the ground state, from lower energy levels (creating holes), and moving them to higher unoccupied levels (excited states).

\begin{figure}[ht!]
		\centering
		\includegraphics[scale=0.6]{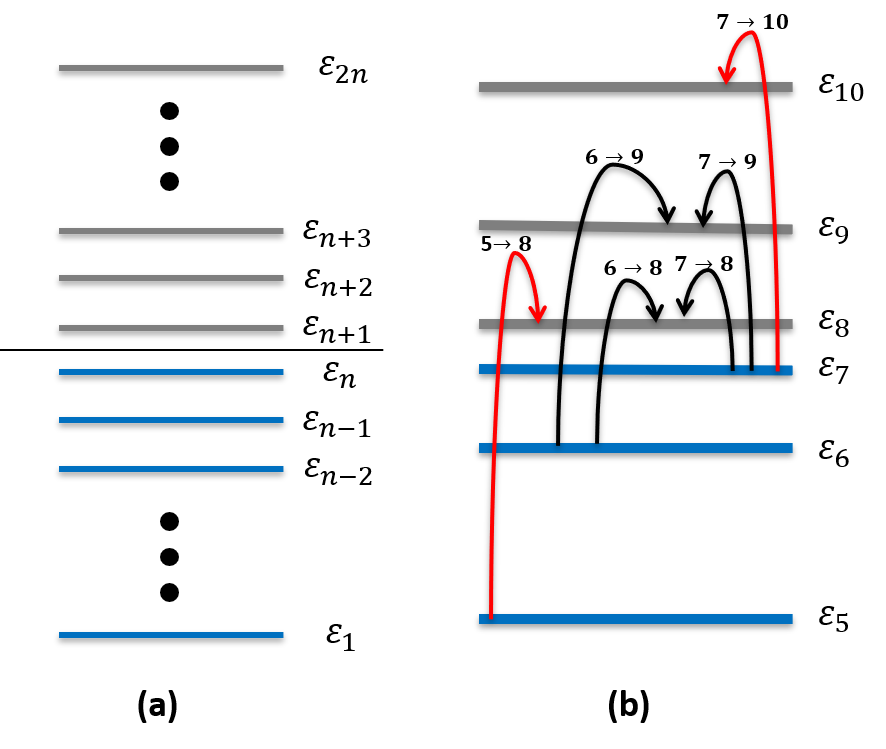}
		\caption{\footnotesize {(a) Energy levels of independent particle solutions of the Hamiltonian \eqref{TP:1}, where $n = \left(\frac{N}{2}+1\right)$, for an even $N$. (b) Construction of the first excited states for $n=7$, the transitions $7 \rightarrow 10$ and $5 \rightarrow 8$, shown in red, have higher energies compared to the others. \Sou}}
		\label{E_levels}
\end{figure}

Using the values $\varepsilon_d = -5$, $U=0$, $V=1.25$, $W_1=1.00$, and $W_2=0.25$, normalized by the bandwidth $D$, and the discretization parameter $\Lambda = 6$ with a maximum iteration $N=12$, we can exactly diagonalize the Hamiltonian \eqref{TP:1} for each of these values of $W_i$ ($i=1,2$), once the non-interacting Hamiltonian is quadratic, and find the 14 eigenenergies $\{\epsilon_l^{(i)}\}$ and the 14 eigenstates $\{g_l^{(i)}\}$ of a single particle, defining the two distinct bases (1) and (2). This calculation was performed by numerically diagonalizing the 14x14 matrix of the Hamiltonian in the basis of operators $c_d$, $f_0$, $f_1$, ..., $f_{12}$, and the results will be referred to as 'exact', as they involve no truncations.

The excited many-body states $\{ \lvert e_n \rangle \}$ with charge 0 (total of 14 electrons) and spin 0, for the bases (1) and (2) defined by the values of $W_{1,2}$, are obtained by creating electrons in the excited states $g^\dagger_{l>7}$ and annihilating electrons (creating holes) $g_{l \le 7}$ from the ground state $ \lvert e_0 \rangle = \prod_{l=1}^{7} g^\dagger_{l\uparrow}g^\dagger_{l\downarrow} \lvert 0 \rangle$ (composed of the double occupation of the 7 lowest-energy states $g_l$). To compare the values obtained by NRG and the exact values, we construct the first ten many-body states and calculate the exact energies of these states and projections between some of them for both bases (1) and (2).

For NRG calculations, we use an energy truncation at a maximum value $E_c$ scaled by $D_N$. Using the NRG diagonalization procedure, we solve the same Hamiltonian with the same values of constants and parameters for $W_1$ and $W_2$, obtaining a spectrum of many-body energies with charge 0 and spin 0, and the projections for both bases (1) and (2). To observe if the NRG numerical procedure is working correctly, we can compare these obtained energy values and projections with the exact values obtained by direct diagonalization of the Hamiltonian in the basis of single-particle operators.

The results are obtained for the ground state energies in Table \ref{Estado Fundamental}, for the energies of the first excited states of base (1) in Table \ref{Excitados1}, and for the energies of the first excited states of base (2) in Table \ref{Excitados2} sequentially.

\begin{table}[h!]
\caption{{Comparing the ground state energies $E_0$, calculated exactly and obtained by NRG, normalized by the bandwidth, for both values of $W_{1}$ and $W_2$.}}
\label{Estado Fundamental}
\begin{center}
\begin{tabular}{| c | c | c | c | c |}
\hline
 $i$ & $E_0$ exact & NRG ($E_c = 20$) & NRG ($E_c = 30$) & NRG ($E_c = 100$) \\
\hline
$1$ & $-11.43865719$ & $-11.43865719$ & -11.43865719 & -11.43865719 \\
$2$ & $-11.76464132$ & $-11.76464130$ & -11.76464132 & -11.76464132 \\
\hline
\end{tabular}
\end{center}
\begin{flushleft}
		Source: By the author. \
\end{flushleft}
\end{table}
The results of the comparison between the projections of the two bases (1) and (2), solutions of the Hamiltonian \eqref{TP:1} with $W_{1~\text{or}~2}$, exact (calculated via diagonalization of the Hamiltonian in the basis of a single particle) and calculated via NRG (using cutoff energies $E_c = 20,30\text{and}~100$) are shown in Table \ref{Proj.}. The projections calculated by NRG with a cutoff energy of 100 are not shown in this table, as they have the same value as the exact result with an accuracy of more than eight decimal places.

We can conclude firstly that the NRG is obtaining results for the energies of many-body states and projections between these bases close to the exact values, for both cutoff energies and values of $W_{1,2}$. Furthermore, the results improve as the cutoff energy increases, until, for $E_c \geq 100$, the values of energies and projections of the first excited states are practically the same as the exact ones. The time required for the NRG calculation also increases with the cutoff energy, taking only a few seconds for $E_c = 20$ and a few minutes for $E_c = 100$. This is because, the higher the $E_c$, the more states are being considered, reducing the error associated with truncation and increasing computational cost.

\begin{table}[t!]
\ABNTEXfontereduzida
\caption{Energies of the first ten excited states, scaled by the term $D_{12}$, calculated exactly and obtained by NRG for different cutoff energy values, using the base (1) defined by the value of $W_1$.}
\begin{center}
\begin{tabular}{| c |c | c | c | c | c |}
\hline
States (1) & $n$ & $E^{(1)}$ exact & $\Mrm{NRG}_{E_c=20}$  & $\Mrm{NRG}_{E_c=30}$  & $\Mrm{NRG}_{E_c=100}$ \\
\hline
$ g_{8\upA}^{\dagger}g_{7\upA} \lvert e_0 \rangle_1$ & 1 & $1.84433637$  & $1.84433636$ & 1.84433637 & $1.84433637$ \\
$ g_{8\upA}^{\dagger}g_{8\downA}^{\dagger}g_{7\upA}g_{7\downA} \lvert e_0 \rangle_1$ & 2 & $3.68867274$ & $3.68867271$ & 3.68867274 & $3.68867274$\\
$ g_{9\upA}^{\dagger}g_{7\upA} \lvert e_0 \rangle_1$ &3 & $6.14384895$ & $6.14384914$ & 6.14384896 & $6.14384895$\\
$ c_{8\upA}^{\dagger}c_{6\upA} \lvert e_0 \rangle_1$ &4 & $7.89996203$ & $7.89996400$ & 7.89996202 & $7.89996203$\\
$ g_{9\upA}^{\dagger}g_{8\downA}^{\dagger}g_{7\upA}g_{7\downA} \lvert e_0 \rangle_1$ &5 & $7.98818533$ & $7.98818580$ & 7.98818533 & $7.98818533$ \\
$ g_{8\upA}^{\dagger}g_{8\downA}^{\dagger}g_{7\upA}g_{6\downA} \lvert e_0 \rangle_1$ &6 & $9.74429840$ & $9.74430066$ & 9.74429840 & $9.74429840$ \\
$ g_{9\upA}^{\dagger}g_{6\upA} \lvert e_0 \rangle_1$ &7 & $12.19947461$ & $12.19947555$ & 12.19947461 & $12.19947461$ \\
$ g_{9\upA}^{\dagger}g_{9\downA}^{\dagger}g_{7\upA}g_{7\downA} \lvert e_0 \rangle_1$ &8& $12.28769791$ & $12.28769734$ & 12.28769791 & $12.28769791$ \\
$ g_{9\upA}^{\dagger}g_{8\upA}^{\dagger}g_{7\upA}g_{6\upA} \lvert e_0 \rangle_1$ &9& $14.04381098$ & $14.04381219$ & 14.04381098 & $14.04381098$\\
$ g_{9\upA}^{\dagger}g_{8 \downA}^{\dagger}g_{7\upA}g_{6\downA} \lvert e_0 \rangle_1$ &10& $14.04381098$ & $14.04381219$ & 14.04381098 & $14.04381098$\\
\hline
\end{tabular}
\end{center}
\label{Excitados1}
\begin{flushleft}
		Source: By the author. \
\end{flushleft}
\end{table}
\begin{table}[t!]
\ABNTEXfontereduzida
\caption{Energies of the first ten excited states, scaled by the term $D_{12}$, calculated exactly and obtained by NRG for different cutoff energy values, using the base (2) defined by the value of $W_2$.}
\begin{center}
\begin{tabular}{| c |c | c | c | c | c |}
\hline
States (2) & $n$ & $E^{(1)}$ exact & $\Mrm{NRG}_{E_c=20}$  & $\Mrm{NRG}_{E_c=30}$  & $\Mrm{NRG}_{E_c=100}$ \\
\hline
$ g_{8\upA}^{\dagger}g_{7\upA} \lvert e_0 \rangle_2$ & 1 & $1.92014286$  & $1.92014278$ & 1.92014280  & $1.92014286$\\
$ g_{8\upA}^{\dagger}g_{8\downA}^{\dagger}g_{7\upA}g_{7\downA} \lvert e_0 \rangle_2$ & 2 & $3.84028573$ & $3.84028556$ & 3.84028560 & $3.84028573$\\
$ g_{9\upA}^{\dagger}g_{7\upA} \lvert e_0 \rangle_2$ &3 & $5.54560884$ & $5.54561078$ & 5.54560916 & $5.54560884$\\
$ g_{9\upA}^{\dagger}g_{8\downA}^{\dagger}g_{7\upA}g_{7\downA} \lvert e_0 \rangle_2$ &4 & $7.46575170$ & $7.46575577$ & 7.46575195 & $7.46575170$ \\
$ c_{8\upA}^{\dagger}c_{6\upA} \lvert e_0 \rangle_2$ &5 & $9.22538836$ & $9.22539330$ & 9.22538742 & $9.22538836$\\
$ g_{8\upA}^{\dagger}g_{8\downA}^{\dagger}g_{7\upA}g_{6\downA} \lvert e_0 \rangle_2$ &6 & $11.09121767$ & $11.09121894$ &11.09121831 & $11.09121767$\\
$ g_{9\upA}^{\dagger}g_{9\downA}^{\dagger}g_{7\upA}g_{7\downA} \lvert e_0 \rangle_2$ & 7& $11.14553122$ & $11.14553844$ & 11.14553022 & $11.14553122$ \\
$ g_{9\upA}^{\dagger}g_{6\upA} \lvert e_0 \rangle_2$ &8 & $12.85085433$ & $12.85085642$ & 12.85085379 & $12.85085433$  \\
$ g_{9\upA}^{\dagger}g_{8\upA }^{\dagger}g_{7\upA}g_{6\upA } \lvert e_0 \rangle_2$ & 9& $14.77099719$ & $14.77100134$ & 14.77099658 & $14.77099719$\\
$ g_{9\upA}^{\dagger}g_{8 \downA}^{\dagger}g_{7\upA}g_{6\downA} \lvert e_0 \rangle_2$ & 10 & $14.77099719$ & $14.77100134$ & 14.77099658 & $14.77099719$\\
\hline
\end{tabular}
\end{center}
\label{Excitados2}
\begin{flushleft}
		Source: By the author. \
\end{flushleft}
\end{table}

\begin{table}[t!]
\ABNTEXfontereduzida
\caption{Projections between the two bases (1) and (2) solutions of the Hamiltonian \eqref{TP:1} for different values of $W_1$ and $W_2$, exact and calculated via NRG, using cutoff energies $E_c = 20,30\text{and}~100$. The projections calculated by NRG with a cutoff energy of 100 are the same as the exact ones with an accuracy of eight decimal places.}
\label{Proj.}
\begin{center}
\begin{tabular}{| c |c | c | c |}
\hline
$\Mrm{~}_1\langle{n'}\lvert n\rangle_2$ & Proj. exact & $\Mrm{NRG}_{E_c=20}$ & $\Mrm{NRG}_{E_c=30}$\\
\hline
$0|0$ & $0.86635934$ & $0.86636382$ & $0.86636219$ \\
$0|1$ & $-0.15740949$& $-0.15741195$& $-0.15740990$\\
$0|2$ & $0.01429992$ & $0.01430070$ & $0.01429995$ \\
$0|3$ & $0.09225636$ & $0.09225876$ & $0.09225661$ \\
$0|4$ & $-0.01185261$& $-0.01185567$& $-0.01185263$\\
$0|5$ & $0.09543813$ & $0.09547710$ & $0.09543840$ \\
$0|6$ & $0.00491207$ & $0.00491244$ & $0.00491208$ \\
$0|8$ & $-0.11597268$& $-0.11596814$& $-0.11597304$\\
$1|3$ & $0.10392632$ & $0.10392808$ &  $0.10392660$\\
$1|8$ & $-0.02985961$& $-0.02985835$& $-0.02985973$\\
$7|8$ & $0.85117398$ & $0.85118840$ & $0.85117683$ \\
$2|2$ & $0.85798359$ & $0.85798733$ & $0.85798642$ \\
$8|2$ & $0.00926520$ & $0.00925770$ & $0.00926521$ \\
\hline
\end{tabular}

\begin{tabular}{| c |c | c | c | }
\hline
$\Mrm{~}_1\langle{n'}\lvert n\rangle_2$ & Proj. exact & $\Mrm{NRG}_{E_c=20}$ & $\Mrm{NRG}_{E_c=30}$\\
\hline
$1|0$ &  $0.15314866$ & $0.15315073$ &$0.15314906$\\
$2|0$ &  $0.01353625$ & $0.01353665$ &$0.01353627$\\
$3|0$ &  $-0.11706927$& $-0.11707054$&$-0.11706927$\\
$4|0$ &  $-0.06699971$& $-0.06702027$&$-0.06699989$\\
$5|0$ &  $-0.01463333$& $-0.01463386$&$-0.01463334$\\
$7|0$ &  $0.11048082$ &  $0.11047978$  &$0.11048115$\\
$8|0$ &  $0.00790966$ &  $0.00790809$  &$0.00790968$\\
$1|1$ &  $0.84824845$ &  $0.84825180$  &$0.84825125$\\
$1|5$ &  $-0.07670606$& $-0.07668241$ &$-0.07670628$\\
$3|3$ &  $0.85910233$ &  $0.85910353$  &$0.85910516$\\
$4|5$ &  $0.86305893$ &  $0.86306046$  &$0.86306177$\\
$2|6$ &  $0.01058718$ &  $0.01058901$ &$0.01058719$\\
$8|6$ &  $0.86431293$ &  $0.86432436$ &$0.86431574$\\
\hline
\end{tabular}
\end{center}
\begin{flushleft}
		Source: By the author. \
\end{flushleft}
\end{table}

\chapter{Transition: neutral-ionized sectors}\label{Wif-Neutral-Ionized}

Prior to continuing, let us ignore the nuclear distribution a little bit and discuss only about the electronic part of the Hamiltonian in Eq. \eqref{H_Colision}, which can be expressed as
\begin{align}\label{H_Colision_ele}
H_e(z) = H_d +\sum_{k}\varepsilon_{k}\tilde{a}^\dagger_{k}\tilde{a}_{k} + V(z)\left(\cd{d}\fn{0}+\hc \right)+ W(z)(n_{d}-1)^2\fd{0}\fn{0},
\end{align}
where $H_d = \varepsilon_d d^\dagger d + U n_{d\upA} n_{d,\downA}$ represents the atomic levels. 

It is straightforward to show that, in situations where the atom is far from the surface and the hybridization is very small ($V(z) \sim 0$), we can express the Hamiltonian in Eq. \eqref{H_Colision_ele} in terms of three independent sectors depending on the occupation of the atom as
\begin{align}\label{Setores_He(z)}
(n_d = 0):~
&\sum_{k}\varepsilon_{k}\tilde{a}^\dagger_{k}\tilde{a}_{k} + \frac{W(z)}{N}\sum_{k,k'}\tilde{a}^\dagger_{k}\tilde{a}_{k'}, \nonumber \\
(n_d = 1):~
& \varepsilon_d +\sum_{k}\varepsilon_{k}\tilde{a}^\dagger_{k}\tilde{a}_{k}, \\
(n_d = 2):~
& 2\varepsilon_d + U +\sum_{k}\varepsilon_{k}\tilde{a}^\dagger_{k}\tilde{a}_{k} + \frac{W(z)}{N}\sum_{k,k'}\tilde{a}^\dagger_{k}\tilde{a}_{k'} \nonumber. 
\end{align}
Here we note that the ionized sector ($n_d = 2/0$) has exactly the same Hamiltonian as the final Hamiltonian in the photoemission problem in Eq. \eqref{H_photo}, and the neutral sector ($n_d = 1$), which represents the initial electronic configuration, has the same metallic electronic configuration as the initial ground state in the photoemission problem. Therefore, the photoemission problem not only can qualitatively mimic the collision process, as discussed in the introduction, but also shares a similar Hamiltonian.

Now, let us consider initially that $ V(z) = 0 $. In this situation, for the neutral sector, the levels of the band remain unperturbed and the many-body levels are the same. However, for the ionized sector, the presence of the scattering potential changes the levels proportionally to the phase shift $\epsilon_l(z) = \varepsilon_l - \frac{\delta_z}{\pi} \Delta\varepsilon$ as in the photoemission problem. Note that the phase shift here $\delta_z = \mathrm{atan}\left(- \pi \rho W_z \right)$ (using Eq. \eqref{phase_shift}) depends on the $W_z$, and for this reason it depends on the atom position $z$, and it grows as the atom approach the surface.

However, as the atom approaches the surface, the hybridization slowly increases. When it is still small, we can use the Fermi golden rule \cite{Sakurai_Napolitano_2020}. Then, we aim to roughly estimate the main contributions to the transitions from the initial neutral electronic configuration $\ket{\Psi_i} \equiv \ket{n_d = 1} \ket{\phi_0}$ of the atom to a final ionized configuration $\ket{\Psi_f} \equiv \ket{n_d = 0/2} \ket{\varphi_f}$, where $\ket{\phi_0}$ represents the initial ground state of the metal and $\ket{\varphi_f}$ represents the final configuration of the electrons in the band in the presence of the scattering potential.

When the hybridization stills small, we can use the Fermi golden rule 
\begin{eqnarray}\label{A-I}
W_{if}^{1 \rightarrow 0/2} =4|\langle H_{\mathrm{hyb}} (z) \rangle|_{i,f}^2  \frac{\sin^2\left( \frac{\Delta E_{if}}{2\hbar} t \right)}{\Delta E_{if}^2}
\end{eqnarray}
to estimate the transition probability $W_{if}$ from the initial neutral electronic configuration $\ket{\Psi_i}$ to a specific final ionized electronic configuration $\ket{\Psi_f}$. The $\Delta E_{if}$ is the energy difference between the initial and final states.

Using the hybridization term defined in Eq \eqref{H_Colision_ele}, the transition probability estimated by the Eq. \eqref{A-I} for the transition from $n_d = 1$ to $n_d = 2$ sectors we become
\begin{eqnarray}
W_{if}^{1 \rightarrow 2} = \frac{4V^2(z)}{N} \frac{\sin^2\left( \frac{\Delta E_{if} (z) }{2\hbar} t \right)}{\Delta E_{if}^2 (z)} \left|\bra{\phi_0}\bra{n_d =1 }\sum_k \tilde a_k^\dagger d \ket{n_d = 2}  \ket{\varphi_f(z) } \right|^2,
\end{eqnarray}
and for the  transition from $n_d = 1$ to $n_d = 0$ sectors
\begin{eqnarray}
W_{if}^{1 \rightarrow 0} = \frac{4V^2(z)}{N} \frac{\sin^2\left( \frac{\Delta E_{if} (z) }{2\hbar} t \right)}{\Delta E_{if}^2 (z)} \left|\bra{\phi_0}\bra{n_d =1 }\sum_k \tilde a_k d^\dagger \ket{n_d = 0}  \ket{\varphi_f(z) } \right|^2.
\end{eqnarray}

It is straightforwards to show that
\begin{eqnarray}
\left|\bra{\phi_0}\bra{n_d =1 }\sum_k \tilde a_k^\dagger d\ket{n_d = 2}  \ket{\varphi_f(z) } \right|^2 = \left|\sum_k \bra{\phi_0} \tilde a_k^\dagger \ket{\varphi_f(z) } \right|^2 \sim  N_e^{-2\left(\frac{\delta_z}{\pi}\right)^2},
\end{eqnarray}
\begin{eqnarray}
\left|\bra{\phi_0}\bra{n_d =1 }\sum_k \tilde a_k d^\dagger \ket{n_d = 0}  \ket{\varphi_f(z)} \right|^2 = \left|\sum_k \bra{\phi_0} \tilde a_k \ket{\varphi_f(z)} \right|^2 \sim  N_e^{-2\left(\frac{\delta_z}{\pi}\right)^2}.
\end{eqnarray}
Here we consider the quantities $\left|\bra{\phi_0}\tilde{a}_k \ket{\varphi_f(z)} \right|$ and $\left|\bra{\phi_0}\tilde{a}_k^\dagger \ket{\varphi_f(z)} \right|^2$ as a sum over many-body projections between the initial unperturbed many-body ground state and some final many-body state configuration in the presence of the scattering potential. Due to the presence of the Anderson orthogonality catastrophe in these projections, as already discussed for the photoemission problem, these quantities must be proportional to $N_e^{-2\left(\frac{\delta_z}{\pi}\right)^2}$.

Taking the last discussion account, to simplified the expression for the transition probability, we can defined the two new quantities
\begin{eqnarray}
 \mathcal{G}_{i,f}^{1 \rightarrow 2} (z) = \frac{ \left|\sum_k \bra{\phi_0} \tilde a_k^\dagger \ket{\varphi_f(z) } \right|^2}{N} N_e^{2\left(\frac{\delta_z}{\pi}\right)^2},
\end{eqnarray}
and
\begin{eqnarray}
\mathcal{G}_{i,f}^{1 \rightarrow 0}(z)  = \frac{\left|\sum_k \bra{\phi_0} \tilde a_k \ket{\varphi_f(z)} \right|^2}{N} N_e^{2\left(\frac{\delta_z}{\pi}\right)^2},
\end{eqnarray}
then the transition probability can be written as 
\begin{eqnarray} 
W_{if}^{1 \rightarrow 2/0 } (z) = \frac{4V^2(z) N_e^{-2\left(\frac{\delta_z}{\pi}\right)^2} }{\left(\Delta E_{if}^{2/0}(z)\right)^2} \sin^2\left( \frac{\Delta E_{if}^{2/0}(z) }{2\hbar} t \right) \mathcal{G}_{i,f}^{1 \rightarrow 2/0} (z). 
\end{eqnarray}

We will not try to estimate the quantities $\mathcal{G}_{i,f}^{1 \rightarrow 2/0}(z)$ since we only need them for qualitative discussions about the transitions from the initial neutral sector to the ionized sectors. Our focus is on identifying the most important factors influencing the transition probability, and the above equation do this job qualitatively well.

As shown above, to obtain the transition probability from an initial neutral state to a final ionized sector, and these transitions are proportional to
\begin{eqnarray}\label{TP_neutral_ionized}
W_{if}^{1 \rightarrow 2/0 }(z) \sim \frac{4V^2(z) N_e^{-2\left(\frac{\delta_z}{\pi}\right)^2} }{\left(\Delta E_{if}^{2/0}(z)\right)^2} \sin^2\left( \frac{\Delta E_{if}^{2/0}(z) }{2\hbar} t \right). 
\end{eqnarray}
Here $\Delta E_{if}^{2/0}(z)$ is the energy difference between the initial and final states.

The equation \eqref{TP_neutral_ionized} provides a rough approximation for the transition probability, but it allows us to qualitatively understand some aspects of the collision process. Initially, when the atom is far from the surface, the hybridization is almost zero, making transitions from the initial neutral configuration to another sector improbable. However, as the atom approaches the surface and the hybridization gradually increases, small energy transitions become possible, especially if the atom remains in this region for a considerable time. Due to the scattering potential, which its amplitude also grows as the atom nears the surface, the energy of the ionized levels decreases and may cross the initial level. When this energy level crossovers occurs, these levels can hybridize, and the transition probability between these two levels can become significant as $\Delta E_{if}^{2/0}(z) \rightarrow 0$.

One point to notice is the factor $N_e^{-2\left(\frac{\delta_z}{\pi}\right)^2}$ that appears in the Eq. \eqref{TP_neutral_ionized} for the transition probability, which indicates the Anderson orthogonality. Another interesting point is that direct transitions from the initial state to another neutral configuration are improbable when hybridization is small, even for small-energy excitations from the ground state in the neutral sector. Instead, transitions typically can only occur through intermediate ionized states, resulting in very small transitions amplitudes for this case.

These qualitative discussions do not capture the entire picture of atom-surface collisions, as they consider small hybridization and disregard nuclear motion and distribution. Atomic motion introduces non-adiabatic effects that contribute to these transitions, while the non-point-like distribution of the atom, accurately obtained by solving the Schrödinger equation for both electronic and atomic part of the wave function, introduces additional complexities. Also, if the hybridization is strong, strongly correlated effects can emerge, complicating the qualitative arguments and disqualifying expressions obtained by perturbation theory. Nonetheless, the crossover mechanism discussed here remains crucial for understanding atomic behavior, specially in the quasi-adiabatic situation, when the atom is approaching the surface and $V(z)$ is small.

\end{apendicesenv}